---

# Chasing Individuation: Mathematical Description of Physical Systems

---

## par Federico ZALAMEA

**Thèse de doctorat en Histoire et Philosophie des Sciences**

**Dirigée par Gabriel CATREN**

Présentée et soutenue publiquement à l'Université Paris Diderot le 23 Novembre 2016

| | |
|---|---|
| Président du jury: | **M. Olivier DARRIGOL**, Directeur de recherche au CNRS |
| Rapporteurs: | **M. James LADYMAN**, Professeur à l'Université de Bristol |
| | **M. Nicolaas P. LANDSMAN**, Professeur à l'Université Radboud de Nijmegen |
| Examinateurs: | **M. Marc LACHIÈZE-REY**, Directeur de recherche au CNRS |
| | **M. Thomas RYCKMAN**, Professeur de l'Université de Stanford |
| Directeur de thèse: | **M. Gabriel CATREN**, Chargé de recherche au CNRS |

---

# Chasing Individuation: Mathematical Description of Physical Systems

---

## by Federico ZALAMEA

### Ph.D. in Philosophy of Physics

### Under the supervision of Gabriel CATREN

Presented and defended publicly at Paris Diderot University on November 23[rd] 2016

| | |
|---|---|
| President of the jury: | **M. Olivier DARRIGOL**, Directeur de recherche at CNRS |
| Rapporteurs: | **M. James LADYMAN**, Professor at the University of Bristol |
| | **M. Nicolaas P. LANDSMAN**, Professor at Radboud University Nijmegen |
| Examinators: | **M. Marc LACHIÈZE-REY**, Directeur de recherche at CNRS |
| | **M. Thomas RYCKMAN**, Professor at the University of Stanford |
| Doctoral supervisor: | **M. Gabriel CATREN**, Chargé de recherche at CNRS |

*"All great insights and discoveries are not only usually thought by several people at the same time, they must also be re-thought in that unique effort to truly say the same thing about the same thing."*

Martin Heidegger

# Acknowledgments

The four years of research culminating with this dissertation were almost entirely funded by the European Research Council, under the European Community's Seventh Framework Programme (Project *Philosophy of Canonical Quantum Gravity*, FP7/2007-2013 Grant Agreement n° 263523).

First, and foremost, I must thank my supervisor, Gabriel Catren, for providing me with the opportunity to be part of his project. During these past years, I have had the chance of evolving in the blurry borders between Physics, Mathematics and Philosophy. To be sure, the demand of navigating at ease within this triangle has oftentimes been a source of difficult challenges, and even of anguishing questions of identity. But it has also significantly enlarged my curiosity and consolidated my thinking. I particularly appreciate the freedom I have enjoyed to explore many different fields and develop my own questions. I know this freedom to be the sign of a rare confidence in my work, for which I am the more grateful.

I also would like to thank warmly James Ladyman and Klaas Landsman for accepting to be the *rapporteurs* of my dissertation, for their careful reading and for the detailed comments which will greatly help me in improving my work; Olivier Darrigol, Marc Lachièze-Rey and Thomas Ryckman for accepting to be part of my jury, for their comments, suggestions and encouragements in the various occasions I have had the luck of meeting them. It is for me an honor to have my work evaluated by such a high-level group of thinkers whose several works I admire.

Because my research would have been exceedingly more difficult without a vibrant institution favoring exchanges and discussions, I am indebted to the laboratory

SPHERE that hosted me during my graduate studies. In particular, I thank the former and current directors, David Rabouin and Pascal Crozet, as well as Roberto Angeloni for our discussions on the history of Quantum Mechanics, Karine Chemla for her inspiring and indefatigable enthusiasm, Nadine de Courtenay, Michel Paty and Jean-Jacques Szczeciniarz for their numerous advises and their generosity in sharing their valuable knowledge.

The assistance of all the administrative staff has also been crucial all along the way. I thank Sandrine Pellé and Patricia Philippe for their help, Nad Fachard for her joyful presence and for her interest in all aspects of life, and Virginie Maouchi for her kindness and her unshakeable efficiency which rendered simple the otherwise labyrinthine administrative procedures.

My colleagues of the ERC Project *Philosophy of Canonical Quantum Gravity* deserve a special mention. In these four years, we spent countless hours together in informal discussions, meetings, seminars and workshops. My work would be significantly poorer had it not been confronted with their critical eye. I am extremely grateful to Alexandre Afgoustidis and Mathieu Anel for their patience and their generosity in guiding me through some regions of Mathematics which seemed inaccessible to me, to Christine Cachot for the many conceptual confusions she helped me dissolving and for her meticulous correction of the various drafts of my dissertation, to Julien Page for all the roads he has been suggesting me in order to broaden my investigations, and to Dimitri Vey for his encouragements and friendship in the most difficult times.

Of course, I also think of all my fellow companions who shared with me the vicissitudes of the daily life of a graduate student. It has been a truly rewarding experience to share so many lunches and coffees discussing our widely different experiences. In particular, I would like to thank: Nacéra Bensaou, Pascal Bertin, Pierre Chaigneau, João Cortese, Vincent Daudon, Simon Decaens, Sarah Erman, Fernando Gálvez, Fabien Grégis, Roberto Hastenreiter-Cruz, Zeinab Karimian, Ramzi Kebaili, Jonathan Regier, Eleonora Sammarchi, Philippe Stamenkovic, Sergio Valencia and Xiaofei Wang.

During my PhD, I have had the chance of traveling and participating in several workshops which have lead to important encounters. Particularly inspiring for me was the meeting of Johannes Kleiner, Robin Lorenz and the Basic Research Community for Physics of which I am now a proud member. I also thank Charles Alunni and Jean-Pierre Marquis for their interest in my work and their support, and Michael Wright for his help with my babbling English. Finally, I am beholden to the whole stackexchange community without which the present document would simply not exist.

Last, my thoughts go to my closest friends—Alexandre, Ivan, Philippe, Lou and Vera—which laughed with me about my obstinate existential doubts, and to my parents. They have managed to create a serene atmosphere of insatiable curiosity and openness towards all regions of knowledge which has deeply influenced me. Their constant support and feedback during these four years, and beyond, has been invaluable.

And, with all my heart, I thank Camila, who accompanied me in every step of this long journey, silently feeling helpless next to me without realizing she brought me the most essential facet of them all: beauty.

# Contents

# List of Figures

# List of Tables

# Introduction

Almost one hundred years after the birth of Quantum Mechanics, the relation in which it stands to Classical Mechanics is still not completely well understood. The problem can take several forms: for instance, one may start with the classical description of some given system and then attempt to generate the corresponding quantum description; on the other hand, one may as well decide to start with the quantum description of a system and then try to recover, through some controlled limiting procedure, a classical description. The former is called the problem of *quantization*; the latter is the problem of the *classical limit*. But, on top of these questions, which investigate the possible *transitions* between both theories, it is also possible to be interested in detecting the core ideas which distinguish the Quantum from the Classical. This last question—which is the question of this dissertation—seems simpler to address; yet, it is a subtle matter.

One cannot expect to have an acute understanding of the differences between Classical and Quantum Mechanics without a thorough analysis of the mathematics underlying both theories. Klaas Landsman, whose many writings are an invaluable reading for anyone interested in such questions, emphasizes the importance of mathematical rigor to achieve conceptual clarity:

> [...] the problem [of the relation between classical and quantum physics] is so delicate that in this area sloppy mathematics is almost guaranteed to lead to unreliable physics and conceptual confusion.[1]

---

[1]N. P. Landsman. "Between Classical and Quantum". In: *Philosophy of Physics (Handbook of the Philosophy of Science) 2 volume set*. Ed. by J. Butterfield and J. Earman. Vol. 1. Amsterdam:

This thesis belongs to the field of *Foundations of Physics*, and, more specifically, of Foundations of Mechanics. This is not to say that its goal is to find the logical structure and the first principles upon which the whole theoretical edifice rests. Rather, the characteristic task of the foundations I wish to practice is the clarification of the *conceptual content* of a physical theory, and its means is the discussion of precise *technical results*. In style, it is closer to the idea described by the mathematician William Lawvere:

> A foundation makes explicit the essential general features, ingredients, and operations of a science, as well as its origins and general laws of development. The purpose of making these explicit is to provide a guide to the learning, use, and further development of the science. A 'pure' foundation that forgets this purpose and pursues a speculative 'foundation' for its own sake is clearly a nonfoundation.[2]

The object of true wonder that launches our investigation is the role symplectic geometry has come to play in the formulation of Mechanics. Although one may argue that symplectic geometry has been present in Classical Mechanics since the seminal work of Joseph-Louis Lagrange at the beginning of the 19th century[3], it is undeniable that the subject has had to wait another hundred and fifty years before acquiring the importance it has today. For in the last fifty years, through the works of, first, Vladimir Arnold, Bertram Kostant and Jean-Marie Souriau, and, afterwards, Victor Guillemin, Jerrold Marsden, Shlomo Sternberg and Alan Weinstein—to name just a few—, symplectic geometry has become an indispensable ingredient in the contemporary understanding of Mechanics. The situation is such that it allows Patrick Iglesias-Zemmour, one of the leading specialists in the subject, to declare:

> Symplectic geometry has become the framework *per se* of mechanics, up to the

North-Holland Publishing Co., 2007, pp. 417–554. URL: http://arxiv.org/abs/quant-ph/0506082, p. 418.

[2]F. W. Lawvere and R. Rosebrugh. *Sets for Mathematics*. Cambridge: Cambridge University Press, 2003, p. 235.

[3]For a short study (in French) on the origins of symplectic geometry, see P. Iglesias-Zemmour. *Aperçu des origines de la géométrie symplectique*. Actes du colloque "Histoire des géométries", vol. 1. 2004.

point one may claim today that these two theories are the same. Symplectic geometry is not just the language of mechanics, it is its essence and its matter.[4]

Now, whenever a new tool acquires such an importance in a large field, it should be expected that not only does this tool constitute a leap forward in the solution of technical problems, but, moreover, that it highlights ideas which were previously unnoticed and that it brings about new ways of structuring the field. This motto, which is at the core of the "philosophical history" that Timmermans tries to develop in his beautiful book *Histoire philosophique de l'algèbre moderne – Les origines romantiques de la pensée abstraite*[5], is also the main methodological premise of this work.

It is my opinion that the collective efforts undertaken to investigate the conceptual lessons to be drawn from the 'symplectic-geometrization' of Mechanics have still to match the tremendous impact this has had on the development of Mechanics itself. In the course of the last century, there have been numerous attempts to capture ever more precisely some of the fundamental features distinguishing the quantum world. Perhaps the two most widely discussed traits are *entanglement* and the *state reduction*: the first deals with the description of *composite* systems and expresses the fact that a state of a composite system may not always be decomposed into states of the various parts[6]; the second regards the *dynamics* of physical systems and is meant to encapsulate the fundamental influence of the measurement process on the time evolution of a system. But in most cases these analyses rely upon a careful examination of the mathematical formulations of Quantum Mechanics. Seldom has it been the case that the novelties introduced by this theory into our conception of Mechanics have been examined under the light of the aforementioned developments in the foundations of *Classical* Mechanics. The present thesis is an attempt to push into that direction of research.

---

[4] "La géométrie symplectique est devenue le cadre par excellence de la mécanique à tel point que l'on peut dire aujourd'hui que ces théories se confondent. La géométrie n'est pas seulement le langage de la mécanique, elle en est l'essence et la matière." (Ibid., p. 2.)

[5] Timmermans' exact formulation is: "Lorsqu'un nouvel outil prend ainsi autant d'importance dans un large domaine [...], on s'attend à ce qu'il n'apporte pas seulement des recettes permettant de régler des questions, mais qu'il détermine aussi des façons nouvelles de poser les problèmes, d'envisager le monde." (Whenever a new tool thus acquires such an importance in a large field, it should be expected that not only does this tool bring about new recipes enabling to solve questions, but, moreover, that it also determines new ways of formulating problems, of looking at the world.) B. Timmermans. *Histoire philosophique de l'algèbre moderne – Les origines romantiques de la pensée abstraite.* Paris: Classiques Garnier, 2012, p. 10.

[6] More precisely, given physical systems $R$ and $S$ with space of states $\mathcal{R}$ and $\mathcal{S}$, it is not the case, in quantum mechanics, that the space of states $\mathcal{R} \otimes \mathcal{S}$ of the composite system $R \sqcup S$ is the cartesian product $\mathcal{R} \times \mathcal{S}$. This means that, unlike in classical mechanics, a state of $R \sqcup S$ cannot in general be thought as a pair $(r, s)$ with $r \in \mathcal{R}$ and $s \in \mathcal{S}$.

The broad goal is therefore to compare Classical and Quantum Mechanics by focusing as much as possible on the concepts highlighted by symplectic geometry. In the midst of the dominant narrative that smugly exclaims over the 'extraordinary' differences between Classical and Quantum Mechanics, symplectic geometry acts as a powerful countercurrent. For indeed, if there is one lesson to take from this geometrization of Mechanics, it is the impressive merger of the two Mechanics. In the setting offered by symplectic manifolds, Poisson algebras, Hamiltonian vector fields and all the surrounding machinery, one clearly realizes how much these two theories actually share. By the same token, precisely because it offers a *common ground* in which to discuss both theories, symplectic geometry allows to pinpoint some of the most fundamental differences between the Classical and the Quantum.

One example of this coming closer together is the treatment of the equations of motion. In Quantum Mechanics, these take the form of the famous Schrödinger equation

$$i\hbar \frac{d\Psi(t)}{dt} = H\Psi(t),$$ (1)

where $H$ is the hermitian operator representing the Hamiltonian and $\Psi$, the wavefunction, denotes a (normed) element of the Hilbert space. (1) is almost systematically perceived as a breakaway from Classical Mechanics. However, through the work of Werner Heisenberg, whose emphasis was on the time-evolution of observables rather than states, and Paul Dirac, who emphasized the role of the Hamiltonian formulation of Classical Mechanics, the equations of motion of Classical and Quantum Mechanics were very early shown to be similar in their form. Indeed, Schrödinger's equation could equivalently be written as

$$i\hbar \frac{dF(t)}{dt} = -[H, F(t)],$$ (2)

where $F$ is any hermitian operator and $[\cdot, \cdot]$ denotes the commutator, not without reminding Hamilton's *classical* equations for the time-evolution of an observable

$$\frac{df(t)}{dt} = \{h, f\},$$ (3)

where $f$ is now any smooth real-valued function over the space of states of the classical system and $\{\cdot, \cdot\}$ denotes the Poisson bracket. Despite this, it was only with the increasing interest for the geometry underlying Classical Mechanics, and the remark that both the space of states of a classical system and the (projective) Hilbert spaces used in Quantum Mechanics were examples of symplectic manifolds, that it was realized equations (2) and (3) were not just similar but, in fact, two instances of *exactly the same equation*:

$$\omega(\frac{d}{dt}, \cdot) = dh, \qquad (4)$$

where $\omega$ is the symplectic 2-form[7].

With the discovery that the time-evolution of any physical system, be it classical or quantum, is governed by the same very simple geometrical equation, one barely begins to scratch the surface of the situation. At the heart of the symplectic geometric formulation of Mechanics lies the idea that physical properties are tied in an essential way to state *transformations*. For some particular physical properties, this idea has been for long included in the folklore of theoretical physics—for instance by regarding angular momentum as the generator of spatial rotations. Nonetheless, that a completely general property–transformation link may constitute a key feature in the conceptual interpretation of both Classical and Quantum Mechanics has remain somehow dormant. This is to be contrasted with the endless efforts to accommodate the quantum theory, despite such no-go theorems as the famous Kochen-Specker, to the prevalent conception of properties as *quantities*, which stresses their numerical character and so smoothly appears to fit the classical formalism. Through the geometric formulation, one clearly perceives the importance of these two aspects of physical properties, and understanding the articulation between properties-as-transformations and properties-as-quantities in the Classical and the Quantum becomes a central issue of our analysis.

This last point is of purely *kinematical* nature. In the contemporary usage, the 'kinematical description' of a physical system has come to signify a characterization of the algebra of observables and of the space of all possible states in which a system may be found. In the light of equation (4), the complete mechanical picture involves, in addition to the kinematical description, choosing a preferred observable (which is then regarded as the Hamiltonian of the system) and writing explicitly the flows generated by it on the space of states. In Quantum Mechanics, one also needs to postulate a second dynamical process that covers the state reduction or wavefunction collapse[8]. In this work, the choice is made to circumscribe our comparison of Classical and Quantum Mechanics to the most basic kinematical level, thus ignoring all interpretational problems related to Dynamics and, in particular, the measurement problem. From Schrödinger's cat[9] to Bell's "and/or" objection[10], the measurement problem has been the focus of much attention in the philosophy of physics—and rightly so: no satisfactory understanding of Quantum Mechanics can avoid addressing this issue—but, as a consequence, it has overshadowed the important conceptual differences between Classical and Quantum Kinematics. After all, one should not forget the major role of purely kinematical considerations in Heisenberg's breakthrough[11].

---

[7]For the precise derivation of this, see N. P. Landsman. *Mathematical Topics Between Classical and Quantum Mechanics.* New York: Springer, 1998, pp. 71–76, and in particular the comment preceding equation (2.39). See also Chapter II of the present work.

Moreover, the concept of a 'kinematical description of a physical system' furnishes us with a perspective from which to question the mathematical formalisms of Classical and Quantum Kinematics. Abstract symplectic manifolds and abstract projective Hilbert spaces are, in a sense to be rendered precise later, _homogeneous_—it is impossible to differentiate their points. This evidently clashes with the idea that one should be able to distinguish states of a physical system by means of physical properties. Thus, when such homogeneous mathematical structures are taken as the starting points in the kinematical descriptions, there must inevitably be a constitution of a labelling scheme which allows to identify each particular state and each specific property of the physical system being described. The progressive constitution of this labelling scheme within the mathematical frameworks of Classical and Quantum Kinematics is what I have wished to call the *Chase for Individuation* which gives the title to this work.

---

[8] Roger Penrose calls these two different evolutions of Quantum Mechanics the U-quantum procedure (which is governed by the Schrödinger equation and is hence unitary) and the R-quantum procedure (which is governed by the state reduction postulate and is hence non-unitary). See R. Penrose. _The Road to Reality: A Complete Guide to the Laws of the Universe._ New York: Alfred A. Knopf, 2005, Chapter 22.

[9] E. Schrödinger. "Die gegenwärtige Situation in der Quantenmechanik". In: _Naturwissenschaften_ 23.48 (1935), pp. 807–812 (English translation: E. Schrödinger. "The Present Situation in Quantum Mechanics". Trans. by J. D. Trimmer. In: _Proceedings of the American Philosophical Society_ 124.5 (1980), pp. 323–338).

[10] J. S. Bell. "Against 'measurement'". In: _62 Years of Uncertainty: Erice, 5-14 August 1989._ Plenum Publishers, 1990. (Reprinted in: J. S. Bell. _Speakable and Unspeakable in Quantum Mechanics._ 2nd ed. Cambridge: Cambridge University Press, 2004, pp. 213–231).

[11] W. Heisenberg. "Über quantentheoretische Umdeutung kinematischer und mechanischer Beziehungen". In: _Zeitschrift für Physik_ 33 (1925), pp. 879–893 (English translation: W. Heisenberg. "Quantum-theoretical Re-interpretation of Kinematic and Mechanical Relations". In: _Sources of Quantum Mechanics._ Ed. by B. Van der Waerden. New York: Dover Publications, Inc., 1967, pp. 261–276).

Hence, two main themes pervade throughout the dissertation and determine its structure:

1) Thinking the difference between Classical and Quantum Kinematics through the looking glass of symplectic geometry and, in particular, through the double role of physical properties;

2) Looking for the techniques, explicitly or implicitly at work in the mathematics of Kinematics, which break the homogeneity of the initially given abstract structures and help constituting a labelling scheme for states and properties.

Chapter I is a detailed discussion of the conceptual background in which the 'Chase for Individuation' is to be understood. We will argue that, in practice, the mathematical objects involved in the kinematical description of a physical system are *abstract structures*. For this, we will look at the example of how a unified framework for Quantum Mechanics was brought out from Göttingen's Matrix Mechanics and Schrödinger's Wave Mechanics. Then, we will try to elucidate the notion of an 'abstract mathematical structure', first by addressing the question of what it means for a mathematical object to be conceived 'abstractly'—making heavy use, in particular, of Jean-Pierre Marquis' account of the abstract method—, and then by discussing mathematical structuralism as developed by Stewart Shapiro and James Ladyman among others. Only in this setting will it become clear in which sense one can say that the elements of a symplectic manifold or a projective Hilbert space are indiscernible. The second and third sections of this first chapter are therefore close in style and matter to the philosophy of mathematics.

In Chapter II, the proper conceptual analysis of the mathematical structures used in Classical and Quantum Kinematics begins. From then on, the style becomes closer to that of mathematical physics: although there is no question of proving theorems, an effort is made to give the exact technical definition of all the new elements that are introduced. Again, the combination of mathematical analysis and conceptual synthesis is crucial. In this chapter, we deal only with the kinematical arenas—that is, with the homogeneous structures which are taken as a starting point in the kinematical description of classical and quantum systems, and where no labelling scheme is still present—and we investigate the articulation between the transformational and the quantitative role of properties. This investigation is performed within three different frameworks: first, the standard kinematical formulation which uses symplectic manifolds for Classical Mechanics and Hilbert spaces for Quantum Mechanics; second, the geometric formulation which emphasizes the notion of 'state' and develops Quantum Mechanics in terms of Hermitian symmetric spaces; and third, the algebraic formulation which emphasizes the notion of 'property' and develops the theory in terms of Jordan-Lie-Banach algebras.

The first two chapters are almost completely independent from each other. From the perspective of the narrative offered by the 'Chase for Individuation' (question 2 above), Chapter II is perhaps an unnecessarily lengthy preparation. However, with regard to the conceptual comparison of Classical and Quantum Kinematics (question 1 above), it surely is the central part of the dissertation. On the other hand, the third and final chapter cannot be understood without the preceding two. Therein, we finally investigate how Lie groups, and their infinitesimal versions Lie algebras, are used to introduce a notion of *difference* within both kinematical arenas, breaking in this way the homogeneity of the initial mathematical structures and providing a sketch of a labelling scheme. This approach—which is in sharp contrast with the traditional view on groups as implementing symmetries, and hence a notion of 'sameness'—illuminates in a new way the role of groups in the mathematics of Mechanics.

By the mathematical notions it deals with—symplectic manifolds, of course, but also (strongly) Hamiltonian actions and the momentum map, $C^*$-algebras and Jordan-Lie algebras, Hermitian symmetric spaces, Poisson spaces with a transition probability, etc.—, the detailed mathematical level to which these notions are discussed and the questions it asks, this work is an unconventional one for the philosophy of physics. But all these recent developments and all these new mathematical formulations show 'basic' Kinematics to be, in fact, a vast and complex territory which ought to be explored once and again, thought and "re-thought in that unique effort to truly say the same thing about the same thing"[12]. The promenade taken here is resolutely against the idea of there being 'technicalities'—supposedly small and harmless mathematical details, important for the correct statement of a theorem or the smooth rolling out of a proof, but insignificant from a conceptual standpoint. In general, 'technicalities' are perceived as such only because of the failure to find the approach from which their true role and meaning is exposed. Finally, it tries to carry as few preconceptions as possible, striving to listen to what these formalisms are expressing instead of attempting to include them in some prefabricated interpretational framework. In this, I am strongly influenced by the words Alexander Grothendieck once wrote:

> One cannot 'invent' the structure of an object. The most we can do is to patiently bring it to the light of day, with humility—in making it known, it is 'discovered'. If there is some sort of inventiveness in this work, and if it happens that we find ourselves the maker or indefatigable builder, we are in no sense 'making' or 'building' these 'structures'. They have not waited for us to find them in order to exist, exactly as they are! But it is in order to express, as faithfully as possible, the things that we have been detecting or discovering, the reticent structure which we are trying to grasp at, perhaps with a language no better than babbling. Thereby are we constantly driven to 'invent' the language most appropriate to express, with increasing refinement, the intimate structure of the mathematical object, and to 'construct' with the help of this language, bit by bit, those 'theories' which claim to give a fair account of what has been apprehended and seen. There is a continual coming and going, uninterrupted, between the apprehension of things, and the means of expressing them by a language in constant state improvement [...].

---

[12]M. Heidegger. *What is a Thing?* Trans. by W. B. Barton and V. Deutsch. Indiana: Gateway Editions, Ltd., 1967, p. 80.

The sole thing that constitutes the true inventiveness and imagination of the researcher is the quality of his attention as he listens to the voices of things.[13]

---

[13]"La structure d'une chose n'est nullement une chose que nous puissions "inventer". Nous pouvons seulement la mettre à jour patiemment, humblement en faire connaissance, la "**découvrir**". S'il y a inventivité dans ce travail, et s'il nous arrive de faire œuvre de forgeron ou d'infatigable bâtisseur, ce n'est nullement pour "façonner", ou pour "bâtir", des "structures". Celles-ci ne nous ont nullement attendues pour être, et pour être exactement ce qu'elles sont ! Mais c'est pour **exprimer**, le plus fidèlement que nous le pouvons, ces choses que nous sommes en train de découvrir et de sonder, et cette structure réticente à se livrer, que nous essayons à tâtons, et par un langage encore balbutiant peut-être, à cerner. Ainsi sommes-nous amenés à constamment **"inventer" le langage** apte à exprimer de plus en plus finement la structure intime de la chose mathématique, et à "construire" à l'aide de ce langage, au fur et à mesure et de toutes pièces, les "théories" qui sont censées rendre compte de ce qui a été appréhendé et vu. Il y a là un mouvement de va-et-vient continuel, ininterrompu, entre **l'appréhension** des choses, et **l'expression** de ce qui est appréhendé, par un langage qui s'affine et se re-crée au fil du travail, [...].

Ce qui fait la qualité de l'inventivité et de l'imagination du chercheur, c'est la **qualité de son attention**, à l'écoute de la voix des choses."

(A. Grothendieck. *Récoltes et semailles – Réflexions et témoignage sur un passé de mathématicien.* 1985, 2.9. Forme et structure - ou la voie des choses. Bold typeface is Grothendieck's. Partial English translation by Roy Lisker, available at **http://uberty.org/wp-content/uploads/2015/12/RS-grothendeick1.pdf**.)

# Chapter I

# Mathematical Description of Physical Systems

In the practice of theoretical and mathematical physics, it has become customary to consider general assignments which, to any given physical system $S$ of a certain type $T_{phys}$ (e.g., non-relativistic systems with finitely-many degrees of freedom), associate a particular mathematical object $\mathcal{D}(S)$ of type $T_{math}$. In general, the object $\mathcal{D}(S)$ is intended to describe the physical system in some way and the map $\mathcal{D}: T_{phys} \longrightarrow T_{math}$ is accordingly called the *mathematical description of a generic physical system* (of the type $T_{phys}$).

This is particularly salient in the foundations of quantization where the main problem could roughly be stated as follows: given the classical description $\mathcal{D}_C(S)$ of a physical system $S$, can we construct its corresponding quantum description $\mathcal{D}_Q(S)$? In the early stages of the quantum theory, this problem was addressed *separately* for each particular system which was of interest at the time, and advances towards the solution were the result of great heuristic physical insights. A paradigmatic example of this situation is Bohr's quantum model of the hydrogen atom of 1913[1]. However, with the progressive development of the mathematical foundations of Classical and Quantum Mechanics, the situation changed and it became possible to attempt to regard quantization as a *systematic procedure* which could be applied to any physical system whose classical description was known. The important point here is that this all-encompassing conception of quantization as a systematic procedure is not possible if there is not, moreover, a definition of general assignments $\mathcal{D}_C$ and $\mathcal{D}_Q$ which associate to *any* physical system $S$ of the type $T_{phys}$ its classical and quantum descriptions $\mathcal{D}_C(S)$ and $\mathcal{D}_Q(S)$. Schematically, the perspective on quantization evolved from the conception of a single assignment

$$\mathcal{D}_C(S) \xmapsto{\;\mathcal{Q}_S\;} \mathcal{D}_Q(S)$$

where the method of quantization $\mathcal{Q}_S$ strongly depended on the particular system $S$ being handled—and thus could hardly be called a *method*—, to the more complex diagram

$$\begin{array}{ccc} & T_{phys} & \\ {\scriptstyle \mathcal{D}_C}\swarrow & & \searrow{\scriptstyle \mathcal{D}_Q} \\ T_{math}^{CM} & \xrightarrow[\;\mathcal{Q}\;]{} & T_{math}^{QM} \end{array} \qquad (\mathrm{I.1})$$

which covers at once the quantization of all the different physical systems $S$ of the type $T_{phys}$.

---

[1] N. Bohr. "On the Constitution of Atoms and Molecules". In: *Philosophical Magazine* 26.151 (1913), pp. 1–25. (Reprinted in: N. Bohr. *Collected Works.* Ed. by U. Hoyer. Vol. 2. Amsterdam: Elsevier, 2008, pp. 161–185); N. Bohr. "On the Constitution of Atoms and Molecules (Part II)". in: *Philosophical Magazine* 26.153 (1913), pp. 476–502. (Reprinted in: N. Bohr. *Collected Works.* Ed. by U. Hoyer. Vol. 2. Amsterdam: Elsevier, 2008, pp. 188–214).

In this setting, the general mathematical problem of quantization is thus: given the maps $\mathcal{D}_C$ and $\mathcal{D}_Q$, to construct a map $\mathcal{Q}$ such that the diagram (I.1) commutes. The choice of the mathematical objects constituting $T_{math}^{CM}$ and $T_{math}^{QM}$ may vary slightly between different approaches to quantization. For example, in his great textbook on geometric quantization, Nick Woodhouse writes:

> The first problem of quantization concerns the kinematic relationship between the classical and quantum domains. At the quantum level, the states of a physical system are represented by the rays in a Hilbert space $\mathcal{H}$ and the observables by a collection $\mathcal{O}$ of symmetric operators on $\mathcal{H}$, while in the limiting classical description, the state space is a symplectic manifold $(M, \omega)$ and the observables are the smooth functions on $M$. The kinematic problem is: given $M$ and $\omega$, is it possible to reconstruct $\mathcal{H}$ and $\mathcal{O}$?[2]

According to this account, in geometric quantization $T_{math}^{CM}$ would then be the type of symplectic manifolds, whereas $T_{math}^{QM}$ would be the type of pairs $(\mathcal{H}, \mathcal{O})$ of Hilbert spaces with collections of symmetric operators. Another example is the so-called "strict deformation quantization": therein, the mathematical problem of quantization is casted in the language of real Poisson algebras (which form $T_{math}^{CM}$) and $C^*$-algebras (which form $T_{math}^{QM}$)[3]. Of course, deformation quantization and geometric quantization are closely related to each other since the smooth real-valued functions over a symplectic manifold form a Poisson algebra, and bounded operators over a Hilbert space form a $C^*$-algebra[4].

Now, the idea of considering assignments $\mathcal{D} : T_{phys} \longrightarrow T_{math}$ does not appear solely in the discussion of quantization: it is explicitly or implicitly present in many works on the foundations of Mechanics. In the present chapter, our primary interest will lie on this notion of 'mathematical description of a generic physical system'. Because of this, let me list a few other places where this idea shows up:

a) In Eduard Prugovečki's textbook *Quantum Mechanics in Hilbert Space*:

> [...] we associate a Hilbert space with any given system which we intend to describe quantum mechanically. For instance [...], if the systems consists of $n$ particles, which are of different kinds, without spin and moving in three dimensions, then [...] we associate with that system the Hilbert space $L^2(\mathbb{R}^{3n})$.
>
> Let us assume now that we are dealing with a particular quantum mechanical problem in which a certain system has been specified (e.g., a hydrogren atom) and with which a certain Hilbert space $\mathcal{H}$ is associated (e.g., $L^2(\mathbb{R}^6)$). It is then postulated that to each observable corresponds in the formalism

---

[2]N. Woodhouse. *Geometric Quantization*. 2nd. Oxford: Clarendon Press, 1991, p. 155.

[3]See for example M. A. Rieffel. "Deformation Quantization and Operator Algebras". In: *Proceedings of Symposia in Pure Mathematics* 51 (1990), pp. 411–423.

[4]See Chapter II for the precise definition of all these structures (symplectic manifolds, Poisson algebras, $C^*$-algebras).

a unique self-adjoint operator acting on $\mathcal{H}$ [...].[5]

b) In Abraham and Marsden's *Foundations of Mechanics*:

A *(simple) mechanical system with symmetry* is a quadruple $(\boldsymbol{M}, \boldsymbol{K}, \boldsymbol{V}, \boldsymbol{G})$ where:

    i. $\boldsymbol{M}$ is a Riemannian manifold [...] called the *configuration space* [...].

    ii. $\boldsymbol{K} \in \mathcal{F}(T^*M)$ is the *kinetic energy* of the system [...].

    iii. $\boldsymbol{V} \in \mathcal{F}(M)$ is the potential energy.

    iv. $\boldsymbol{G}$ is a connected Lie group acting on $\boldsymbol{M}$ [...].[6]

c) In Carlo Rovelli's monograph *Quantum Gravity*:

A [classical] dynamical system is determined by a triple $(\Gamma_0, \omega_0, H_0)$, where $\Gamma_0$ is a manifold, $\omega_0$ is a symplectic two-form and $H_0$ is a function on $\Gamma_0$. [...]

A [classical] dynamical system is thus completely defined by a presymplectic space $(\Sigma, \omega)$. [...]

A given quantum system is defined by a family (generally an algebra) of operators $A_i$, including $H_0$ [the Hamiltonian operator corresponding to the energy], defined over an Hilbert space $\mathcal{H}_0$.[7]

And the examples could be multiplied *ad infinitum*...

The first important question regarding this notion is to understand its *purpose*: What information about the physical system $S$ do we expect to capture through its mathematical description $\mathcal{D}(S)$? Put differently, given the knowledge of $\mathcal{D}(S)$, how much ambiguity do we expect to have about the system $S$ being thus handled? This question may be easily translated into a more precise one in the following fashion: given a certain choice of a map $\mathcal{D} : T_{phys} \longrightarrow T_{math}$, define the equivalence relation "two systems $S$ and $S'$ of type $T_{phys}$ are $\mathcal{D}$–*equivalent*, denoted by $S \underset{\mathcal{D}}{\sim} S'$, if and only if $\mathcal{D}(S)$ and $\mathcal{D}(S')$ are equal", and call $T_{phys}\big/\mathcal{D}$ the collection of all such equivalence classes. The above question wishes to understand the difference between $\mathcal{D}$–equivalence and physical identity. In other words, it asks how faithful a picture of $T_{phys}$ the quotient $T_{phys}\big/\mathcal{D}$ is.

---

[5]E. Prugovečki. *Quantum Mechanics in Hilbert Space.* 2nd ed. New York: Academic Press, 1981, p. 258.

[6]R. Abraham and J. E. Marsden. *Foundations of Mechanics.* 2nd ed. Redwood City: Addison-Wesley Publishing Company, 1978, p. 341.

[7]C. Rovelli. *Quantum Gravity.* Cambridge: Cambridge University Press, 2004, pp. 100, 101, 165.

If taken literally, the language of some authors suggest quite an ambitious view on the power of these mathematical descriptions. Thus, when Rovelli writes that a dynamical system is "*determined*" or "*completely* defined" by the mathematical object $\mathcal{D}(S)$, he seems to claim that *the data of $\mathcal{D}(S)$ fully and unambiguously characterizes the system $S$*. According to this reading, if a physicist is given the mathematical description of a given physical system *and nothing else*, he will nonetheless be able to recognize which physical system is being described. Let me call this view the *descriptive perspective* towards the mathematical description of a physical system. It is defined by the wish of constructing a map $\mathcal{D}: T_{phys} \longrightarrow T_{math}$ such that any difference between two physical systems is reflected in their respective mathematical descriptions. In other words, it demands the following faithfulness requirement:

> **Faithfulness requirement (descriptive perspective):** consider two physical systems $S$ and $S'$ of type $T_{phys}$ described by the mathematical objects $\mathcal{D}(S)$ and $\mathcal{D}(S')$ of type $T_{math}$. We must have:
>
> $$\mathcal{D}(S) =_M \mathcal{D}(S') \text{ if and only if } S =_P S'.$$

Of course, this is tantamount to requiring the map $\mathcal{D}$ to be injective. But it seems preferable to write the condition explicitly: the two different signs of equality, $=_M$ and $=_P$, are there to stress that we are in fact dealing with two different criteria of identity, one for physical systems of the type $T_{phys}$ and one for mathematical objects of the type $T_{math}$.

To be sure, the descriptive perspective may appear as a very naive answer to the question of the relation between $T_{phys}$ and $T_{phys}\big/\mathcal{D}$. Rather, it is certainly more natural to adopt some variant of a *formalist perspective* towards these mathematical descriptions. The mathematical object $\mathcal{D}(S)$ is then perceived as a *formal framework* in which it is possible to develop certain techniques, useful for the theoretical analysis of physical systems. But the full description of the specificities of a particular system is never reached by the sole study of $\mathcal{D}(S)$: to unambiguously refer to one specific system it is necessary to consider additional information, extraneous to the mathematical formalism, that conveys to $\mathcal{D}(S)$ its complete physical meaning. There are then two separate levels of meaning: a first one, encoded in the Mathematics of Mechanics, that captures the general features common to certain physical systems; and a second one added on top, the so-called *physical interpretation*, invariably situated beyond the grasp of mathematical formalization. This idea appears quite clearly in the writings of many authors, and an explicit example is found in Franco Strocchi's textbook on Algebraic Quantum Mechanics. He writes:

> In the mathematical literature, given a $C^*$-algebra $\mathcal{A}$, any normalized positive linear functional on it is by definition a state; here we allow the possibility that the set $\mathcal{S}$ of *states with physical interpretation* (briefly called physical states) is full but *smaller* than the set of all the normalized positive linear functionals on $\mathcal{A}$.[8]

There is in this quote a sharp contrast between what is declared by definition in Mathematics and what is to be interpreted in Physics. According to this formalist perspective, the physical interpretation is not determined by the mathematical formalism. Therefore, the same object $\mathcal{D}(S)$ may describe a wide range of different physical systems and the quotient $T_{phys}\big/\mathcal{D}$ only provides a rough picture of $T_{phys}$.

Although one may suspect that it is simply not possible for the mathematical object $\mathcal{D}(S)$ to capture *every*thing about the physical system $S$, it certainly encapsulates *some*thing about it. In other words, although one probably has $T_{phys}\big/\mathcal{D} \not\cong T_{phys}$, it is certainly the case that $T_{phys}\big/\mathcal{D} \not\cong *$. Despite the soundness of the formalist perspective, it then becomes interesting to adopt the descriptive perspective as a *working hypothesis*—as an ideal asymptotic situation one should try to approach. In this way, by attempting to fully and unambiguously characterize a physical system through its mathematical description, one is forced to study in detail the mechanisms through which some physical information is encoded in the mathematics of Mechanics.

Given a certain type of physical systems $T_{phys}$, it cannot be the task of our work

---

to determine 'the correct' map $\mathcal{D} : T_{phys} \longrightarrow T_{math}$ which succeeds in performing this characterization—otherwise, it would be a research program in mathematical physics. Rather, our main task in this chapter will be to clarify some general characteristics the mathematical objects of type $T_{math}$ should have in order to be considered as a priori *'acceptable'* candidates for describing physical systems.

In general, the object $\mathcal{D}(S)$ is intended to describe either the *space of states* of the system $S$—sometimes also called the "phase space"—or its *algebra of observables*. Now, an experimental physicist must be able to clearly identify the state in which a physical system $S$ is prepared. But if the mathematical object $\mathcal{D}(S)$ is to capture all the physical information of the system, then one should expect this ability of the experimentalist to be reflected within $\mathcal{D}(S)$. In other words, one is confronted with the following requirement:

> **Requirement of individuation:** it must be possible, in practice, to *individuate* any specific element of the mathematical object $\mathcal{D}(S)$ used to described the physical system $S$.

As innocent-looking as the requirement of individuation may appear, it nonetheless imposes some stringent conditions on the mathematical formalisms to be used for the development of Classical and Quantum Mechanics. To see this, we need however to clarify further the content of the two main requirements of the descriptive perspective. The goal of the present chapter is to do so by addressing the following three questions:

♣ How does one conceive the mathematical entities involved in the kinematical description of physical systems?

♥ Which is the *criterion of identity* $=_M$ used in practice for the mathematical objects $\mathcal{D}(S)$?

♠ What exactly does it mean to *individuate* an element within the mathematical entity $\mathcal{D}(S)$?

# I.1   A case study: the birth of Quantum Mechanics (1925–1932)

The turning years, from 1925 to 1932, during which a *unified* conceptual and mathematical framework was built for the theory of Quantum Mechanics, appear as a particularly well suited example to launch our investigations. With the advent of both Göttingen's matrix mechanics and Schrödinger's wave mechanics, the quantum theory passed, in only one year (June 1925 to June 1926), from lacking any systematic scheme to having two seemingly different foundations. The sudden rise of two empirically equivalent theories which however seemed conceptually at odds from each other baffled the Physics community. The perplexity felt at the time is clearly expressed by Schrödinger himself:

> Considering the extraordinary differences between the starting-points and the concepts of Heisenberg's quantum mechanics and of the theory which has been designated "undulatory" or "physical" mechanics, it is very strange that these two new theories agree with one another with regard to the known facts, where they differ from the old quantum theory. [...] That is really very remarkable, because starting-points, presentations, methods, and in fact the whole mathematical apparatus seem fundamentally different.[9]

This exceptional situation[10] inevitably brought to the forefront the question of the relation between the mathematical characteristics of the theories being developed and their physical content. It offers a particularly well-suited context to perceive how different physicists dealt with the questions we have posed.

In the following subsections, I wish to investigate what can be learned with respect to these questions from the attempts of the founders of Quantum Mechanics in clarifying the relation between wave and matrix mechanics. I will therefore summarily present the key points in these developments, from Heisenberg's *Umdeutung* paper until von Neumann's introduction of abstract Hilbert spaces. The historical account does not pretend to any originality, even if, I hope, its synthetic and reflexive approach might be useful. It is based on some much more detailed works, which are now standard. I think in particular of Jammer's *The Conceptual Development of Quantum Mechanics*, Darrigol's *From c-Numbers to q-Numbers: The Classical Analogy in the History of Quantum Theory* and Mehra and Rechenberg's *The Historical Development of Quantum Theory. Volumes 1 – 6*. Along the way, I will pay particular attention to the places where emerges a reflection on the physical content of the mathematical apparatus used in Quantum Mechanics.

---

[9]E. Schrödinger. "On The Relation Between The Quantum Mechanics of Heisenberg, Born, and Jordan, and That of Schrödinger". In: *Collected Papers on Wave Mechanics*. Trans. by J. Shearer and W. Deans. London: Blackie & Son, 1928, pp. 45–61, p. 45.

[10]As Jammer puts it: "It is hard to find in the history of physics two theories designed to cover the same range of experience, which differ more radically than these two." (M. Jammer. *The Conceptual Development of Quantum Mechanics.* 2nd ed. Los Angeles: Tomash Publishers, 1989, p. 270)

### I.1.1  Matrix and Wave Mechanics

Prior to 1925, the quantum theory, although empirically very successful, lacked any clarity from the conceptual and methodological point of view. In the introduction of the book which represented the culmination in the establishing of quantum mechanics—von Neumann's *Mathematische Grundlagen der Quantenmechanik*—, the author recalled the situation as follows:

> In spite of the claim of quantum theory to universality, which had evidently been vindicated, there was lacking the necessary formal and conceptual instrument; there was a conglomeration of essentially different, independent, heterogeneous and partially contradictory fragments.[11]

Along the same lines, Jammer describes the state of the art as "a lamentable hodge-podge of hypotheses, principles, theorems, and computational recipes rather than a logical consistent theory"[12]. However, during the second semester of that year, the situation widely changed with the development of what later came to be known as 'matrix mechanics'[13]. This new quantum theory of Mechanics was essentially established by four physicists: the three-man Göttingen group formed by Werner Heisenberg, Max Born and Pascual Jordan, and the Cambridge student Paul Adrien Maurice Dirac.

---

[11] J. von Neumann. *Mathematical Foundations of Quantum Mechanics*. Trans. by R. T. Beyer. Princeton: Princeton University Press, 1955, p. 4.

[12] Jammer, op. cit., p. 208.

### I.1.1.a  Matrix Mechanics

The seminal work that cleared the fog and set up the path was Heisenberg's article *"Über quantentheoretische Umdeutung kinematischer und mechanischer Beziehungen"*, received on July 29 1925, written while he was part of Max Born's group at Göttingen[14]. His main motivation is clearly stated in the introduction of his paper:

> Instead it seems more reasonable to try to establish a theoretical quantum mechanics, analogous to classical mechanics, but in which only relations between observable quantities occur.[15]

There are two important ideas here. On the one hand, there is the emphasis on "*observable* quantities". In particular, this meant refusing any attempt to describe the trajectory in space-time of the electrons of an atom. Instead, the central phenomenon upon which the theory had to be built was the emission of radiation[16]. Through this move, frequencies and energy were conferred a preferred role amongst physical

---

[13]The term of 'matrix mechanics' does not appear in any of the first papers on the theory. Mehra and Rechenberg track back this expression to a review written in late 1926 by Edwin C. Kemble. (See J. Mehra and H. Rechenberg. *The Historical Development of Quantum Theory. Volume 3: The Formulation of Matrix Mechanics and Its Modifications*. New York: Springer-Verlag, 1982, footnote 74, pp. 61–62.)

[14]W. Heisenberg. "Über quantentheoretische Umdeutung kinematischer und mechanischer Beziehungen". In: *Zeitschrift für Physik* 33 (1925), pp. 879–893 (English translation: W. Heisenberg. "Quantum-theoretical Re-interpretation of Kinematic and Mechanical Relations". In: *Sources of Quantum Mechanics*. Ed. by B. Van der Waerden. New York: Dover Publications, Inc., 1967, pp. 261–276).

[15]Ibid., p. 262.

[16]"[I]t is necessary to bear in mind that in quantum theory it has not been possible to associate the electron with a point in space, considered as a function of time, by means of observable quantities. However, even in quantum theory it is possible to ascribe to an electron the emission of radiation." (Ibid., p. 263.)

Jean Petitot—following Alain Connes—considers this change of "regional object", from motion to spectra, as the fundamental move which characterizes the transition from Classical to Quantum Mechanics (see, for example, A. Connes. *Noncommutative Geometry*. Trans. by S. Berberian. London: Academic Press, 1994, pp. 33–39 and J. Petitot. "Noncommutative Geometry and Transcendental Physics". In: *Constituting Objectivity. Trascendental Perspectives on Modern Physics*. Ed. by M. Bitbol, P. Kerszberg, and J. Petitot. Springer, 2009, pp. 415–455).

quantities, and this had two important consequences on the form matrix mechanics was going to take. First, as we will shortly see, in the first papers on the theory one almost invariantly considered states with a definite value of energy—i.e., stationary states. Second, the theory of quantum mechanics was to be built by close analogy with classical Fourier analysis, which allowed to represent any physical quantity—in particular, position—in terms of frequencies and amplitudes.

On the other hand, Heisenberg insisted on the fact that quantum mechanics had to be "*analogous* to classical mechanics". The young german physicist was in fact deeply influenced by Bohr's correspondence principle and wished to give a rigorous formulation of it[17]. Therefore, he was aiming, as the title of the article underlines, at a systematic method of re-interpreting or *translating* the classical laws which related classical properties into similar laws relating quantum quantities[18]. In the road to this systematic translation scheme, the first fundamental problem became to translate into quantum mechanics the square of a given physical quantity:

---

[17]On the paper he co-authored with Born and Jordan, Heisenberg would be more explicit about the relation between his work and Bohr's. In the introduction of this famous "three-man paper", he wrote:

> If one reviews the fundamental differences between classical and quantum theory, differences which stem from the basic quantum theoretical postulates, then the formalism proposed [...], if proved to be correct, would appear to represent a system of quantum mechanics as close to that of classical theory as could reasonably be hoped. [...] This similarity of the new theory with classical theory also precludes any question of a separate correspondence principle outside the new theory; rather, the latter can itself be regarded as an exact formulation of Bohr's correspondence considerations.
>
> (M. Born, W. Heisenberg, and P. Jordan. "On Quantum Mechanics II". in: *Sources of Quantum Mechanics*. Ed. by B. Van der Waerden. New York: Dover Publications, Inc., 1967, pp. 321–384, p. 322)

(Although the paper was conjointly written by the three man, the introduction was written by Heisenberg, as it is explained in Mehra and Rechenberg, op. cit., pp. 92–102.)

[18]This point was clearly understood by Dirac, who, only a few months after the publication of the Heisenberg's work, concisely captured its essence:

> In a recent paper Heisenberg puts forward a new theory which suggests that it is not the equations of classical mechanics that are in any way at fault, but that the mathematical operations by which physical results are deduced from them require modification. *All* the information supplied by the classical theory can thus be made use of in the new theory.
>
> (P. A. M. Dirac. "The Fundamental Equations of Quantum Mechanics". In: *Proceedings of the Royal Society of London* A109 (1925), pp. 642–653. (Reprinted in: P. A. M. Dirac. *The Collected Works of P.A.M. Dirac: 1924–1948*. Ed. by R. Dalitz. Cambridge: Cambridge University Press, 1995, pp. 65–78), p. 642)

> This point has nothing to do with electrodynamics but rather — and this seems to be particularly important — is of purely kinematical nature. We may pose the question in its simplest form thus: If instead of a classical quantity $x(t)$ we have a quantum-theoretical quantity, what quantum-theoretical quantity will appear in place of $x^2(t)$?[19]

In other words, he asked about the product governing the algebra of quantum properties. As it is well-known, basing his considerations on the frequency conditions for the emission of radiation and using the analogy with the decomposition of a product in Fourier analysis, he concluded the quantum product needed to be non-commutative: "Whereas in the classical theory $x(t)y(t)$ is always equal to $y(t)x(t)$, this is not necessarily the case in the quantum theory"[20]. Having determined this new product, he was able to solve the quantum anharmonic oscillator, by re-interpreting the classical equations of motion $\ddot{x} + \omega_0^2 x + \lambda x^2 = 0$. This was the first success of Göttingen's quantum mechanics.

Immediately after Heisenberg had finished writing his article, Max Born recognized that the "*law of multiplication* of quantum-theoretical quantities [introduced by Heisenberg] was none other than the well-known mathematical rule of *matrix multiplication*"[21]. This realization launched a collaboration with Pascual Jordan, and in a few months they properly rewrote the mathematical aspects of Heisenberg's work in terms of infinite matrices, in a paper received on September 27 1925[22]. The major contribution of this work was, of course, the understanding that in this newly developed theory of quantum mechanics physical quantities were to be represented by matrices:

---

[19]Heisenberg, loc. cit.

[20]Ibid., p. 266. Although the introduction of this noncommutative product constitutes probably the most important breakthrough of the paper, Heisenberg was not satisfied with it, as he later acknowledged in a interview with Kuhn: 'In my paper, the fact that $XY$ was not equal to $YX$ was very disagreeable to me. I felt this was the only point of difficulty in the whole scheme, otherwise I would be perfectly happy.' (Cited in Mehra and Rechenberg, op. cit., p. 94)

[21]M. Born and P. Jordan. "On Quantum Mechanics". In: *Sources of Quantum Mechanics*. Ed. by B. Van der Waerden. New York: Dover Publications, Inc., 1967, pp. 277–306, p. 278, author's emphasis.

[22]M. Born and P. Jordan. "Zur Quantenmechanik". In: *Zeitschrift für Physik* 34 (1925), pp. 858–888 (English translation: M. Born and P. Jordan. "On Quantum Mechanics". In: *Sources of Quantum Mechanics*. Ed. by B. Van der Waerden. New York: Dover Publications, Inc., 1967, pp. 277–306).

The infinite square array (with discrete or continuous indices) which appears at the start of the next section, termed *matrix*, is a representation of a physical quantity which is given in classical theory as a function of time. The mathematical method of treatment inherent in the new quantum mechanics is thereby characterized through the employment of *matrix analysis* in place of the usual number analysis.[23]

More precisely, the matrix

$$\boldsymbol{q}(t) = \big(q(nm)e^{2i\pi\nu(nm)t}\big)$$

which represented the dynamical quantity $\boldsymbol{q}(t)$ (e.g., the position) was considered to be the quantum re-interpretation of the Fourier series expansion of the classical quantity $q(t)$:

$$q(t) = \sum_{n=-\infty}^{+\infty} q_n e^{2i\pi n\nu t}.$$

Although in principle the indices of the matrix coefficients could also be continuous, in practice they confined themselves from the outset to systems whose motion was *periodic*, a fact which restricted them to only considering discrete indices[24].

On top of this major insight, they introduced for the first time the canonical commutation relations for a system with one degree of freedom

$$\boldsymbol{pq} - \boldsymbol{qp} = \frac{h}{2\pi i}\boldsymbol{1}$$

(which they called at the time "stronger quantum conditions" since they were related to the quantization of angular momentum required by Bohr), proposed the probabilistic interpretation of the amplitude $|q(nm)|^2$ and wrote Hamilton's equations of motion in

$$q(t) = \int_{-\infty}^{+\infty} q(\alpha)e^{2i\pi\alpha\nu t}d\alpha.$$

terms of the commutator with the Hamiltonian:

$$\begin{cases} \dot{\boldsymbol{q}} = \dfrac{2\pi i}{h}(\boldsymbol{Hq} - \boldsymbol{qH}) \\ \dot{\boldsymbol{p}} = \dfrac{2\pi i}{h}(\boldsymbol{Hp} - \boldsymbol{pH}). \end{cases}$$

---

[23]Ibid., p. 278.

[24]Indeed, for a non-periodic function, the Fourier series is replaced by a Fourier *integral*:

Independently of the work of Born and Jordan at Göttingen, Paul Dirac, who was at the time a student at Cambridge, also managed to put Heisenberg's *Umdeutung* paper on a firmer mathematical ground in his article *"The Fundamental Equations of Quantum Mechanics"*, received on November 7 1925[25]. The whole paper is devoted to the study of the mathematical operations which can be performed on the physical quantities in the new quantum theory: section 2, "*Quantum algebra*", restates Heisenberg's law of multiplication for quantum variables; section 3, "*Quantum differentiation*", deals with "the most general quantum operation"[26] satisfying Leibniz's rule (multiplication and differentiation were the two main operations needed to find the quantum 're-interpretation' of the classical laws); section 4, "*The quantum conditions*", introduces the quantum Poisson bracket which allows him to write the canonical commutation relations for an arbitrary number of degrees of freedom

$$\begin{cases} q_r p_s - p_s q_r = \delta_{rs}\dfrac{ih}{2\pi} \\ q_r q_s - q_s q_r = 0 \\ p_r p_s - p_s p_r = 0; \end{cases}$$

and section 5, "*Properties of the quantum Poisson bracket expression*", introduces Hamilton's equations of motion. Hence, he essentially recovered (and slightly extended), in a beautifully concise style, all the basic results of Born and Jordan[27]. And he did so without mention of the just-found matrix representation (which he was unaware of). Instead, Dirac preferred to work at a purely symbolic level, disregarding any particular representation of what he later decided to call the "q-numbers" and focusing on the algebraic relations they satisfied. This methodology was clearly described in his second paper on the subject:

> This [matrix] representation was taken as defining a q-number in the previous papers [of Heisenberg, Born and Jordan] on the new theory. It seems preferable though to take the above algebraic laws and the general conditions (1) [the canonical commutation relations] as defining the properties of q-numbers, and to deduce from them that a q-number can be represented by c-numbers in this manner when it has the necessary periodic properties.[28]

---

[25] P. A. M. Dirac. "The Fundamental Equations of Quantum Mechanics". In: *Proceedings of the Royal Society of London* A109 (1925), pp. 642–653. (Reprinted in: P. A. M. Dirac. *The Collected Works of P.A.M. Dirac: 1924–1948.* Ed. by R. Dalitz. Cambridge: Cambridge University Press, 1995, pp. 65–78).

[26] Ibid., p. 311.

[27] Dirac's paper is only ten pages long. Compare this with the thirty pages of Born and Jordan's first paper!

Finally, simultaneously with Dirac's first paper, Born, Jordan and Heisenberg wrote yet another article that further developed the new quantum theory of mechanics: *"On Quantum Mechanics II"*, received on November 16 1925[29]. For a system with an arbitrary number of (periodic) degrees of freedom, they proposed the same generalized commutation relations found by Dirac. But the main novelty of this sixty pages long 'three man paper' was their introduction of the concept of "canonical transformations" (leaving invariant the quantum commutation relations) in order to translate the dynamical problem of integrating the equations of motions into the problem of diagonalizing the energy matrix $\boldsymbol{H(p,q)}$[30]. In this way, the theory of eigenvalues of hermitian forms became of central importance to quantum mechanics.

At this point, essentially all the fundamental ingredients of what became to be known as 'matrix mechanics' had been properly laid down. All the relevant information of a quantum-theoretical dynamical system was to be completely described by the form of the Hamiltonian and the set of observable quantities, which were represented

---

[28]P. A. M. Dirac. "Quantum Mechanics and a Preliminary Investigation of The Hydrogen Atom". In: *Proceedings of the Royal Society of London* A110 (1926), pp. 561–579. (Reprinted in: P. A. M. Dirac. *The Collected Works of P.A.M. Dirac: 1924–1948*. Ed. by R. Dalitz. Cambridge: Cambridge University Press, 1995, pp. 85–105), p. 563.

[29]M. Born, W. Heisenberg, and P. Jordan. "Zur Quantenmechanik II". in: *Zeitschrift für Physik* 35 (1926), pp. 557–615 (English translation, M. Born, W. Heisenberg, and P. Jordan. "On Quantum Mechanics II". in: *Sources of Quantum Mechanics*. Ed. by B. Van der Waerden. New York: Dover Publications, Inc., 1967, pp. 321–384)

[30]"Then, the dynamic problem, e.g., the determination of the $\boldsymbol{p_k, q_k}$ can be formulated as: A transformation $\boldsymbol{p_k^0, q_k^0 \to p_k, q_k}$ is to be found which leaves [the basic commutation relations] invariant and at the same time reduces the energy to a diagonal matrix." (Ibid., p. 349)

by infinite dimensional matrices satisfying a certain number of algebraic relations. Pauli's application of this matrix scheme to solve the hydrogen atom was the first truly impressive success of the theory[31].

Therefore, only six months after the publication of Heisenberg's seminal paper, the new theory of 'matrix mechanics' seemed to be expanding full steam ahead, especially carried by the Göttingen group. But this swift development was to be echoed, during the first semester of 1926, by the equally sudden and expeditious construction of 'wave mechanics'.

### I.1.1.b   Wave Mechanics

Contrary to the situation of the quantum theory of 'matrix mechanics', which progressively took its form thanks to the contribution of various physicists, 'wave mechanics' was single-handedly developed by Erwin Schrödinger, who at the time worked in Zurich. He did so in a series of four papers written and published in only six months, from end of December 1925 to end of June 1926:

 – *"Quantisierung als Eigenwertproblem (I)"*, received on January 27 1926[32],

 – *"Quantisierung als Eigenwertproblem (II)"*, received on February 23 1926[33],

 – *"Quantisierung als Eigenwertproblem (III)"*, received on May 10 1926[34],

---

[31]W. Pauli. "Über das Wasserstoffspektrum vom Standpunkt der neuen Quantenmechanik". In: *Zeitschrift für Physik* 36 (1926), pp. 336–363, received on January 17 1926 (English translation: W. Pauli. "On The Hydrogen Spectrum From The Standpoint of The New Quantum Mechanics". In: *Sources of Quantum Mechanics*. Ed. by B. Van der Waerden. New York: Dover Publications, Inc., 1967, pp. 387–415).

[32]E. Schrödinger. "Quantisierung als Eigenwertproblem (I)". in: *Annalen der Physik* 79 (1926), pp. 361–376 (English translation: E. Schrödinger. "Quantisation as a Problem of Proper Values. Part I". in: *Collected Papers on Wave Mechanics*. Trans. by J. Shearer and W. Deans. 2nd ed. London and Glasgow: Blackie & Son, Ltd, 1928, pp. 1–12).

[33]E. Schrödinger. "Quantisierung als Eigenwertproblem (II)". in: *Annalen der Physik* 79 (1926), pp. 489–527 (English translation: E. Schrödinger. "Quantisation as a Problem of Proper Values. Part II". in: *Collected Papers on Wave Mechanics*. Trans. by J. Shearer and W. Deans. 2nd ed. London and Glasgow: Blackie & Son, Ltd, 1928, pp. 13–40).

[34]E. Schrödinger. "Quantisierung als Eigenwertproblem (III)". in: *Annalen der Physik* 80 (1926), pp. 437–490 (English translation: E. Schrödinger. "Quantisation as a Problem of Proper Values. Part III". in: *Collected Papers on Wave Mechanics*. 2nd ed. London and Glasgow: Blackie & Son, Ltd, 1928, pp. 62–101).

– *"Quantisierung als Eigenwertproblem (IV)"*, received on June 21 1926[35].

This four-parts paper has been qualified by Jammer as "one of the most influential contributions ever made in the history of science"[36]. As Schrödinger explains, they were not first conceived as a whole and then written; rather, the writing process accompanied the author's progressive understanding of the theory he was building[37]. As a consequence, they do not constitute the best exposition for a reader who would be approaching the subject for the first time. Far more enlightening are the presentations of the same material he wrote in English and French a few months later[38].

As it had been the case with Heisenberg, Schrödinger wanted to ban the notion of "trajectory" from the theory of Mechanics. But his reasons for this widely differed from those of the German physicist. As we saw in the previous section, Heisenberg's dismissal of trajectories originated in the impossibility found at the time to observe the position of the electrons in an orbit. Because of this, Heisenberg rejected any spatial picture of atomic phenomena. Instead, Schrödinger's main motivation was to get rid of the picture of *material points* and build a theory of Mechanics entirely based upon the notion of "wave":

> The theory which is reported in the following pages is based on the very interest-
> ing and fundamental researches of L. de Broglie on what he called "phase-waves"
> ("ondes de phase") and thought to be associated with the motion of material
> points, especially with the motion of an electron or proton. The point of view

---

[35]E. Schrödinger. "Quantisierung als Eigenwertproblem (IV)". in: *Annalen der Physik* 81 (1926), pp. 109–139 (English translation: E. Schrödinger. "Quantisation as a Problem of Proper Values. Part IV". in: *Collected Papers on Wave Mechanics*. Trans. by J. Shearer and W. Deans. 2nd ed. London and Glasgow: Blackie & Son, Ltd, 1928, pp. 102–123).

[36]Jammer, op. cit., p. 266.

[37]"[T]he papers now combined in one volume were originally written *one by one* at different times. The results of the later sections were largely unknown to the writer of the earlier ones." (E. Schrödinger. *Collected Papers on Wave Mechanics*. Trans. by J. Shearer and W. Deans. 2nd ed. London and Glasgow: Blackie & Son, Ltd, 1928, p. v.)

[38]E. Schrödinger. "An Undulatory Theory of the Mechanics of Atoms and Molecules". In: *The Physical Review* 28 (1926), pp. 1049–1070. (Reprinted in: E. Schrödinger. *Gesammelte Abhandlungen / Collected Papers. Volume 3*. Vienna: Austrian Academy of Science, 1984, pp. 280–301), and E. Schrödinger. "La mécanique des ondes". In: *Electrons et Photons: Rapports et Discussions du Cinquième Conseil de Physique, tenu à Bruxelles du 24 au 29 Octobre 1927*. Paris: Gauthiers-Villars, 1928, pp. 185–213. (Reprinted in: E. Schrödinger. *Gesammelte Abhandlungen / Collected Papers. Volume 3*. Vienna: Austrian Academy of Science, 1984, pp. 302–323).

taken here [...] is rather that material points consist of, or are nothing but, wave-systems.[39]

The urge Schrödinger felt to build this "undulatory" or "wave" Mechanics was indeed strongly influenced by the recent work of the French Louis de Broglie, which allowed to associate a wavelength to material particles. But there was also the work of William Hamilton, who had constructed his analytical Mechanics by analogy with geometrical optics. This exerted a decisive influence on Schrödinger's ideas. In fact, during the process of firmly establishing his quantum theory, Schrödinger was driven by the fundamental idea that quantum mechanics should be to classical mechanics what undulatory optics had been to geometrical optics:

> [...] *we know to-day, in fact, that our classical mechanics fails for very small dimensions of the path and for very great curvatures.* Perhaps this failure is in strict analogy with the failure of geometrical optics, i.e., "the optics of infinitely small wave lengths", that becomes evident as soon as the obstacles or apertures are no longer great compared with the real, finite, wave length. Perhaps our classical mechanics is the *complete* analogy of geometrical optics and as such is wrong and not in agreement with reality [...]. Then it becomes a question of searching for an undulatory mechanics, and the most obvious way is the working out of the Hamiltonian analogy on the lines of undulatory optics.[40]

---

[39] Idem, "An Undulatory Theory of the Mechanics of Atoms and Molecules", p. 1049.

[40] Idem, "Quantisation as a Problem of Proper Values. Part II", p. 18, author's emphasis. Another paradigmatic paragraph of Schrödinger's thought is the following:

> The true mechanical process is realised or represented in a fitting way by the *wave process* in *q*-space, and not by the motion of *image points* in this space. The study of the image points, which is the object of classical mechanics, is only an approximate treatment, and has, as such, just as much justification as geometrical or "ray" optics has, compared with the true optical process. A macroscopic mechanical process will be portrayed as a wave signal [...], which can approximately enough be regarded as confined to a point compared with the geometrical structure of the path. [...] This manner of treatment, however, loses all meaning where the structure of the path is no longer very large compared with the wave length or indeed is comparable with it. Then we *must* treat the matter strictly on the wave theory, *i.e.* we must proceed from the *wave equation* [yet to be found] and not from the fundamental equations of mechanics [...]. These latter equations are just as useless for the elucidation of the micro-structure of the mechanical processes as geometrical optics is for explaining the *phenomena of diffraction*. (Ibid., p. 25.)

From a mathematical point of view, Schrödinger's program of going from material-point Mechanics to undulatory Mechanics meant the passage from *total* differential equations whose only parameter is time—such as the equation of motion of a point-like harmonic oscillator $\left(\frac{d^2}{dt^2}+\omega^2\right)q(t)=0$—to *partial* differential equations where both time and space coordinates are involved—such as the usual one-dimensional wave equation $\left(\frac{\partial^2}{\partial t^2}-\frac{1}{v^2}\frac{\partial^2}{\partial x^2}\right)u(x,t)=0$.

The first article is solely dedicated to treating the simplified non-relativistic hydrogen atom, described by the Keplerian central potential $V=-\frac{e^2}{r}$. For this, he proposed the partial differential equation

$$\nabla^2\psi(x)+K(E-V)\psi(x)=0 \qquad\qquad (I.2)$$

and re-obtained Bohr's energy levels for the stationary states $E_n\propto-\frac{e^4}{n^2}$ $(n=1,2,3,\dots)$ as the only possible negative values of the parameter $E$ for which the equation admitted a finite, continuous solution. In this way, Schrödinger claimed, the postulation of whole numbers, which was the core of the old quantum theory, acquired a more natural justification:

> The essential thing seems to me to be, that the postulation of "whole numbers" no longer enters into the quantum rules mysteriously, but that we have traced the matter a step further back, and found the "integralness" to have its origin in the finiteness and single-valuedness of a certain space function.[41]

Despite this success, the justification for Equation I.2, which Schrödinger had essentially guessed[42] and later came to be known as the 'time-independent Schrödinger equation', was not very convincing (to say the least)[43]. By extensively discussing the already mentioned analogy between Mechanics and Optics, the second paper tried to establish it on firmer grounds, but the result was not much more transparent. Notwithstanding this, he continued developing his theory and considered further applications of his wave equation: in the second article, he managed to solve other simple mechanical systems such as the harmonic oscillator and the rigid rotator, while in the third he developed the time-independent perturbation theory which allowed him to compute the Stark effect in the hydrogen atom. Finally, in the fourth article, Schrödinger dealt with non-conservative systems, i.e. systems for which the external potential $V$ varied with time, and was lead to introduce the full Schrödinger equation:

$$\nabla^2 \Psi - \frac{8\pi^2}{h^2} V \Psi \mp \frac{4\pi i}{h} \frac{\partial \Psi}{\partial t} = 0 \tag{I.3}$$

which yielded the previous time-independent equation (I.2) whenever the wave-function was stationary: $\Psi(x,t) = \psi(x)e^{2i\pi Et/h}$.

---

[41] Idem, "Quantisation as a Problem of Proper Values. Part I", p. 9.

[42] For a detailed reconstruction of Schrödinger's complex route of "educated guesses" towards his wave equation, see J. Mehra and H. Rechenberg. *The Historical Development of Quantum Theory. Volume 5: Erwin Schrödinger and the Rise of Wave Mechanics.* New York: Springer-Verlag, 1987, Chapter III.

[43] In the first paragraph of his second paper, Schrödinger himself described his previous derivation of the time-independent equation as "unintelligible" and "incomprehensible"!

Therefore, by the end of the fourth paper, Schrödinger had written down mostly all of quantum mechanics as it is known today. Table I.1 (page 31) summarizes the situation at that point, with the fundamental traits of the two different quantum theories of Mechanics[44].

## I.1.2   Extracting the physics (1): Schrödinger's mathematical equivalence

The state of perplexity of the Physics community, after having witnessed in such a short lapse of time the development of both Matrix and Wave Mechanics, could very well be described by the following words, which Einstein supposedly addressed to a crowded room of physicists in the University of Berlin:

> Now listen! Up to now we had no exact quantum theory, and today we suddenly
> have two of them. You will agree with me that these two theories exclude one

---

| **Matrix Mechanics** | **Wave Mechanics** |
|---|---|
| Description of the mechanical system: by its **observable quantities**, which are represented by infinite square **matrices**. | Description of the mechanical system: by its **stationary states**, which are described by **complex-valued functions** over $q$-space. |
| Dynamical problem: to find matrices $\boldsymbol{q}$ and $\boldsymbol{p}$ satisfying the **algebraic relation** $$\boldsymbol{pq} - \boldsymbol{qp} = \frac{\hbar}{i}\boldsymbol{1}$$ and for which the Hamiltonian matrix $H(\boldsymbol{p},\boldsymbol{q})$ is **diagonal**. | Dynamical problem: to find the wave functions $\psi$ and the values $E$ satisfying the **partial differential equation** $$\nabla^2\psi(q) + \frac{2m}{\hbar^2}(E - V)\psi(q) = 0$$ |
| Energy levels: eigenvalues of the matrix $H(\boldsymbol{p},\boldsymbol{q})$. | Energy levels: eigenvalues of the time-independent Schrödinger equation. |

**Table I.1** – Fundamental ingredients of Matrix and Wave Mechanics.

another. Which theory is the correct one? Perhaps neither of them is correct![45]

This surprise was the combination of two facts. On the one hand, Matrix and Wave Mechanics were 'extraordinarily different in their starting-points and concepts' (Schrödinger), up to the point that they seemed to 'exclude one another' (Einstein). On the other hand, however, the two theories had lead to the same predictions for the energy levels of the systems they had considered—coincidence which was specially spectacular for the harmonic oscillator since the predictions differed from those of the old quantum theory. To be sure, the *observation* of their empirical equivalence raised the question of whether or not the 'extraordinary differences' were nothing but a misleading *impression*. In other words, it raised the question of the *identity* of the two theories. The importance of reflecting on the notion of identity was immediately recognized by Schrödinger, who was the first to unveil the profound relation between the two new quantum theories in the article *"On The Relation Between The Quantum*

---

[45]H. Kallmann. "Von der Anfängen der Quantentheorie—Eine persönliche Rückschau". In: *Physikalische Blätter* 22 (1966), pp. 489–500. Cited in: Mehra and Rechenberg, op. cit., p. 636. Mehra and Rechenberg raise doubts on the faithfulness of Kallmann's recollection of Einstein's words, but this need not worry us here.

*Mechanics of Heisenberg, Born, and Jordan, and That of Schrödinger*"[46].

But before reviewing this work, let us look more closely at the reasons Schrödinger could have had for considering these two theories to be so radically *different*. In the introduction of his paper, the Austrian physicist explains:

> Above all, however, the departure from classical mechanics in the two theories seems to occur in diametrically opposed directions. In Heisenberg's work the classical continuous variables are replaced by systems of discrete numerical quantities (matrices), which depend on a pair of integral indices, and are defined by *algebraic* equations. The authors themselves describe the theory as a "true theory of the discontinuum". On the other hand, wave mechanics shows just the reverse tendency; it is a step from classical point-mechanics towards a *continuum-theory*. In place of a process described in terms of a finite number of dependent variables occurring in a finite number of total differential equations, we have a continuous *field-like* process in configuration space, which is governed by a single *partial* differential equation [...].[47]

Hence, for him, the couple of opposite notions Continuum/Discrete characterized the essential trait distinguishing Wave and Matrix Mechanics.

Undoubtedly, Schrödinger could not be wrong when claiming that, in the development of undulatory mechanics, the pursue of continuity played a major role—after all, he was the author! However, the idea that 'discreteness' went hand in hand with Matrix Mechanics—idea which has been repeated ever so often (for example in Jammer's book, p. 270)—is less obvious. From the above-quoted passage, it would seem that this association stemmed from the use of "integral indices" in Göttingen's theory. However, although in all *practical* examples considered by the end of 1925, the indices labelling the matrix coefficients were discrete—and this surely conveyed an impression of an essential discreteness in the theory—, this trait arose only because the systems considered were periodic and hence the coordinates' domain of variation was bounded. *In principle*, Born, Jordan and Heisenberg insisted, these coefficients could be also continuous[48]. Even more, this *cohabitation of the continuum and the discrete* was emphasized as a major feature of their theory:

> [...] a particularly important trait in the new theory would seem to us to consist of the way in which both continuous and line spectra arise in it on an equal footing, i.e. as solutions of one and the same equation of motion and closely connected with one another mathematically; obviously, in this theory, any distinction between 'quantized' and 'unquantized' motion ceases to be at all meaningful [...].[49]

---

[46]E. Schrödinger. "Über das Verhältnis der Heisenberg-Born-Jordanschen Quantumemchanik zu der meinen". In: *Annalen der Physik* 79 (1926), pp. 734–756 (English translation: E. Schrödinger. "On The Relation Between The Quantum Mechanics of Heisenberg, Born, and Jordan, and That of Schrödinger". In: *Collected Papers on Wave Mechanics*. Trans. by J. Shearer and W. Deans. London: Blackie & Son, 1928, pp. 45–61).

[47]Ibid., pp. 45-46, author's emphasis.

Thus, this circumstantial use of integral indices could certainly not be what rendered Matrix Mechanics a "true theory of the discontinuum".

Now, among all the seminal papers on matrix mechanics, there was indeed *one* sentence in which discreteness was designated as fundamental to the theory: the sentence from Born and Jordan that Schrödinger paraphrases in the above passage—and which is also used by Jammer. In their paper *"On Quantum Mechanics"*, they wrote:

> The new mechanics presents itself as an essentially discontinuous theory [...].

But, as far as I can tell, the reasons which led the two authors to this conclusion have never been discussed. These had in fact little to do with the use of discrete indices to label the matrix coefficients. Indeed, by analyzing the properties of the equations they had just established, Born and Jordan realized that their matrix representation of physical quantities was ambiguous. For given any pair of matrices $\big(\boldsymbol{q}(t), \boldsymbol{p}(t)\big)$ solution to the equations of motion, the permutation of some chosen rows and columns—that is, the permutation of the the the order of two indices $n_0$ and $n_1$—allowed to generate a different pair of matrices $\big(\boldsymbol{q}'(t), \boldsymbol{p}'(t)\big)$ which would also solve the equations:

> [...] one can see right away that the exact form of the matrix can never be deduced

---

[48]Thus, in the first section of Heisenberg's *Umdeutung* article, all formulas are given in two versions: one in terms of discrete sums and one in terms of integrals. This only ceases to be the case when he turns to the consideration of the anharmonic oscillator.

[49]Born, Heisenberg, and Jordan, op. cit., pp. 322–323. The three authors insisted in this point with a similar quote found thirty pages later in the same article: "The simultaneous appearance of both continuous and line spectra as solutions to the same equations of motion and the same commutation relations seemed to us to represent a particularly significant feature of the new theory." (Ibid., p. 358)

from the fundamental equations, since if rows and columns be subjected to the same permutation, the canonical equations and the quantum condition remain invariant and thereby one obtains a new and apparently different solution. But all such solutions naturally differ only in the notation, i.e., in the way the elements are numbered.[50]

The two solutions $\big(\boldsymbol{q}(t), \boldsymbol{p}(t)\big)$ and $\big(\boldsymbol{q}'(t), \boldsymbol{p}'(t)\big)$, although different in their exact numerical form, described the same physical situation. The remark of this indeterminacy in the form of the matrices had an important conceptual consequence for them: it showed that the order in which were arranged the labels $n_0, n_1, \ldots$, used to distinguish the different frequencies $\nu(n_0, n_1)$ was completely arbitrary, and therefore physically irrelevant. And it was this last fact which, in turn, was read as a manifestation of an essential discontinuity of the quantum theory:

The classically calculated orbits merge into one another continuously; consequently the quantum orbits selected at a later stage have a particular sequence right from the outset. The new mechanics presents itself as an essentially discontinuous theory in that herein there is no sequence of quantum states defined by the physical process, but rather of quantum numbers which are indeed no more than distinguishing indices which can be ordered and normalized according to any practical standpoint whatsoever.[51]

Therefore, *for Born and Jordan, the 'essential discontinuity' of matrix mechanics did not lie in the fact that the quantum numbers were discrete, but rather in the fact there was no preferred order in which to arrange these numbers*[52].

This may appear as strange a conclusion from the modern point of view. For the ambiguity in the matrix representation of a physical quantity observed by Born and Jordan is now understood as the indication that physical quantities are better described by *operators* rather than by matrices (the latter being a representation of the former in a particular system of coordinates). But this was impossible for them to foresee, since at the time they were not acquainted with this more abstract notion[53]. Hence, the invariance of the fundamental equations of Göttingen's Mechanics under permutation of indices was completely unrelated with any sort of fundamental discontinuity.

---

[50]Born and Jordan, op. cit., p. 298.

[51]Ibid., pp. 300–301.

[52]The intuition behind their idea of relating *continuity* and *order* seems to be somewhat along the following lines: given a discrete set of objects $\bullet$, $\bullet$, $\bullet$, $\ldots$, their labelling by natural numbers is arbitrary: one may choose $\bullet = 1$, $\bullet = 2$, $\bullet = 3$, $\ldots$, but one can also perform the permutation $\bullet \longleftrightarrow \bullet$ which yields the labelling $\bullet = 1$, $\bullet = 2$, $\bullet = 3$, $\ldots$ If, however, these objects are part of a larger continuous set (e.g., a straight line), then the picture changes. Now, we have ●━━●━━● and their labelling by given real numbers $\alpha < \beta < \gamma < \ldots$ is no longer arbitrary: the number $\beta$ must be used to label $\bullet$ since there is no *continuous* transformation which, starting from ━━●━━●━━, would yield the picture ━━●━━●━━. In other words, by introducing continuity into the picture, the notion of 'being the object in the middle' becomes meaningful.

Besides the couple Continuum/Discrete, another notion which can help us understand the seemingly irreducible differences between Wave and Matrix Mechanics is that of *visualizability* in space-time. In the introduction to one of their papers, the Göttingen group had written that the new quantum theory was "not directly amenable to a geometrically visualizable interpretation, since the motion of electrons cannot be described in terms of the familiar concepts of space and time"[54], and had called their theory a "symbolic quantum geometry" in contrast with the "visualizable classical geometry"[55]. On the contrary, Schrödinger was clearly striving for an intuitive spatial picture of atomic processes:

> [...] it has even been doubted whether what goes on in the atom could ever be described within the scheme of space and time. From the philosophical standpoint, I would consider a conclusive decision in this sense as equivalent to a complete surrender. For we cannot really alter our manner of thinking in space and time, and what we cannot comprehend within it we cannot understand at all. There *are* such things—but I do not believe that atomic structure is one of them.[56]

---

[53]The idea that physical quantities were described by operators rather than matrices emerged from the collaboration of Born with the mathematician Norbert Wiener at the beginning of 1926. However, they considered operators acting on functions depending solely on time. As we will later explain, it was Schrödinger who first considered operators acting on functions of configuration space. See M. Born and N. Wiener. "A New Formulation of The Laws of Quantization of Periodic and Aperiodic Phenomena". In: *Journal of Mathematics and Physics (MIT)* (1925–1926), pp. 84–98. (Reprinted in: N. Wiener. *Norbert Wiener: Collected Works. Volume III.* ed. by P. Masani. Cambridge: The MIT Press, 1981, pp. 427–441).

[54]Born, Heisenberg, and Jordan, op. cit., p. 322.

[55]Ibid., p. 322.

[56]Schrödinger, "Quantisation as a Problem of Proper Values. Part II", pp. 26-27.

And this difference seemed to be at the origin of the mutual dislike Heisenberg and Schrödinger felt towards each other's theories[57].

Now, Schrödinger's claims on the visualizability of his theory strongly depended on his tentative to attach a physical meaning to the wave-function $\Psi$ he had introduced, and which he expected to "represent the true mechanical process in a fitting way"[58]. However, this intended interpretation was made difficult by two mathematical characteristics of the theory: first, the wave-function was not a function of space but rather a function of the configuration space ("the $q$-space")[59]. Hence, only when dealing with one electron could one readily interpret the atomic process as a "vibration in real three-dimensional space"[60]. Second, after the introduction of the time-dependent Schrödinger equation (Equation I.3, page 30), complex numbers had become a seemingly unavoidable and essential ingredient, somewhat against Schrödinger's will[61]. To some extent, these two problems—of the interpretation of the wave-function and of the

---

[57]Thus, Heisenberg, in a letter to Pauli, wrote: "The more I ponder about the physical part of Schrödinger's theory, the more horrible I find it. [...] What Schrödinger writes on the visualizability of his theory [...] I find rubbish. The great achievement of Schrödinger's theory is the calculation of matrix elements". (Cited in: Mehra and Rechenberg, op. cit., p. 821.)

On the other side, Schrödinger confessed in a letter to Wien: "I firmly hope, of course, that the matrix method, after its valuable results have been absorbed by the eigenvalue theory, will disappear again. [...] Because the mere thought makes me shudder, if I later had to present the matrix calculus to a young student as describing the true nature of the atom." (Cited in: ibid., p. 639.)

[58]Schrödinger, "Quantisierung als Eigenwertproblem (II)", p. 25.

[59]"$\Psi$ itself is in the general case a function of the generalized coordinates $q_1 \cdots q_n$ and the time,— not a function of ordinary space and time as in ordinary wave-problems. This raises some difficulty in attaching a physical meaning to the wave-function." (Idem, "An Undulatory Theory of the Mechanics of Atoms and Molecules", p. 1066.)

[60]Idem, "Quantisation as a Problem of Proper Values. Part II", p. 28.

[61]In the French translation of his collected papers on wave mechanics, Schrödinger added the following comment:

> In his desire to consider at any cost the propagation phenomenon of the waves $\psi$ as something real in the classical sense of the word, the author had refused to acknowledge that the whole development of the theory increasingly tended to highlight the essential complex nature of the wave function.
>
> (*Dans son désir d'envisager à tout prix le phénomène de la propagation des ondes $\psi$ comme quelque chose de réel dans le sens classique du terme, l'auteur s'était refusé de reconnaître franchement que tout le développement de la théorie mettait de plus en plus clairement en évidence le caractère essentiellement complexe de la fonction d'onde.* E. Schrödinger. *Mémoires sur la mécanique ondulatoire.* Trans. by A. Proca. Paris: Librairie Alcan, 1933, p. 166.)

His apprehension towards complex numbers is also apparent in the concluding paragraph of his fourth and last part of the article "Quantisation as a Problem of Proper Values":

role of complex numbers in Quantum Mechanics—are still discussed nowadays[62].

Thus—we see—the impression of radical conceptual and physical differences owed much to the tendency of attaching a meaning to certain *particularities of the mathematical form* of one theory or the other: the use of discrete indices, the invariance of the equations under permutation of indices, the appearance of a function depending on the space-time coordinates, etc. Yet, despite having observed this and having explained how "the whole mathematical apparatus [of Matrix and Wave Mechanics] seem fundamentally different", Schrödinger made the following claim:

> From the formal mathematical standpoint, one might well speak of the *identity* of the two theories.[63]

From what has been explained so far, this statement is surely surprising. One would have perhaps expected an argument along these lines: i) the mathematical apparatus of the two theories are different; ii) however, the *noticed* empirical equivalence stifles a naive and direct reading of these mathematical apparatus; iii) therefore, although mathematically different, the two theories must be *physically* identical. In fact, in the article he did also introduce the notion of "physical identity" and discussed its relation with the notion of "mathematical identity":

> To-day there are not a few physicists who, like Kirchhoff and Mach, regard the task of a physical theory as being merely a mathematical description (*as economical as possible*) of the empirical connections between observable quantities [...]. On this view, mathematical equivalence has almost the same meaning as

---

Meantime, there is no doubt a certain crudeness in the use of a *complex wave function. If it were unavoidable *in principle*, and not merely a facilitation of the calculation, this would mean that there are in principle *two* wave functions, which must be used *together* in order to obtain information on the state of the system. [...] Our inability to give more accurate information about this is intimately connected with the fact that, in the pair of equations (I.3), we have before us only the *substitute*—extraordinarily convenient for the calculation, to be sure—for a real wave equation of probably the fourth order, which, however, I have not succeeded in forming for the non-conservative case." (Idem, "Quantisation as a Problem of Proper Values. Part IV", p. 123, author's emphasis.)

[62]Cf. for example the modern debate between Ψ-epistemic models and Ψ-ontologists. For a review of this, see M. S. Leifer. "Is the Quantum State Real? An Extended Review of ψ-ontology Theorems". In: *Quanta* 3 (2014), pp. 67–155.

[63]Schrödinger, "On The Relation Between The Quantum Mechanics of Heisenberg, Born, and Jordan, and That of Schrödinger", p. 46, author's emphasis.

physical equivalence. [...]

[However] the validity of the thesis that mathematical and physical equivalence mean the same thing, must itself be qualified. Let us think, for example, of the two expressions for the electrostatic energy of a system of charged conductors, the space integral $\frac{1}{2}\int E^2 d\tau$ and the sum $\frac{1}{2}\Sigma e_i V_i$ taken over the conductors. The two expressions are completely equivalent [...]. Nevertheless we intentionally prefer the first [...].

We cannot yet say with certainty to which of the two new quantum theories preference should be given, from *this* point of view. As natural advocate of one of them, I will not be blamed if I frankly [...] bring forward the arguments in its favour.[64]

Thus, it is clear from this passage that Schrödinger was considering the possibility that, despite the empirical equivalence and mathematical identity of the two theories, his Wave Mechanics was perhaps physically superior to Göttingen's Quantum Mechanics. Hence, the above outlined argument was surely *not* what Schrödinger was claiming. What, then, did he mean by the "*mathematical* identity" of these two theories?

Here, the best is to look at the way in which he attempted to prove his claim. He first proceeded to show how his wave functions could be used to construct matrices obeying the algebraic relations written by Heisenberg, Jordan and Born. This he achieved by a two-step procedure that is now well-known:

1. <u>From functions to operators:</u> to the $2n$ quantities $\{q_1, \ldots, q_n; p_1, \ldots, p_n\}$ (position and canonically conjugate momentum co-ordinates), associate linear operators acting on functions of the first $n$ variables as follows:

$$q_k \longmapsto q_k \cdot \big[\,-\,\big] \qquad \text{(multiplication by } q_k\text{)}$$
$$p_k \longmapsto \frac{\hbar}{i}\frac{\partial}{\partial q_k}\big[\,-\,\big] \qquad \text{(differentiation by } q_k\text{)}.$$

This allows to associate a linear operator $[F, \cdot]$ to any given (well-ordered) function $F(p, q)$ of '$pq$-space'.

---

2. <u>From operators to matrices</u>: given the choice of a complete orthogonal system of functions of 'q-space' $\{u_1(q), u_2(q), \ldots\}$, associate to the operator $[F, \cdot]$ a matrix whose coefficients are defined by:

$$F^{kl} := \int dq\, u_k(q)[F, u_l(q)].$$

In this way, Heisenberg's 'mysterious' commutation relations for the quantum representatives $\{\boldsymbol{q_1}, \ldots, \boldsymbol{q_n}; \boldsymbol{p_1}, \ldots, \boldsymbol{p_n}\}$ revealed themselves to be nothing else but the trivial identity in ordinary analysis for linear differential operators: $\frac{\partial}{\partial q_k}q_k - q_k\frac{\partial}{\partial q_k} = 1$. Moreover, the complete orthogonal system of functions which associated to the energy function $H(p, q)$ a diagonal matrix, as required by Göttingen's quantum mechanics, were precisely Schrödinger's wave-functions $\psi_E(q)$, solutions to his time-independent equation (which was the only one he had introduced at the time of the writing of the article).

In other terms, one could say that, by the above construction, Schrödinger showed how any statement of Matrix Mechanics could be interpreted in the language of Wave Mechanics. But he did not stop here, for he knew this relation between both theories was insufficient to prove the identity:

> [...] the equivalence *actually* exists, and it also exists *conversely*. Not only the matrices can be constructed from the proper functions as shown above, but also, conversely, the functions can be constructed from the numerically given matrices. Thus the functions do not form, as it were, an *arbitrary* and *special* "fleshly clothing" for the bare matrix skeleton, provided to pander to the need for intuitiveness. This really would establish the superiority of the matrices, from the epistemological point of view.[65]

The last two sentences are enlightening. In his worry of being "as economical as possible" and excluding any "arbitrary and special "fleshly clothing"", Schrödinger was precisely trying to avoid the mistake of conceding a key role to a feature of the theory that could turn out to be completely irrelevant. And to protect himself from this, he needed to show how *any* statement expressed in terms of waves functions could be found to be expressible as well in terms of matrices (and, of course, vice versa). It thus appears that, for Schrödinger, the task of proving the 'mathematical identity' consisted in finding a perfect *translation*, in building an *dictionary* between the two theories. In other words, Matrix and Wave Mechanics were to be identical, "from the formal mathematical standpoint", because one could make explicit a *bi-directional* correspondence between what could be expressed within one language and what could be expressed within the other.

---

[65]Ibid., p. 58, author's emphasis

Contrary to his claims however, Schrödinger did not really manage to rigorously prove this 'backwards translation'[66]. It was only with the work of Dirac and Jordan first, and von Neumann afterwards, that the situation became completely understood[67].

### I.1.3 Extracting the physics (2): Dirac's transformation theory

In December 1926, as he was visiting Niels Borh's group at Copenhagen, Dirac finished the writing of a much celebrated article that brought a completely new perspective on the relation between Wave and Matrix Mechanics: *"The Physical Interpretation of*

---

[66]For a much more detailed analysis Schrödinger's notion of 'mathematical equivalence' and the reasons why Schrödinger's work should not be considered as a rigorous proof of the equivalence, see the excellent two-part article F. A. Muller. "The Equivalence Myth of Quantum Mechanics—Part I". in: *Studies in History and Philosophy of Science Part B: Studies in History and Philosophy of Modern Physics* 28 (1997), pp. 35–61; F. A. Muller. "The Equivalence Myth of Quantum Mechanics—Part II". in: *Studies in History and Philosophy of Science Part B: Studies in History and Philosophy of Modern Physics* 28 (1997), pp. 219–247.

[67]In the road towards the proof of the equivalence between the two quantum mechanics, I should also mention the names of other physicists whose works represented an advance in the completion of the proof, but which I will not comment here. These are:
- Wolfgang Pauli, in particular a letter from April 12th 1926 he wrote to Jordan in which he sketched the proof of the equivalence. An English translation of the full letter can be found in Van der Waerden, op. cit., pp. 278–282.
- Carl Eckart, in particular his article C. Eckart. "The Solution of the Problem of the Single Oscillator by a Combination of Schrödinger's Wave Mechanics and Lanczos' Field Theory". In: *Proceedings of the National Academy of Science* 12 (1926), pp. 473–476.
- Fritz London, in particular his article F. London. "Über die Jacobischen Transformationen der Quantenmechanik". In: *Zeitschrift für Physik* 37 (1926), pp. 383–386.

*the Quantum Dynamics*"[68]. As it was the case in Schrödinger's approach to the equivalence between his theory and Göttingen's, the idea that indeed both theories were but two different *descriptions* of the same situation was central to Dirac's article. However, in his own approach, the English physicist introduced a crucial conceptual shift that it is important to highlight from the outset. As we have just seen, Schrödinger's original conception may be fairly well understood with the following claim: 'Wave and Matrix Mechanics are different descriptions of the same situation because these descriptions are written in *different languages*'. On the contrary, as we are about to see, Dirac's conception is more accurately captured by saying: 'Wave and Matrix Mechanics are different descriptions of the same situation because these descriptions are performed from *different points of view*'. Therefore, from Schrödinger's article to Dirac's article, we pass from a 'linguistic' analogy to a 'relativistic' analogy.

Let me unpack, in more detail, what I mean by this last statement. The problem of the relation between the two quantum theories, as tackled by Schrödinger, appears to be quite similar to the problem of the relation between Lagrangian and Hamiltonian Classical Mechanics. We are confronted with two, *and only two*, formulations of apparently the same theory. Surely, in order to secure that they really correspond to the same theory, it is important to find a correspondence/dictionary connecting them. But the main interest of the theoretical physicist lies in the languages, not in the dictionary: most efforts will be spent in *working within one fixed formulation* and exploring all its implications. Now, the problem of the relation between Wave and Matrix Mechanics, as tackled by Dirac in 1927, should instead be compared with the problem of relating two descriptions of the same motion by different frames of reference. The two descriptions appear then to be just two possible choices *among a multitude of other possible ones*. And it is precisely the realization of the existence of this multitude which produces the conceptual shift introduced by Dirac: the two particular formulations of Schrödinger and Göttingen become less worthy of study in themselves; rather—this was perhaps the great methodological lesson of Einstein's relativity theory—the primary object of investigation becomes the *transformations* between the whole variety of points of view. Presumably, it was in this spirit that Dirac famously wrote in the preface of his book *The Principles of Quantum Mechanics*:

> The growth of the use of transformation theory, as applied first to relativity and later to the quantum theory, is the essence of the new method in theoretical physics. Further progress lies in the direction of making our equations invariant under wider and still wider transformations.[69]

---

[68]P. A. M. Dirac. "The Physical Interpretation of the Quantum Dynamics". In: *Proceedings of the Royal Society of London* 113 (1927), pp. 621–641. (Reprinted in: P. A. M. Dirac. *The Collected Works of P.A.M. Dirac: 1924–1948*. Ed. by R. Dalitz. Cambridge: Cambridge University Press, 1995, pp. 207–229) (reprinted in: idem, *The Collected Works of P.A.M. Dirac: 1924–1948*, pp. 207–229).

In this way, not only did Dirac manage to render more precise Schrödinger's equivalence between Wave and Matrix Mechanics, but he wildly extended it.

With this in mind, let us now turn to the precise content of Dirac's article, in which the transformation theory was presented for the first time[70]. As we have recalled, Dirac played a major role in the development of Göttingen's Matrix Mechanics from its very beginning. It is then not surprising that the starting point of his considerations was the general dynamical problem as it had been formulated by Heisenberg and his collaborators. In this setting, the key concept was that of a "*scheme of matrices*"—that is, a particular set $\mathcal{G}$ of infinite square matrices, each of which represented a dynamical variable. Dirac concisely summarized the problem as follows:

> The solving of a problem in Heisenberg's matrix mechanics consists in finding a scheme of matrices to represent the dynamical variables, satisfying the following conditions:
>
> (i) The quantum conditions, $q_r p_r - p_r q_r = i\hbar$, etc.
>
> (ii) The equations of motion, $gH - Hg = i\hbar \dot{g}$, or if $g$ involves the time explicitly
> $$gH - Hg + i\hbar \frac{\partial g}{\partial t} = i\hbar \dot{g}.$$

---

(iii) The matrix representing the Hamiltonian $H$ must be a diagonal matrix.

(iv) The matrices representing real variables must be Hermitian.[71]

However, the solution was not thereby uniquely determined: given a scheme of matrices $\mathcal{G}$ and any invertible matrix $b$, one could define a new matrix scheme through the transformation

$$\mathcal{G} \longrightarrow \mathcal{G}' := b\mathcal{G}b^{-1} = \left\{ bgb^{-1} \middle| g \in \mathcal{G} \right\}. \tag{I.4}$$

The result would also solve the Heisenberg problem if

(2) the coefficients of $b$ did not depend explicitly on time,

(3) $b$ commuted with $H$,

(4) $b^{-1}$ was equal to the conjugate transposed of $b$[72],

where the conditions (2)–(4) insured that $\mathcal{G}'$ met the requirements (ii)–(iv) respectively (requirement (i) being met for *any* choice of matrix $b$).

As we have seen, the remark of this fundamental indeterminacy in the solution of the general dynamical problem of the quantum theory had first lead Born and Jordan to claim that the theory was "essentially discontinuous" (cf. subsection I.1.2, page 34), but it had also allowed them to introduce the general theory of canonical transformations in the three-man paper. This work was of course known to Dirac from the beginning—he had already discussed it in his paper on the hydrogen atom. At the time, he had considered these transformations of "no great practical value"[73]. But this was before the emergence of Schrödinger's undulatory quantum theory and Bohr's statistical interpretation.

Now, Dirac returned with renewed energy to understanding the physical significance of these transformations of the scheme of matrices used to represent the dynamical variables. In his approach, the focus was set on the meaning of the indices which labelled the matrix coefficients. Associated to a given a matrix scheme, he said, one should find a set of $n$ real dynamical variables ($n$ being the number of degrees of freedom of the system under consideration) which were represented by *diagonal* matrices: the values of the diagonal elements could then be used as labels for the rows and columns. In this way, *a particular scheme of matrices appeared to be closely related to a preferred choice of dynamical quantities*, the indeterminacy in the scheme of matrices being just a manifestation of a certain freedom in the choice of the preferred dynamical quantities.

---

[71] Dirac, "The Physical Interpretation of the Quantum Dynamics", p. 627. I have here changed Dirac's notation slightly: in the original paper, Dirac uses the letter $h$ to refer to the *reduced* Planck constant (which is hence sometimes also called the Dirac constant). I have here instead followed the modern conventions, where $h$ denotes the Planck constant and the reduced Planck constant is referred to with the symbol $\hbar := \frac{h}{2\pi}$.

[72] In modern terms, this last condition means that $b$ was required to be a unitary matrix.

[73] Idem, "Quantum Mechanics and a Preliminary Investigation of The Hydrogen Atom", p. 565.

Let us explain this important idea more explicitly[74]. In the scheme of matrices $\mathcal{G}$, the physical quantity g is represented by the matrix $g$ with coefficients $g_{ab}$. Here, the labels $a$ and $b$ are $n$-tuples of real values: $a = (a_1, \ldots, a_n)$ where $a_r \in I_r$ and each $I_r$ is the set of possible values of a certain quantity $\mathtt{f}_r$. As Dirac remarks, each $I_r$ may be a discrete and/or continuous set of real values. Therefore, in the scheme of matrices $\mathcal{G}$, the quantity $\mathtt{f}_r$ is represented, *by definition of the matrix scheme* $\mathcal{G}$, through the diagonal matrix $f_r$ whose coefficients are

$$(f_r)_{ab} = a_r\, \delta(a - b). \tag{I.5}$$

In order for this latter and similar expressions to make sense even in the case of continuously-varying indices, Dirac famously introduced the "Delta function" $\delta(x)$ defined by the two properties[75]

$$\delta(x) = 0 \text{ for } x \neq 0 \quad \text{and} \quad \int_{\mathbb{R}} \delta(x) dx = 1.$$

Similarly, in the matrix scheme $\mathcal{G}'$, the same physical quantity g is represented by the matrix $g'$ with coefficients $g'_{\alpha\beta}$, where now $\alpha, \beta \in I'_1 \times \ldots \times I'_n$ and each $I'_r$ refers to the set of possible values of a *different* physical quantity $\mathtt{k}_r$ which is represented in the scheme $\mathcal{G}'$ through a diagonal matrix $k'_r$: $(k'_r)_{\alpha\beta} = \alpha_r\,\delta(\alpha - \beta)$.

---

[74]Despite the importance that Dirac attached to notation and his undisputed mastery in the invention of new useful ones, I will in the following somehow depart from the original notation of his paper, which I sometimes find rather confusing for the modern reader. I am thinking in particular of his choice of using unprimed symbols for the variables $(g, \xi, \ldots)$ and primed symbols for the values of these variables $(g', \xi', \ldots)$.

[75]Although this is sometimes called "Dirac's delta function" and its use certainly became widespread through the work of Dirac, he was *not* the inventor of it. The function had already been thus defined by Kirchhoff in 1882 and was also used at least by Heaviside (see O. Darrigol. *From c-Numbers to q-Numbers: The Classical Analogy in the History of Quantum Theory.* Berkeley: University of California Press, 1992, footnote 84, p. 339 and also Jammer, op. cit., p. 316).

In this way, the transformation (I.4) is understood as a transformation from a point of view which confers a preferred role to the quantities $\mathtt{f}_1, \ldots, \mathtt{f}_n$ to a point of view which confers a preferred role to the quantities $\mathtt{k}_1, \ldots, \mathtt{k}_n$. With all the labels written down explicitly, (I.4) becomes, for a given quantity $\mathtt{g}$:

$$g'_{\alpha\beta} = b_{\alpha a}\, g_{ab}\, b^{-1}_{b\beta} \tag{I.6}$$

where I have used Einstein's summation convention for repeated indices. In particular, we see that "the new parameters $[\alpha, \beta$ of the new matrix scheme $\mathcal{G}']$ are quite unconnected" with the parameters $a, b$ of the matrix scheme $\mathcal{G}$, as Dirac insisted[76]. There may even happen that there is "no one-one correspondence between the rows and columns of the new matrices and those of the original matrices"[77] (the spectra of the quantities $\mathtt{f}_1, \ldots, \mathtt{f}_n$ may happen to be all discrete whereas the spectra of the quantities $\mathtt{k}_1, \ldots, \mathtt{k}_n$ may happen to be all continuous).

Finally, Dirac noticed, since by definition the coefficients $b_{\beta a}$ and $b^{-1}_{a\beta}$ of the transformation matrices satisfy the equations

$$\begin{cases} b^{-1}_{a\beta} = \overline{b_{\beta a}} \\ b_{\alpha a}\, b^{-1}_{a\beta} = \delta(\alpha - \beta) \\ b^{-1}_{b\alpha}\, b_{\alpha a} = \delta(b - a), \end{cases}$$

one could consider them as a complete family of mutually orthogonal complex-valued functions of the parameters $\alpha$ (with $a$ fixed), or of the parameters $a$ (with $\alpha$ fixed). As we will see in a moment, this apparently trivial remark was quite important for Dirac's understanding of Wave Mechanics.

Which was then the relation between this theory of transformations, Heisenberg's Matrix Mechanics and Schrödinger's Wave Mechanics? The answer was in fact quite simple: *the two quantum theories were simply two matrix schemes which differed only*

[76]Dirac, "The Physical Interpretation of the Quantum Dynamics", p. 628.

[77]Ibid., p. 628.

*in their choice of the preferred quantities* $\mathtt{f}_1, \ldots, \mathtt{f}_n$. Indeed, Göttingen's quantum theory had always been set in a matrix scheme $\mathcal{G}'$ in which the Hamiltonian or energy function $H$ was a preferred quantity—it was required to be represented by a diagonal matrix (cf. condition (iii), page 43). The coefficients $\alpha, \beta$ labelled stationary states and contained among the numbers $\alpha_r$ the value of energy $E$. On the other hand, Schrödinger's Wave Mechanics was set in a matrix scheme $\mathcal{G}$ in which the configuration variables $\mathtt{q}_1, \ldots, \mathtt{q}_n$ were chosen as preferred quantities. The labels $a, b$ were simply the possible values $\boldsymbol{q^0} = (q_1^0, \ldots, q_n^0)$ of these quantities, which were represented by the matrices $(q_r)_{\boldsymbol{q^0 q^1}} = q_r^0 \, \delta(\boldsymbol{q^0} - \boldsymbol{q^1})$ (compare this with Equation I.5, page 44)[78]. Moreover, Schrödinger's wave-functions $\psi_E(q)$ were nothing but the complex coefficients $b_{\beta a}$ of the transformation $b$ which enabled to switch between these two particular matrix schemes.

Hence, from the heights of Dirac's general theory of "scheme of matrices" and transformations between them, one recovered both Heisenberg's and Schrödinger's quantum mechanics as two particular examples or points of view. This theory relied on two crucial concepts. The first was that of *a set of preferred quantities*, which was somehow the analogue of a reference frame, thereby furnishing a certain point of view from which to describe the quantum system. It allowed to define a matrix scheme by providing labels for the rows and columns. In modern terminology, this "set of preferred quantities" is called a "complete set of commuting observables" and corresponds also to the better formalized notion of a "maximal abelian von Neumann subalgebra"[79].

---

[78]For this last formula, Dirac's original notation is more transparent, so let me rewrite it with his notation. Consider the variables $\mathtt{q}_1, \ldots, \mathtt{q}_n$, represented by the matrices $q_1, \ldots, q_n$. Let $q' = (q_1', \ldots, q_n')$ be a certain $n$-tuple of possible values of the quantities $\mathtt{q}_1, \ldots, \mathtt{q}_n$. Then $(q_r)_{q'q''}$ denotes the numerical coefficient of the matrix $q_r$ which is on the row labelled by the $n$-tuple of values $q'$ and on the column labelled by $q''$. In this matrix scheme, we therefore have

$$(q_r)_{q'q''} = q_r' \delta(q' - q'').$$

This is exactly the statement that, in Schrödinger's Wave Mechanics representation, the position operator acts by multiplication. In more familiar notation involving operators, this would be written as $\widehat{q_0} = q_0 \, \delta(q - q_0)$.

[79]A "complete set of commuting observables" or CSCO is a set of commuting self-adjoint operators which admit a unique (up to phase factors) orthonormal basis of common eigenvectors. This is a standard definition which is found in any textbook on Quantum Mechanics (see for example C. Cohen-Tannoudji, B. Diu, and F. Laloë. *Mécanique quantique*. Paris: Hermann, Collection Enseignement des Sciences, 1973, p. 144). However, this definition does not quite work for infinite-dimensional Hilbert

The second notion was that of a *transformation between matrix schemes*, which was the central object of the theory and contained all the physical information: Schrödinger's fundamental wave equation appeared to be an equation on the coefficients $b_{\beta a}$ of these transformations[80]. The coefficients, in turn were interpreted as "transition amplitudes". Thus, the knowledge of $b_{\beta a}$ allowed to answer what Dirac considered to be "the only question to which the physicist requires an answer"[81]: If a system is assumed to have values $\beta_1, \ldots, \beta_n$ of the compatible quantities $\mathtt{k}_1, \ldots, \mathtt{k}_n$, what is the probability that a measurement of the compatible quantities $\mathtt{f}_1, \ldots, \mathtt{f}_n$ will yield the values $a_1, \ldots, a_n$?

Although it is very tempting, from the vantage modern perspective, to interpret all Dirac's machinery in terms of states (a transformation being a change in the basis of states), it is important to insist on the fact that this notion—of "state"—was completely absent in his 1927 article. Darrigol lucidly highlights this when discussing Dirac's Quantum Mechanics:

> There is one feature of Dirac's original transformation theory that is likely to surprise the modern quantum physicist: the notion of state vector is completely absent. [...] Perhaps modern-day interpreters of quantum mechanics should nevertheless remember that there exists a formulation of quantum mechanics

---

spaces, where self-adjoint operators may not admit any eigenvector. The best strategy is then to recast it in an algebraic form. For finite-dimensional Hilbert spaces, the existence of a CSCO is equivalent to the existence of a *maximal abelian von Neumann subalgebra* of operators, and this latter notion is still well-defined in the infinite-dimensional case (a 'maximal abelian subalgebra' $\mathcal{U}$ is an algebra such that $\mathcal{U} = \mathcal{U}'$, where $\mathcal{U}'$ denotes the commutant; a 'von Neumann subalgebra' $\mathcal{U}$ is an algebra such that $\mathcal{U} = \mathcal{U}''$). For the precise definitions and relations between these notions, see J.-M. Jauch. "Systems of Observables in Quantum Mechanics". In: *Helvetica Physica Acta* 33 (1960), pp. 711–726.

[80] Indeed, Schrödinger's equation arises from Equation I.6 (page 45) by taking $\mathtt{g}$ to be the Hamiltonian $H$. To see this, suppose you want to find a matrix scheme $\mathcal{G}'$ such that $\mathtt{g}$ is represented by a diagonal matrix. In other words, you want $g'_{\alpha\beta} = \alpha_r\, \delta(\alpha - \beta)$ for some $r \in \{1, \ldots, n\}$. In this case, Equation I.6 becomes $\alpha_r\, \delta(\alpha - \beta) = b_{\alpha a}\, g_{ab}\, b_{b\beta}^{-1}$, which is equivalently written as

$$g_{ab}\, b_{b\alpha}^{-1} = \alpha_r\, b_{a\alpha}^{-1}.$$

Now, with a simple change of notation: $g \to H$, $\alpha_r \to E$, $a \to q$, $b^{-1} \to \psi$, one gets

$$H\, \psi_E(q) = E\, \psi_E(q)$$

which is Schrödinger's time-independent equation. There is however one caveat to this "proof" which will be commented on the next section (cf. page 49).

[81] Dirac, op. cit., p. 623.

without state vectors, and with transition amplitudes (transformations) only.[82]

In the rapidly developing field of quantum mechanics, it was the arrival of the young Hungarian mathematician John von Neumann which brought again to the fore-front the notion of state and completed the understanding of the relation between Heisenberg's and Schrödinger's approaches.

## I.1.4   Extracting the physics (3): von Neumann's process of abstraction

In 1926, around the same time during which Dirac, in Copenhagen, and Jordan, in Göttingen, were finishing the development of the transformation theory, John von Neumann came to Göttingen to be David Hilbert's assistant at only twenty-two years of age. By then, he had already shown outstanding mathematical skills and Hilbert called him in order to contribute to his (Hilbert's) new foundations of Mathematics programme. However, as was often the case, Hilbert had decided to give a series of lectures on the mathematical methods of Physics, and the topic of the 1926/1927 winter semester was—as it could not have been otherwise—the quantum theory. In this way, by helping in the preparation of these lectures, von Neumann found himself involved with the most recent developments in the foundations of Quantum Mechanics[83]. In May 1927, he presented his first important work on the subject, a sixty-pages long

---

[82]Darrigol, op. cit., p. 344.

[83]For at least twenty five years, Hilbert lectured every so often at Göttingen on the mathematics of Physics and covered an impressively broad spectrum of topics: Mechanics, Special Relativity, Kirchhoff's laws of radiation, Boltzmann's kinetic theory of gases, General Relativity, etc. This is all the more impressive if one thinks of it as being just a peripheral activity compared to Hilbert's involvement with the whole of Mathematics. For an interesting analysis of the influence of Hilbert's lectures in the foundations of Physics, see U. Majer. "The Axiomatic Method and the Foundations of Science: Historical Roots of Mathematical Physics in Göttingen (1900-1930)". In: *John von Neumann and the Foundations of Quantum Physics*. Ed. by M. Rédei and M. Stöltzner. Dordrecht: Kluwer Academic Publishers, 2001, pp. 11–31.

For a more detailed historical account of von Neumann's arrival in Göttingen and the importance of the 1926/1927 winter lectures on the quantum theory, see J. Mehra and H. Rechenberg. *The Historical Development of Quantum Theory. Volume 6: The Completion of Quantum Mechanics. 1926–1941.* New York: Springer-Verlag, 2000, pp. 392–411.

article entitled *Mathematical Foundations of Quantum Mechanics*[84], which culminated five years later in his book with the same title, a masterpiece and invaluable reading still today for anyone interested in the foundations of Quantum Mechanics[85].

As he explains in the opening words of his book, von Neumann had mixed feelings towards the work accomplished by Dirac. On the one hand, he considered the transformation theory to be "presumably the definitive form" of the new quantum mechanics and Dirac's own book on the subject to be "a representation of quantum mechanics which is scarcely to be surpassed in brevity and elegance"[86]. On the other hand, he was dissatisfied with the mathematical treatment of the theory, which, he said, "in no way satisfies the requirements of mathematical rigor—not even if these are reduced in a natural and proper fashion to the extent common elsewhere in theoretical physics"[87]. Therefore, when von Neumann approached the subject of quantum mechanics, his main goal was to find a clean mathematical reformulation of Dirac's work. He wrote:

> It should rather be pointed out that the quantum mechanical "transformation theory" can be established in a manner which is just as clear and unified, but which is also without mathematical objections. It should be emphasized that the correct structure need not consist in a mathematical refinement and explanation of the Dirac method, but rather that it requires a procedure differing from the very beginning, namely, the reliance on the Hilbert theory of operators.[88] [89]

---

[84]J. von Neumann. "Mathematische Begründung der Quantenmechanik". In: *Nachr. Ges. Wiss. Göttingen* (1927), pp. 1–57.

[85]J. von Neumann. *Mathematische Grundlagen der Quantenmechanik*. Heidelberg: Springer-Verlag, 1932 (English translation: J. von Neumann. *Mathematical Foundations of Quantum Mechanics*. Trans. by R. T. Beyer. Princeton: Princeton University Press, 1955).

[86]Ibid., p. i and viii.

[87]Ibid., p. ix.

[88]Ibid., p. ix.

[89] Although the mathematical reformulation of the transformation theory is certainly von Neumann's original motivation, this fails to do justice to everything he accomplished with his book. For it is certainly much more than a simple rigorous "reformulation" of Dirac's work! In this regard, I prefer another characterization by von Neumann of the achievements of his book, found in a letter to the president of Dover publications (which would later publish the English translation):

> The subject-matter is partly physical-mathematical, partly, however, a very involved conceptual critique of the logical foundations of various disciplines (theory of probability, thermodynamics, classical mechanics, classical statistical mechanics, quantum mechanics). This philosophical-epistemological discussion has to be continuously tied in and In the eyes of von Neumann, Dirac's method was unacceptable for at least two reasons. First, it clearly relied on the ability of finding, for any given quantity g, a matrix scheme in which the chosen quantity would be represented by a diagonal matrix $g$. But this was equivalent to supposing any self-adjoint operator to be diagonalizable, which was a "mathematical fiction". Second, as we saw in the previous section, Dirac insisted in regarding Schrödinger's wave equation

$$H\psi_E(q) = E\,\psi_E(q) \qquad (q \in \Omega),$$

which is a differential equation, as the analogue of the linear transformation equation

$$H_{ab}(\psi_E)_b = E\,(\psi_E)_a \qquad (a, b = 1, 2, \ldots)$$

(cf. page 47 and specially footnote 80). In this way, because he wished to regard differential operators acting on Schrödinger's wave functions as infinite matrices labelled with a continuously varying index, Dirac's unification of Matrix and Wave Mechanics was somewhat grounded in an analogy between the continuous configuration space of the physical system $\Omega$ and the discrete space $Z = \mathbb{N}^*$. And this intended analogy was the deep source of all the "violence to the formalism and to mathematics"[90] inflicted by the transformation theory.

Dirac had arrived at this analogy as a consequence of his focus on the labels and their meaning. The crucial move of von Neumann, which allowed him to get round all the mathematical difficulties of Dirac's theory, was a shift of attention. Indeed, he noticed that all the relevant information of both Wave and Matrix Mechanics was not contained in the spaces $\Omega$ and $Z$ but rather in the *functions* over these spaces. At this level, the Fischer-Riesz theorem claimed $F_\Omega := L^2(\Omega)$ and $F_Z := l^2(Z)$ were

---

quite critically synchronised with the parallel mathematical-physical discussion. It is, by the way, one of the essential justifications of the book, which gives it a content not covered in other treatises, written by physicists or by mathematicians, on quantum mechanics.
(Letter to H. Cirker, October 3, 1949. In: J. von Neumann. *John von Neumann: Selected Letters.* Ed. by M. Rédei. History of Mathematics. American Mathematical Society, 2005, p. 92)

[90]Idem, *Mathematical Foundations of Quantum Mechanics*, p. 28.

isomorphic[91]:

> $Z$ and $\Omega$ are very different, and to set up a direct relation between them must
> lead to great mathematical difficulties. On the other hand, $F_Z$ and $F_\Omega$ are
> isomorphic [...]—and since they (and not $Z$ and $\Omega$ themselves!) are the real
> analytical substrata of the matrix and wave theories, this isomorphism means
> that the two theories must always yield the same numerical results.[92]

Had von Neumann only wished to prove the *equivalence* of Matrix and Wave Mechanics, as Schrödinger wanted, this could have very well been his final remark on the subject. Once it had become clear that only the elements of $F_Z$ or $F_\Omega$ were involved in any calculation of Matrix or Wave Mechanics, the explicit construction of the isomorphism given by Fischer and Riesz was exactly the bi-directional correspondence or dictionary that Schrödinger had been after. With this result, the equivalence between the two approaches to the quantum theory was firmly established. However, for von Neumann, the Fischer-Riesz theorem was not the end point but rather the *starting point* of his investigations. For indeed the Hungarian mathematician shared with Dirac the idea of *unifying* Heisenberg's and Schrödinger's mechanics—that is, he wished to find an overhanging perspective from which both theories would appear as particular cases of a more general framework. But, as he had already warned, von Neumann's method required "a procedure differing from the very beginning" from Dirac's.

Let us see how he described it:

> The following may be said regarding the method employed in this mode of treatment: as a rule, calculations should be performed with the operators themselves (which represent physical quantities) and not with the matrices, which after the introduction of a (special and arbitrary) coordinate system in Hilbert space, result from them. This "coordinate free", i.e. invariant method, with its geometric language, possesses noticeable formal advantages.[93]

---

[91] Here, $L^2(\Omega)$ is the space of complex-valued square integrable functions over $\Omega$, and $l^2(Z)$ is the space of square summable sequences (i.e., sequences $(x_1, x_2, \ldots)$ such that $\Sigma |x_i|^2$ is convergent). The theorem was proven independently by Frigyes Riesz and Ernst Sigismund Fischer in 1907.

[92] Ibid., p. 31.

[93] Ibid., p. viii.

In some regards, this may sound conceptually very similar to Dirac's 1927 transformation theory. Recall the relativistic analogy used to grasp Dirac's approach: the difference between Matrix and Wave Mechanics could be conceived as similar to the difference between two descriptions of the same motion from two different frames of reference. Thus, for both Dirac and von Neumann, Göttingen's and Schrödinger's theories arose from particular and arbitrary choices among many other possible ones. And the choice being made was that of a frame of reference or coordinate system.

Yet, Dirac's 1927 article and von Neumann's 1932 book differ radically in how to deal with this multiplicity of particular and arbitrary choices. Indeed, von Neumann's whole methodology consists in *avoiding such choices*: to elevate oneself and work in a "coordinate free" fashion with the "operators *themselves*"—idea which is absent in Dirac's original conception[94]. This move towards a "coordinate free", "invariant" or "intrinsic" formulation of Mechanics I call the *process of abstraction*. It characterizes the core of von Neumann's approach and led him to introduce the fundamental notion of an "***abstract*** Hilbert space" in the following all-important passage which ended the introductory chapter of his book:

> Since the systems $F_Z$ and $F_\Omega$ are isomorphic, and since the theories of quantum mechanics constructed on them are mathematically equivalent, it is to be expected that a unified theory, independent of the accidents of the formal framework selected at the time, and exhibiting only the really essential elements of

---

[94]In fact, if the relativistic analogy is taken seriously, one could even argue that, for the Dirac of 1927, it was impossible *not* to make a choice: it is indeed far from clear what it would mean to free oneself from all frames of reference and describe the 'motion itself'...

This reading seems further supported by a passage Dirac wrote in the preface of the first edition of his 1930 book:

> The growth of the use of transformation theory, as applied first to relativity and later to the quantum theory, is the essence of the new method in theoretical physics. [...] This state of affairs is very satisfactory from a philosophical point of view, as implying an increasing recognition of the part played by the observer in himself introducing the regularities that appear in his observations [...].
>
> (Dirac, *The Principles of Quantum Mechanics*, p. v)

Under this light, von Neumann's "coordinate free" method could appear as a step backwards in the "recognition of the part played by the observer" since it avoids describing the physical system from the point of view any particular frame of reference whatsoever.

Of course, Dirac's attitude towards the multiplicity of choices is much more complex, and this claim only applies to what stems from his article *"The Physical Interpretation of the Quantum Dynamics"*. As a matter of fact, Di quantum mechanics, will then be achieved if we do this: Investigate the intrinsic properties (common to $F_Z$ and $F_\Omega$) of these systems of functions, and choose these properties as a starting point.

The system $F_Z$ is generally known as "Hilbert space". Therefore, our first problem is to investigate the fundamental properties of Hilbert space, independent of the special form of $F_Z$ or $F_\Omega$. The mathematical structure which is described by these properties (which in any special case are equivalently represented by calculations within $F_Z$ or $F_\Omega$, but for general purposes are easier to handle directly than by such calculations), is called "abstract Hilbert space".

We wish then to describe the abstract Hilbert space, and then to prove rigorously the following points:

1. That the abstract Hilbert space is characterized uniquely by the properties specified, i.e., that it admits of no essentially different realizations.

2. That its properties belong to $F_Z$ as well as $F_\Omega$. [...] When this is accomplished, we shall employ the mathematical equipment thus obtained to shape the structure of quantum mechanics.[95]

As announced, in Chapter II: "Abstract Hilbert Space", von Neumann went on to extract the intrinsic properties to be used as a starting point in the definition of this new mathematical structure $\mathcal{R}$. He proposed the five well-known axioms:

**A.** $\mathcal{R}$ is a complex linear space.

**B.** An Hermitian inner product is defined in $\mathcal{R}$.

**C.**$^{(\infty)}$ There are arbitrarily many linear independent vectors.

**D.** $\mathcal{R}$ is complete.

**E.** $\mathcal{R}$ is separable.[96]

---

[95] Von Neumann, op. cit., pp. 32–33.

[96] These are the axioms for an infinite-dimensional Hilbert space. For a $n$-dimensional Hilbert space, axiom **C.**$^{(\infty)}$ is replaced by:

**C.**$^{(n)}$ There are exactly $n$ linear independent vectors.

Axioms **D.** and **E.** then follow from the first three. See, respectively, ibid., pp. 36, 38, 45, 46.

* * * * *

In our survey of the earlier developments in the foundations of Quantum Mechanics, we have finally arrived at an important methodological precept, which can be summarized as follows:

> **Abstract methodological precept.** In order to find the physical information contained in the mathematics of Mechanics and avoid the "accidents of the formalism chosen at the time", *work directly with the relevant abstract mathematical structures "themselves"* and not with the systems that represent it.

I have voluntarily phrased this advice in more general terms, so that it is not thought to be confined solely to Quantum Mechanics and abstract Hilbert spaces, nor is it associated only to von Neumann. Indeed, although I have motivated it in this particular context, the abstract methodological precept should be considered as much more general, and owes as much to Dirac and Weyl as it does to von Neumann[97].

With regard to our initial inquiry—namely, to clarify which mathematical objects can be considered as acceptable candidates for characterizing physical systems—the main lesson to take from these years of development of Quantum Mechanics is the following: *the mathematical objects involved in the descriptions of physical systems need to be conceived <u>abstractly</u>.* If a quantum-mechanical system is to be described by a Hilbert space at all, then it ought to be described by an *abstract* Hilbert space. In the same way, one needs *abstract* symplectic manifolds in Classical Mechanics.

---

[97]Taking aside his 1927 *"The Physical Interpretation of the Quantum Dynamics"*, Dirac indeed emphasised in many other places the advantages of working at a purely symbolic level. Here is for example a quote from his book in 1930:

> One does not anywhere specify the exact nature of the symbols employed, nor is such a specification at all necessary. They are used all the time in an abstract way, the algebraic axioms that they satisfy and the connections between equations involving them and physical conditions being all that is required.

(Dirac, op. cit., p. 18)

The awareness of this feature is perhaps not as clear in the Classical setting as it is in the Quantum. As we have just seen, the particular way in which Quantum Mechanics was developed forced it, from the outset, into higher levels of abstraction. But in Classical Mechanics, one still tended to adopt, in the practice, a naive realistic reading of the mathematical formalism and to get attached to the 'materiality' of certain mathematical constructions—just recall Schrödinger's appeal to the "visualizability" of Classical Mechanics. For instance, when dealing with the Hamiltonian description of a given classical system, one generally had the impression of working with a particular, 'concrete' symplectic manifold and sticking to it. Therein, there seemed to be no analogue of a 'change of representation' from a certain 'Heisenberg representation' to a 'Schrödinger representation'. Nonetheless, when one considers the developments in the mathematical foundations of Classical Mechanics of the last sixty years, it becomes clear this is an impression we must abandon: to describe the space of states of a free massive non-relativistic particle, one can use the cotangent bundle $T^*\mathbb{R}^3$ with its natural symplectic structure or decide to work in the space $\mathbb{R}^3 \times \mathbb{R}^3$ and impose $\omega = dq^i \wedge dp_i$ as an *ad hoc* definition. Of course, the two descriptions are 'the same', but only insofar as the two symplectic spaces are isomorphic[98]. The sense of working "up to isomorphism" is hence present in both Classical and Quantum Mechanics.

However, von Neumann's call to work at the level of the abstract structures *themselves*, and not at the level of their particular representations or coordinatizations, surely requires some important elucidations. What exactly does the adjective 'abstract' stand for when we apply it to a mathematical object? What exactly is an *abstract* Hilbert space $\mathcal{R}$, and in which way does it differ from the Hilbert spaces $F_\Omega$ and $F_Z$? In this regard, the above quoted passages from his book mention many different ideas which need to be clarified and which will be discussed at length in the

---

[98] A less trivial—but mathematically more sophisticated—example can be found in the classical analogue of the theory of systems of imprimitivity developed by George Mackey. In this theory, the classical phase space associated to a free massive (spin-zero) non-relativistic particle is the dual of the action Lie algebroid $\mathbb{R}^3 \ltimes \mathbb{R}^3$, and one knows this theory to be equivalent to the usual point of view because this space can be proven to be isomorphic to the cotangent bundle $T^*\mathbb{R}^3$. For a more detailed exposition of this description, see N. P. Landsman. "Between Classical and Quantum". In: *Philosophy of Physics (Handbook of the Philosophy of Science) 2 volume set.* Ed. by J. Butterfield and J. Earman. Vol. 1. Amsterdam: North-Holland Publishing Co., 2007, pp. 417–554. URL: http://arxiv.org/abs/quant-ph/0506082, pp. 461–462.

following sections. Let us list them:

1. *Hierarchy of levels.* There seems to be (at least) two different levels: the one of abstract mathematical structures and the one of particular "systems", which are "realizations" of the abstract ones.

2. *One-to-many relation.* Given an abstract structure, there are many isomorphic systems which realize it. Conversely, given many isomorphic systems, there is a unique abstract structure of which they are realizations.

3. *Uniqueness of kind.* Both the abstract and the systems are of the same kind (e.g., both the abstract $\mathcal{R}$ and the system $F_\Omega$ are Hilbert spaces).

4. *Commonality.* The "intrinsic" properties of the "abstract" are "common" to the different systems.

5. *Schematic nature.* The properties of the abstract Hilbert space are "independent of the accidents" of the systems that realize it.

Moreover, the concept of "isomorphism" plays a fundamental role: it is the starting point of von Neumann's process of abstraction.

Undoubtedly, these features are not particular to the discussion of abstract Hilbert spaces. The same ideas may be found in, e.g., Hermann Weyl's writings on abstract groups—to choose another example highly relevant for the mathematics of Mechanics. He says:

> A group $\Gamma$ is a set of correspondences containing the identity $E$, the inverse $S^{-1}$ of any $S$ in $\Gamma$ and the composite $TS$ of any two correspondences $S$ and $T$ in $\Gamma$. Considered as an *abstract group* **scheme** $\gamma$, our set $\Gamma$ consists of elements $s$ (**of irrelevant nature**) for which a composition $st$ is defined satisfying [the usual group axioms]. The given transformation group $\Gamma$ is a *faithful realization* of the abstract group scheme $\gamma$.[99]

And also:

> In the study of groups of transformations one does well to stress the mere structure of such a group. This is accomplished by attaching arbitrary labels to its elements and then expressing in terms of these labels for any two group elements $s$, $t$ what the result $u = st$ of their composition is. If the group is finite one could tabulate the composition of elements. **The group scheme or abstract group thus obtained is itself a structural entity**, its structure represented by the law or table of composition for its elements, $st = u$.[100]

---

[99]H. Weyl. *The Classical Groups - Their Invariants and Representations*. 2nd ed. Princeton: Princeton University Press, 1946. (First edition: 1939), p. 14. Italics are Weyl's, bold typeface is mine.

Or, to choose a couple of more modern mathematicians, we can read Makkai claiming that

[...] two groups that are isomorphic share all structural properties; they are structurally indistinguishable.[101]

Mac Lane explaining that

All infinite cyclic groups are isomorphic, but this infinite group appears over and over again — in number theory, in ornaments, in crystallography, and in physics. Thus, the 'existence' of this group is really a many-splendored matter.[102]

and Lawvere emphasizing that

In the mathematical development of recent decades one sees clearly the rise of the conviction that the relevant properties of mathematical objects are those which can be stated in terms of their abstract structure rather than in terms of the elements which the objects were thought to be made of.[103]

To sum up, it has emerged that physical systems are described by using some sort of 'abstract mathematical structures'. Yet, we are missing a precise account of what abstract mathematical structures are. The task of such an account is to provide an explanation of the above features 1 through 5. Of course, it is a priori possible to have more than one account. Our goal is simply to find one proposal that does the job and that might fruitfully be used in the remaining parts of our analysis of Mechanics. Methodologically, it is important to distinguish the work done by the notions of "abstraction" and "structure": which of the features can be explained by the fact that an abstract mathematical structure is an *abstract* entity? Which necessitate the fact that this entity is, in particular, a *structure*? Therefore, the strategy to follow to address the conceptual clarification of von Neumann's methodological precept is the following:

   i) Confer a meaning to "abstraction" and "mathematical structuralism".

  ii) Understand the key role played by isomorphisms in relation with these two concepts.

 iii) Explain, in this conceptual scheme, the remaining features (1 through 5).

---

[100]H. Weyl. *Symmetry*. Princeton: Princeton University Press, 1952 (reprinted in 1989), p. 145. Bold is mine.

[101]M. Makkai. "Towards a Categorical Foundation of Mathematics". In: *Logic Colloquium '95 (Haifa)*. Vol. 11. Lecture Notes Logic. Berlin: Springer, 1998, pp. 153–190, p. 161.

[102]S. Mac Lane. "Structure in Mathematics". In: *Philosophia Mathematica* 4.2 (1996), pp. 174–183, p. 182 (Cited in: A. Rodin. "Categories Without Structures". In: *Philosophia Mathematica* 19 (2011), pp. 20–46, p. 22).

[103]F. W. Lawvere. "The Category of Categories as a Foundation for Mathematics". In: *Proceedings of the Conference on Categorical Algebra*. Springer. Berlin, 1966, pp. 1–20, p. 1 (cited in R. Krömer. *Tool and Object: A History and Philosophy of Category Theory*. Vol. 32. Historical Studies. Berlin: Springer Science & Business Media, 2007, p. 211).

This is certainly a large problem which goes well beyond the physical interpretation of the mathematics of Mechanics. As so, we will not restrain ourselves from treating it in as general terms as possible. But in doing so, we should not forget that, ultimately, we are interested in the mathematical objects involved in the foundations of Mechanics, and in the conceptual problems that arise in the *practice* of this field. Hence, although we will try to deal with abstraction and structuralism in full generality, we will nonetheless systematically ignore problems that are seemingly irrelevant for the foundations of *Mechanics* (cf. Lawvere's quote, section ). We will start by first considering abstraction, to understand which of the features 1. through 5. can or cannot be accounted for (section I.2). We will then move to the study of mathematical structuralism (section I.3).

## I.2   Abstract mathematical entities

There is in the analytical tradition of philosophy a vivid debate on the status of abstract objects. It usually discusses three clearly distinguished questions: one *ontological*—what kind of *objects* are abstract objects?—, one *epistemological*—how do we gain knowledge on these objects?—and one *terminological*—how should be precisely drawn the distinction between abstract and concrete objects? The debate, in its contemporary analytical setting, is often seen as originating from the 1947 article *"Steps Toward a Constructive Nominalism"* by Nelson Goodman and Willard V.O. Quine[104], and the main reference on the subject is perhaps the work of Michael Dummett[105]. Most famous are also Lewis' four ways of attempting to explain the abstract/concrete distinction: the Way of Example, which simply lists paradigmatic examples of concrete objects, such as donkeys, and of abstract objects, such as numbers; the Way of Conflation, which relates the abstract/concrete distinction to other metaphysical distinctions, such as the particular/universal distinction; the Negative Way, which attempts to characterize abstract objects by their lacking of some property, such as their lack of causal interaction or of spatiotemporal location; and finally the Way of Abstraction, which rests on having already an account of the process of abstraction[106].

However, as it should be clear, our quest for an explanation of the adjective 'abstract', as it appears in expressions such as 'abstract Hilbert spaces', 'abstract groups' or 'abstract $C^*$-algebras', has little to do with the general debate on abstract objects that occupies many analytical philosophers. For indeed the adjective 'abstract' applied to such mathematical structures is used here to introduce a distinction between entities which are all abstract on any account of the abstract/concrete distinction. The Hilbert space $F_Z := l^2(\mathbb{N}^*)$, although it is not an abstract Hilbert space in the sense of von Neumann, would certainly not belong to the category of concrete objects in the sense of Lewis or Dummett. Therefore, if we are to address the problem of the nature of abstract mathematical entities, it is better to understand 'abstract' not as opposed to 'concrete', but rather to 'particular'.

---

[104]N. Goodman and W. V. O. Quine. "Steps Toward a Constructive Nominalism". In: *Journal of Symbolic Logic* 12 (1947), pp. 105–122.

[105]In particular, M. Dummett. *Frege: Philosophy of Mathematics*. London: Duckworth, 1991. For a nice introduction to Dummett's ideas, see also G. Duke. *Dummett on Abstract Objects*. History of Analytical Philosophy. Hampshire: Palgrave MacMillan, 2012

[106]See D. Lewis. *On the Plurality of Worlds*. New York: Basil Blackwell, 1986, pp. 81–86.

## I.2.1   Ontology: the abstract/particular as hierarchy of objects

In order to give an account of the abstract/particular distinction, one first possible strategy is to try to interpret the hierarchy of levels—abstract mathematical structures on one side, particular systems on the other—as a hierarchy between two clearly distinguished kinds of mathematical objects. Von Neumann's wish of "describing *the* abstract Hilbert space" and its "intrinsic properties" certainly seems to point in the direction of reifying the abstract entities. The goal is then to find a precise definition of these abstract mathematical objects, such as $\mathcal{H}$[107].

To this end, a natural move is to attempt to use the so-called *abstraction principles* of the neo-Fregean logicism programme[108]. In their general form, these principles are written as

$$\forall S, S' \in V_1 \left( \triangle(S) = \triangle(S') \longleftrightarrow S \sim S' \right) \tag{I.7}$$

where $S, S'$ are elements of a certain universe $V_1$, $\sim$ is an equivalence relation defined on $V_1$ and $\triangle$ is an operator applicable to elements of $V_1$. It is precisely this operator that introduces the hierarchy of levels: $S, S'$ are the particular objects and $\triangle(S)$ is intended to be a new abstract object found as a result of the abstraction principle (I.7). Some paradigmatic examples of this abstraction principle which are often cited are:

– the *direction* of a line, where $V_1$ is the set of all lines contained in, say, the Euclidean two-dimensional plane, $\triangle(-)$ is 'the direction of' and $\sim$ is the equivalence

---

[107] Now that we have widened the discussion and the focus is no longer only on von Neumann's work, I will change the notation referring to abstract Hilbert spaces: instead of von Neumann's $\mathcal{R}$, I will use the nowadays more standard $\mathcal{H}$.

[108] This programme—sometimes dubbed Neo-Fregeanism, Neo-Logicism or Abstractionism—is usually taken to initiate with Crispin Wright's *Frege's Conception of Numbers as Objects* in 1983, which revived some of Frege's central ideas in his attempt to provide a foundation for arithmetic. Since then, the philosophical literature on the subject has abysmally grown. Attempting a survey of this would certainly take us too far afield. For an introduction, I refer the reader to F. MacBride. "Speaking with Shadows: A Study of Neo-logicism". In: *British Journal for the Philosophy of Science* 54 (2003), pp. 103–163 or to B. Hale and C. Wright. "Logicism in The Twenty-First Century". In: *The Oxford Handbook of Philosophy of Mathematics and Logic*. Ed. by S. Shapiro. New York: Oxford University Press, 2005, pp. 166–202. A more in-depth approach is provided by the collection of articles R. T. Cook, ed. *The Arché Papers on the Mathematics of Abstraction*. Dordrecht: Springer, 2007.

relation of being parallel (this is Frege's original example);

– the *shape* of a figure, where $V_1$ is the set of all possible figures contained in, say, the Euclidean two-dimensional plane, $\triangle(-)$ is 'the shape of' and $\sim$ is similarity between figures (this is Weyl's example)[109];

– the *cardinal number* of a finite set, where $V_1$ is a fixed universe of finite sets, $\triangle(-)$ is 'the cardinal number of' and $\sim$ is the existence of a bijection between two sets.

Already at this early stage of the discussion, some basic features of the process of abstraction, as understood by the neo-Fregean logicists, are apparent. First, the particular systems $S, S'$ are at least epistemologically prior to the abstract object $\triangle(S)$: one starts by having a fixed domain of particulars on which there is defined an equivalence relation, and *only then* one has access to the abstract. This appears to be in agreement with—or at least does not blatantly contradict—von Neumann's road towards abstract Hilbert spaces which starts by studying the properties of two particular systems ($F_\Omega$ and $F_Z$). Second, according to this account of abstraction, the fundamental ingredient which underlies the process of abstraction is that of an *equivalence relation*. Without the notion of 'parallelism of lines', one cannot even conceive the concept of 'direction'. Hence, under this view, the definition of the equivalence relation is also prior to the definition of the abstract object. Third, in the process of abstraction, as conceived by the neo-Fregeans, the essential trait of isomorphisms is that they induce the equivalence relation "being isomorphic". Following Andrei Rodin, we will call this conception "isomorphism–*qua*–equivalence"[110].

---

[109] H. Weyl. *Philosophy of Mathematics and Natural Science*. Trans. by O. Helmer. Princeton: Princeton University Press, 1949, p. 9. Recall that two figures $F$ and $F'$ in the Euclidean plane $E_2$ are similar if there exists an automorphism of $E_2$ which transforms $F$ into $F'$.

[110] In his article *"Categories Without Structures"*, Rodin distinguishes three different ways of thinking about isomorphisms: isomorphism–*qua*–equivalence, in which one only retains that 'being isomorphic' is en equivalence relation; isomorphism–*qua*–correspondence, where an isomorphism between $E_1$ and $E_2$ is thought as a one-to-one (and onto) correspondence between the *elements* of the two objects; and isomorphism–*qua*–transformation, where one stresses the *direction* of the isomorphism (say, from $E_1$ to $E_2$) and where the idea that $E_1$ and $E_2$ are composed of elements is not essential. Thus, "the same isomorphism–*qua*–correspondence gives rise to two isomorphisms–*qua*–transformations" (Rodin, op. cit., p. 25). In the next section, we will add to these three conceptions a fourth one: isomorphism–*qua*–possible-identification.

But, as Angelelli rightly remarks in his discussion of Frege's account of abstraction, the sole principle (I.7) does not suffice, for it does not give an explicit and unambiguous definition of what $\triangle(S)$ is[111]. One simple possibility is to read it as saying that, in fact, $\triangle(S)$ is defined as the equivalence class of the system $S$. Symbolically:

$$\triangle(S) := [S]_{eq}. \tag{I.8}$$

For the above paradigmatic examples, this reading works perfectly fine. There is indeed no fundamental problem in defining the finite cardinal numbers to be equivalence classes of finite sets[112]. Furthermore, this view is explicitly endorsed by some important mathematicians. Let us see, for example, how Irving Segal explains the difference between concrete and abstract $C^*$-algebras:

> [...] we define as a concrete $C^*$-algebra $\mathcal{A}$, an algebra of bounded linear operators on a real or complex Hilbert space [...]. Now two concrete $C^*$-algebras may be algebraically isomorphic (in one-to-one correspondence in a fashion making sums, products, and adjoints correspond) without there being any simple connection whatsoever between the Hilbert spaces on which the respective operators act. The relevant object here is an *abstract $C^*$-algebra*, **which may be defined as an equivalence class of $C^*$-algebras under algebraic isomorphism**. The set of all self-adjoint elements of an abstract $C^*$-algebra forms then a physical

---

[111] I. Angelelli. "Frege and Abstraction". In: *Philosophia Naturalis* 21 (1984), pp. 453–471, pp. 463–464.

[112] See for example S. Mac Lane. *Mathematics, Form and Function*. New York: Springer, 1986, p. 59. Nevertheless, with that definition, a finite cardinal number would not be a set but rather a *proper class*. This implies that one cannot talk about the *set* of all finite cardinal numbers, since in axiomatic set theory it is not allowed to form a set whose elements are proper classes. Bertrand Russell did not consider this point to be a serious objection:

> Thus a cardinal number is the class of all those classes that are similar to a given class. This definition leaves unchanged the truth-values of all propositions in which cardinal numbers occur, and avoids the inference to a set of entities called 'cardinal numbers', which were never needed except for the purpose of making arithmetic intelligible, and are no longer needed for that purpose.
>
> (B. Russell. "Logical Atomism". In: *Contemporary British Philosophers*. Ed. by J. Muirhead. London: Allen and Unwin, 1924, pp. 356–383. (Reprinted in: B. Russell. *Logic and Knowledge*. Ed. by R. Marsh. London: Allen and Unwin, 1956, pp. 323–343), p. 327.)

However, precisely in order to avoid this 'objection' and be able to consider the set of all finite cardinal numbers, von Neumann proposed another definition of them (as initial ordinals).

system.[113]

This simple view has a certain evident appeal. It certainly establishes a two-level hierarchy of well-defined objects and it illuminates the one-to-many relation: there is one unique abstract object of which $S$ is a realization because there is one unique equivalence class to which $S$ belongs, and different "realizations" of the abstract object are just different 'representatives'—that is, different members—of the equivalence class. It therefore explains features 1 and 2 (page 56) and also seems to deal with feature 5—the particular nature of the members of the class is completely irrelevant to the properties of the equivalence class.

But, despite all this, and despite Segal's explicit definition, there is an *obvious* problem with definition (I.8). This is best seen by taking an example: according to (I.8), we should have

$$\mathcal{H} := [L^2(\mathbb{R})]_{eq}.$$

Now, since, *by definition*, $L^2(\mathbb{R})$ and $l^2(\mathbb{N}^*)$ are among the elements of the equivalence class, and since the combination $L^2(\mathbb{R}) - l^2(\mathbb{N}^*)$ is meaningless, we see there is no sense in writing arbitrary linear combinations of elements of $\mathcal{H}$. Thus, this newly defined abstract 'Hilbert space' clearly fails to be a complex Hilbert space and any talk about linear operators defined on $\mathcal{H}$ reveals to be complete nonsense![114]

In fact, in the attempt to explain notions such as 'abstract groups', 'abstract Hilbert spaces' or 'abstract $C^*$-algebras', definition (I.8) is such a blunder that one cannot seriously consider that such fine a mathematician as Segal literally endorsed it. On the contrary, one is pressed to find a more sophisticated account that avoids the

---

[113]I. E. Segal. "Mathematical Problems of Relativistic Physics". In: *Proceedings of the Summer Conference, Boulder, Colorado*. Ed. by M. Kac. American Mathematical Society, 1960, pp. 8–9, the italics are Segal's, the bold type emphasis is mine.

[114]This critique is pointed out by John Burgess in his review of Stewart Shapiro's book. He writes:

> Sometimes they [abstract structures] are confused with isomorphism types, but this is a mistake: An isomorphism type is no more a special kind of system than a direction is a special kind of line.

(J. P. Burgess. "Review of *Philosophy of Mathematics: Structure and Ontology* by Stewart Shapiro". In: *Notre Dame Journal of Formal Logic* 40.2 (1999), pp. 283–291, p. 287) definition of the abstract object as the isomorphism class of its realizations and yet remains somewhat faithful to this idea. Here, it is important to remark a main difference between the three examples given above—direction, shape and cardinal number—and the type of abstract objects we are after—groups, Hilbert spaces, $C^*$-algebras. It is the following: whereas we think of an abstract group as composed of various abstract elements, we do not think in the same way of one given cardinal number. To put it succinctly, the difference is exactly that between the notion of an 'abstract set with three elements' and the notion of the 'cardinal number 3'. The abstraction principle (I.7) and the definition (I.8) allow to grasp the latter but not the former.

This last remark immediately suggests a new attempt to define the abstract objects by means of an abstraction principle. The main idea is simply to somewhat reverse the procedure of definition (I.8): instead of using an abstraction principle to directly define the abstract object as a whole—strategy which fails, as we just saw, since it does not give a proper account of the abstract elements: the elements of $\mathcal{H} := [L^2(\mathbb{R})]_{eq}$ were not the abstract vectors we were expecting—, use an abstraction principle to *first define each abstract element separately* and then define the abstract object as the set of all the abstract elements. A flavor of this is clearly found in the work of the 19th century German mathematician Heinrich Weber, one of the first mathematicians to work on abstract groups. He says:

> We can ... combine all isomorphic groups into a single class of groups that is itself a group *whose elements are the generic concepts obtained by making one general concept out of the corresponding elements of the individual isomorphic groups.* The individual isomorphic groups are then to be regarded as different representatives of the generic concept, and it makes no difference which representative is used to study the properties of the group.[115]

---

[115]H. Weber. "Die allgemeinen Grundlagen der Galois'schen Gleichungstheorie". In: *Mathematische Annalen* 43 (1893), pp. 521–549, p. 524 (English translation cited in H. Wussing. *The Genesis of the Abstract Group Concept.* Trans. by A. Shenitze. Cambridge, MA: Dover Publications, Inc., 1984, p. 248, my emphasis).

Now, a precise elaboration of this idea is found in the paper *"Two Types of Abstraction for Structuralism"* written by Linnebo and Pettigrew[116]. The proposal runs as follows: for any two systems $S, S'$ and any elements $x \in S$ and $x' \in S'$, define the abstraction principle

$$\triangle(x) = \triangle(x') \longleftrightarrow \exists f\big(f : S \xrightarrow{\sim} S' \text{ and } f(x) = x'\big). \tag{I.9}$$

Here, $\triangle(x)$ denotes the abstract element of which $x$ is a representative. In other words, the abstract element is the equivalence class of elements of particular systems which are connected by an isomorphism:

$$\triangle(x) := [x]_{eq}. \tag{I.10}$$

With this definition in hand, the abstract object $\triangle(S)$ is defined by

$$\triangle(S) := \{\triangle(x) \,|\, x \in S\}. \tag{I.11}$$

As a side remark, one should notice that now the fact that an isomorphism is a bijective function is essential. We have thus moved from the conception of isomorphism–*qua*–equivalence of the previous definition to the conception of isomorphism–*qua*–correspondence (cf. footnote 110, page 61).

Under this second view, when Segal writes that 'an abstract $C^*$-algebra is an equivalence class of $C^*$-algebras under algebraic isomorphism', he rather meant: 'an *element* of an abstract $C^*$-algebras is an equivalence class of *elements* of $C^*$-algebras under algebraic isomorphism'. The particular example the two authors have in mind is the abstract field of real numbers $\mathbb{R}$—in contradistinction with the various particular realizations of this field. The system $S$ would be Dedekind's model of the real numbers in terms of cuts, and $S'$ would be Cantor's model in terms of equivalence classes of Cauchy sequences. In this case, definitions (I.10) and (I.11) lead to defining the abstract

real number field by

$$\mathbb{R} := \{[x]_{eq} \mid x \in S\}. \tag{I.12}$$

This a perfectly satisfactory definition which succeeds in explaining all the features of the abstract/particular relation (page 56). In particular, it is indeed the case that the abstract object $\mathbb{R}$ so defined is a field isomorphic to the system $S$ which realizes it.

However, as Linnebo and Pettigrew do not miss to notice, the success of this second approach to the definition of abstract mathematical objects is concomitant of the example chosen. It works for the real number field but fails in general. It even fails spectacularly for other well-chosen examples. Consider for instance the notion of an 'abstract $n$-dimensional vector space $\mathcal{V}$'—which one easily finds in the mathematical literature—and attempt to define it explicitly according to definitions (I.10) and (I.11) so that

$$\mathcal{V} := \{[x]_{eq} \mid x \in \mathbb{R}^n\}. \tag{I.13}$$

The problem here is that for any two non-zero vectors of $\mathbb{R}^n$—call them $x_1$ and $x_2$—there exists an isomorphism of vector spaces of $\mathbb{R}^n$ with itself—call it $f$—such that $f(x_1) = x_2$. Hence, we see that the abstract $n$-dimensional 'vector space' $\mathcal{V}$ defined by (I.13) consists of only two elements: the class $[0]_{eq}$ of the zero vectors and the class $[\neq 0]_{eq}$ of non-vanishing vectors. Thus, $\mathcal{V}$ is certainly not an $n$-dimensional vector space!

It therefore appears that, after all, this second, somewhat more sophisticated approach to understanding abstract entities by means of abstraction principles faces exactly the same problems as the naive approach of (I.8) (page 62), exception made of some very particular examples[117]. If the abstract/particular hierarchy is to be understood as a hierarchy of mathematical objects, then the logicists abstraction principles do not seem enough to grasp the situation. As Linnebo and Pettigrew conclude,

> [...] this cannot be the sort of abstraction championed by Frege and developed by the neo-Fregeans: for this sort of abstraction yields an unacceptable result for many non-rigid structures.[118]

But then, in the face of such difficulties, it is natural to wonder whether one has not taken the wrong road from the very beginning. To be successful, any ontological account of the abstract/particular hierarchy needs to define two types of mathematical objects such that:

i) both the abstract and the particular are of the same kind,

ii) the abstract is defined in such a way to be clearly distinguished from the particular.

By adopting first definition (I.8) (page 62) and then definition (I.11) (page 65), the latter requirement was directly addressed, but the former was seen not to be met. In fact, one clearly perceives the difficulty of satisfying both requirements, since they tend to pull in opposite directions: the one wishes a *unifying* framework for the abstract object and its realizations, while the other demands a *distinguishing* framework. The striving for an ontological distinction may thus seem artificial and unnecessarily restrictive. Perhaps, one should better abandon altogether this ontological interpretation of the abstract/particular hierarchy, and rather explore a more natural *epistemological* explanation. To this other possible strategy we now turn.

---

[117]It is easy to characterize precisely when this approach is going to be successful, as it was the case with the real numbers. The key feature that distinguished the example of the real number field from the example of abstract vector spaces was the presence, in the latter case, of *non-trivial automorphisms*. Mathematical objects which admit no non-trivial automorphisms, such as the real number field $\mathbb{R}$ or the rational number field $\mathbb{Q}$, are called *rigid*. For such objects, it will never be the case that two elements of a certain particular realization belong to the same equivalence class (as defined by equation (I.9)), and the abstract object (as defined by (I.11)) will then be isomorphic to each of its realizations (see ibid., Proposition 5.1., p. 276).

[118]Ibid., p. 283.

## I.2.2 Epistemology: the abstract/particular as hierarchy of information

As we have just explained, the 'hierarchy of levels' and the 'uniqueness of kind' (features 1 and 3, page 56) seem to pull in opposite directions if they are both interpreted ontologically. This was recognized as the core difficulty of the ontological approach to abstract entities. A natural move is then to circumvent the problem by simply interpreting the abstract/particular distinction as a purely epistemological issue and understanding the hierarchy of levels as a *hierarchy of information*.

An account along these lines is readily found in the writings of the mathematician Saunders Mac Lane. In his book *Mathematics, Form and Function*, the author gives a short account of his views on abstraction. After giving a very general idea of what abstraction amounts to[119], he proceeds to distinguish between three forms of abstraction, which he calls abstraction by *deletion*, abstraction by *analogy* and abstraction by *shift of attention*. The three are certainly worth discussing by themselves. However, in relation to our present investigation, the process of abstraction by deletion appears to be the most relevant one: Mac Lane indeed regards 'abstract groups' as a paradigmatic example of a mathematical concept attained through abstraction by deletion. Thus we concentrate solely on this specific form of abstraction, which is probably also the simplest of them all. The author describes it as follows:

> *Abstraction by deletion* is a straightforward process: One carefully omits part of the data describing the mathematical concept in question to obtain the more "abstract" concept. [...] For example, if one starts with the notion of a transformation, one may delete the elements being transformed but retain the associative, identity and inverse laws for the composition of transformations. The result is the notion of an "abstract" group.[120]

---

[119]He writes: "An "abstraction" is intended to pick out certain central aspects of the prior instances and to free them from aspects extraneous to the purpose at hand. Thus abstraction is likely to lead to the description and analysis of new and more austere or more "abstract" mathematical concepts." (Mac Lane, op. cit., p. 436)

[120]Ibid., p. 436. Here is a rough idea of the other two forms of abstraction. First, abstraction by shift of attention occurs when, in the development of a certain mathematical theory, some concepts which were first ignored are realized to be the key ingredients. The example used by Mac Lane to

As simple and "straightforward" as this process may be, it surely requires some further clarifications. But the main point should not be missed: from this point of view, the process of abstraction (by deletion) is not conceived as a process allowing to attain entities of a dubious ontological nature, but just as a *methodological decision* to disregard a certain amount of data. All mathematical objects are on equal footing, and an abstract object is nothing more than a particular object for which some data is *forgotten*, left unspecified[121]. In other words, any Hilbert space is some specific Hilbert space, and one talks about an "abstract" Hilbert space when there is not enough available information to ascertain which specific Hilbert space one is referring to. Now, as Mac Lane explains, trying to recover the missing information—thus trying to decide to which particular entity the abstract is referring to—is usually an important mathematical problem:

> This [abstraction by deletion] often leads to a reverse process, in which it is shown that all (or some) of the abstract objects can have the deleted data restored, perhaps in more than one way. Such a restoration is then called a "representation theorem".[122]

But one should not necessarily strive for this restoration: the lack of information characteristic of the abstract entity should by no means be thought as a drawback. Instead, it may appear as a welcomed methodological simplification allowing to focus on some chosen features of the theory—remain at a certain *level of 'unspecification'* and study which knowledge can be gained without further assumptions. Consider for example the following two theorems:

---

illustrate his point is the development of topology, where the notion of '*open subsets*' slowly evolved to become the central concept of the theory: one passed from sets embedded in Euclidean space to abstract metric spaces and finally to abstract topological spaces. Second, abstraction by analogy arises when a strong similarity is recognized between two different theories. Mac Lane here cites as example the introduction of the notion of a *modular* lattice, which is the key ingredient to prove both the Jordan-Hölder theorem for finite groups (two composition series of a finite group are of same length) and the fact that two bases of a vector space have same cardinality.

[121]The idea that *abstracting* is not possible without *forgetting*—for one would constantly be immersed in an infinite sea of details—is beautifully put in Jorge Luis Borges' short story *"Funes the Memorious"*.

[122]Ibid., p. 436.

**Theorem I.1.** *Let $\mathcal{H}$ be an abstract Hilbert space and $A$ any linear operator on $\mathcal{H}$. Then $A$ is everywhere continuous if it is continuous at the point $f = 0$.*[123]

**Theorem I.2.** *Let $x$ be a real number, then $x^2 + 1$ admits a square root.*[124]

According to the present view of abstraction by deletion, there is no conceptual difference between these two theorems. In the same way that Theorem I.2 is clearly not perceived as referring to some strange mathematical object 'the abstract real number' of which all particular real numbers would be realizations, Theorem I.1 should not be interpreted as a claim about a strange mathematical structure called "the" abstract Hilbert space, but simply as an assertion valid for *any* particular Hilbert space. And precisely because of this—because abstraction is here seen as a linguistic shortcut allowing to express claims which are valid for a variety of particulars—the specificities of the particular become irrelevant. If abstraction is the methodological decision of 'omitting part of the data' and remaining at the chosen level of unspecification, then, *by definition*, abstract entities such as 'abstract Hilbert spaces' are schematic. Therefore, features 1, 3, 4 and 5 are easily understood in this setting.

Now, by dissolving the abstract/particular as an ontological distinction, eliminating the idea of abstract Hilbert spaces as objects in their own right, and declaring that 'any Hilbert space is a particular Hilbert space', this point of view has come to rest on the notion of 'particular entities'. And a clarification of this notion is needed. This point seems especially pressing since we are here dealing with the realm of Mathematics and it is by no means clear what a particular (as opposed to abstract) mathematical object might be. If we were here concerned by general abstraction, the notion of 'particular objects' would perhaps not be so suspicious. To take a common example in the literature, the claim 'The white queen is allowed to move in any direction of the chess board' can be thought as a claim about an abstract entity (the white queen)[125]. Here, one can adopt exactly the same epistemological account of abstraction and understand this as a claim valid for *any* white queen. 'The white queen' is not a platonic abstract object and any white queen is some particular white queen, having a specific shape, made of some specific material, etc. Whether it is a plastic queen or a wooden queen is irrelevant to the claim, and the information is judiciously omitted. Presumably, the concept of 'particular objects' does not seem to pose any trouble in this case because there is an *underlying ontology* presupposed[126]. Similarly, it appears that the above epistemological view on mathematical abstraction needs to presuppose an analogous underlying ontology—one allowing to define what it means to be a 'particular mathematical object'. One possible solution is to adopt the extensional view that dominated Mathematics during the first half of the 20th century and that still overwhelmingly dominates the Philosophy of Mathematics. According to it, all mathematical objects are sets, explicitly constructed from the null set, and to know all the information about a particular object is to know exactly which are the elements that compose it[127].

---

[123]I take this theorem from von Neumann, op. cit., p. 99.

[124]This example is taken from S. Awodey. "An Answer to Hellman's Question: 'Does Category Theory Provide a Framework for Mathematical Structuralism?'" In: *Philosophia Mathematica* 12.1 (2004), pp. 54–64, p. 59.

[125]Analogies with the game of chess are widespread in the work of Stewart Shapiro, one of the main defenders of mathematical structuralism.

This being said, let us turn to the understanding, in this account, of the role of isomorphisms and the one-to-many relation. What does it mean to say that "$L^2(\mathbb{R}^3)$ and $l^2(\mathbb{N})$ are different realizations of $\mathcal{H}$"? Remember, $\mathcal{H}$ and $L^2(\mathbb{R}^3)$ are of the same nature—they are both particular Hilbert spaces and what distinguishes them is the amount of information *we have* about them: we know more about $L^2(\mathbb{R}^3)$ than about $\mathcal{H}$. So, in the ongoing understanding of abstraction, it would make no sense to interpret 'realization' as 'embodiment'. Given $\mathcal{H}$, $L^2(\mathbb{R}^3)$ and $l^2(\mathbb{N})$, consider the following important question: is $\mathcal{H}$ *equal* to $L^2(\mathbb{R}^3)$? As it should be clear, there is no way to

---

[126]In fact, even here one could argue that the notion of a 'particular white queen' is not so easily dealt with. Consider for instance a game of chess on the internet between two players in different parts of the world. To explain what kind of particular object is 'the white queen' to which both players refer is certainly not an easy task...

[127]This extensional view has been recently criticized, both for mathematical and philosophical reasons. For a nice description—geared to a philosophically-oriented audience—of some developments in main stream contemporary mathematics pushing away from this view, see J.-P. Marquis. "Mathematical Forms and Forms of Mathematics: Leaving the Shores of Extensional Mathematics". In: *Synthese* 190.12 (2013), pp. 2141–2164.

However, the criticisms against the set-theoretic ontology of Mathematics need not worry much the defenders of the 'abstraction by deletion' account. Indeed, as long as one in not wary of the idea of an underlying ontology in Mathematics, this epistemological account of abstraction is independent of *what* this ontology might be.

answer this in a definite fashion. We simply are lacking enough information about $\mathcal{H}$ to do so! Since all we know about $\mathcal{H}$ is that it is a Hilbert space, the best answer which can be provided at this point is to say that, given the available information, it is *possible* for $\mathcal{H}$ to be equal to $L^2(\mathbb{R}^3)$. But of course, it is also possible for $\mathcal{H}$ to be equal to $l^2(\mathbb{N})$. And—it seems to me—it is precisely in this *modal* sense that the one-to-many relation between the abstract and the particular entities should be interpreted in this epistemological account of what is meant by 'abstract groups', 'abstract Hilbert spaces' or 'abstract $C^*$-algebras'. Following Mac Lane, we could say that a realization of any of those abstract mathematical entities is one of the many possible *restorations* of the missing information.

Under this light, isomorphism appears as the technical tool that captures this possibility of equality. But to understand this, it seems preferable at this point to distinguish three different types of isomorphisms:

- *Isomorphism between an abstract and a particular*: in the ongoing conceptual setting, this inter-level type of isomorphism is the most natural type to consider. As we have just explained, the assertion $\mathcal{H} \simeq L^2(\mathbb{R}^3)$ is interpreted as 'it is possible for $\mathcal{H}$ to be equal to $L^2(\mathbb{R}^3)$'.

- *Isomorphism between two particulars*: obviously, this intra-level type of isomorphism cannot be interpreted in the same manner. We certainly have $L^2(\mathbb{R}^3) \neq l^2(\mathbb{N})$ (the elements of the former are continuous functions whereas the elements of the latter are infinite series), so there is no possibility of them being equal. Nevertheless, the claim $L^2(\mathbb{R}^3) \simeq l^2(\mathbb{N})$ remains of interest, since it is equivalent to the claim: if $\mathcal{H} \simeq L^2(\mathbb{R}^3)$ then $\mathcal{H} \simeq l^2(\mathbb{N})$. Thus, an isomorphism between two particulars may be regarded as the statement that they are two different possible restorations of the unspecified information of an abstract entity.

- *Isomorphism between two abstract entities*: in the neo-Fregean process of abstraction through equivalence, the loss of isomorphisms between abstract entities was a major caveat. It is easy to see that this is no longer the case: the claim $\mathcal{H}_1 \simeq \mathcal{H}_2$ is readily interpreted in the same way as $\mathcal{H}_1 \simeq l^2(\mathbb{N})$. Again, isomorphic abstract entities are entities which are possibly equal.

This last point seems to be so important, that it is worthwhile to restate it somewhat differently. In the conception of isomorphism–*qua*–correspondence, there is no fundamental difference between the three types of isomorphisms just discussed. They are all thought of as bijective functions allowing to translate statements about one entity into analogous statements about another entity. Thus conceived, isomorphisms–*qua*–correspondences give rise to an equivalence relation. However, one can adopt a different conception of isomorphisms—namely, isomorphism–*qua*–possible-identification. Now, two objects are isomorphic if they might be equal. As I see it, there are two main differences between the two conceptions. First, *strictu sensu*, isomorphisms–*qua*–possible-identifications give rise to a relation that is *not* an equivalence relation, for it fails to be transitive: $\mathcal{H}$ might be equal to $L^2(\mathbb{R}^3)$ and $\mathcal{H}$ might be equal to $l^2(\mathbb{N})$, but $L^2(\mathbb{R}^3)$ cannot be equal to $l^2(\mathbb{N})$. This is explained by the fact that *the conception of isomorphism-*qua*-possible–identification can only be adopted when abstract, partly unspecified entities are involved*. Second, given two abstract entities $\mathcal{H}_1$ and $\mathcal{H}_2$, in the isomorphism–*qua*–correspondence point of view, it is perfectly alright to claim '$\mathcal{H}_1 \simeq \mathcal{H}_2$ but $\mathcal{H}_1 \neq \mathcal{H}_2$'. But this claim is contradictory if one adopts the isomorphism–*qua*–possible-identification point of view. Literally, this claim would mean: 'it is possible for $\mathcal{H}_1$ and $\mathcal{H}_2$ to be equal, but they are different'. This poses the question of why one would adopt a point of view that restricts the sort of claims allowed? Is it not as artificial and unnecessarily restrictive as the attempt of conferring a distinct ontological status to the abstract?

The answer is simple: the motivation stems from the *practice* of mathematics and mathematical physics. Even though a question about the equality of $\mathcal{H}_1$ and $\mathcal{H}_2$ may have *in principle* a definite answer (it may indeed be a matter of fact whether $\mathcal{H}_1$ is equal to $\mathcal{H}_2$), this is not so in practice: as long as one remains at the level of unspecification characteristic of the abstract objects one is dealing with, there is simply not enough information to ascertain the validity of such an equality claim! Therein, *isomorphism appears as the strongest possible claim about the equality of two abstract entities*. Since, when considering abstract entities of a certain type, equality claims cannot be ascertained, one might forget equality altogether and use only isomorphisms. The talk 'up to isomorphism', ubiquitous in the practice of contemporary mathematics, reveals itself a natural and fundamental feature of this epistemological account of the process of abstraction.

The identity conditions of these abstract entities is a quite subtle matter. It would be a mistake to understand the above replacement as a move pushing to conflate the notions of isomorphism and identity. There are two main reasons why it would be so. On the one hand, this would be missing the point of the *modal* conception of isomorphism: it is isomorphism–*qua*–*possible*-identification, not isomorphism–*qua*–identification. In the same way that two $n$-dimensional Hilbert spaces $\mathcal{H}_1$ and $\mathcal{H}_2$ might be identical, they might also be different, so one should better keep them distinct. And this is indeed akin to what mathematicians do[128]. On the other hand, there may be a whole myriad of different isomorphisms between two isomorphic objects, and it is important to retain this multiplicity, for it contains relevant information about the mathematical situation being handled. If one does not properly distinguish 'identification' from 'identity', one obfuscates the existence of this multiplicity. Given two objects $\mathcal{H}_1$ and $\mathcal{H}_2$, it seems indeed rather difficult to understand how they can be *identical to* each other in different ways (they are either identical or not), but it is much easier to imagine several distinct manners in which they can be *identified with* each other. There can be many possible different processes of identifications that reveal the same identity.

To understand this last distinction between identity and identification, it is useful to consider one single object $\mathcal{H}$. If one reduces isomorphisms to just 'possible equality', the notion of automorphism appears to be redundant—if not quite obscure. Since $\mathcal{H}$ is *certainly* equal to itself, what would be the use of saying that, moreover, $\mathcal{H}$ is *possibly* equal to itself? On the contrary, if one conceives a given isomorphism as an explicit identification, the situation changes: even though $\mathcal{H}$ is equal to itself, there is an *ambiguity* in this claim, which is precisely captured by the existence of several different possible self-identifications or automorphisms.

---

[128]Here is a subtle point that often creates confusion: isomorphic objects are not identified but *canonically* isomorphic objects are. One speaks of "the" terminal object of a category (if it exists) because between any two such terminal objects there is a canonical isomorphism. Given a finite-dimensional vector space $V$, it is isomorphic to the dual $V^*$ and canonically isomorphic to $V^{**}$. Accordingly, one distinguishes the first from the second but not from the third. This remark shows that, in fact, isomorphism is not the 'strongest possible claim about the equality of two abstract entities', as I have affirmed above. It is canonical isomorphism. For a beautiful discussion of this point, see B. Mazur. "When is One Thing Equal to Some Other Thing?" In: *Proof and Other Dilemmas: Mathematics and Philosophy.* Ed. by B. Gold and R. A. Simons. Spectrum Series. Mathematical Association of America, 2008, pp. 221–242.

This is best illustrated by an explicit example. Take for $\mathcal{H}$ a set of two elements and consider two copies of it[129]. If $\mathcal{H}$ is an abstract set, then, although we know the two copies to be equal to each other, there are two possible identifications of them and there is simply no possible way to determine which one is the correct:

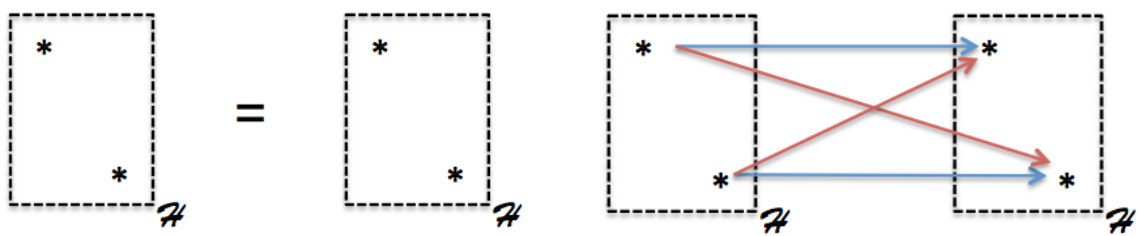

**(a)** Picture of $\mathcal{H} \simeq \mathcal{H}$ if isomorphism is conceived as identity, for $\mathcal{H}$ an abstract entity.

**(b)** Picture of $\mathcal{H} \simeq \mathcal{H}$ if isomorphism is conceived as identification, for $\mathcal{H}$ an abstract entity.

**Fig. I.1** – Difference between identity and identification for abstract entities.

This ambiguity finds its roots in the fact that, among the omitted data turning $\mathcal{H}$ into an abstract entity, is the information allowing us to distinguish the two elements of $\mathcal{H}$. If this lacking information was somehow completely restored, so that $\mathcal{H}$ became a particular set of two elements, there would no longer be a need to distinguish between identity and identification: indeed, in this case the elements could be distinguished and one could determine the unique correct identification of the two copies of $\mathcal{H}$:

---

[129]For expository reasons, I have taken the simple example of a set with two elements to illustrate my point. Isomorphisms are then bijective functions. But the argument does not depend on mathematical objects being sets (with possibly extra-structure). One might as well reason in terms of transformations, as defined in category theory. However, this seemed an unnecessary complication at this point.

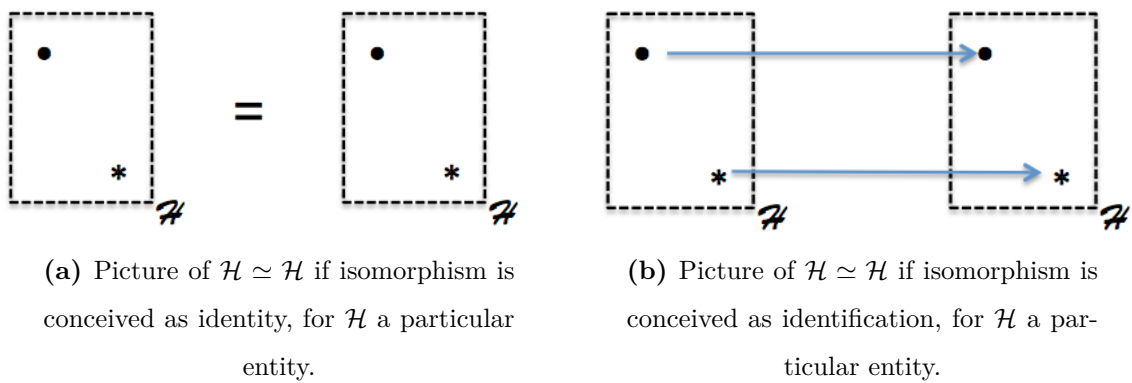

**(a)** Picture of $\mathcal{H} \simeq \mathcal{H}$ if isomorphism is conceived as identity, for $\mathcal{H}$ a particular entity.

**(b)** Picture of $\mathcal{H} \simeq \mathcal{H}$ if isomorphism is conceived as identification, for $\mathcal{H}$ a particular entity.

**Fig. I.2** – Difference between identity and identification for particular entities.

Thus, *it is characteristic of the abstract to have a complex structure of self-identifications which, in turn, is a manifestation of an essential ambiguity.*

As a side remark, note that in this view of abstraction by deletion the notion of isomorphism does not come first in the process of abstraction. Rather, one starts by choosing certain properties that are to be retained—i.e. abstracted—, and *only then* the relevant notion of isomorphism follows. In other terms, one first defines the abstract entity $\mathcal{H}$—for example, through an axiomatic presentation—and then isomorphism[130]. This is the opposite of what the process of abstraction through equivalence suggested.

Let us sum up what has emerged during this discussion of the epistemological account of the abstract/particular distinction as a hierarchy of information. An abstract entity is here understood as a particular entity for which only partial information is at our disposal. Given this omission of information, there can be many particular entities that may be the abstract entity handled. Realizations are possible restorations of the unknown. This explains the one-to-many relation, which is understood thus as a modal relation, best described using isomorphisms-*qua*-possible-identifications. With the information at our disposal, one can try to deduce some properties the abstract entity must have. These properties are necessarily independent of the omitted data— which explains the schematic nature of abstract entities. Finally, since any particular realization of the abstract entity is possibly equal to it, all known properties of the abstract must also hold for the particular. It is in this sense that the 'properties of the abstract are common to all the different realizations'. So, we see, this account of abstraction can explain on its own the importance of isomorphisms and all features 1. through 5. which were all emphasized by both von Neumann and Weyl (page 56).

---

[130]However, it would be a mistake to believe that the relevant notion of isomorphism follows *immediately* once the choice of properties to be retained is fixed. The mathematicians' struggle to find the good notion of homeomorphism in point-set topology is a nice historical example of this. See, G. H. Moore. "The Evolution of the Concept of Homeomorphism". In: *Historia Mathematica* (2007), pp. 333–343.

Nonetheless, this view of how to make sense of notions such as 'abstract Hilbert spaces' is not free of problems. The move of dissolving the abstract/particular as an ontological distinction, eliminating the idea of abstract Hilbert spaces as objects in their own right, and declaring that 'any Hilbert space is a particular Hilbert space', has two main drawbacks. First, the whole point of view has now come to rest on the notion of 'particular entities'. And a clarification of what exactly is meant by this is still lacking. This point seems especially pressing since we are here dealing with the realm of Mathematics and it is by no means clear what a particular (as opposed to abstract) mathematical object might be. If we were here concerned by general abstraction, the notion of 'particular objects' would perhaps not be so suspicious. To take a common example in the literature, the claim 'The white queen is allowed to move in any direction of the chess board' can be thought as a claim about an abstract entity (the white queen)[131]. Now, one can adopt exactly the same epistemological account of abstraction and understand this as a claim valid for *any* white queen. 'The white queen' is not a platonic abstract object and any white queen is some particular white queen, having a specific shape, made of some specific material, etc. Whether it is a plastic queen or a wooden queen is irrelevant to the claim, and the information is judiciously omitted. Presumably, the concept of 'particular objects' does not seem to pose any trouble in this case because there is an *underlying ontology* presupposed[132]. Similarly, it appears that the above epistemological view on mathematical abstraction needs to presuppose an analogous underlying ontology—one allowing to define what it means to be a 'particular mathematical object'. But this is a serious problem only if one is wary of the idea of an underlying ontology in Mathematics. For if one is ready to accept this possibility, all that has been said is in fact independent of *what* this ontology might be[133].

---

[131]Analogies with the game of chess are widespread in the work of Stewart Shapiro, one of the main defenders of mathematical structuralism.

[132]In fact, even here one could argue that the notion of a 'particular white queen' is not so easily dealt with. Consider for instance a game of chess on the internet between two players in different parts of the world. To explain what kind of particular object is 'the white queen' to which both players refer is certainly not an easy task...

Second, by interpreting all statements about abstract entities as statements quantified over a class of particular entities—symbolically: 'abstract' = 'any particular'—, this epistemological approach emphasizes the notion of *generalization* rather than abstraction. In fact, the process of abstraction by deletion appears simply to be the creation of a language useful for expressing general statements. In this new language, all superfluous details have indeed been deleted, but this is perceived as a useful way of *referring to many particulars* at a single stroke, not as a way of *conceiving* a new entity deprived from all these details. In short, abstraction is here considered to be a mere form of generalization. Moreover, it is not difficult to find, in the writings of some 19th century and early 20th century mathematicians, passages which seemingly support this claim. For example, we can read Stephan Banach explaining that:

> The aim of the present work is to establish certain theorems valid in different functional domains, which I specify in what follows. Nevertheless, in order not to have to prove them for each particular domain, I have chosen to take a different route [...]; I consider sets of elements about which I postulate certain properties; I deduce from them certain theorems, and I then prove for each particular functional domain that the postulates adopted are true for it.[134]

---

[133]For instance, one might adopt the extensional view that dominated Mathematics during the first half of the 20th century and that still overwhelmingly dominates the Philosophy of Mathematics. According to it, all mathematical objects are sets, explicitly constructed from the null set. To know all the information about a particular object is then to know exactly which are the elements that compose it. But this view has been recently criticized, both for mathematical and philosophical reasons. For a nice description—geared to a philosophically-oriented audience—of some developments in main stream contemporary mathematics pushing away from this view, see Marquis, op. cit.

[134]Quoted in J.-P. Marquis. "Mathematical Abstraction, Conceptual Variation and Identity". In: *Logic, Methodology and Philosophy of Science, Proceedings of the 14th International Congress (Nancy).* Ed. by P. E. Bour et al. London: College Publications, 2014, pp. 299–322, p. 315.

This short passage may not do full justice to Banach's conception of abstraction, but it does stress how abstraction involves a form of generalization. Another example is found in the work of Maurice Fréchet:

> In modern times it has been recognized that it is possible to elaborate full mathematical theories dealing with elements of which the nature is not specified, that is, with abstract elements. A collection of these abstract elements will be called an *abstract set.* [...]

The trouble with this claim is that it is clearly orthogonal to von Neumann's *intentions* when he introduced abstract Hilbert spaces in the context of Quantum Mechanics. As we saw in subsection I.1.3 and subsection I.1.4, Dirac and von Neumann conceived working with a specific realization of a Hilbert space to be tantamount to working with a specific choice of coordinates. Now, we may ask: What does it mean then to work with an abstract Hilbert space? A defender of the eliminative account of abstraction would answer: it means to work with an arbitrary choice of coordinates. However, this was not von Neumann's answer: for him, it meant to work in a *coordinate-free* formulation of the theory[135]. In other words, the eliminativist conceives these abstract mathematical entities as abstract particulars, whereas the physicist, following von Neumann, wants to conceive them as *universals*.

These remarks suggest that, despite its evident merits, the purely epistemological account of the abstract/particular distinction as a hierarchy of information fails to do justice to what is achieved through the introduction of these abstract mathematical structures, and that one should try to develop yet another alternative account.

### I.2.3  Mixture: the abstract/particular as hierarchy of identities

In a series of recent papers, Jean-Pierre Marquis has developed a description of the abstract method, as used in modern and contemporary mathematics, which aims

---

It is necessary to keep in mind that these notions are *not of a metaphysical nature*; that when we speak of an abstract element we mean that the nature of the element is indifferent, but *we do not mean at all that this element is unreal*. Our theory will apply to all elements; in particular, applications of it may be made to the natural sciences.

(M. R. Fréchet. "Abstract Sets, Abstract Spaces and General Analysis". In: *Mathematics Magazine* 24.3 (1951), pp. 147–155, p. 147, author's emphasis.)

[135]Here is a perhaps more modern and precise rephrasing of this last point. Given an $n$-dimensional abstract Hilbert space $\mathcal{H}_n$, the choice of an orthogonal basis $\mathcal{B}_n = \{e_1, \ldots, e_n\}$ induces a canonical isomorphism $\Phi_{\mathcal{B}_n} : \mathcal{H}_n \xrightarrow{\sim} \mathbb{R}^n$. This hints to the idea of perceiving isomorphisms between abstract Hilbert spaces and particular realizations as choices of a basis. In fact, it is perfectly alright to *define* a basis of $\mathcal{H}_n$ as such an isomorphism. Although technically more involved, the situation is essentially the same for an infinite-dimensional separable Hilbert space $\mathcal{H}$. There, one can always choose a countable orthonormal basis $\mathcal{B}$, and this choice will induce an isomorphism $\Phi_{\mathcal{B}} : \mathcal{H} \xrightarrow{\sim} l^2(\mathbb{N})$. In this way, one has precisely the 'equation'

particular Hilbert space = abstract Hilbert space + choice of a basis.

to point at the inherent features of abstraction[136]. His detailed account presents the abstract method as a process which is epistemological in nature but which nonetheless culminates in the creation of new mathematical entities. It can therefore be perceived as a mixture of the previous two approaches. In particular, Marquis emphasizes what I take to be the main lesson of the isomorphism–*qua*–possible-identification conception: the importance—often understated if not unnoticed—of reflecting on the criteria of identity for abstract entities. But his merit is to do so without mention of any type of modality and without such strong a reliance on the notion of 'particulars'. In this way, he is able to avoid the main drawback of the previous account: the reduction of abstraction to a particular form of generalization. Therefore, the essence of Marquis' point of view may be captured by the following three fundamental claims:

1. *Abstraction is epistemological.*

   > It is my profound belief that abstraction in mathematics is solely an episte-
   > mological issue and that the abstract character of mathematics is *not* an on-
   > tological property but rather derives from epistemological features of math-
   > ematical knowledge itself. I am not so much concerned with abstract objects
   > than with the *process* of abstraction and the abstract *method*. Some math-
   > ematical objects, or rather mathematical concepts, are *abstracted*. They do
   > not inherit a dubious ontological status for that reason.[137]

2. *Abstraction is not generalization*: the distinction lies in the creation of new enti-
   ties in the process of abstraction.

   > One could [...] work in a purely formal fashion or consider that one is doing
   > algebra in the classical sense of that word, that is working on generalized

---

[136] J.-P. Marquis. "Mathematical Abstraction, Conceptual Variation and Identity". In: *Logic, Methodology and Philosophy of Science, Proceedings of the 14th International Congress (Nancy).* Ed. by P. E. Bour et al. London: College Publications, 2014, pp. 299–322; J.-P. Marquis. "Stairway to Heaven – The Abstract Method and Levels of Abstraction in Mathematics". In: (forthcoming).

Besides these papers, which deal directly with the abstract method, at least two other articles are relevant to the question of what 'abstract' means in the realm of mathematical entities. These are: J.-P. Marquis. "Categorical Foundations of Mathematics – Or how to provide foundations for *abstract* mathematics". In: *The Review of Symbolic Logic* 6.1 (2013), pp. 51–75 and J.-P. Marquis. "Categories, Sets and the Nature of Mathematical Entities". In: *The Age of Alternative Logics: Assessing Philosophy of Logic and Mathematics Today.* Ed. by J. van Benthem et al. Dordrecht: Springer, 2006, pp. 181–192.

[137] Idem, "Mathematical Abstraction, Conceptual Variation and Identity", p. 300, author's emphasis.

arithmetic operations. [...] The situation changes radically once it is clear that it is possible to consider a new type of entities supporting these properties and relations. [...]

[F]rom an epistemological point of view, to focus on generality is to miss the point of the conceptual difference between the two notions [of generalization and abstraction].[138]

3. *It is characteristic of abstract entities to have a complex identity structure.*

[O]n any view of abstract mathematics, the notion of identity has a rich, complex structure which is not prior to the abstract objects present.[139]

With this in mind, let us now review in some detail Marquis' work. To describe the process of abstraction, the author brings out three basic components that are to constitute mathematical abstraction[140]:

– a domain of significant variation,

– a method of presentation and development,

– the extraction of a new criterion of identity.

The first two components are fairly intuitive. For there to be abstraction, one has to encounter several objects which appear somehow radically different—they constitute the domain of *significant* variation. Yet, one recognizes some invariant properties in the domain under consideration. It is "this interplay of variation and invariance [that] opens the door to the possibility of abstracting"[141]. In the main example we have been considering, these significantly different objects are $L^2(\mathbb{R}^3)$ (functions on a continuous space) and $l^2(\mathbb{N})$ (infinite series) and, among the invariant properties is, for instance, the existence of a Hermitian product. Once these invariant properties have been recognized,

---

[138]Idem, "Stairway to Heaven – The Abstract Method and Levels of Abstraction in Mathematics", pp. 7–8 and 13.

[139]Idem, "Categorical Foundations of Mathematics – Or how to provide foundations for *abstract* mathematics", p. 27.

[140]The number and the description of these components varies slightly from one work to another, but the ideas remain the same. This is the terminology of *"Stairway to Heaven – The Abstract Method and Levels of Abstraction in Mathematics"*, p. 3. See also *"Mathematical Abstraction, Conceptual Variation and Identity"*, pp. 308–309.

[141]Idem, "Mathematical Abstraction, Conceptual Variation and Identity", p. 309.

"the epistemic attention has to shift from certain pregnant features of the objects under study to the invariant elements involved"[142]. In other words, for a given element $f \in L^2(\mathbb{R}^3)$, one needs to ignore all questions about it being continuous or derivable—which make sense because it is a function—and consider it solely as point on a Hilbert space. Thus, one needs a method of presentation and development—a "*systematic ignorance* of the specific properties of the objects" leading to an "appropriate language" in which to present and investigate in an autonomous fashion the invariant properties[143]. In von Neumann's work, this method of presentation corresponds to the axiomatic definition of a Hilbert space[144].

Up to now, there is little novelty in this description of abstraction. Marquis' "systematic ignorance" is essentially the same as Mac Lane's "deletion". Nonetheless, this presentation shows already its usefulness, for it allows us to disentangle three concepts that are often conflated: abstraction, the axiomatic method and formalism[145]. As claimed by Marquis, the latter two should not be identified with abstraction, nor are they essential elements of the process:

> [...] in order to see the invariant features [...], one has to *forget* or *ignore* essential aspects of the objects and their properties involved. One has to ignore key properties of functions, of series, of the complex numbers, etc. **One of the ways** to succeed this operation is to concentrate on the formalism, the symbols and the operations on these symbols.[146]

A formalist approach to mathematics facilitates the access to invariant properties. Nonetheless, it is but one possible methodological decision enabling to perform one basic step of abstraction. In the same manner, the axiomatic method can be used as an implementation of the method of presentation and development characteristic of abstraction. Again, it is just one possible method among others:

> [...] axioms and the axiomatic method did play a key role in the rise of the abstract method. The axioms capture the invariant features of the theories under investigation. Once these have been identified, the axiomatic method allows for the systematic and rigorous development of the consequences of these features. One *could* use a different method of presentation of the invariant features. It depends on the linguistic means available. For instance, nowadays, it would be possible to use a graphical language to present a new theory by using what are called *sketches*.[147]

---

[142]Idem, "Stairway to Heaven – The Abstract Method and Levels of Abstraction in Mathematics", p. 7.

[143]Ibid., p. 3, my emphasis.

[144]Von Neumann, op. cit., pp. 35–45.

[145]For example, the abstract and axiomatic method are used as if interchangeable concepts by Mac Lane himself in the following passage: "The abstract or postulational development of these systems must then be supplemented by an investigation of their "structure"." (S. Mac Lane. "Some Recent Advances in Algebra". In: *American Mathematical Monthly* 46 (1939), pp. 3–19, pp. 17–18, cited in: Marquis, op. cit., p. 8.)

[146]Ibid., p. 5, italics are from the author, bold typeface is mine.

Moreover, as the historical example of Euclidean geometry suffices to illustrate, this is not the only use axioms may have in Mathematics. There are other contexts in which axioms may be perceived as evident, basic truths. In fact, it is a feature of abstraction that, whenever axioms appear, their role is not to assert but to present—axioms as defining conditions, not as evident assertions[148].

But, Marquis insists over and over, the process of abstraction is unfinished if one stops here and ignores the third, fundamental component: the extraction of a criterion of identity—this "blind spot in the mathematicians' journey through the abstract method"[149]. It is the central point which distinguishes his approach from the one sketched in the previous section. He explains this as follows:

> [...] it is not until the proper criterion of identity has been identified and applied systematically that the theory acquires an autonomy, both epistemological and ontological. Notice also that it is the presence of a new criterion of identity that allows to say that we are indeed in the presence of a new type of abstraction, for as we have seen, the usage of the axiomatic method in itself does not entail the need of a new criterion of identity. [...] The identification of the proper criterion of identity is of fundamental importance, since it allows to sift the properties of the resulting theory from the properties of the previous theories. In other words,
*it captures the process of abstraction itself.*[150]

---

[147]Ibid., p. 10, author's emphasis.

[148]For an analysis of the different uses of axioms in the practice of Mathematics, see D. Schlimm. "Axioms in Mathematical Practice". In: *Philosophia Mathematica* 21 (2013), pp. 37–92.

[149]Marquis, op. cit., p. 11.

The fundamental thesis of Marquis' account is then the following: to perceive the new abstract entities as emerging, not from the definition of a new object, but rather from the definition of a new criterion of identity. In this way, identity ceases to be a universal and unrestricted notion, admitting one single form which applies to the entire realm of Mathematics. Rather, the process of abstraction is better described using a multi-sorted or typed logic, in which *identity is contextual*. For Marquis, "this is precisely where certain aspects of abstract mathematics escape the standard analysis or explication in terms of ZF-sets, or any other notion of sets based on extensionality"[151]. In turn, type theory illuminates the complexity of identity: instead of describing identity of two objects as a property, the theory describes it as a structure[152]. The multiplicity of possible identifications between two abstract objects, encountered in the account of abstraction by deletion (Section I.2.2), is precisely the sort of structure here mentioned. The realm of Mathematics is seen to be governed by a complex net of different criteria of identity and the abstract/particular distinction appears as the consequence of a *hierarchy of identities*. More precisely, the abstract/particular opposition is dissolved and replaced by a ladder of *levels of abstraction*: a mathematical entity is not *either* abstract *or* non-abstract, it is simply more or less abstract *than another entity*.

This phenomenon of there being several different criteria of identity, associated to different levels of abstraction, can very well be perceived in our main example with Hilbert spaces. So far, the situation had been presented as having only two levels: the one of 'particular' Hilbert spaces, such as $L^2(\mathbb{R})$ and $l^2(\mathbb{N})$, and the one of the abstract Hilbert space $\mathcal{H}$. Now, notice that in order to talk about *the* Hilbert space $L^2(\mathbb{R})$ of (Lebesgue equivalence classes of) square-integrable functions over $\mathbb{R}$, or about

---

[150] Ibid., p. 12.

[151] Idem, "Categorical Foundations of Mathematics – Or how to provide foundations for *abstract* mathematics", p. 58.

[152] For given a type $T$ and objects $X : T, Y : T$, the proposition $X =_T Y$ is better captured as a type $\mathrm{Id}_T(X, Y)$. In fact, a central idea of type theory is to view any proposition as a type. Some of the main differences between set and type theory will be discussed in more detail in Subsection I.3.3.a.

*the* Hilbert space of square-summable functions over $\mathbb{N}$, one needs to be able to talk about *the* field of real numbers $\mathbb{R}$ and about *the* natural number system $\mathbb{N}$. But of course, from a set-theoretical point of view, we know that $\mathbb{R}$ could equally well refer to Dedekind's model $\mathbb{R}_D$ or to Cantor's model $\mathbb{R}_C$, in the same way that $\mathbb{N}$ could refer to von Neumann's ordinals $\mathbb{N}_N$ or to Zermelo's numerals $\mathbb{N}_Z$. In other words, one can only perceive $L^2(\mathbb{R})$ as being a well-defined specific entity if one decides to regard the different models of $\mathbb{R}$ as being equal. Therefore, the situation we have been dealing with is more accurately described as having (at least) three different levels of abstraction and three criteria of identity:

(i) First level of abstraction, governed by identity $=_{T_1}$. We have four entities:

$$l^2(\mathbb{N}_N) \neq_{T_1} l^2(\mathbb{N}_Z) \neq_{T_1} L^2(\mathbb{R}_D) \neq_{T_1} L^2(\mathbb{R}_C).$$

The identity $=_{T_1}$ may be based for example in the usual extensionality for sets. Then, $\mathbb{N}_N$ and $\mathbb{N}_Z$ are different sets and $l^2(\mathbb{N}_N)$ and $l^2(\mathbb{N}_Z)$ are different Hilbert spaces.

(ii) Second level of abstraction, governed by identity $=_{T_2}$. We have two entities:

$$l^2(\mathbb{N}) =_{T_2} l^2(\mathbb{N}_N) =_{T_2} l^2(\mathbb{N}_Z) \neq_{T_2} L^2(\mathbb{R}_D) =_{T_2} L^2(\mathbb{R}_C) =_{T_2} L^2(\mathbb{R}).$$

With this second criterion of identity, one can conceive *the* abstract natural number system $\mathbb{N}$ and *the* abstract field of real numbers.

(iii) Third level of abstraction, governed by identity $=_{T_3}$. We have one entity:

$$l^2(\mathbb{N}) =_{T_3} l^2(\mathbb{N}_N) =_{T_3} l^2(\mathbb{N}_Z) =_{T_3} L^2(\mathbb{R}_D) =_{T_3} L^2(\mathbb{R}_C) =_{T_3} L^2(\mathbb{R}) =_{T_3} \mathcal{H}.$$

With this third criterion of identity, there is only the one abstract Hilbert space. This remark simply points to the fact that in any mathematical situation there is always an implicit threshold level of abstraction below which one will not descend.

We still have to explain the precise manner in which the new criterion of identity is extracted. As Marquis emphasizes, the main reason for defining this new identity is to switch from a point of view where properties are *invariant* to a point of view where properties are *intrinsic*. Given a domain of variation, and a collection of invariant properties that are to be retained, this move is achieved by the following two-step procedure. First, for any property $P_0$ that is to be retained, and any two objects $X, Y$ of the domain of variation, the new identity $=_{T_2}$ should be constructed so that

$$\text{if } P_0(X) \text{ and } X =_{T_2} Y, \text{ then } P_0(Y) \tag{I.14}$$

Second, one reverses the approach and uses $=_{T_2}$ to define what qualifies as an abstract property $P$:

$$\text{if } P(X) \text{ and } X =_{T_2} Y, \text{ then } P(Y) \tag{I.15}$$

The two steps (I.14) and (I.15) are not to be confused: in (I.14) the properties $P_0$ are given and this allows us to define the new identity, whereas in (I.15) $=_{T_2}$ is given and this allows to define the abstract properties $P$. Of course, the initially chosen properties $P_0$ become *by construction* abstract properties. It is precisely this move that accomplishes the "shift of attention" mentioned by Marquis. Under the light of the old criterion of identity, one could have $X \neq_{T_1} Y$ but $P_0(X)$ and $P_0(Y)$. $P_0$ was then *invariant* by the change from $X$ to $Y$. But under the light of the new criterion of identity, *there is no such variation* from $X$ to $Y$ since $X =_{T_2} Y$: $P_0$ shows now to be an intrinsic property of the abstract entity emerged through the newly defined identity.

Through this account, it appears (again) that it is "impossible to think of the abstraction process in terms of an equivalence relation"[153]. Indeed, the choice of an identity criterion strongly depends on the properties one is wishing to abstract:

> One has to have the properties that will be abstracted in order to define the criterion of identity between the abstract entities. In other words, the criterion of identity can not be given a priori but is derived from the theory.[154]

One then needs to have an idea of the new abstract entity *before* considering the definition of the equivalence relation, unlike what is claimed by the neo-Fregean account of abstraction.

Now, the reader should be struck by the fact that the definition of the new abstract identity $=_{T_2}$ coincides exactly with the definition of *isomorphisms*. These are constructed precisely in order to preserve certain properties, as stated in definition (I.14)—e.g., an isomorphism of finite-dimensional vector spaces is a morphism preserving the linear structure and the dimension. In this way, one understands that isomorphism is nothing but the identity governing the new level of abstraction. Here is Marquis explaining this crucial point:

> In fact, it might be wise to replace the term "isomorphism" by a more neutral term that evokes a type of identity. Notice that one could stipulate that once the proper criterion of identity has been discovered, then the meaningful properties are precisely those that satisfy Leibniz's principle. I would even dare suggest that the latter is a key property of what it means to be abstract for mathematical objects. [...]
>
> For many mathematicians, *being isomorphic is precisely what being abstract amounts to. This means that X and Y are, from an abstract point of view, essentially the same.*[155]

---

[153]Idem, "Mathematical Abstraction, Conceptual Variation and Identity", p. 312.

[154]Ibid., p. 312.

Marquis is certainly not alone in claiming this. In fact, in *"Towards a Categorical Foundation of Mathematics"*, the logician Michael Makkai has proposed to elevate (I.15) to the rank of a principle, which he calls the "Principle of Isomorphism"[156]. He considers it a crucial tenet of what Abstract Mathematics are. To wit:

> [...] the Principle of Isomorphism itself appears to be a generally accepted idea in Abstract Mathematics. [...] The basic character of the Principle of Isomorphism is that of a *constraint* on the language of Abstract Mathematics; a welcome one, since it provides for the separation of sense from nonsense. But the fact that **isomorphism is the real equality in Abstract Mathematics** is also an *experience.*[157]

---

[155]Idem, "Categorical Foundations of Mathematics – Or how to provide foundations for *abstract* mathematics", p. 58–59, my emphasis.

[156]The similarity between these two works is not surprising since Makkai's attempts to develop a foundations of Mathematics based on category theory has exerted a strong influence on Marquis' ideas, as the latter has acknowledged in many occasions (see, for instance, ibid., p. 72).

[157]Makkai, op. cit., p. 161, author's italics, boldtype face is mine.

In view of the central role it plays in the process of abstraction, I propose to name it the *"Principle of Abstraction"*[158]:

> **Principle of Abstraction:** all grammatically correct properties of abstract objects are to be invariant under the relevant isomorphism type.

Through Marquis' account of the method of abstraction, we have reached a better understanding of what it means to conceive some mathematical entities *abstractly*. By the same token, we now know which criterion of identity to use for the mathematical description $\mathcal{D}(S)$ of the physical system $S$. Therefore, out of the three initial questions (cf. page 17), it only remains to elucidate the content of the requirement of individuation. For this, we need to turn to an analysis of mathematical structuralism.

## I.3    Abstract mathematical structures

In none of the last two accounts of abstraction the word 'structure' has appeared. Yet, they both manage to explain the essential features about abstract Hilbert spaces, abstract symplectic manifolds, abstract groups, etc. which were highlighted when analyzing von Neumann's work. These were: the hierarchy of levels, the one-to-many relation, the uniqueness of kind, the commonality of properties and the schematic nature of these new abstract entities (cf. pages 56–56). This poses the question of why

---

[158] Makkai's precise formulation is "**Principle of Isomorphism:** all grammatically correct properties *of objects of a fixed category* are to be invariant under isomorphism" (ibid., p. 161, my emphasis).

A major caveat of my terminological decision—besides the confusion it may create with the neo-Fregean abstraction principles—is that the expression "Principle of abstraction" has already been used before in a quite different sense. In *The Principles of Mathematics*, Bertrand Russell introduces an axiom with this precise name and defined as follows:

> "Every transitive symmetrical relation, of which there is at least one instance, is analyzable into joint possession of a new relation to a new term, the new relation being such that no term can have this relation to more than one term, but that its converse does not have this property." This principle amounts, in common language, to the assertion that transitive symmetric relations arise from a common property. (p. 220)

For a discussion on Russell's principle of abstraction, see I. Angelelli. "Adventures of Abstraction". In: *Poznań Studies in the Philosophy of the Sciences and the Humanities* 82 (2004), pp. 11–35.

the concept of 'structure' is present. Von Neumann talks of Hilbert spaces as "mathematical structures"; Weyl refers to an abstract group as "a structural entity". Now, if abstract Hilbert spaces and groups are best described as mathematical structures and not simply as abstract mathematical entities, it is essential to understand which features are intended to be captured by the concept of 'structure' that cannot be accounted for by solely appealing to the concept of 'abstraction'. In fact, it is my belief that the whole philosophical discussion on mathematical structuralism often suffers from not systematically distinguishing these two concepts.

A good example where this confusion is particularly evident is the role attributed to isomorphisms. As we have seen in some detail in the previous section, from the above description of the process of abstraction emerged the understanding that isomorphisms are the pertinent identity criterion for abstract entities. Makkai wrote that "isomorphism is the real equality in Abstract Mathematics" and Marquis went almost as far as claiming that "being isomorphic is precisely what being abstract amounts to". But this view appears to be quarrelsome, for it is not difficult to find, in the philosophy of mathematics literature, statements pulling in another direction. For instance, Andrei Rodin explains how "the idea that isomorphic objects can be treated as equal is, in [his] view, crucial for structuralism"[159]. Moreover, Steve Awodey has recently proposed that the statement 'isomorphic objects are identical'—that is, essentially the same statement we have wished to call "Principle of Abstraction"—should be called the *Principle of Structuralism*[160]. Thus, Awodey would be inclined to say that "isomorphism is the real equality in *Structural* Mathematics"...

Our immediate goal becomes thus to clarify what is to be meant by 'mathematical structuralism' and to understand what the concept of 'structure' adds to the above account of the process of abstraction.

---

[159] Rodin, op. cit., p. 23.

[160] S. Awodey. "Structuralism, Invariance, and Univalence". In: *Philosophia Mathematica* 22.1 (2014), pp. 1–11, p. 2.

### I.3.1   Mathematical structuralism

In any view of mathematical structuralism, mathematical structures are a specific kind of abstract mathematical entities. As such, the whole discussion of Section I.2 immediately applies to the question of interpreting what mathematical structures are. In the spirit of the epistemological approach to abstraction (subsection I.2.2), some consider statements about mathematical structures to be implicit general statements about all the systems which realize them. Following Charles Parsons, this view which negates the existence of abstract mathematical structures is called *eliminative structuralism*[161]. In contrast, in the spirit of the views on abstraction developed in subsections I.2.1 and I.2.3, others insist on conceiving mathematical structures as entities in their own right, independent from their realizations. This view is called *ante rem structuralism* by Stewart Shapiro[162] and *sui generis structuralism* by Geoffrey Hellman[163].

Before discussing the differences between these different approaches to mathematical structuralism, we must however focus on what should be the first task in any discussion of the subject: to characterize the *common core* which allows to consider the two versions as being two versions *of structuralism*—in other words, to locate the essential features that distinguish structuralism from the general abstract method described in Section I.2.

---

[161]"A reading [...] that seems to me to accord reasonably well [... holds] that statements about natural numbers are implicitly general, about *any* simply infinite system. [...] It clearly avoids singling out any one simply infinite system as the natural numbers and expresses the general conception I have in mind in speaking of the structuralist view. [...] Such a program I will call *eliminative structuralism*." (C. Parsons. "The Structuralist View of Mathematical Objects". In: *Synthese* (1990), pp. 303–346, p. 307, author's emphasis.)

[162]"Any usual array of philosophical views on universals can be adapted to structures. One can be a Platonic *ante rem* realist, holding that each structure exists and has its properties independent of any systems that have that structure. On this view, structures exist objectively, and are ontologically prior to any systems that have them (or at least ontologically independent of such systems). Or one can be an Aristotelian *in re* realist, holding that structures exist, but insisting that they are ontologically posterior to the systems that instantiate them." (S. Shapiro. "Mathematical Structuralism". In: *Internet Encyclopedia of Philosophy*. URL: http://www.iep.utm.edu/m-struct/, pp. 2–3.)

[163]See for example G. Hellman. "Structuralism". In: *The Oxford Handbook of Philosophy of Mathematics and Logic*. Ed. by S. Shapiro. New York: Oxford University Press, 2005, pp. 536–562, p. 541 or G. Hellman. "Three Varieties of Mathematical Structuralism". In: *Philosophia Mathematica* 9.3 (2001), pp. 184–211, p. 188.

### I.3.1.a   Characterizing structuralism

This task is not always carefully carried out and some philosophers tend to attribute to mathematical structuralism virtues that belong to Abstract Mathematics in general. A striking example of this is the work of Steve Awodey on what he calls "categorical structuralism". Thus, in the very first paragraphs of his article *"An Answer to Hellman's Question: 'Does Category Theory Provide a Framework for Mathematical Structuralism?"'*, the author describes categorical structuralism as follows:

> As a first, very rough, approximation, we may say that the point of view that we are going to describe emphasizes form over content, descriptions over constructions, specification of assumptions over deductive foundations, characterization of essential properties over constitution of objects having those properties.
>
> [...] The 'categorical-structural' [view] we advocate is based instead on the idea of specifying, for a given theorem or theory only the required or relevant degree of information or structure, the essential features of a given situation, for the purpose at hand, without assuming some ultimate knowledge, specification, or determination of the 'objects' involved.[164]

Now, as a first description of what structuralism amounts to, this account should be surprising. For, as the reader will immediately notice, this quote has little to do with structuralism *per se*: the concept of structure does not seem to be doing any work here. In fact, Awodey's description applies admirably well to the abstract method in general! However, this is but a "first, very rough, approximation" and one can hope that, later on, mathematical structuralism—which is "a *certain*, now typical, 'abstract' way of practicing mathematics"[165]—will be clearly distinguished from all other types of abstraction. But, in my opinion, Awodey fails to do so. A major part of his paper concentrates on showing that "mathematical theorems are schematic", characteristic which is "clearly essential to this approach"[166]. He certainly makes a good point here, but the worry is he still seems to be talking about abstraction, not about structuralism. Consider for instance the following passage:

> The proof of a theorem involves the structures mentioned, and perhaps many others along the way, together with some general principles of reasoning like those collected up in logic, set theory, category theory, etc. But it does not involve the specific nature of the structures, or their components, in an absolute sense. That is, there is a certain degree of 'analysis' or specificity required for the proof, and beyond that, it does not matter what the structures are supposed to be or to 'consist of'—the elements of the group, the points of the space, are simply *undetermined*.[167]

---

[164]Awodey, "An Answer to Hellman's Question: 'Does Category Theory Provide a Framework for Mathematical Structuralism?'", p. 55–56.

[165]Ibid., p. 54, my emphasis.

[166]Ibid., p. 62.

Again, one can replace 'structures' by 'abstract entities' and the description continues to be correct. So, perhaps, Awodey uses 'structure' as just another name for 'abstract entities'.

To find a definition of the former notion, one has to go back to his first article on the subject, *"Structure in Mathematics and Logic: A Categorical Perspective"*. He has the merit of clearly stating what is to be meant by 'structure':

> The categorical notion of isomorphism may now serve as a *definition* of 'having the same structure of a given kind'.[168]

The idea here is that a category defines a kind of structure, and the morphisms are by definition 'structure-preserving maps'. Then, two objects of the category bear the same structure if they are isomorphic. Hence, as Awodey himself explains in a much more recent article, he is determining the concept of structure through a Fregean abstraction principle of the form (I.7) (page 60):

$$\forall A, B \in \mathcal{C}_0, \big(\mathrm{str}(A) = \mathrm{str}(B) \longleftrightarrow A \cong B\big)$$

---

[167] Ibid., p. 59, author's emphasis.

[168] S. Awodey. "Structure in Mathematics and Logic: A Categorical Perspective". In: *Philosophia Mathematica* 4 (1996), pp. 209–237, p. 214.

where $\mathcal{C}_0$ is the collection of objects of a given category $\mathcal{C}$. In words: "The structure of $A$ is the same as the structure of $B$ just in case $A$ and $B$ are isomorphic"[169].

Now, as a terminological decision, there is nothing to be objected, as long as the author takes good care to distinguish his use of the word from what 'structure' may mean in other contexts. But Awodey's whole defense of structuralism, as a philosophical position, is entirely based on the above definition—and this is a problem. One needs to justify in which way the concept of structure so defined manages to capture the main intuitions behind mathematical structuralism, and not simply those behind abstraction. Thus, we are back to the question: What is the core of mathematical structuralism, which distinguishes it from other methods of abstraction?

Let us return once more to Awodey. In the concluding paragraph of *"Structure in Mathematics and Logic: A Categorical Perspective"*, he writes:

> The structural perspective on mathematics codified by categorical methods might be summarized in the slogan: The subject matter of pure mathematics is invariant form, not a universe of mathematical objects consisting of logical atoms. [...] The tension between mathematical form and substance can be recognized already in the dispute between Dedekind and Frege over the nature of the natural numbers, the former determining them structurally, and the latter insisting that they be logical objects.[170]

In my view, the key to clarifying the situation emerges here. Indeed, there are three concepts at play: "(invariant) form", "structure" and "logical atoms". As we have seen, the objects of abstract mathematics always come with a level of unspecification—part of their "substance" is omitted—and, in this sense, they are not constituted by logical atoms. If one defines "form" as what is generally obtained by the abstraction principle,

$$\forall A, B\big(\text{form}(A) = \text{form}(B) \longleftrightarrow A \cong B\big)$$

then, together with Awodey, we can safely say: 'the subject of Abstract Mathematics is form, not substance'. This is the definition of Abstract Mathematics, not a statement

---

[169]Idem, "Structuralism, Invariance, and Univalence", p. 3.

[170]Idem, "Structure in Mathematics and Logic: A Categorical Perspective", p. 235.

about them. The important point becomes then to understand the relation between form and structure.

For Awodey and Rodin, these two words are synonyms[171]. This explains why much of what they say fits very well with the method of abstraction in general. However, by this decision, the concept of 'structure' becomes unnecessary... It seems the situation is better understood if one keeps the three concepts distinct and claims that *structure is a particular kind of form*. In this way, mathematical structuralism appears indeed to be a particular method of abstraction and one may say: "*the subject of Abstract Mathematics is form, not substance; the subject of Structural Mathematics is structure, not any form*".

What kind of process of abstraction would structuralism be? To see this, let us contrast the different ways in which some of the main philosophers endorsing mathematical structuralism have attempted to introduce and motivate the subject[172]:

– *Geoffrey Hellman*: "[C]ertain views called "structuralist" have become commonplace. Mathematics is seen as the investigation, by more or less rigorous deductive means, of "abstract structures", systems of objects fulfilling certain structural relations among themselves and in relation to other systems, without regard to the particular nature of the objects themselves."[173]

– *Stewart Shapiro*: "The theme of mathematical structuralism is that what matters to a mathematical theory is not the internal nature of its objects, such as its numbers, functions, sets, or points, but how those objects relate to each other. In a sense, the thesis is that mathematical objects (if there are such objects) simply have no intrinsic nature."[174]

---

[171]Prior to the last paragraph of his paper, Awodey never uses the concept of 'form'. Hence, it is not clear how he understands this. But the quote seems to suggest he indeed considers them as synonyms. For Rodin, the situation is far clearer, since he declares that "invariant [form] in the given context is just another word for structure" (Rodin, op. cit., p. 29) and that "the desired 'purely structural' mathematics would deal only with the 'invariant Form' (ibid., p. 31).

[172]Again, for the moment we are just trying to pinpoint the main intuitions attached to structuralism. Thus we are interested in how the subject is introduced, and not that much in the detailed attempts to articulate it.

[173]Hellman, "Structuralism", p. 536.

[174]Shapiro, op. cit., p. 1.

– *Charles Parsons*: "By the 'structuralist view' of mathematical objects, I mean the view that reference to mathematical objects is always in the context of some background structure, and that the objects involved have no more to them than can be expressed in terms of the basic relations of the structure."[175]

– *Michael Resnik*: "The underlying philosophical idea here is that in mathematics the primary subject-matter is not the individual mathematical objects but rather the structures in which they are arranged. The objects of mathematics, that is, the entities which our mathematical constants and quantifiers denote, are themselves atoms, structureless points, or positions in structures. And as such they have no identity or distinguishing features outside a structure."[176]

Evidently, there are some recurrent themes in these quotes. The first, most important one, is the *emphasis on relations*. Above all, structuralism is the *methodological decision* of never studying an entity in isolation but rather of considering a collection of entities and focusing on the relations in which they stand. Then, one investigates how much knowledge can be gained of these various entities through the sole consideration of structural properties—i.e., those arising from the relations. Thus, structuralism shows itself as one certain process of abstraction: one in which a decision is made to retain only structural properties and to ignore the particular nature of the entities. By this act of abstraction, the collection becomes an abstract structure.

As a method of study, structuralism appears to be interested in two types of properties. On the one hand, it is wishes to investigate *properties of the structure.* Hellman speaks of "investigating the abstract structures" and Resnik emphasizes that the "primary subject-matter is the structures". For example, one asks whether a given group $G$ admits finite linear representations, or whether a given field $K$ admits proper algebraic extensions, and these are questions that could even be positively investigated without ever considering the "things" that constitute the structure. Let us call this first type of properties '*global properties*'. On the other hand, even though the structuralist is interested in the structure as an object of study, his interest also turns towards what lies *within* the structure. This is particularly clear in the terminology of Parsons and Shapiro, where the notion of 'mathematical object' is not used to denote a structure but rather a place within a structure. The natural number structure may be a structure; yet, it is made of natural numbers, and one asks whether any even number can be written as the sum of two primes, or whether 8 and 9 are the only consecutive numbers which are pure powers of non-zero integers[177]. Hence, structuralism also wishes to investigate properties of the 'things' constituting the structure. Let us call this second type of properties '*internal properties*'. In my opinion, both the holistic point of view— which takes the structure as main object of study—and the internal point of view— which takes the 'things' inside a structure as main object of study—are constitutive of structuralism: the whole is always thought as composed of elements and, in return, the elements can only be understood if conceived as part of a whole.

---

[175]Parsons, op. cit., p. 303.

[176]M. D. Resnik. *Mathematics as a Science of Patterns.* New York: Oxford University Press, 1997, p. 201.

The holistic approach is always emphasized: do not study objects in isolation; study relations between objects. But the importance of the internal point of view is sometimes understated if not completely ignored. For all four philosophers, a structure is always a collection. For Hellman, it is "a system of objects fulfilling certain structural relations"[178]. For Shapiro, "a structure is the abstract form of a system, highlighting the interrelationships among the objects, and ignoring any features of them that do not affect how they relate to other objects in the system"[179]. For Parsons, "[w]hat is meant by a structure is usually a domain of objects together with certain functions and relations on the domain, satisfying certain given conditions"[180]. And Resnik takes "a pattern to consist of one or more objects, which [Resnik] call[s] positions that stand in various relationships"[181]. I thus join Feferman in saying that the notions of collection

---

[177]The first problem is the famous Goldbach's conjecture, which is still an open problem. The second is the less known 'problem of Catalan', which was asked in 1842 and only solved recently by Preda Mihăilescu (see Y. Bilu, Y. Bugeaud, and M. Mignotte. *The Problem of Catalan*. Springer, 2014).

[178]Hellman, loc. cit.

[179]S. Shapiro. *Philosophy of Mathematics: Structure and Ontology*. New York: Oxford University Press, 1997, p. 74.

[180]Parsons, op. cit., p. 305.

[181]Resnik, op. cit., p. 203.

and relation are the fundamental primitive notions on which structuralism rests[182].

Hence, I take the core of mathematical structuralism to be the following:

> **Mathematical method of structuralism:** to consider always collections of things, among which some relations have been defined, and to investigate which knowledge can be gained, about both the whole collection and the individual things, by studying these relations and the properties derived from them.

Collections, relations and the global/local two-fold level of study. These I take to be the three essential components of structuralism as a method of study. They shape what Structural Mathematics are, and I believe any version of structuralism, as a *philosophical position* about Mathematics, should include them[183].

---

[182]"The point is simply that *when explaining* the general notion of structure and of particular kinds of structures such as groups, rings, categories, etc., we implicitly *presume as understood* the ideas of *operation* and *collection* [...]." (S. Feferman. "Categorical Foundations and Foundations of Category Theory". In: *Logic, Foundations of Mathematics, and Computability Theory (Proc. Fifth Internat. Congr. Logic, Methodology and Philos. of Sci., Univ. Western Ontario)*. Philos. Sci. Dordrecht, The Netherlands: University Western Ontario, 1977, pp. 149–169, p. 150, author's emphasis.)

[183] In the light of this, let me continue commenting Awodey's work. In his answer to Hellman, he explains:

> [...] the essential difference between the position being sketched here and old-fashioned, relational structuralism is the idea of a top-down description, which presupposes no bottom-up hierarchy of things. For Russell, every relation had to be a relation on some *things* which, even if they were themselves analyzable into relations, had to be among some other *things*, ... , and either this process had to stop somewhere (atoms), or an account had to be given of infinite analysis.
>
> The difficulty arises in the preoccupation with *relations* as the fundamental notion of 'structure'; for a relation presupposes its relata, and off we go into the descent of Russellean analysis. If we take instead the perfectly autonomous notion of a morphism in a category, we can build structures out of them to our heart's content, **without ever having to ask what might be in them**.
>
> (Awodey, "An Answer to Hellman's Question: 'Does Category Theory Provide a Framework for Mathematical Structuralism?'", p. 61, author's emphasis, bold typeface is mine.)

Awodey's move of completely ignoring the elements of the structure, of 'never asking what might be in them', appears to be completely at odds from what other philosophers consider to be the essence of structuralism, and it is not at all clear in which sense one can still talk about structuralism if the notion of collection—and together with it, the internal level of interest—is evacuated, ignored. For instance, Resnik goes as far as claiming:

> [P]ositing *mathematical objects* that are not themselves taken as positions in a pattern is to give up a basic structuralist thesis. (Resnik, op. cit., p. 205, author's emphasis.)

For Resnik, the primordial interest for a structuralist is to be found in what lies within the structure. And, if one is giving up collections, relations and any interest in the interior of the structure, what is there left of structuralism? What difference is there left to distinguish the structural method from the abstract method? I would say there is none and that Awodey's position is not a structuralist one.

Let me be clear: I am not criticizing Awodey for giving up the concepts of collection and relation. I am criticizing him for giving them up and, despite of it, claiming he holds a *structuralist* view of Mathematics. Rodin reaches a similar conclusion by distinguishing structural abstraction from *categorical* abstraction, which

> [...] forgets the fact that [these abstract mathematical entities] have elements and considers only how they map to (i.e., transform into) one another. (Rodin, op. cit., p. 35.)

In fact, if the aim is to embrace the whole of Mathematics, I agree with Awodey there are very good reasons to depart from those concepts—that simply means *there are very good reasons to abandon structuralism as a possible foundation for **all** of Mathematics*. Indeed, contemporary Mathematics has furnished objects that cannot be properly understood if they are conceived as 'sets plus extra structure'. In more technical terms, these correspond to the so-called *not-concretizable categories*. A famous example is given by homotopy types: the category *hTop*, with topological spaces as objects and homotopy classes of functions as morphisms, was proven to be non-concretizable by Peter Freyd in 1970 (cf. his article *"Homotopy is Not Concrete"*). Marquis, in his article *"Mathematical Forms and Forms of Mathematics: Leaving the Shores of Extensional Mathematics"*, discusses many other examples, such as ∞-categories and stacks.

As another witness that some parts of Mathematics are 'leaving the shores of structuralism', let me close this digression with a quote of Yuri Manin:

> [...] after Cantor and Bourbaki, no matter what we say, set theoretic mathematics resides in our brains. [...] I cannot do otherwise. If I'm thinking of something completely new, I say that it is a set with such-and-such a structure [...].
>
> But fundamental psychological changes also occur. Nowadays these changes take the form of complicated theories and theorems, through which it turns out that the place of old forms and structures, for example, the natural numbers, is taken by some geometric, right-brain objects.
>
> Instead of sets, clouds of discrete elements, we envisage some sorts of vague spaces, which can be very severely deformed, mapped one to another, and all the while the specific space is not important, but only the space up to deformation. If we really want to return to discrete objects, we see continuous components, the pieces whose form or even dimension does not matter. [...]
>
> I am pretty strongly convinced that there is an ongoing reversal in the collective consciousness of mathematicians: the right hemispherical and homotopical picture of the world becomes the basic intuition, and if you want to get a discrete set, then you pass to the set of connected components of a space defined only up to homotopy.
>
> That is, the Cantor points become continuous components, or attractors, and so on—almost from the start. Cantor's problems of the infinite recede to the background: from the very start, our images are so infinite that if you want to make something finite out of them, you must divide them by another infinity.
>
> (I. Gelfand. "We Do Not Choose Mathematics as Our Profession, It Chooses Us: Interview with Yuri Manin". Trans. by M. Saul. In: *Notices of the AMS* 56.10 (2009), pp. 1268–1274, p. 1274.)

### I.3.1.b   Eliminative vs. *ante rem* structuralism

An advantage of having already discussed the abstract method is that, since we now understand structuralism as a particular process of abstraction, the different approaches to structuralism may be seen as a consequence of the many views on abstraction.

- Eliminative structuralism

In particular, the main ideas of eliminative structuralism exactly correspond with the above account of abstraction by deletion (subsection I.2.2). According to this point of view, abstraction is but a method of generalization: to abstract is to leave unspecified. All mathematical entities are particular entities and an abstract entity is one for which some of the data has been omitted. This—we concluded—necessitates the choice of a background ontology, which determines what is to be meant by a particular object.

When applied to the particular case of mathematical structures, abstraction by deletion yields an account of eliminative structuralism which is very similar to the one provided by Shapiro, who starts by distinguishing two different perspectives one can adopt towards the 'things' constituting a structure—which he calls "places":

> There are, in effect, two different orientations involved in discussing structures and their places [...]. Sometimes the places of a structure are discussed in the context of one or more systems that exemplify the structure. Call this the *places-are-offices* perspective. This office orientation presupposes a background ontology that supplies objects that fill the places of the structures. [...]
>
> In contrast to this office orientation, there are contexts in which the places of a given structure are treated as objects in their own right, at least grammatically. That is, sometimes items that denote places are bona fide singular terms. [...] Call this the *places-are-objects* perspective. Here, the statements are about the respective structure as such, independent of any exemplifications it may have.[184]

---

[184]Shapiro, op. cit., pp. 82–83, author's emphasis.

and then adds:

> For the eliminativist, the surface grammar of places-are-objects statements does
> not reflect their underlying logical form, since, from that perspective, there are
> no structures and there are no places to which one can refer. [...] The eliminative
> structuralist holds that places-are-objects statements are just ways of expressing
> the relevant generalizations [...].[185]

Hence, "the eliminative structuralism program paraphrases places-are-objects state-
ments in terms of the places-are-offices perspective"[186].

A customary choice for the background ontology is to take an ontology of sets.
We then get the following standard account of structures as structured sets:

(i) All the mathematical objects considered are specific sets, explicitly constructed
    from the null set.

(ii) Given an object $E$, an *n-ary relation* $\mathcal{R}$ is simply a subset of $E^n$. Functions
    and $n$-ary operations are particular relations: a function $f : E \rightarrow E$ is a binary
    relation that is one-to-one in the first variable; an $n$-ary operation on $E$ is a
    function from $E^n$ to $E$, thus it is a particular kind of $(n+1)$-ary relation.

(iii) A *particular structure*, or system, is a set $E$ for which some relations $\mathcal{R}_1, \mathcal{R}_2, \ldots$
    have been defined.

(iv) Given two particular structures $(E, \mathcal{R})$ and $(E', \mathcal{R}')$, an *isomorphism* is a bijective
    function $\phi : E \rightarrow E'$ such that for any $e_1, e_2, \ldots$ elements of $E$, $\mathcal{R}(e_1, e_2, \ldots) \Leftrightarrow$
    $\mathcal{R}'(\phi(e_1), \phi(e_2), \ldots)$.

(v) An *abstract structure* is a particular structure for which the explicit construction
    has been omitted and only the relations have been retained.

To take an example, consider the following three structures:

– $\mathcal{S}_1 \equiv (E_1, \mathcal{R}_1)$, with $E_1 = \left\{ \emptyset, \{\emptyset\}, \{\{\emptyset\}\} \right\}$ and $\mathcal{R}_1 \subset E_1^2$ the binary relation
  defined by $\mathcal{R}_1 = \left\{ \big(\emptyset, \{\emptyset\}\big), \big(\{\emptyset\}, \{\{\emptyset\}\}\big), \big(\emptyset, \{\{\emptyset\}\}\big) \right\}$,

---

[185] Idem, "Mathematical Structuralism", p. 6.

[186] Idem, *Philosophy of Mathematics: Structure and Ontology*, p. 86.

– $\mathcal{S}_2 \equiv (E_2, \mathcal{R}_2)$, with $E_2 = \left\{ \emptyset, \{\emptyset\}, \{\emptyset, \{\emptyset\}\} \right\}$ and $\mathcal{R}_2 \subset E_2^2$ the binary relation defined by $\mathcal{R}_2 = \left\{ (\emptyset, \{\emptyset\}), (\{\emptyset\}, \{\emptyset, \{\emptyset\}\}), (\emptyset, \{\emptyset, \{\emptyset\}\}) \right\}$,

– $\mathcal{S} \equiv (E, \mathcal{R})$ with $E = \{e_0, e_1, e_2\}$ and $\mathcal{R} = \{(e_0, e_1), (e_1, e_2), (e_0, e_2)\}$.

All three structures are isomorphic (they can all be thought as describing an ordered set of three numbers, with the relation $\mathcal{R}(x, y)$ being conceived as "$y$ is greater than $x$"). All of them are particular structures but only for $\mathcal{S}_1$ and $\mathcal{S}_2$ all the information has been given. For $\mathcal{S}$, the explicit construction of the elements is not specified, and it is because this knowledge is missing that $\mathcal{S}$ appears to be an abstract structure. The distinction abstract/particular is not ontological but epistemological[187].

As Shapiro explains, one of the main differences between the places-are-offices perspective and the places-are-objects perspective lies in the way a statement like '$e_2$ is the biggest place of $\mathcal{S}$' is perceived. In the former, the copula *is* denotes a *predication*: the statement is expressing one property amongst many others $e_2$ may have. In the latter, the copula *is* denotes an *identity*: the statement defines $e_2$. It is in this sense that Shapiro says places are treated "grammatically" as objects[188].

However, as it is to be presently seen, the eliminative account is in fact a very mild version of mathematical structuralism, one which is not quite faithful to the main initial motivations. As we said earlier, the structuralist needs to articulate a clear conception of what the places of an abstract structure are. By the double move of first claiming there are only particular structures and then choosing a background ontology upon which to base the construction of all these structures, the eliminativist has addressed this issue directly. Indeed, the "things" constituting a given structure are here specific sets. But, through this, it is now possible to conceive the places *by themselves*, independently of the structures they may be part of. Returning to the above example, it is possible to study $e_2$ in isolation: as a place of the structure $\mathcal{S}$, $e_2$ acquires some structural properties (e.g., it is the greatest "thing" of $\mathcal{S}$), but it also has other properties which are completely independent of it being part of $\mathcal{S}$ (e.g., it makes sense to ask which is the cardinality of $e_2$). Therefore, in this view, the structuralist emphasis on relations is just a matter of interest. He cannot claim structural properties are all there is to an object: the latter may have—and indeed has!—non-structural, 'intrinsic' properties, but the structuralist takes the decision to focus on relations and not to care about the remainder properties—which he leaves unspecified.

---

[187]This view is very similar to that described by Hellman under the name of *Structuralism in Set Theory* (STS), although he does not emphasize the epistemological interpretation of it. See Hellman, op. cit., pp. 538–541.

[188]"It is common to distinguish the "is" of identity from the "is" of predication. The sentence "Cicero is Tully" does not have the same form as "Cicero is Roman". When in the places-are-objects perspective [...], we use the "is" of identity. We could just as well write "=" or "is identical to". In contrast, when we invoke the places-are-offices perspective [...], we use something like the "is" of predication [...]." (Shapiro, op. cit., p. 83.)

One may then wonder what knowledge of the individual places is gained through the study of the structural properties. And the answer is worrisome. On the one hand, any set whatsoever can be part of a particular structure which is isomorphic to $\mathcal{S}$. The statement "$e_2$ is the greatest place of $\mathcal{S}$" does not yield any information whatsoever about $e_2$, since $e_2$ could be any set. On the other hand, *the same* $e_2$ may be part of infinitely many other structures—among which, for instance, a structure $\mathcal{S}' \equiv (E, \mathcal{R}')$ with $\mathcal{R}' = \big\{(e_2, e_1), (e_1, e_0), (e_2, e_0)\big\}$, isomorphic to $\mathcal{S}$, and where $e_2$ is now the smallest place. The structural properties an object may acquire as a place of a structure are therefore completely extraneous to the object: they are accidental properties which reveal nothing of its essence.

After this remark, it is hard to still retain the interpretation of the structural statement "$e_2$ is the greatest place of $\mathcal{S}$" as a predication *about* $e_2$. Rather, since whatever $e_2$ turns out to be is completely irrelevant, one would like to interpret it now as a predication about *the structure*. In other words, to read the statement "$e_2$ is the greatest place of $\mathcal{S}$" as rather being "$\mathcal{S}$ has a greatest place, which we call $e_2$". But in this way the copula *is* becomes again a copula of identity and we seem to be back to the places-are-objects perspective. The background ontology appears now to be "an arbitrary and special fleshly clothing provided to pander to the need for intuitiveness"[189] which has nothing to do with the structure itself. There is then a strong temptation of developing a framework in which to reify abstract structures and set them free from the hypothetical and perhaps artificial background ontology.

---

[189]Schrödinger, op. cit., p. 58 (cf. footnote 65, page 39).

- The idea of an *ante rem* structure

To be sure, the eliminative move, tying all statements to an underlying ontology, appears to be at opposite ends from the philosophical motivations behind the structuralist approach. The fundamental problem of this move is that it allows to conceive the places of the particular structures by themselves, in isolation from all other places. The elements become ontologically prior to the structure, and this undermines much of the holistic ideas the structuralist wanted to emphasize. For Parsons, "reference to mathematical objects is always in the context of some background structure"[190]. Shapiro is also very clear about this point:

> For us [structuralists], a real number *is* a place in the real-number structure. It makes no sense to "postulate one real number", because each number is part of a large structure. It would be like trying to imagine a shortstop independent of an infield, or a piece that plays the role of the black queen's bishop independent of a chess game. [... I]t is nonsense to contemplate numbers independent of the structure they are part of.[191]

It thus becomes clear that the main goal of any non-eliminative structuralist is to articulate an account of the structural method of abstraction in which the structure is (onto)logically prior to the entities it contains. Such a conception of *ante rem* structures adopts a places-are-objects perspective: a structure is not a particular arrangement of objects which have an internal composition and can be conceived independently of the structure they are part of. Rather, as Resnik puts it, these objects within a structure "are structureless points [... which] have no identity or distinguishing features outside a structure"[192]. All their properties stem from the relations defining the given structure.

However, already at this first conceptive level of an *ante rem* account of the structural method of abstraction, one can clearly perceive some serious difficulties such an articulation will have to face. To eventually have a good conceptual hold of these sought for *sui generis* structures, I believe it is important to have these *prima facie* problems in mind. So, for the moment, let us give a succinct account of what these objections are, and expand on them later. We here closely follow the presentation of Hellman[193].

---

[190]Parsons, op. cit., p. 303.

[191]Shapiro, op. cit., p. 76.

[192]Resnik, op. cit., p. 201.

– *Identity of structural indiscernibles.* The idea that places of structures are to have no identity or distinguishing features outside a structure suggests some sort of Principle of Identity of Indiscernibles, where only relational properties are taken into account. Thus, "any items bearing exactly the same intrastructural relations to other items should be not many but one"[194]. But there are many familiar mathematical situations where this seems to be false. The complex numbers $i$ and $-i$ inside the structure $\mathbb{C}$, or any two points of the Euclidean space are the most famous examples. This can be seen as a proof that *ante rem* structures fail to describe the objects they were designed for.

– *Ontological priority of relations over relata.* That structures are to be ontologically prior to the elements constituting it seems to necessitate that relations are prior to relata. But "a relation presupposes its relata"[195] and hence the "notion of an *ante rem* structure seems to involve a vicious circularity"[196]. This is thus a threat to the fundamental basic notion characteristic of Sui Generis Structuralism.

– *Multiple reductions.* One of the main goals of ante rem structuralism is to make sense of the discourse referring to, say, *the* infinite-dimensional separable Hilbert space, in such a way that the explanation is consistent with the use of the definite article "the". However, given an *ante rem* structure $(\mathcal{S}, \mathcal{R})$, it seems that, by simply permuting the places of the structure and redefining the relations, one can define *another* structure $(\mathcal{S}', \mathcal{R}')$ that is just as valid a candidate for being *the* structure.

Strong physico-mathematical and philosophical reasons tempted us towards the idea of an abstract structure as an independent entity. Yet, some conceptual difficulties have immediately appeared. Any sound version of *sui generis* structures must give clear answers to these three objections. Now, among the different points raised by Hellman, the one which questions the notion of identity within an abstract structure appears to be most relevant to our inquiries, for it is directly related to the requirement of individuation (page 17). Indeed, in the mathematical description of classical and quantum mechanical systems, points of an abstract symplectic manifold are used to describe states of a classical system, rays of an abstract Hilbert space represent states of a quantum system, elements of an abstract $C^*$-algebra refer to properties, etc. But if these 'things' that constitute an abstract structure "have no intrinsic nature" and are "structureless points with no identity or distinguishing features outside a structure" (Resnik), how does one manage to identify the specific 'thing' which describes this particular state or that given property?

---

[193]Hellman, op. cit., pp. 544–546.

[194]Ibid., p. 544.

[195]Awodey, op. cit., p. 61.

[196]Hellman, op. cit., p. 545.

For our investigation, it is therefore essential that we turn towards reflecting on the interior of these mathematical structures and that we articulate a way of thinking about the 'things' that constitute them.

## I.3.2 Identity within an abstract mathematical structure

### I.3.2.a The problem of the identity of indiscernibles

The problem of the identity of structural indiscernibles has raised an important debate in the last decade. It is usually attributed to Jukka Keränen[197] and John Burgess[198] who, in two independent papers, pointed to this fundamental objection against non-eliminative structuralism. In the philosophy of physics literature, specially when discussing the nature of space-time, this problem is sometimes called the "abysmal embarrassment argument" in reference to the critique raised by Christian Wüthrich[199].

---

[197] J. Keränen. "The Identity Problem for Realist Structuralism". In: *Philosophia Mathematica* 9.3 (2001), pp. 308–330.

[198] Burgess, op. cit.

[199] C. Wüthrich. "Challenging the Spacetime Structuralist". In: *Philosophy of Science* 76 (2010), pp. 1039–1051.

Keränen's own exposition of the problem is extremely clear, so I will just describe the basic steps of his argument.

First. The structuralist "must furnish an account of the identity for places"[200]. This follows simply from the fact that "within a given theory, language, or framework, there should be a definite criteria for identity among its objects [... and t]here is no reason for structuralism to be the single exception to this"[201], as Shapiro acknowledges. Such an account amounts to completing the following 'identity schema':

$$\text{for any } x, y \text{ places of } \mathcal{S}, \ (x = y \iff \text{----})$$

Second. Keränen explains there are two ways of completing the identity schema: "the account of identity will be either a general-property account or haecceity account"[202]. To understand the difference between both options, it is necessary to grasp the concept of haecceity, or *primitive thisness*. This is nicely explained in the paper *"Primitive Thisness and Primitive Identity"* of Robert Adams:

> Intended to be a synonym or translation of the traditional term "haecceity", [...] a thisness is the property of being identical with a certain particular individual.[203]

Hence, some thing has a thisness as long as it is an individual. Intuitively, this means that the thing has a property that allows one to *point* at it in a precise way—to say, in a meaningful way, '*this* thing'. On the other hand,

> a property is purely qualitative—a suchness—if and only if it could be expressed, in a language sufficiently rich, without the aid of such referential devices as proper names, proper adjectives and verbs (such as 'Leibnizian' and 'pegasizes'), indexical expressions, and referential uses of definite descriptions.[204]

Now, the thisness of an object may or may not be reducible to a set of qualitative properties. When it is not the case, one talks about a '*primitive thisness*'. This is, I believe, the distinction Keränen has in mind: if one explains the identity of an object by appealing to its primitive thisness, then one is using a haecceity account of identity; if one explains the identity through the sole use of qualitative properties, then it is a general-property account of identity. But for the non-eliminative structuralist, places "have no more to them than can be expressed in terms of the basic relations of the structure"[205]. Hence, "any haecceity account directly conflicts with the spirit and motivations of realist structuralism"[206] and we are left only with the second option.

---

[200]Keränen, op. cit., p. 314.

[201]Shapiro, op. cit., p. 92, cited in Keränen, loc. cit.

[202]Ibid., p. 313.

[203]R. M. Adams. "Primitive Thisness and Primitive Identity". In: *The Journal of Philosophy* (1979), pp. 5–26, p. 6.

[204]Ibid., p. 7.

Third. At this point, the identity schema the non-eliminative stucturalist needs to provide appears to be of the following form:

$$\text{for any } x, y \text{ places of } \mathcal{S}, \ \big(x = y \Longleftrightarrow (\text{for any property } P, \ P(x) \Leftrightarrow P(y))\big).$$

But, as Keränen explains, "we need to be careful about which properties we admit"[207]. The question becomes then to determine precisely which qualitative properties the non-eliminative structuralist is allowed to use. To this, Keränen answers: "the places of the structure $\mathcal{S}$ must be individuated by properties that are *invariant under the automorphisms of* $\mathcal{S}$"[208].

Given this three steps, one finally arrives at the identity problem:

> We are now ready to state the identity problem. Given a structure $\mathcal{S}$, the schema
> says that two singular terms denoting places of $\mathcal{S}$ denote the same place precisely

---

[205]Parsons, loc. cit. Cf. also Shapiro's slogan: "There is no more to the individual numbers "in themselves" than the relations they bear to each other" (Shapiro, op. cit., p. 73).

[206]Keränen, op. cit., p. 314.

[207]Ibid., p. 316.

[208]Ibid., p. 318. In fact, this is not the way Keränen first characterizes the grammatically correct properties. He says:

> We maintain that there are two crucial constraints [about which properties we admit]:
> (1) No property the specification of which essentially involves an individual constant denoting an element in $\mathcal{S}$ may be admitted. [...]
> (2) No property the specification of which essentially involves an individual constant denoting an element in $\mathcal{S}$ may be admitted. [...]
> In sum, only the properties that can be specified by formulae in one free variable and without individual constants may be admitted. (Ibid., pp. 316–317)

(Here $\mathcal{S}$ is the abstract structure and $\mathcal{S}$ is any system exemplifying it.) The author then proves that any such properties are necessarily invariant under automorphisms and that any invariant property satisfies the above two crucial constraints.

> when their referents have the same intra-structural relational properties that can
> be specified without using individual constants. [...] The problem is that it
> does not at all square with the use of the identity predicate in mathematical
> discourse. For example, since 1 and −1 in any system $(Z, +)$ have the same
> intra-systemic relational properties, the realist structuralist must view '**1**' and
> '**−1**' in the language of the structure $(\mathbf{Z}, +)$ as *co-referential* terms.[209]

The structuralist must conclude that 1 and −1 are equal. In the same way, he must
conclude that $i$ and $-i$ are not many but one, and that the Euclidean space contains
only one point, not infinitely many. These conclusions are clearly absurd—Wüthrich
would say: they are an abysmal embarrassment!

The reader may perhaps wonder how Keränen can be so confident about the fact
that 1 and −1 have the same relational properties within $(\mathbb{Z}, +)$. However, with the
notions of automorphism at our disposal, it is not hard to be convinced that this is
indeed so. For consider the general case of a structure $\mathcal{S}$ and two places $x$ and $y$ such
that there exists an automorphism $\phi : \mathcal{S} \to \mathcal{S}$ relating them—i.e., $\phi(x) = y$. Since any
qualitative property is invariant under automorphisms, we have, for any property $P$: if
$P(x)$ then $P(y)$. So, if the *ante rem* structuralist accepts the identity schema suggested
by Keränen, he must indeed conclude that any two places related by an automorphism
are equal. In other terms, Keränen's Principle of Identity entails a structure should not
admit any automorphisms besides the trivial one. Thus, the identity problem arises for
any non-rigid structure[210]. In the case of the structure $(\mathbb{Z}, +)$, the transformation $\phi$
defined by $\phi(z) = -z$ for any place $z$ of $\mathbb{Z}$ is a non-trivial automorphism (this is the case
for any non-trivial abelian group). Notice that this identity problem applies precisely
to those mathematical structures involved in the Mathematics of Mechanics: groups,
Hilbert spaces and symplectic manifolds admit many non-trivial automorphisms! In
sum, if he accepts the three steps of Keränen's argument, the structuralist should
indeed be embarrassed.

---

[209]Ibid., p. 317. Again, bold characters refer to the structure whereas plain characters refer to the systems.

[210]A structure $\mathcal{S}$ is said to be *rigid* if it its group of automorphisms $Aut(\mathcal{S})$ is trivial. $\mathbb{R}$, seen as *topological* field, is an example of such a rigid structure.

### I.3.2.b First attempts at a solution

So much for the identity problem. Let us now look at how the *ante rem* structuralist may answer this objection. Certainly, all qualitative properties should be invariant under automorphisms. I know of no structuralist that rejects this point, which has been stressed many times: it can already be found in Rudolf Carnap

> The structural properties are so to speak the invariants under isomorphic transformation.[211]

in Hermann Weyl

> A point relation is said to be objective if it is invariant with respect to every automorphism.[212]

or, to take a recent example, it can also be found in F.A. Muller

> [...] structuralism should be taken to include that all and only automorphic subsets represent properties.[213]

From our perspective, since the structural method is a particular case of the abstract method, this constraint of invariance under automorphism is simply the restatement of the Principle of Abstraction (cf. page 88): the allowed, grammatically correct, qualitative properties are those properties invariant under isomorphisms. Therefore, if

---

[211] R. Carnap. *Untersuchungen zur allgemeinen Axiomatik*. Darmstadt: Wissenschaftliche Buchgesellschaft, 2000, p. 74 (cited in G. Schiemer and J. Korbmacher. "What Are Structural Properties?" Preprint available at `http://www.jkorbmacher.com/`, p. 8).

[212] This is taken from the passage where Weyl describes the problem of relativity. An extended version of the quote would be

> Our knowledge stands under the norm of *objectivity*. He who believes in Euclidean geometry will say that all points in space are objectively alike, and that so are all possible directions. [...] Whereas the philosophical question of objectivity is not easy to answer in a clear and definite fashion, we know exactly what the adequate mathematical concepts are for the formulation of this idea. [...] An automorphism is a one-to-one mapping [...] which leaves the basic relations undisturbed. [...] A point relation is said to be objective if it is invariant with respect to every automorphism. In this sense, the basic relations are objective, and so is any relation defined in terms of them.
>
> (Weyl, op. cit., pp. 71–73.)

Replace "objective" by "structural" and you get a modern version of structuralism!

[213] F. A. Muller. "How to Defeat Wüthrich's Abysmal Embarrassment Argument against Space-Time Structuralism". In: *Philosophy of Science* 78.5 (2011), pp. 1046–1057, p. 1051.

the first two steps of the above argument are granted, the problem cannot be avoided.

One first strategy for the structuralist seems to be the rejection of the *dichotomy* presented by Keränen: identity need not be grounded on primitive thisness or on qualitative properties, understood as formulae in *one* free variable. For MacBride, this dichotomy is "the most questionable feature of the argument"[214]. To overcome the problem, the structuralist may try to furnish a third possible manner of completing the identity schema. Now, since the whole point of structuralism is to put the emphasis on *relations* between the places of a structure, the structuralist has a very natural place where to start looking for a third alternative account of identity.

This is indeed the strategy followed by James Ladyman in his first attempt to overcome the identity problem. In his short article *"Mathematical Structuralism and The Identity of Indiscernibles"*, the author reactivates Quine's distinction between three different levels of discernibility. Two objects are said to be:

– *absolutely discernible* if there exists a one-place predicate that is true of one object but not of the other,

– *relatively discernible* if there exists a two-place relation that is true of them in one order but not in the other,

– *weakly discernible* if there exists a two-place relation, irreflexive for the pair and that is true of them[215].

Keränen considers only formulae in one variable and is therefore building his identity schema based on absolute discernibility. By "demanding only weak, and not strong or relative, discernibility of numerically distinct individuals"[216], the *ante rem* structuralist can expect to solve the identity problem. And this is indeed the case for all the examples discussed above. Therein the wished-for relation that is irreflexive for the problematic pairs can readily be found: for the structure $(\mathbb{Z}, +)$, one can choose the relation $\mathcal{R}(x, y) \equiv$ '$x$ is the additive inverse of $y$' ($-1$ is the inverse of 1, but 1 is not its own inverse); this same relation allows to distinguish $i$ from $-i$ in the complex field structure; and the Euclidean distance allows to distinguish any two points on the space[217].

---

[214]"The most questionable feature of [Keränen's] argument is its most basic assumption, the thesis that necessary and sufficient conditions for the identity of objects can and should be states in exclusively property-theoretic terms." (F. MacBride. "Structuralism Reconsidered". In: *The Oxford Handbook of Philosophy of Mathematics and Logic.* Ed. by S. Shapiro. New York: Oxford University Press, 2005, pp. 563–589, p. 582.)

[215]In *Word and Object*, Quine only introduced the distinction between absolute and relative discernibility. The third term was introduced later, in "Grades of Discriminability", but there he changed "relative discernibility" into "moderate discernibility". However, Ladyman seems to be following the terminology adopted by Simon Saunders (S. Saunders. "Physics and Leibniz's Principles". In: *Symmetries in Physics: Philosophical Reflections.* Ed. by K. Brading and E. Castellani. Cambridge University Press, 2003, pp. 289–308).

The introduction of the symmetric/asymmetric and reflexive/irreflexive distinctions among relations may appear as a rather *ad hoc* move from the structuralist who is trying to overcome the identity problem. One can attempt to avoid any appeal to this distinction by adopting a strategy slightly different from Ladyman's: to treat all relations indistinctly and build the identity schema from *all* of them. Indeed, one can propose the following:

**Relational Principle of Identity**. Given a structure $\mathcal{S}$, for any $x, y$ places of $\mathcal{S}$, $x$ is identical to $y$ if and only if, for any $n$-ary relation $\mathcal{R}$ defined on $\mathcal{S}$ and any $z_1, \ldots, z_{n-1}$ places of $\mathcal{S}$, we have:

$$
\begin{cases}
(x, z_1, \ldots, z_{n-1}) \in \mathcal{R} \longleftrightarrow (y, z_1, \ldots, z_{n-1}) \in \mathcal{R} \\
(z_1, x, \ldots, z_{n-1}) \in \mathcal{R} \longleftrightarrow (z_1, y, \ldots, z_{n-1}) \in \mathcal{R} \\
\quad \vdots \\
(z_1, \ldots, z_{n-1}, x) \in \mathcal{R} \longleftrightarrow (z_1, \ldots, z_{n-1}, y) \in \mathcal{R}
\end{cases}
$$

If no relation can perceive the difference between two places—in other terms, if two places are *relationally indiscernible*—then, these places must not be many but one. This certainly appears to be a faithful implementation of the structuralist intuition that there is no more to the places than the relations they bear to each other. This relational account of identity within a structure is the precise answer given by Muller

[216] J. Ladyman. "Mathematical Structuralism and The Identity of Indiscernibles". In: *Analysis* 65.3 (2005), pp. 218–221, p. 220.

[217] Ibid., p. 220.

in the context of spacetime structuralism[218].

Again, with this choice of identity schema, Keränen's objection no longer holds in the general case. Given $x$ and $y$ places of $\mathcal{S}$ related by an automorphism $\phi$, one can no longer conclude $x = y$ from the data $y = \phi(x)$. Indeed, grammatically correct relations are certainly also invariant under automorphisms, but this now means

$$(x, z_1, \ldots, z_{n-1}) \in \mathcal{R} \leftrightarrow (\phi(x) = y, \phi(z_1), \ldots, \phi(z_{n-1})) \in \mathcal{R}$$

and not at all

$$(x, z_1, \ldots, z_{n-1}) \in \mathcal{R} \leftrightarrow (\phi(x) = y, z_1, \ldots, z_{n-1}) \in \mathcal{R}$$

It is thus a priori possible to find a relation $\mathcal{R}$ such that $(x, z_1, \ldots, z_{n-1}) \in \mathcal{R}$ and yet $(y, z_1, \ldots, z_{n-1}) \notin \mathcal{R}$. If such a relation exists, one is forced to conclude $x \neq y$, unless one is willing to abandon the usual axiom of substitution of identicals (sometimes also called Principle of Indiscernibility of Identicals).

Ladyman's and Muller's approaches have in common the fundamental idea of building identity based on relations, not on properties. There are however some important differences as well. The latter seems to be more general than the former: for all cases where weak discernibility applies, Muller's relational principle of identity will do the work as well, but it is not obvious that the converse is also true. Despite this, there is at least one reason why one could prefer to stick to the Ladyman-Quine strategy: the introduction of the two distinctions among relations, which are absent in Muller's account, allows Ladyman to keep the question of the discernibility of two places a *local* matter. Given the places $x$ and $y$, one can consider them *in isolation* and answer the question of their identity without having to ever consider the remainder places of the structure. This is far from true with Muller's notion of relational indiscernibles—which does not coincide with neither relative nor weak indiscernibles. As it is easy to see, with the Relational Principle of Identity, the identity of $x$ and $y$ involves *all* other places of the structure. Thus, Muller's identity scheme is explicitly *holistic*; it involves the whole structure! But, from the structuralist perspective, this holism should not be seen as a drawback. On the contrary, it could be perceived as a welcomed feature, since—as I explained in I.3.1.a, page 97—the local/global two-fold level of study is characteristic of the structural method. After all, Shapiro does state that "an *ante-rem* structure is a *whole* consisting of, or constituted by, its places and relations"[219].

---

[218]Muller, op. cit., p. 1054. I have generalized slightly Muller's account. The author only considers the $n = 1$ and $n = 2$ cases (that is, predicates and binary relations).

Either way, by considering not only properties but also relations as a means to discern between two places, the *ante rem* structuralist seems to have at his disposal a satisfactory and general answer to the objection. The situation is however more involved and the structuralist cannot escape the identity problem so lightly. One first way to see this is to notice the existence of mathematical objects which fail to meet both the weak version of the Principle of Indiscernibles and the Relational Principle of Identity! Button finds in graph theory two such examples[220]:

– **G1:** $b \longleftarrow a \longrightarrow c$

– **G2:** $\circlearrowright b \longleftrightarrow c \circlearrowleft$

Therein, any relation that holds of $(b, b)$ will be satisfied as well by $(b, c), (c, b), (c, c)$, and there is hence no hope of finding an (irreflexive) relation distinguishing $b$ from $c$. Leitgeb, Ladyman and Shapiro discuss very similar examples[221].

These examples seem to condemn the idea of accounting for the identity of places within a structure solely in terms of intra-structure relations and with no appeal to primitive identity. But one can still try to "save" some version of *ante rem* structuralism from accepting primitive identity facts by rejecting that Button's examples be named structures. That some mathematical objects should not be considered structures is not an option for many philosophers endorsing mathematical structuralism. This is because they regard mathematical structuralism as an attempt to build a *foundation* for the

---

[219]Shapiro, "Mathematical Structuralism", p. 2, my emphasis.

[220]T. Button. "Realistic Structuralism's Identity Crisis: A Hybrid Solution". In: *Analysis* 66 (2006), pp. 216–222, p. 218.

[221]Button's graph **G2** is very similar to the unlabelled graph with two nodes and no edges considered by Leitgeb and Ladyman (H. Leitgeb and J. Ladyman. "Criteria of Identity and Structuralist Ontology". In: *Philosophia Mathematica* 16.3 (2008), pp. 388–396). This, in turn, corresponds precisely to the 'finite cardinal patterns' considered by Shapiro (Shapiro, *Philosophy of Mathematics: Structure and Ontology*, p. 115).

*whole* of Mathematics. However, as I have emphasized several times already, these foundational aspirations are extraneous to our purpose. We are trying to learn how to conceive abstract mathematical structures because some of them play a fundamental role in the foundations of Mechanics, and the question of whether all mathematical objects are structures is of no matter to us[222]. Therefore, one could attempt to define a structure as a collection of places and relations for which the Relational Principle of Identity (or the Principle of Identity for weak Indiscernibles) holds. This move seems to be alright as long as only artificially constructed objects—like the graphs **G1** and **G2**—are denied the status of abstract structures[223].

However, even these precautions do not suffice. As Shapiro remarks, it is not clear in which way some of the relations used to discern two places are different from a brute non-identity relation. And this is a real worry for this strategy:

> [...] if non-identity does count as an irreflexive relation for these metaphysical purposes, then the distinguishing task is trivial, and unilluminating. The thesis is just that distinct objects must be distinct. Notice that identity, or non-identity, is presupposed in the very formulation of some of the requirements and the examples.[224]

The problem is very well perceived if one considers vector spaces. Indeed, let $V$ be an

---

[222]In fact, I believe the foundational aspiration of structuralism to be hopeless. To me, it has been outdated by many of the objects pure mathematics has introduced in the last sixty years which do not have an underlying set.

[223]Essentially, this is Button's proposal. Structures satisfying the identity of weak indiscernibles, which he calls "basic structures" are to be interpreted realistically; those structures which fail to meet such an identity criterion, called "constructed structures" are treated eliminativistically (Button, op. cit., p. 220). Nonetheless, this way out of the problem seems dubious. As noted by Ladyman,

> "Graphs such as $G'$ [not satisfying weak discernibility] are not exceptional; all other unlabelled graphs that contain at least two isolated nodes (for example, 11 out of the 156 possible unlabelled graphs with 6 nodes) include nodes that are not even weakly discernible. Furthermore, an analogous point can be made about all unlabelled graphs which include at least two distinct but isomorphic and unconnected components."

> (J. Ladyman. "Scientific Structuralism: On The Identity and Diversity of Objects in a Structure". In: *Aristotelian Society Supplementary Volume*. Vol. 81. 1. Wiley Online Library. 2007, pp. 23–43, p. 35.)

Thus, Button's proposal seems to exclude all graph theory from an *ante rem* structural account.

[224]S. Shapiro. "Identity, Indiscernibility, and *ante rem* Structuralism: The Tale of $i$ and $-i$". In: *Philosophia Mathematica* 16.3 (2008), pp. 285–309, footnote 2, p. 288.

abstract vector space and $\Psi_1, \Psi_2$ two distinct vectors. If one uses Ladyman's identity schema, what is the irreflexive relation distinguishing them? I claim it would have to be something of the sort "$-\Psi_2$ is not the additive inverse of $\Psi_1$". Of course, this can also be stated as "the additive inverse of $\Psi_2$ is not the additive inverse of $\Psi_1$", which also means "$\Psi_2$ is not $\Psi_1$". Symbolically, the argument is to conclude that $\Psi_1 \neq \Psi_2$ *because* $\Psi_1 - \Psi_2 \neq 0$[225]. It can hardly get less illuminating than that...

On the other hand, Muller's Relational Principle of Identity runs into similar difficulties. To discern the two vectors $\Psi_1$ and $\Psi_2$, it would be enough to find any (not necessary irreflexive) binary relation $\mathcal{R}$ for which there exists a *third* vector $\Psi_3$ such that $(\Psi_1, \Psi_3) \in \mathcal{R}$ and $(\Psi_2, \Psi_3) \notin \mathcal{R}$[226]. Now, the existence of this sought-for $\Psi_3$ would have to be proven by a *demonstrative act*: it would have to be exhibited, explicitly constructed. This third vector cannot not be proven to exist *in principle*, for that would presuppose we already knew $\Psi_1 \neq \Psi_2$. But it is hard to see how one could effectively exhibit one particular element of an *abstract* structure, which, by its unspecified, freestanding nature cannot be laid 'in front of our eyes'.

### I.3.2.c  The solution: primitive *typed* identity

All in all, it very well seems the *ante rem* structuralist cannot avoid committing to some primitive identity facts. This conclusion is explicitly endorsed by Ladyman and Leitgeb:

> [...] the identity relation for positions in a structure is a relation that ought to be viewed as an integral component of a structure in the same way as, for example, the successor relation is an integral component of the structure of natural numbers. [...]

---

[225] The situation is exactly the same for the important example of Hilbert spaces. Given an abstract Hilbert space $\mathcal{H}$ and two elements $\Psi_1, \Psi_2$ of same norm, one would mimic the case of the Euclidean space and use the metric relation to distinguish them: $d(\Psi_1, \Psi_2) \neq 0$, and this seems alright. But the notation is hiding the triviality of such a statement, since $d(\Psi_1, \Psi_2) = ||\Psi_1 - \Psi_2||$ and thus $d(\Psi_1, \Psi_2) \neq 0 \iff \Psi_1 - \Psi_2 \neq 0$.

[226] In his article, when discussing the identity of space-time points, Muller gives the following example: consider the light cone relation $\mathcal{R}(x, r) \equiv$ "$r$ lies inside the light-cone of $x$". Then, if for every point $r$ we have both $\mathcal{R}(x, r)$ and $\mathcal{R}(y, r)$, we can conclude that $x = y$. (Muller, op. cit., p. 1056).

> The fact that [the graph] **G2** consists of precisely two nodes is simply part
> of what **G2** is; it is 'built into' its graph-theoretic structure. Adapting the
> structuralist slogan on natural numbers [...], we are still allowed to say that
> 'There is no more to the individual nodes "in themselves" than the relations
> they bear to each other', the only addition that we have to make is that we have
> to count identity and difference of nodes among the very relations that the nodes
> in a graph bear to each other.[227]

This is also the position adopted by Shapiro, who "wholeheartedly rejects the identity
of indiscernibles"[228]. When discussing his 'finite cardinal structures'—which are finite
abstract sets—he says:

> The cardinal-four structure is the worst offender of (IND) [absolute discernibility]
> possible. Since there are no relations to preserve, every bijection of the domain is
> an automorphism. Each of the four places is structurally indiscernible from the
> others and yet, by definition, there are four such places, and so not just one.[229]

It is question-begging to demand the structuralist to justify why a graph with two nodes
has two nodes and not three nor one. *The cardinality of a structure is an information
given a priori, not an information one acquires a posteriori.* A graph with two nodes
has two nodes *by definition*.

It is easy to see this idea of a primitive cardinality is indeed faithful to the way
mathematicians work. As Leitgeb and Ladyman emphasize[230], in graph theory one
never asks the question of how many nodes a given graph $G$ has. Rather, the pertinent
question is to find how many different graphs with a given number of nodes there are.
Moreover, almost any description of an abstract structure includes an axiom about
its cardinality. To give some examples, to uniquely characterize the field of complex
numbers, one has to describe it as an algebraically closed field of characteristic zero *and
of cardinality the continuum*. If this choice of cardinality is not made, then the structure is not fixed[231]. Second, in his *Mathematical Foundations of Quantum Mechanics*, when von Neumann defines Hilbert spaces, he includes an axiom about the cardinality of a family of linearly independent vectors, besides the axioms of a vector space and of a Hermitian inner product:

> The properties **A.**, **B.** [of linearity and Hermitian product] permit us, as we see, to state a great deal about $\mathcal{R}$ [an abstract Hilbert space], yet *they are not sufficient to enable us to distinguish* the $\mathcal{R}_n$ from each other and from $\mathcal{R}_\infty$. This concept is clearly associated with the maximum number of linearly independent vectors. If $n = 0, 1, 2, \ldots$ is such a maximum, then we may state for this $n$:
>
> – [Axiom] **C.**$^{(n)}$ There are exactly $n$ linearly independent vectors. [...]
>   If there exists no maximum number, then we have:
>
> – [Axiom] **C.**$^{(\infty)}$ There are arbitrarily many linearly independent vectors.
>
> [...] We obtain a different space $\mathcal{R}$, depending on which we *decide* upon.[232]

---

[227]Leitgeb and Ladyman, op. cit., pp. 390 and 392–393.

[228]Shapiro, op. cit., p. 292.

[229]Ibid., p. 287.

[230]Leitgeb and Ladyman, op. cit., p. 392.

Von Neumann could not be more transparent: the structural relations retained in the process of abstraction are not enough to force upon us the number of places of the structure. This cardinality can only be fixed by a decision we make.

In the light of all this, we conclude: *Identity within an abstract structure is primitive*: it cannot be grounded on structural properties nor relations. This the structuralist can no longer deny. The worry is to understand whether this return of primitive identity within a structure reduces the structuralist identity schema to a haecceity account. If, indeed, the identity of the places of a structure is to be grounded on a primitive *intrinsic* self-identity, the whole *ante rem* perspective on abstract structures would be undermined. For it then would make perfect sense to consider abstract places in isolation—and this would be to abandon one of the main tenets of structuralism: places are nothing in themselves. Recall Shapiro: "It makes no sense to postulate *one* real

---

[231] This follows from Steinitz's theorem which proves that, for every characteristic $p \geqslant 0$ and uncountable cardinal $\kappa$, there is, up to isomorphism, exactly one algebraically closed field of characteristic $p$ and cardinality $\kappa$ (E. Steinitz. "Algebraische Theorie der Körper". In: *Journal für die reine und angewandte Mathematik* 137 (1910), pp. 167–309).

[232] Von Neumann, op. cit., p. 45, my emphasis.

number"[233]. Fraser MacBride encapsulates well the danger the structuralist is facing:

> [The places of a structure] cannot simply be bundles of structural relations; they
> are a separate, irreducible category of existent. So the structuralist must admit
> (at least) a two-category ontology of objects and relations. The failure of property
> reductionism indicates that mathematical objects [here: places of a structure] are
> also the bearers of properties and relations that take them outside their parent
> structure [...].[234]

If the last sentence is true, then *sui generis* structuralists need to postulate some
background ontology of abstract objects from which structures will be constituted.
But this was precisely what *ante rem* structuralism was trying to avoid in the first
place!

Luckily, the return to primitive identity within structures need not mean the
appeal to haecceity. Consider an abstract structure $\mathcal{S}$ and two places $x, y$. To ground
identity on haecceity means to claim:

1. that there exists a grammatically correct property $H_x \equiv$ 'being identical with $x$',
2. that $x$ is different from $y$ *because* $H_x(x)$ is true and $H_x(y)$ is false.

This move certainly goes against structuralism but also conflicts with the Principle
of Abstraction, for the haecceity $H_x$ is in general not invariant under automorphisms.
However, this is not what *ante rem* structuralists are committed to. Rather, their claim
is:

i) that there exists a basic binary relation on $\mathcal{S}$ 'being identical with' and denoted
   $=_S$ (in other words, such that $z_1 =_S z_2 \iff$ '$z_1$ is identical with $z_2$'),
ii) that $x$ is different from $y$ *because* $x =_S y$ is false.

At first sight, this may appear to be a trivial restatement of the haecceity account,
but in fact it is not. The first thing to notice is that now the conflict with the Principle
of Abstraction has evaporated: $=_S$ is manifestly invariant under any automorphism
and is hence a grammatically correct relation. But to stop at this remark would be to
miss the really crucial point. What makes the structural account of identity radically different from a haecceity account is that this primitive identity $=_S$ is a relation that *only makes sense within the given structure*. Whereas a haecceity account of identity may suggest—to say the least—the existence of an independent entity named $x$, the structuralist once more insists on the importance of remembering the *context* inside which the identity claims are being made. Identity is primitive, but identity is also contextual. Therefore, the *ante rem* structuralist escapes MacBride's conclusion that 'places can be taken outside of their structure'. Ladyman rightly insists on this point, by distinguishing intrinsic and contextual individuality—distinction which he borrows from Stachel[235]:

> [...] primitive contextual individuality is different to primitive intrinsic individuality [...], for only the latter and not the former implies haecceitism. If individuation is intrinsic, and not grounded in qualitative properties but is either ungrounded or grounded in haecceities, then the identity of an individual objects is determinate in other counterfactual situations [...]. On the other hand, *if individuality is contextual then there is in general no reason to regard talk of the same object in another relational structure as intelligible.*[236]

---

[233]Shapiro, *Philosophy of Mathematics: Structure and Ontology*, p. 76, my emphasis.

[234]MacBride, op. cit., p. 584.

In sum, we have arrived to the following conclusion:

> **Identity within a structure:** For an *ante rem* structuralist, identity of places within an abstract structure is primitive. But it is not *absolute*, unrestricted primitive identity; rather, it is primitive *typed* identity.

## I.3.3 Individuation within an abstract mathematical structure

The previous section considered the question of what grounds identity within an abstract mathematical structure, and the conclusion was that, in fact, this internal identity is in general *ungrounded*: given places $s_1, s_2$ of an abstract structure $\mathcal{S}$, their

---

[235]J. Stachel. "Structural Realism and Contextual Individuality". In: *Hilary Putnam.* Ed. by Ben-Menahem. Cambridge: Cambridge University Press, 2005, pp. 203–219.

[236]Ladyman, op. cit., p. 37, my emphasis.

*difference* is a primitive statement which is *stipulated* and needs not be justified. Yet, the existence of a primitive typed identity does not render superfluous all discussion about discernibility within an abstract structure. It still remains an important question to understand precisely which are the available *descriptive resources* inside a given abstract structure. For indeed our interest lies not so much in determining whether two places ares different but, foremost, in determining whether a particular place (describing a specific state or a given property of the system) can be objectively singled out among all abstract places. In other words, we ask about the possibilities of *individuation* within a structure, or, in Hermann Weyl's more elegant language, about the possibility of a "conceptual fixation of points [...] that would enable one to reconstruct any point when it has been lost"[237].

To capture precisely what is at stake, we need to take a small detour and fully pursue the consequences of what emerged in the last few pages.

### I.3.3.a  Abstract structures as structured types

The need for *ante rem* structuralists to appeal to a primitive *typed* identity points to the crucial idea that type theory, and not set theory, is the natural home for conceiving *abstract* structures. Once this has been hinted at, it may appear as a blunder not having considered it from the start. For the tokens-to-type relation has the perfect characteristics to capture exactly the places-to-structure relation that *ante rem* structuralist advocates. To understand why this is so, it is important to recall the main conceptual differences between set theory and type theory.

As the Univalent Foundations Program explains in the introduction of its book, "[o]ne problem in understanding type theory from a mathematical point of view, however, has always been that the basic concept of *type* is unlike that of *set* in ways that have been hard to make precise"[238]. Nonetheless, some pages later they describe a first fundamental difference:

---

[237]Weyl, op. cit., p. 75.

[238]The Univalent Foundations Program. *Homotopy Type Theory: Univalent Foundations of Mathematics*. Institute for Advanced Study: http://homotopytypetheory.org/book, 2013, p. 2, authors' emphasis.

> [...] if the type $A$ is being treated more like a set than like a proposition [...], then
> "$a : A$" may be regarded as analogous to the set-theoretic statement "$a \in A$".
> However, there is an essential difference in that "$a : A$" is a *judgment* whereas
> "$a \in A$"" is a *proposition.* [...]
>
> A good way to think about this is that in set theory, "membership" is
> a relation which may or may not hold between two pre-existing objects "$a$"
> and "$A$", while **in type theory we cannot talk about an element "$a$" in
> isolation: every element by *its very nature* is an element of some type,**
> and that type is (generally speaking) uniquely determined. Thus, when we say
> informally "let $x$ be a natural number", in set theory this is shorthand for "let
> $x$ be a thing and assume that $x \in \mathbb{N}$", whereas in type theory "let $x : \mathbb{N}$ is an
> atomic statement: we cannot introduce a variable without specifying its type.**[239]**

The relation between a token $a$ and its type $A$ is not a proposition because it is not a
statement susceptible of being proven. Rather, the statement $a : A$ is a *definition* that
allows to render explicit a *context*, and the set-theoretic statement $\neg(a \in A)$ simply
cannot be transposed into type theory.

An immediate consequence of the difference between "$a : A$" in type theory and
"$a \in A$" in set theory is the difference in the treatment of identity. Since the types are
inseparable from the entities, identity statements must always be considered within
a given type. Given $a : A$ and $b : B$, it makes no sense in general to consider the
proposition $a = b$. Equality statements can only make sense for tokens of the same
type: only for $a, b : A$, one can ask whether $a =_A b$. Unlike in set theory, in type
theory *the equality sign always comes with a subscript.* This means that identity is
a *dependent* type: in addition to the invariant element $=$, one always has to add a
variable specifying the context in which equalities are being predicated.

In sum, as Makkai succinctly puts it:

> "[In type theory] both equality and membership are denied the free reign they
> enjoyed in the standard [set theoretical] foundation."**[240]**

---

**[239]** Ibid., p. 18, authors' italics, bold emphasis is mine.

**[240]** Makkai, op. cit., p. 156.

Now, compare this quote with Shapiro's:

> In mathematics, at least, the notions of "object" and "identity" are unequivocal but thoroughly relative. Objects are tied to the structures that contain them.[241]

It should strike how good a fit this is. The type-theoretical insistence that tokens should not be considered in isolation is a familiar one for any structuralist. But the rules of type theory are much stronger than just this. It is not that one *should not* talk about tokens in isolation (methodological decision granted by any structuralist); it is that one *cannot* talk about the tokens without at the same time talking about the types (ontological constraint adopted only by non-eliminative structuralists). The token is not a pre-existent object, prior to the type; it is not the case that the type *A* is defined extensionally, by the collection of its tokens. It is the other way around. Here is again Makkai explaining it:

> An entity belonging to a type cannot be discussed without reference to the type; the type logically precedes the entity, and the type is inseparable from the entity.[242]

As we have already seen, this (onto)logical priority of the structure is characteristic of non-eliminative structuralists. Recall the insistence of Shapiro on this point:

> Structures are prior to places in the same sense that any organization is prior to the offices that constitute it. The natural number structure is prior to '6', just as 'baseball defense' is prior to 'shortstop' or 'U.S. Government' is prior to 'Vice President'.[243]

Therefore, by conceiving abstract structures as structured types, many objections against *sui generis* structuralism vanish into thin air. Type theory provides the *ante rem* structuralists with both a relation capturing the places-to-structure relation they advocate, and a treatment of identity allowing them to solve the problem of identity for structural indiscernibles. As indicated by Benacerraf, the structuralist way out to the (in)famous "Julius Caesar problem", posed by Frege, is another good example of the usefulness of type theory:

> To speak from Frege's standpoint, there is a world of objects [...] in which the identity relation had free reign. [...] Hence the complaint at one point of his argument that, thus far, one could not tell whether Julius Caesar was a number.
>
> I rather doubt that in order to explicate the use and meaning of numbers one will have to decide whether Julius Caesar was (is?) or was not the number 43. [...] I propose to deny that all identities are meaningful [...]. Identity statements make sense only in contexts where there exist possible individuating conditions.[244]

---

[241]Shapiro, op. cit., p. 81.

[242]Makkai, loc. cit.

[243]Shapiro, op. cit., p. 9.

As Shapiro stresses, "a good philosophy of mathematics need not answer questions like "Is Julius Caesar = 2?" and "Is $1 \in 4$?" Rather, a philosophy of mathematics should show why these questions need no answers [...]"[245]. And this is precisely what the appeal to type theory does: it simply dissolves such questions by considering them as grammatically incorrect!

In the light of this, it is surprising that, despite the natural match between the language of types and the main ideas of *ante rem* structuralists, the use of type theory in the philosophical discussion of abstract structures still has a feeble existence. Therein, model theory and set theory overwhelmingly dominate. In fact, it seems that when the type-to-tokens relation is invoked, it is usually done in order to capture the structure-to-systems relation and not the structure-to-places relation[246]. To my knowledge, only Michael Makkai has tried to develop a systematic account of abstract structures using type theory. He calls it the Structuralist Foundation of Abstract Mathematics (SFAM), and the central notion allowing to describe abstract structures is that of "*abstract set*":

> In Abstract Mathematics, we find the intuitive idea of *abstract sets* one whose elements are characterless, nevertheless distinct, points. An "abstract" structure is one whose underlying set is an abstract set.[247]

---

[244]P. Benacerraf. "What Numbers Could Not Be". In: *Philosophical Review* 74 (1965), pp. 47–73, p. 64. For that matter, one could also recall one of the main disputes between Benacerraf's "two militant logicists" Ernie and Johnny on whether 3 belonged to 17. Using von Neumann's ordinals, Ernie answers positively whereas Johnny, using Zermelo's numerals, answers negatively. Therein, the discussion was dominated by the set-theoretical conception of membership, and this allowed questions about membership to be considered pertinent, well-posed questions.

[245]Shapiro, op. cit., p. 79.

[246]Examples of this are: Resnik, op. cit., p. 228; Shapiro, op. cit., p. 85; C. Chihara. *A Structural Account of Mathematics.* Oxford: Oxford University Press, 2004, p. 170.

But contrary to what the name seems to indicate, "abstract sets" are not sets in the sense of ZFC set-theory and are better thought as types. The distinction between these abstract sets and the "concrete" ZFC sets is crucial, and many objections to *ante rem* structuralism originate in a failure to see this. It is thus important to focus our attention for a while on this notion.

As Marquis points out, the idea of abstract sets can already be found in the early work of Fréchet[248] and is discussed at length in Lawvere's 1976 article *"Variable Quantities and Variable Structures in Topoi"*:

> The traditional view that membership is primary leads to a mysterious absolute distinction between $x$ and $\{x\}$, to agonizing over whether or not the rational numbers are literally contained in the real numbers, [...] to debates over whether the members of the natural number 5 are 0, 1, 3, 4 or not, and all that is clearly just getting started [...]. I believe the conclusion is that membership-as-primary entails membership as *global and absolute* whereas in practice membership is *local and relative* [...].
>
> These considerations lead one to formulate the following "purified" concept of (constant) abstract set as the one actually used in naive set-theoretic practice of modern mathematics: An abstract set $X$ has elements each of which has no internal structure whatsoever; $X$ has no internal structure except for equality and inequality of pairs of elements, and has no external properties save its cardinality [...].[249]

Notice once again the importance of departing from the usual, absolute set-theoretical notion of membership.

---

[247] Makkai, op. cit., p. 157.

[248] M. R. Fréchet. "Les ensembles abstraits et le calcul fonctionnel". In: *Rendiconti del Circolo Matematico di Palermo (1884–1940)* 30 (1910), pp. 1–26. See also the quote of footnote 134, page 79.

[249] F. W. Lawvere. "Variable Quantities and Variable Structures in Topoi". In: *Algebra, Topology and Category Theory - A collection of Papers in Honor of Samuel Eilenberg.* Ed. by A. Heller and M. Tierney. London: Academic Press, 1976, pp. 101–131, pp. 118-119.

These abstract sets are obtained by applying the method of abstraction, as described by Marquis (subsection I.2.3). In this case, one first has a domain of significant variation which includes "concrete" sets of very different sorts—sets of points, sets of transformations, sets of numbers, etc. Along with the cardinality of these sets, one then decides to retain solely the *relation* of identity and omit all the remainder information. The transformations that preserve this chosen information are the bijections. And one finally arrives to abstract sets, as entities on their own, by declaring that bijection is the relevant criterion of identity for sets: "what can be declared in the given [new] language is that abstract sets can be *isomorphic*"[250]. Hence, one gets to the notion of abstract sets by a process of *structural* abstraction. Clearly, for abstract sets cardinality is given a priori and identity within a set is primitive.

Even though the notion of abstract set is certainly not Makkai's invention, he is, to my knowledge, the only one that captures their nature. The structural process of abstraction applied to concrete ZFC sets, which culminates in the notion of 'abstract sets', does not produce sets but rather produces *types*:

> [An abstract] set $a$ is a *type*, and a variable $x$ may be *declared* to be of type $a$. "$x \in a$" will not be a predicate subject to free manipulation with the connectives and quantifiers as fullfledged predicates will be; for instance, we will never write $\neg(x \in a)$. The statement "$x \in a$" will have the role of defining *contexts* of variables, in the style of P. Martin-Löf and Cartmell. Thus SFAM is a *type theory*.[251]

Hence, following Makkai, we shall now say that an abstract structure is a type equipped with some relations between its tokens. This in particular holds for an abstract Hilbert space, which is a type $\mathcal{H}$ such that the usual Hilbert space axioms are met by its tokens[252].

[250]Marquis, op. cit., p. 62.

[251]Makkai, op. cit., p. 156.

### I.3.3.b   Abstract sets vs. abstract collections

As we have just explained, abstract sets can be seen as the result of an abstraction process that starts from 'concrete', 'material' sets (which include both sets of physical objects and ZFC-sets) and then declares bijection to be the new criterion of identity. Therefore, it would be most natural that, at some point of the discussion on abstract sets, the reader may have had the impression that these abstract entities were nothing more than the usual Frege-Russell numbers. Recall the two logicians' definition of cardinal numbers: these are equivalence classes of sets under the bijection relation. Moreover, were the reader familiar with Shapiro's work, the notion of 'finite cardinal structure' would have also come to mind and strengthened this impression: by their definition—Shapiro defines the $n$ cardinal structure to be the structure common to all collections of exactly $n$ objects[253]—they look very much like our abstract sets, by their name—finite *cardinal* structures—they suggest the link with cardinal numbers.

Nonetheless, the two notions—of 'abstract set' and 'cardinal number'—should be

---

[252]I should mention a post from Michael Schulman in the *n*-category café blog where Makkai's idea that abstract structures are in fact structured types is also discussed. He says:

> Now, the sets in a material set theory are admittedly closer to the natural-language meaning of "set": a set of three sheep can be distinguished from a set of three chairs, and each of the sheep and chairs might also be an element of other sets. However, the claim is that the sets in a structural set theory are closer to the way sets are *used* in *mathematics*. These "structural sets" are also very similar to the *types* in a type theory (regarded as the object-theory, as suggested in the previous post). In fact, Toby [Bartels] has convinced me that it's difficult to decide exactly where to draw the line between type theory and structural set theory, although there are differences in how the words are most commonly used. **It might be better, terminologically speaking, if mathematicians had used a word such as "type" instead of "set" all along. But by now the notion that (for instance) a group is a set equipped with an identity and a multiplication is so firmly entrenched in most mathematicians' consciousnesses that I think there's little point trying to change it**. Anyway, as I mentioned in the previous post, "set" and "type" and "class" are basically fungible words—especially when used structurally. (https://golem.ph.utexas.edu/category/2009/12/syntax_semantics_and_structura_1.html, italics are Schulman's, bold emphasis is mine.)

We here find again the typical terminological problem: 'material sets' correspond to Makkai's 'concrete sets', whereas 'structural sets' are Makkai's 'abstract sets'.

[253]Shapiro, op. cit., p. 115.

distinguished, for abstract sets furnish a perfect example of why the structural process of abstraction produces entities which cannot be thought as equivalence classes.

As Lawvere insists, abstract sets are more complex than cardinal numbers:

> [...] an abstract set is more refined (less abstract) than a cardinal number in that it does have elements while a cardinal number does not. The latter feature makes it possible for abstract sets to support the external relations known as *mappings*, which constitute the second fundamental concept of naive set theory (cardinal numbers would admit only the less refined external relations expressed by one being less than another or not).[254]

A good way to understand Lawvere's quote is to insist on the difference between an abstract set and an *abstract collection*. On the one hand, an abstract set, says Lawvere, "has no internal structure *except for* equality and inequality". But *it is the presence of this identity relation that makes abstract sets structural*. Abstract sets are abstract structures—indeed, the simplest of all—and they are so because they are collections equipped with an identity relation that holds within them. For that reason, abstract sets are also sometimes called "structural sets"[255]. Makkai puts it nicely:

> [An abstract] set is a relatively orderly part of the world in which an *equality predicate* reigns. The elements of a set are individuated *with respect to each other*. However, there is no global equality present for all things simultaneously. An equality predicate is an equivalence relation on the given set. In fact, the set is the underlying collection *together with* its equality predicate.[256]

On the other hand, one can decide to push the abstract method one step further by deciding to omit the identity relation within abstract sets. The more abstract entities thus obtained I call 'abstract collections'. As I will now try to argue, abstract collections behave very much like cardinal numbers.

---

[254] Lawvere, op. cit., p. 119, author's emphasis.

[255] This seems to be the terminology adopted by the community involved in the nLab project (e.g., John Baez, Toby Bartels, Michael Schulman, Urs Schreiber). See for instance the entry *Structural Set Theory* in nLab. http://ncatlab.org/nlab/show/structural+set+theory

[256] Makkai, loc. cit.

Following Lawvere's suggestion, this is best captured by reflecting on the morphisms between these entities. Consider for instance two abstract sets $E, F$ of cardinality $n-1$ and $n$. To the question "How many different injective morphisms from $E$ to $F$ are there?", the mathematician will answer that there are $n!$ many of them. The subsequent reasoning is one possible way to arrive to this answer. Start by stating the usual identity criterion for functions: given two morphisms $f, g : E \longrightarrow F$, we have

$$(f =_{Hom(E,F)} g) \Longleftrightarrow \big(\text{ if } x =_E y \text{ then } f(x) =_F g(y)\big)$$

Notice that, since we are dealing with types, we need to specify the type in which each identity relation is being stated. To explicitly construct one such injective morphism $g : E \longrightarrow F$, choose first a token of $F$ that will not have a preimage—call it $f_n$—and then choose one particular bijection between $E$ and the remainder tokens of $F$—call them $f_1, \ldots, f_{n-1}$. It is now clear why there are $n!$ injective morphisms from $E$ to $F$: there are $n$ ways of choosing the element $f_n$ and $(n-1)!$ different bijections between two abstract sets of cardinality $n-1$. If one defines the relation on abstract sets "'$F$ is bigger than $E$' when there exists an injective morphism from $E$ to $F$", what the above shows is that the 'bigger than' relation has a more complex structure for abstract sets than for numbers: there is a *plurality* of ways in which the set $F$ is bigger than the set $E$, plurality that is clearly absent for numbers. In this sense, the external relation is "less refined" for numbers, as Lawvere claims.

However, if a decision is taken to omit the relations $=_E$ and $=_F$, this plurality is immediately lost as well. Indeed, the determination of the multiplicity of *different* morphisms rested on the identity criterion for functions, which crucially involved the primitive identities within both $E$ and $F$ allowing to decide whether $x =_E y$ and whether $f(x) =_F g(y)$. This phenomenom can also be perceived in the way the monomorphism $g$ was constructed, as "a function which reaches all elements of $F$ except *this* particular one". But this is possible because, once an element or token of the abstract set $F$ has been chosen, the primitive (typed) identity that holds within $F$ allows to distinguish the given element from all others—in virtue of the irreflexive non-identity relation, elements of an abstract set are indeed relationally discernible.

Therefore, with the deletion of primitive identities, it is no longer possible to determine the difference between two functions. Even worse: it is not even possible to define the notion of a function! (For how does one check that it is one-to-one and not one-to-many?)

The situation can also be pictured through the conception of morphisms as transformations. When dealing with abstract sets, the primitive typed identities allow to 'follow' the transformation of each token. The arrow $E \xrightarrow{g} F$ is conceived as a collection of *local* arrows —"*this* token of $E$ gets transformed into *that* token of $F$". On the contrary, by omitting primitive identities the tokens become strongly indistinguishable— that is: they cease to be even weakly discernible. Then, when considering abstract collections, the transformation can only be considered externally, as a whole. In this way, whereas there are many transformations from the abstract set $(E, =_E)$ to the abstract set $(F, =_F)$, there is on the contrary only *one unique* way of transforming the abstract collection $E$ into the abstract collection $F$, namely: "add one element". In this sense, abstract collections and numbers look very much alike.

The distinction between abstract sets and abstract collections is summed up in the following figure[257]:

---

[257]With this distinction in hand, we can now comment in some more detail on Shapiro's "finite cardinal structures". The essential trait of the $n$ cardinal structure is that any concrete set of cardinality $n$ should be seen as a realization of it. Therefore, it is clear that the $n$ cardinal structure and the abstract set of cardinality $n$ are intended to be the same abstract structure. However, I only partially agree with Shapiro's description of this structure. He says:

> The finite cardinal structures have no relations and so are as simple as structures get. (Shapiro, loc. cit.)

True: finite cardinal structures or abstract sets are the simplest of all structures, since any abstract structure is an abstract set on which some further relations have been defined. Wrong: finite cardinal structures or abstract sets do have one relation, namely primitive typed identity. Otherwise, they would be abstract collections and would not be able to support functions on them.

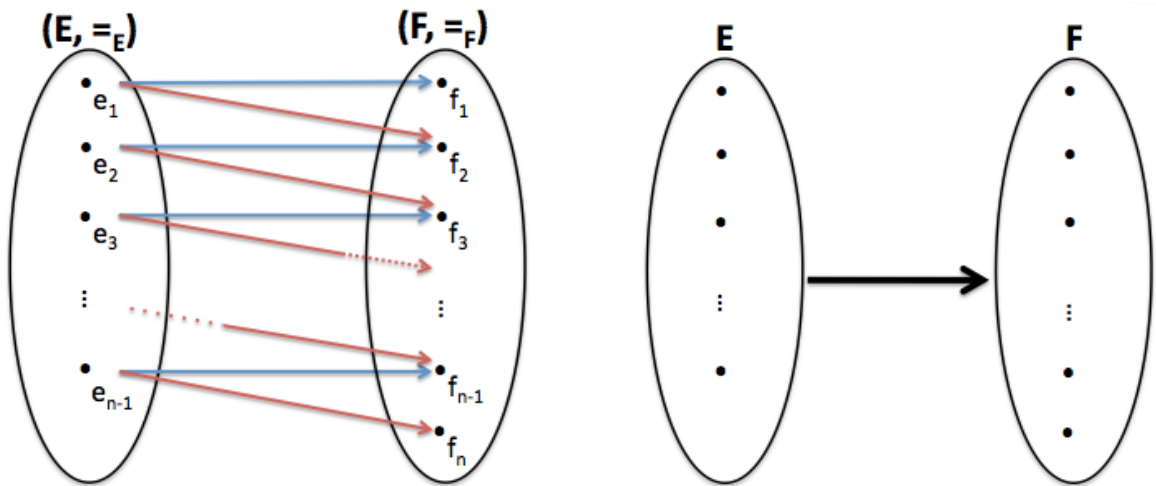

**(a)** Two abstract sets $E$ and $F$ with their primitive identity relations $=_E$ and $=_F$. A transformation from $E$ to $F$ may be depicted by an internal (local) diagram. In red and blue are represented two different injective morphisms from $E$ to $F$.

**(b)** The abstract collections, as a result of the omission of the identity relations. A transformation from $E$ to $F$ can only be seen as an external (global) diagram and there is now only one possible injective morphism from $E$ to $F$.

**Fig. I.3** – Difference between abstract sets and abstract collections.

### I.3.3.c   Four grades of discernibility

The discussion has hinted at the fact that abstract sets differ from abstract collections inasmuch as it is possible to *discern* and *name* the tokens of the former but not of the latter. In other terms, it has allowed to underline the triad:

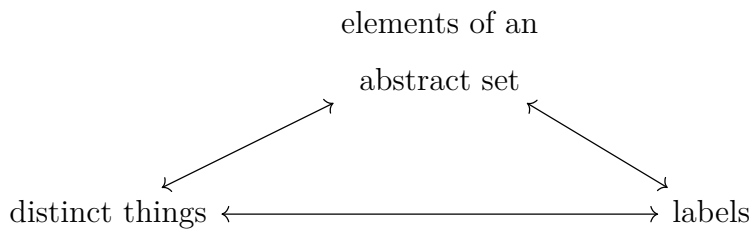

The idea of this triad is certainly not new. Already Cantor had stressed the importance of distinguishability within a set:

> By a 'set' we mean any collection $M$ into a whole of definite, *distinct* objects $m$
> (called 'elements' of $M$) of our perception or our thought.[258]

On the other hand, the intimate relation between abstracts sets and labeling was clearly
put forward by Lawvere:

> The only possible use of abstract sets $T$ is the possibility of indexing or para-
> metrizing things by the elements of $T$ in the hope of clarifying actual relations
> between the things [...].[259]

In fact, although I had not emphasized it at the time, we had already encountered the
idea in Weyl's description of how to obtain abstract groups:

> This [the obtention of an abstract group] *is accomplished by attaching arbitrary*
> *labels to its elements* and then expressing in terms of these labels for any two
> group elements $s$, $t$ what the result $u = st$ of their composition is.[260]

Nonetheless, we still need a better conceptual grasp of the triad. In particular, we
need to further clarify the kind of discernibility which is present in abstract sets but
absent from abstract collections. So far, we have come across four different grades of
discriminability within structures:

i) the case of the real number 1, which can be distinguished from all other numbers
   through the property: $P(x) \equiv$ 'for any $y : \mathbb{R}$, $xy = y$',

ii) the case of two points $x$ and $y$ of a homogeneous metric space $(M, d)$, which can
    be distinguished through the relation $d(x, y) \neq 0$,

iii) the case of two elements of an abstract set (or of an abstract vector space), which
     can be distinguished through a primitive typed identity,

iv) the case of two elements of an abstract collection, which cannot be distinguished
    in any way.

---

[258]"*Unter einer 'Menge' verstehen wir jede Zusammenfassung $M$ von bestimmten wohlunterschiede-
nen Objekten $m$ unserer Anschauung oder unseres Denkens (welche die 'Elemente' von $M$ genannt
werden) zu einem Ganzen*". Both the original German quote and the English translation are cited in
Y. I. Manin. "Georg Cantor and His Heritage". In: *Tr. Mat. Inst. Steklova* 246 (2004), pp. 208–216,
p. 214, my emphasis.

[259]Lawvere, op. cit., pp. 120–121.

[260]Weyl, *Symmetry*, p. 145, emphasis is mine.

The Quine-Muller terminological distinction—absolute and relational discernibility—is not powerful enough to describe the full situation, for it was constructed to capture the distinction between the first case and the second. We thus need more terminology:

**Definition I.1** (Individuals)**.** A collection within which an identity predicate reigns will be called a collection of ***individuals***.[261]

This terminological decision serves a double purpose. On the one hand, it points to the fact that tokens of abstract sets and, more generally, places of any abstract mathematical structure are individuals. And this is so *by definition* of what abstract sets and abstract structures are. In other terms, it points to the fact that, from the very start, abstract sets are not metaphysically neutral: they cannot serve to describe *any*-thing, but only a "relatively orderly part of the world"[262], namely: collections of *individuals.* It automatically extends the conclusion of French and Krause on material set theory:

> [...] standard set theories involve a theory of identity which takes the elements of a set [...] to be *individuals* of a kind.[263]

In fact, the realization of the metaphysical commitments of set theory was already emphasized by the mathematician Yuri Manin[264] and has been one motivation to

---

[261]The concept of "individuality" has received many different characterizations in the philosophical literature. My terminological decision is strongly influenced by the ideas developed by Steven French and Décio Krause (particularly in their joint work *Identity in Physics: A Historical, Philosophical, and Formal Analysis*). Therein, they develop a conception of "individuality in terms of self-identity" (p. 15)—or, what amounts to the same, they

> [...] defend the claim that the notion of non-individuality can be captured [...] by formal systems in which self-identity is not always well defined, so that the reflexive law of identity, namely $\forall x \, (x = x)$, is not valid in general. (pp. 13-14)

This is also similar to the notion of individuals presented by Lowe in his chapter "Individuation" of the *Oxford Handbook of Metaphysics.* He says:

> [...] for many kinds of entity, identity and countability are indeed inseparable—and it is these entities that may properly be described as being 'individuals' or as having 'individuality'. (p. 78)

[262]Makkai, loc. cit. See page 127.

[263]S. French and D. Krause. *Identity in Physics: A Historical, Philosophical, and Formal Analysis.* Oxford: Oxford University Press, 2006, p. 240.

[264]Thus, he says: "The birth of quantum physics [...] made clear that Cantor's famous definition of sets represented only a distilled classical mental view of the material world as consisting of pairwise distinct things residing in space [...]. Once this view was shown to be only an approximation to the

construct alternative theories[265].

On the other hand—and here I follow French and Krause again—this choice of terminology allows to conceptually separate "individuality" from "distinguishability". It thereby introduces new degrees of discernibility. Indeed, since non-equality is a grammatically correct irreflexive relation, individuals of a same structure are, *by definition*, relationally discernible. But there is little point in keeping this lax use of relational discernibility. Rather, we will say that places of a structure are *primitively* discernible. The interesting question is then to investigate whether this primitive discernibility can be described without appealing to the primitive typed identity. That is to say, to investigate whether the non-trivial relations of the structure are powerful enough to account for the discernibility of places[266].

Given an abstract collection $\mathcal{C}$, the question on discernibility is thus : "Is this a collection of discernible individuals?" The above mentioned four cases correspond to the four possible answers:

---

incomparably more sophisticated quantum description, sets lost their direct roots in reality." (Manin, op. cit., p. 9.)

[265]Two examples of these are:

1. *The theory of quasets*, developped by Dalla Chiara and Toraldo di Francia, which "have a cardinal but not an ordinal" and are not determined extensionally but intensionally (see e.g. M. L. Dalla Chiara and G. Toraldo di Francia. "Individuals, Kinds and Names in Physics". In: *Bridging the Gap: Philosophy, Mathematics, Physics*. Dordrecht: Kluwer Academic Publishers, 1993, pp. 261–283).

2. *The theory of quasi-sets*, introduced first by Newton da Costa and then also developed by Krause and French: "It is important to realize that quasi-set theory may be said to be inspired by the idea that the concept of identity might not be applicable to elementary particles, as Schrödinger claimed. The limitation imposed on the concept of identity will offer us the opportunity to elaborate a mathematical theory in which we can talk of indistinguishable but not identical objects, as we will see." (French and Krause, op. cit., p. 241)

[266]A very similar point is made by Muller: "The aim is not, when we begin with a differentiable manifold of infinitely many distinct space-time points, to find out whether there really is more than one space-time point [...] but *the aim is to find out whether the distinctness of the points can be grounded qualitatively*, physically, and structurally, and that has not been assumed tacitly" (Muller, op. cit., p. 1057, my emphasis). What Muller meant by the possibility of "grounding physically" the distinctness of points of a differentiable manifold remains obscure to me. But the point he is making is essentially the same as mine.

**Definition I.2** (Four Grades of Discernibility)**.**

i) A collection of **qualitatively (or absolutely) discernible individuals** is a collection such that

$$\text{for any } x, y : \mathcal{C}, \ \exists P, P(x) \text{ and } \neg P(y)$$

where $P$ is invariant under automorphisms.

ii) A collection of **comparatively discernible individuals** is a collection such that i) fails but one has

$$\text{for any } x, y : \mathcal{C}, \ \exists \mathcal{R}, \exists z : \mathcal{C}, \ \mathcal{R}(x, z) \text{ and } \neg \mathcal{R}(y, z)$$

where $\mathcal{R}$ is an invariant relation other than the primitive typed identity[267].

iii) A collection of **indiscernible individuals** is a collection such that i) and ii) fail but there is a primitive typed identity.

iv) A collection of **non-individuals** (thus strongly indiscernible things) is a collection for which i), ii) and iii) fail.

Of course, as French and Krause very well emphasize, the conceptual distinction between individuality and discernibility is only useful in practice if there are legitimate cases of indiscernible individuals—that is: if there are cases of (contextual) primitive thisness[268]. But we have already seen that mathematics has plenty of those (abstract sets, abstract vector spaces, etc.).

What precedes should dissolve what Lawvere calls the "strong contradiction" of abstract sets: their points are completely distinct, because they are individuals, and yet are indistinguishable, because no relation other than the primitive identity can discern them[269].

---

[267]For simplicity, I have here only considered binary relations. Following Muller's approach, one can extend this definition in the obvious way so to include all *n*-ary relations.

[268]They say: "Our conceptual distinction between individuality and distinguishability can then only be maintained in practice under [the] view that individuality is grounded on something else, 'over and above' properties." (French and Krause, op. cit., p. 16.)

[269]"Yes, the notion of an abstract set (Cantor's Kardinalzahl) is a strong contradiction: its points are completely distinct and yet indistinguishable." (F. W. Lawvere. "Foundations and Applications: Axiomatization and Education". In: *The Bulletin of Symbolic Logic* 9.2 (2003), pp. 213–224, p. 215.)

With these four grades of discernibility within a collection properly distinguished, we can finally render precise the content of the requirement of individuation for physical states and properties (page 17).

**Definition I.3** (Individuation). We say that an element $x$ of an abstract structure $\mathcal{S}$ can be ***individuated*** if, for all $y : \mathcal{S}$ such that $y \neq_{\mathcal{S}} x$, $x$ and $y$ are qualitatively discernible individuals.

By the above definitions, it follows that two elements of an abstract structure which are related by an automorphism cannot be qualitatively discerned from each other. This simple remark is actually quite fruitful, for it furnishes a practical tool to find the 'amount of individuation' that can be achieved within a given mathematical structure $\mathcal{S}$. It now appears that this information can be easily read off from the action of the group of automorphisms $Aut(\mathcal{S})$ on $\mathcal{S}$: the orbits $\mathcal{O} \in \mathcal{S}/Aut(\mathcal{S})$ are the smallest parts of the structure which can be individuated[270].

# I.4   Conclusion

Let me now briefly summarize what has emerged in the course of this chapter and put it in perspective with respect to the main goal of this thesis. To recall: the intention is to provide a clear conception of what quantization means—and, in particular, to grasp the "real difference between Classical and Quantum Mechanics"[271] by plunging as much as possible into the depths of the mathematical formalisms underlying both theories. Now, as Darrigol reminds us, "any application of [quantization] starts with *formally defining* a classical system, and the quantum theoretical level is then reached by applying a precise mathematical procedure"[272]. It thus has seemed to me that any attempt to understand quantization should deal first with the concept of "mathematical description of a physical system". Accordingly, the first part of this work has been devoted to a general reflection on this notion.

---

[270] Given the left action of a group $G$ on a set $E$, the *orbit* $\mathcal{O}_x$ of an element $x \in E$ is the subset $\mathcal{O}_x := \{y \in E \mid \exists g \in G, y = g \cdot x\}$.

[271] This expression is the title of a conference given on February 13th 2014 by Andreas Döring at the Workshop "Philosophy of Mechanics: Mathematical Foundations" held in Paris.

[272] Darrigol, op. cit., p. xvi, my emphasis.

Therein, the main driving question was: Which is the role theoretical physicists expect to confer to these mathematical descriptions? Many different expectations are of course possible (and indeed found amongst the scientific community), but I have chosen to focus on the more ambitious one, which I have dubbed the "*descriptive perspective*": it considers the goal of these mathematical descriptions to provide a full and unambiguous *intrinsic characterization* of the physical systems being thus described. Even though such an aspiration may appear as quite naive an utopia, adopting it as a working hypothesis and seeking to push it to its limits can yield interesting insights in theoretical physics—as I hope the remainder of this work will show. As Gabriel Catren beautifully puts it: "It is necessary to be programmatically ambitious in order to fail in a productive way"[273].

The mathematical description of a generic physical system can be thought as a map $\mathcal{D} : T_{phys} \longrightarrow T_{math}$ from a certain class of physical systems (e.g., non-relativistic systems with finitely many degrees of freedom) towards a specific class of mathematical objects. Given a physical system $S$, the mathematical object $\mathcal{D}(S)$ is usually intended to furnish a description of the geometry of the state space and/or of the algebra of properties. To give a precise content to the expectation of an intrinsic characterization portrayed in the descriptive perspective, I have formulated two requirements these mathematical objects should meet: first, the *faithfulness requirement* (page 15), which demands that the map $\mathcal{D}$ be injective, and thus concerns identity between mathematical objects; second, the *requirement of individuation* (page 17), which demands that both states and properties of the system be qualitative discernible individuals, and therefore focuses on individuation within mathematical entities.

The next step is to understand the particular nature of the mathematical objects customarily involved in the formalisms of both Classical and Quantum Mechanics. For only once this nature has been properly grasped, will one clearly understand the specific conditions imposed by both requirements. After having taken a closer look at the turning years during which the foundations of Quantum Mechanics were first developed, I have proposed that mathematical objects appearing at this fundamental level of Mechanics should be conceived as *abstract structures*. Sections I.2 and I.3 were dedicated to the clarification of what is meant by this. As it appears in the case of von Neumann, one is pushed towards abstraction by the will of finding an *intrinsic* description of physical systems. Abstract mathematics allow to answer the question: *Where*, in the formalism, should we look for the relevant physical information? Indeed, through its systematic use of isomorphisms as equalities—which constitutes the core of abstraction—, it manages to precisely define a level in which the physical information is to be found. Mathematical objects become "schematic" and one no longer worries about the "arbitrary and special 'fleshly clothing'" of the formalism[274]. On the other hand, structural mathematics is the answer to the question: *How*, from the formalism, should we recover the relevant physical information? It is so because, with the requirement of individuation in mind, the look for discernible individuals becomes one of the main goals of the analysis and structuralism proves to be extremely useful in this regard. First, I have shown how it allows to distinguish four different grades of discernibility within a collection—qualitatively discernible individuals, comparatively discernible individuals, indiscernible individuals and non-individuals (page 134). Second, it furnishes practical tools to extricate from within the abstract structure those individuals that can be qualitatively individuated. This last point is accomplished by use of the pivotal group of automorphisms: qualitative discernible individuals correspond exactly to the orbits of the structure under the defining action of this group.

---

[273]G. Catren. "A Throw of the Quantum Dice Will Never Abolish the Copernican Revolution". In: *Collapse: Philosophical Research and Development* 5 (2009), pp. 453–500, p. 470.

I have tried as much as possible to discuss the issues about Abstract and Structural Mathematics in their natural general context. As a result of this, Sections I.2 and I.3 may have seemed to the reader too long a detour from the main subject of the thesis. I hope nonetheless that they present an interest in their own right. Be that as it may, with this conceptual background firmly understood, we are now ready to analyze, in all the technical detail they deserve, both Classical and Quantum Kinematics. The unfolding of this analysis will be the content of the remaining two chapters.

---

[274]Schrödinger, "On The Relation Between The Quantum Mechanics of Heisenberg, Born, and Jordan, and That of Schrödinger", p. 58.

# Chapter II

# The Classical and Quantum Kinematical Arenas

In their standard formulation, the Classical and the Quantum are respectively casted into the language of symplectic (or Poisson) manifolds and Hilbert spaces. Their use is so widespread among theoretical physicists that one may be tempted to write the following definitions:

**Classical System 1.** A classical system is characterized by an abstract symplectic manifold $(S, \omega)$.

**Quantum System 1.** A quantum system is characterized by an abstract Hilbert space $\mathcal{H}$.

The first definition is in fact readily found in many standard textbooks[1], whereas one can easily have the impression that the second is latent in many works on the subject[2]. However, from the perspective that is ours, it is clear that both definitions are wrong: neither the simple data of a symplectic manifold nor that of a Hilbert space can be enough to characterize a physical system. This may be seen in two ways:

– *Failure to meet the faithfulness requirement.* As we have already mentioned, when von Neumann introduced the notion of a Hilbert space, he also proved the categoricity of the axioms: up to isomorphism, there is only one separable Hilbert spaces for a given dimension[3]. Thus, if there is more than one quantum system—which is obviously the case—the faithfulness requirement fails. The classical analogue of this is Darboux's theorem[4]: any two symplectic manifolds of same dimension are *locally* isomorphic. For a given dimension, we thus get infinitely many non-isomorphic symplectic manifolds, but all differences are only of a global nature. Although strictly speaking this does not suffice to prove the failure of the faithfulness requirement in the classical case, it strongly suggests there are not enough differences between symplectic manifolds to account for the actual variety of physical systems. The next point will confirm this impression.

– *Failure to meet the requirement of individuation.* Instead of looking at the identity between various Hilbert spaces or symplectic manifolds, one can focus on the descriptive power within these objects. The group of automorphisms of a Hilbert space is the group $U(\mathcal{H})$ of unitary transformations. Given any two unit vectors $\phi$ and $\psi$, there exists a unitary transformation relating them. Said differently, the action of the group $U(\mathcal{H})$ on the projective Hilbert space $\mathbb{P}\mathcal{H}$ is transitive[5]. Exactly the same result holds in Classical Kinematics: for a given connected symplectic manifold $S$, any two points can be transformed into each other by a symplectomorphism[6].

---

[1]For example, see C. Rovelli. *Quantum Gravity*. Cambridge: Cambridge University Press, 2004, p. 100, N. Woodhouse. *Geometric Quantization*. 2nd. Oxford: Clarendon Press, 1991, p. 155, or M. Puta. *Hamiltonian Mechanical Systems and Geometric Quantization*. Dordrecht, The Netherlands: Kluwer Academic Publishers, 1993, p. 28.

[2]Thus, Livine declares that "[i]n order to talk about the quantum theory [of gravity], we should precisely define the Hilbert space and our quantum states of space(-time) geometry". (E. Livine. "Covariant Loop Quantum Gravity?" In: *Approaches to Quantum Gravity*. Ed. by D. Oriti. Cambridge: Cambridge University Press, 2009, pp. 253–271, p. 262).

---

[3] J. von Neumann. *Mathematical Foundations of Quantum Mechanics.* Trans. by R. T. Beyer. Princeton: Princeton University Press, 1955, section II.2., pp. 46–59.

[4] Cf. R. Abraham and J. E. Marsden. *Foundations of Mechanics.* 2nd ed. Redwood City: Addison-Wesley Publishing Company, 1978, p. 175.

[5] This is fairly obvious for finite dimensions. The proof that this result also holds for the infinite-dimensional case can be found in R. Cirelli, M. Gatti, and A. Manià. "The Pure State Space of Quantum Mechanics as Hermitian Symmetric Space". In: *Journal of Geometry and Physics* 45.3 (2003), pp. 267–284. URL: http://arxiv.org/abs/quant-ph/0202076.

[6] See W. M. Boothby. "Transitivity of the Automorphisms of Certain Geometric Structures". In: *Transactions of the American Mathematical Society* 137 (1969), pp. 93–100, Theorem A, p. 98 or P. W. Michor and C. Vizman. "N–transitivity of Certain Diffeomorphism Groups". In: *Acta Math.*

It thus appears that both (projective) Hilbert spaces and (connected) symplectic manifolds are homogeneous structures: their points cannot be qualitatively discerned and it is therefore impossible to make unambiguous reference to any such point without appealing to primitive thisness. The upshot is that neither of these structures are sophisticated enough to describe, on their own, all the physical information of a mechanical system. This conclusion is of course not new and statements of the like are scattered through the literature, particularly in the context of Quantum Mechanics[7].

Confronted with this, one option would be to discard Hilbert spaces and symplectic manifolds from the outset and start looking for completely different mathematical structures that could do a better job. But this would be to miss the point of the above criticism towards the standard formalisms of Mechanics. The goal is not to drive us away from these frameworks, but rather to urge us to *take a closer look* at them. For indeed the practice of theoretical physics never considers *bare* Hilbert spaces—that is, abstract Hilbert spaces *and nothing else.* Nor does it consider bare symplectic manifolds. Explicitly or implicitly, these structures always come along with other additional mathematical structures. For instance, when Ashtekar and Lewandowski explain that when considering "a 'free' particle on the group manifold of a compact Lie group $G$ [...], the Hilbert space of quantum states can be taken to be the [Hilbert] space $L^2(G, d\mu_H)$ of square integrable functions on $G$ with respect to the Haar measure"[8], they have in mind not only a Hilbert space but, in fact, a particular unitary representation of the group $G$—namely, the so-called regular representation. Thus, they are describing the quantum space of states by an abstract Hilbert space *together with additional data—*here, the choice of a particular morphism of groups $\rho : G \longrightarrow U(\mathcal{H})$. This extension, from the sole data of the abstract $\mathcal{H}$ to the more sophisticated structure of a triple

---

*Univ. Comenianae* 63.2 (1994), pp. 221–225. URL: `http://arxiv.org/abs/dg-ga/9406005#`, Theorem (4), p. 221.

[7]For example, Landsman expresses this in quite the same vein: "all Hilbert spaces of a given dimension are isomorphic, so that one cannot characterize a physical system by saying that 'its Hilbert space of (pure) states is $L^2(\mathbb{R}^3)$'." (N. P. Landsman. "Lecture Notes on $C^*$-algebras, Hilbert $C^*$-modules, and Quantum Mechanics". In: (1998). URL: `http://arxiv.org/abs/math-ph/9807030`, p. 6.)

[8]A. Ashtekar and J. Lewandowski. "Background Independent Quantum Gravity: A Status Report". In: *Classical and Quantum Gravity* 21.15 (2004). URL: `http://arxiv.org/abs/gr-qc/0404018`, pp. 25–26.

$(\mathcal{H}, G, \rho)$, appears naturally in the quantum formalism precisely because of the failure of Hilbert spaces to meet the requirement of individuation. And a similar phenomenon occurs in the Classical realm.

The point here is to emphasize the existence of a fundamental tension between the intended purpose of the mathematical descriptions of physical systems and the basic formalism used in the theory. This translates into a crucial question:

> *How can we break homogeneity and introduce qualitative*
> *discernibility into the mathematical structures underlying*
> *the formalisms of Classical and Quantum Mechanics?*

I claim this should be recognized as a driving force in the Foundations of Mechanics, in the sense that many developments in the field can be retrospectively read as attempts to overcome this tension and answer the question. Surely, attempting to justify this claim will be one major point underlying the remainder of this work.

From this emerges a conceptual scheme, or program of investigation, that I will develop in the next two chapters and have wished to call *The Chase for Individuation*. It is the following. First, Hilbert spaces and symplectic manifolds appear simply as the *starting point* of the formalism: they represent, so to speak, the homogeneous receptacle or the arena in which the kinematical description of physical systems takes place. Notwithstanding their homogeneity and lack of discernibility, this 0-level involves many sophisticated mathematical structures whose interplay reveals many conceptually interesting features. The seeds of the mechanisms that will allow the introduction of discernibility are already present, and this chapter is dedicated to a careful analysis of this.

Once the structure of this 0-level will be properly understood, the next move, to be studied in Chapter III, will be to start constructing candidates for a mathematical description of a physical system by breaking the homogeneity of the kinematical arenas studied in Chapter II. The general strategy, which involves the the introduction into the picture of additional external structures, is explained in section III.1. Then, we study in detail the particular case of one type of such structures: groups. At every stage of the road towards individuation, the mathematical mechanisms introduced in Classical and Quantum Kinematics will be compared, thus shedding some light on the procedure of quantization. At the end, we hopefully will have a better understanding of the real difference between the Classical and the Quantum.

## II.1  The double role of properties in standard Kinematics

The kinematical description of a physical system attempts to completely characterize each possible state and to understand all the possible properties of the system. The two fundamental notions are here "state" and "property". Assuredly, an analytically inclined reader will immediately raise an eyebrow and ask with suspicion: What exactly do these two words mean? And the question would be legitimate for both notions often hide strong metaphysical commitments. For example, depending on one's favourite ontology of objects persisting in time, one will have a different conception of "states": an endurantist may perhaps be inclined to conceive "states" as instantaneous points in the evolution of the system, whereas, on the contrary, a perdurantist may tend to consider "states" as extended processes in time—in which case they are often called "histories"[9]. Moreover, a realist may regard "properties" as true qualities possessed by the system, while an empiricist will insist on thinking them as the result of an interaction with an observer—in which case calling them "observables" seems more convenient. And so it continues, with a plethora of other difficult metaphysical debates underlying the use and meaning of these two fundamental words.

It is therefore possible to feel that, before immersing ourselves in the conceptual analysis of Kinematics, it is necessary to first clarify the spectrum of all metaphysical positions one may adopt towards the basic notions. This is not, however, the path I shall follow. Indeed, my wish is to turn all attention to what can be learned *from the formalism itself*: instead of approaching the mathematics of Kinematics through some firmly constructed looking glass, to carefully and patiently listen the "inner voice" of these structures[10]. Hence, the notions of "state" and "property" should carry as light a metaphysical baggage as possible, and to achieve this I will simply leave the two words somewhat undefined.

---

[9]Although the question of temporal parts goes back at least to the Greeks, the debate between endurantism and perdurantism was introduced, in these terms, by David Lewis in *On the Plurality of Worlds* (p. 202). A good general introduction to the subject is M. J. Loux. *Metaphysics, a Contemporary Introduction.* 3rd. New York: Routledge, 2006, Chapter 8.

Prima facie, the conceptual skeleton of Kinematics may therefore appear to be captured by this simple *duality* diagram[11]:

$$\text{states} \xleftarrow{\quad ? \quad} \text{properties.}$$

And the first obvious relation one can think of is the fact that "states take definite values of properties": given a property $f$ and a well-chosen state $q$, one can assign a number to the pair $(f, q)$. This number, denoted by $f(q)$—or by $\langle f, q \rangle$ if one wants to stress the dual role of properties and states in the assignment of a number—allows to partially characterize the specificities of the state $p$. Under this light, the role of properties is to introduce discernibility and separate states. Typical questions will then concern:

i) the conditions under which such a pairing can be done (e.g.: Given a property $f$, can we assign a number to any state $q$? Given a state $q$, can we assign a number to any property $f$?),

ii) the knowledge we can gain from this pairing (e.g.: Given a state $q$, can we find a set of properties $\{f_1, \ldots, f_n\}$ such that the numbers $\{\langle f_1, q \rangle, \ldots, \langle f_n, q \rangle\}$ fully characterize the state?).

There is however a second role played by properties in Kinematics. Besides their relation to numbers, properties are also related to *transformations*. In my view, the consciousness of this relation, progressively built during the 20th century, is one of the major conceptual achievements of modern physics. One most commented instance of this is of course Noether's theorem, which relates the existence of symmetries to the existence of conserved quantities[12]. To realize the importance this theorem has acquired, it suffices to read what Robert Mills writes:

> It seems to me quite possible that Noether's theorem is the more fundamental fact—that the physical theories that we devise to describe the universe about us have the structure they do because of this fundamental relationship between symmetries and conservation laws. If this is so, then Noether's theorem becomes a principle rather than a theorem.[13]

---

[10]Cf. Grothendieck's quote at the end of the Introduction (page 9).

[11]I here use the term "duality" in the rather loose sense of a "contrast between two notions". The link with the precise mathematical notion of duality will be commented later on.

But the link between properties and transformations does not restrict to this symmetry-conservation relationship. In fact, in modern expositions of Mechanics, transformations are often used to *define* properties. To attest, we learn in Towsend's textbook on Quantum Mechanics that

> [...] the best way to define the momentum operator is as the generator of translations, just as we defined the angular momentum operators as the generators of rotations and the Hamiltonian, or energy operator, as the generator of time translations.[14]

Properties play then a double role: as quantities, they allow to separate states and as transformations, they allow to relate states. This fact is clearly known to physicists and mathematicians[15], but its conceptual significance has been largely ignored. One

---

[12]In fact, Noether proved two theorems and only one of them—the one dealing with *global* symmetries—deals with conserved quantities. The second theorem, dealing with infinite-dimensional groups—and hence with local gauge transformations—is much less known. A good concise review for the precise content of both Noether's theorem is K. Brading and H. R. Brown. "Noether's Theorems and Gauge Symmetries". In: *arXiv preprint* (2000). URL: http://arxiv.org/abs/hep-th/0009058. A more exhaustive discussion of the subject is the excellent Y. Kosmann-Schwarzbach. *Les Théorèmes de Noether. Invariance et lois de conservation au XXème siècle.* Palaiseau: Les éditions de l'école polytechnique, 2004.

[13]R. Mills. "Gauge fields". In: *100 Years of Gravity and Accelerated Frames: The Deepest Insights on Einstein and Yang-Mills.* Ed. by J.-P. Hsu and D. Fine. Vol. 9. Singapore: World Scientific, 2005, pp. 512–526, p. 513.

[14]J. S. Townsend. *A Modern Approach to Quantum Mechanics.* Sausalito: University Science Books, 2000, p. 156.

[15]For example, Guillemin and Sternberg mention it explicitly: "[...] in classical mechanics as in quantum mechanics there is a double role: a function is an observable and it also determines an infinitesimal symmetry of the space of observables [...]." (V. Guillemin and S. Sternberg. *Variations on a Theme by Kepler.* Vol. 42. American Mathematical Soc., 2006, p. 9.)

Also, this double role appears to be one major motivation underlying the book of Alfsen and Shultz on operator algebras, as they explain in the preface: "[...] it is an important feature of quantum notable exception is the work of Gabriel Catren, who has insisted in the importance of the number-transformation double facet of properties:

> The twofold role played by classical observables in mechanics—as functions that can be evaluated on states and as generators of canonical transformations—is considered here a fundamental feature that deserves further attention.[16]

His influence on the conception of my work is substantial, and I here follow Catren in considering the double role of properties a key feature that should be made a cornerstone on which to center the conceptual analysis of Kinematics. To clearly distinguish both roles, I will often talk of "properties-as-quantities" and "properties-as-transformations".

In the light of this, the supposed state-property duality explodes and becomes *the fundamental conceptual triad of Kinematics*:

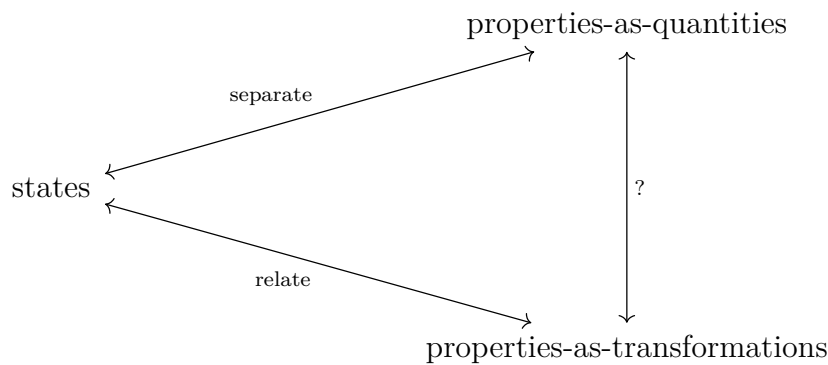

**Fig. II.1** – The fundamental conceptual triad of Kinematics.

The remainder of this chapter is a detailed analysis of this triad for the homogeneous arenas of the Classical and the Quantum. I start with their standard formulations.

---

mechanics that the physical variables play a dual role, as observables and as generators of transformation groups. The observables are random variables with a specified probability law in each state of the quantum system, while the generators determine one-parameter groups of transformations of observables (Heisenberg picture) or states (Schrödinger picture). [...] Both aspects can be adequately dealt with in *C*∗-algebras." (E. M. Alfsen and F. W. Shultz. *State Spaces of Operator Algebras.* Boston: Birkhäuser, 2001, pp. vii-viii.)

[16]G. Catren. "On Classical and Quantum Objectivity". In: *Foundations of Physics* 38.5 (2008), pp. 470–487. URL: http://philsci-archive.pitt.edu/4298/, p. 485.

## II.1.1 Standard Classical Kinematics

Classical Hamiltonian Mechanics is casted in the language of symplectic geometry. There are many excellent expositions of the subject. Some of the most standard references, in which all the technical details of this section can be found, are:

- J.-M. Souriau. *Structure des systèmes dynamiques.* Paris: Dunod, 1970

- P. R. Chernoff and J. E. Marsden. *Properties of Infinite Dimensional Hamiltonian Systems.* Lecture Notes in Mathematics. Heidelberg: Springer-Verlag, 1974

- R. Abraham and J. E. Marsden. *Foundations of Mechanics.* 2nd ed. Redwood City: Addison-Wesley Publishing Company, 1978

- V. I. Arnold. *Mathematical Methods of Classical Mechanics.* Trans. by K. Vogtmann and A. Weinstein. 2nd ed. Vol. 60. New York: Springer-Verlag, 1989

- M. Puta. *Hamiltonian Mechanical Systems and Geometric Quantization.* Dordrecht, The Netherlands: Kluwer Academic Publishers, 1993

- J. E. Marsden and T. S. Ratiu. *Introduction to Mechanics and Symmetry. A Basic Exposition of Classical Mechanical Systems.* 2nd ed. New York: Springer, 1999

The **space of states** is described by a finite-dimensional symplectic manifold $(S, \omega)$. This is a differentiable manifold *S equipped with one extra-structure*: a 2-form $\omega$ that is closed and non-degenerate. This means:

i) $\omega \in \Omega^2(S)$,

ii) $d\omega = 0$,

iii) $\forall p \in S, \forall v \in T_p S \ (v \neq 0), \exists v' \in T_p S, \omega(v, v') \neq 0$.

The dimension $d$ of a symplectic manifold is necessarily even: $d = 2n$. A state of the system is described by a point of the state space.

The **Lie group of global transformations** is the group $Aut(S) = Symp(n)$ of symplectomorphisms[17]. It is the subgroup of diffeomorphisms $\phi : S \longrightarrow S$ leaving invariant the symplectic 2-form: $\phi^* \omega = \omega$, where $\phi^* \omega$ is the pull-back of the symplectic

---

[17]Sometimes, these transformations are also called *canonical transformations*.

form[18].

The **Lie algebra of infinitesimal state transformations** is the Lie algebra associated to the group of global transformations. It is the Lie algebra $\Gamma(TS)_\omega$ of vector fields leaving invariant the symplectic 2-form: $\Gamma(TS)_\omega = \{v \in \Gamma(TS) \,|\, \mathcal{L}_v\omega = 0\}$ where $\mathcal{L}$ denotes the Lie derivative[19].

Finally, the **algebra of properties** is described by a Poisson algebra $(\mathcal{U}_\mathbb{R}, \bullet, \star)$.

**Definition II.1.** A **Poisson algebra** is a real (usually infinite-dimensional) vector space *equipped with two extra-structures*: a Jordan product $\bullet$ and a Lie product $\star$ such that

   i) $\bullet$ is a bilinear symmetric product,

  ii) $\bullet$ is associative,

 iii) $\star$ is a bilinear anti-symmetric product,

 iv) $\star$ satisfies the Jacobi identity: $f \star (g \star h) + g \star (h \star f) + h \star (f \star g) = 0$,

  v) $\star$ and $\bullet$ satisfy the Leibniz rule: $f \star (g \bullet h) = (f \star g) \bullet h + g \bullet (f \star h)$.

The Lie product of a Poisson algebra is very often called a *Poisson bracket* and denoted by $\{\cdot, \cdot\}$.

A property is described by an element of such an algebra. We see that it presents a structure slightly more involved than what we have encountered so far. Let us make a series of comments on this. First, axioms iii) and iv) turn $(\mathcal{U}_\mathbb{R}, \star)$ into a Lie algebra. Second, the Jacobi identity can be read in at least three different ways: using the anti-symmetry of the Lie product, one may write it as

$$f \star (g \star h) - (f \star g) \star h = -(f \star h) \star g$$

---

[18] A diffeomorphism $\phi : S \longrightarrow S$ induces a transformation $\Phi : \mathcal{C}^\infty(S, \mathbb{R}) \longrightarrow \mathcal{C}^\infty(S, \mathbb{R})$ defined by:

$$\forall f \in \mathcal{C}^\infty(S, \mathbb{R}), (\Phi f)(p) = f(\phi(p)).$$

This in turn allows to define the *push-forward* $\phi_*$ of vector fields and the *pull-back* $\phi^*$ of $n$-forms by:

$$\forall v \in \Gamma(TS), \ (\phi_* v)[f] := v[\Phi f]$$
$$\forall \alpha \in \Omega^n(S), \ (\phi^* \alpha)(v_1, \ldots, v_n) := \alpha(\phi_* v_1, \ldots, \phi_* v_n).$$

[19] For a given two-form $\alpha \in \Omega^2(S)$, the Lie derivative with respect to the vector field $v \in \Gamma(TS)$ is given by the so-called "Cartan's magic formula": $\mathcal{L}_v\alpha = (\iota_v d + d\iota_v)\alpha$, where $\iota_v\alpha := \alpha(v, \cdot) \in \Omega^1(S)$.

and it then expresses the non-associativity of the Lie product. But one may also write it as

$$f \star (g \star h) = (f \star g) \star h + g \star (f \star h)$$

and the Jacobi identity then captures the fact that, for any property $f \in \mathcal{U}_{\mathbb{R}}$, the map

$$v_f : \mathcal{U}_{\mathbb{R}} \longrightarrow \mathcal{U}_{\mathbb{R}}$$
$$g \longmapsto v_f(g) := f \star g$$

is a derivation with respect to the Lie product. Third, the Leibniz rule is the only one establishing a relation between the Jordan and Lie structures and it implies the map $v_f$ is a also derivation with respect to the Jordan product. There is hence a map

$$v_- : \mathcal{U}_{\mathbb{R}} \longrightarrow Der(\mathcal{U}_{\mathbb{R}})$$
$$f \longmapsto v_f$$

from the algebra of properties to the derivative operators on that algebra. This remark furnishes us the third possible reading of the Jacobi identity. Indeed, it may be now rewritten as

$$[v_f, v_g](h) = v_f \circ v_g(h) - v_g \circ v_f(h) = v_{f \star g}(h) \tag{II.1}$$

which expresses the fact that the map $v_-$, assigning operators to properties, is a morphism of Lie algebras. If we denote by $Der(\mathcal{U}_{\mathbb{R}})_H$ the image of $\mathcal{U}_{\mathbb{R}}$ by the map $v_-$ (that is, the Lie subalgebra of derivative operators arising from properties), and by $Z(\mathcal{U}_{\mathbb{R}})$ the center of $(\mathcal{U}_{\mathbb{R}}, \star)$, we then have the isomorphism of Lie algebras[20]

$$\mathcal{U}_{\mathbb{R}}/Z(\mathcal{U}_{\mathbb{R}}) \simeq Der(\mathcal{U}_{\mathbb{R}})_H. \tag{II.2}$$

This says that, up to an element of $Z(\mathcal{U}_{\mathbb{R}})$, *a classical property may as well be thought as a derivative operator, <u>as far as the Lie structure is concerned</u>*. If one takes into account the commutative Jordan product, this is no longer true since the algebra of derivative operators is just a Lie algebra and cannot be naturally equipped with a Jordan product[21]. In any case, this shows that, already at the level of *classical* mechanics, there is a close connection between physical properties and linear operators. In fact, this remark may be seen as one of the fundamental points in the rationale of geometric quantization[22].

---

[20]This follows immediately from the fact $Z(\mathcal{U}_{\mathbb{R}})$ is the kernel of the map $v_-$.

So far, I have discussed the Poisson algebra of classical properties without any mention to the space of states in order to stress what can be learned from the sole algebraic structure. Nonetheless, the back and forth between the geometrical picture naturally attached to states and the algebraic picture naturally attached to properties is a fundamental movement to properly understand the Kinematical arena. Of course, classical properties are most commonly described as smooth real-valued functions over the state space:

$$(\mathcal{U}_{\mathbb{R}}, \bullet, \star) \simeq (\mathcal{C}^{\infty}(S, \mathbb{R}), \cdot, \{\cdot, \cdot\}).$$

The commutative and associative Jordan product $\bullet$ is nothing but the usual point-wise multiplication $\cdot$, whereas the anti-commutative Lie product $\star$ is more commonly called in this context the Poisson bracket and denoted by $\{\cdot, \cdot\}$. Now, derivative operators of an algebra of functions over a smooth manifold are nothing but vector fields:

$$Der(\mathcal{U}_{\mathbb{R}}) \simeq \Gamma(TS)$$

and the derivative operator $v_f$ associated to the property $f$ is usually called the *Hamiltonian vector field* associated to $f$[23].

Both the Poisson bracket and the Hamiltonian vector field can alternatively be defined in terms of the symplectic structure. Since $\omega$ is non-degenerate, for any $p \in S$

---

[21]Indeed, neither the composition nor the anti-commutator of derivative operators yield another derivative operator. Thus, the purported analogue of Equation II.1

$$[v_f, v_g]_+(h) = v_f \circ v_g(h) + v_g \circ v_f(h) = v_{f \bullet g}(h)$$

fails to be true.

[22]For a detailed analysis of this last point, see Catren, "On Classical and Quantum Objectivity".

[23]More generally, the adjective "Hamiltonian" will almost systematically mean "arising from a property" (e.g. a Hamiltonian curve, a Hamiltonian vector field, a Hamiltonian $\mathfrak{g}$-action (cf. Definition III.2, page 247), etc.).

it establishes an isomorphism

$$\omega_p(\cdot, \cdot) : T_p S \overset{\sim}{\longrightarrow} T_p^* S$$

$$v \longmapsto \omega_p(v, \cdot)$$

which allows to define the Hamiltonian vector field associated to the property $f$ by

$$\omega(v_f, \cdot) := df.$$

Therefore, the Hamiltonian vector field $v_f$ is sometimes also called the *symplectic gradient* of $f$. Given this, the Poisson bracket of two functions $f$ and $g$ can be now defined as

$$\forall f, g \in \mathcal{C}^\infty(S, \mathbb{R}), \ \{f, g\} := \omega(v_g, v_f).$$

The fact that $\omega$ is a 2-form implies the anti-commutativity of the bracket, whereas the closedness of the symplectic form forces the Jacobi identity.

Conversely, the data of a bracket $\{\cdot, \cdot\}$ turning $(\mathcal{C}^\infty(S, \mathbb{R}), \cdot, \{\cdot, \cdot\})$ into a Poisson algebra is (almost) enough to define a symplectic structure on $S$. Indeed, given such a bracket, one can define a smooth antisymmetric tensor field $B \in \Gamma(\wedge_2(S))$, called the *Poisson tensor*, by the equation $\{f, g\} = B(df, dg)$. In virtue of the Jacobi identity, it will automatically satisfy a condition analogue of $d\omega = 0$[24]. Equipped with this structure, $(S, B)$ is a *Poisson manifold*[25] but not necessarily a symplectic manifold. In fact, a Poisson manifold is symplectic if and only if, locally, any possible transformation of the space is generated by a property. In other words, when we choose the space of states of a system to be symplectic, we are forcing any two states to be related by a locally Hamiltonian curve[26]. This is precisely the feature that conveys to a symplectic manifold its homogeneity and provokes the failure to meet the requirement of individuation. Were the space of states a general Poisson manifold, there would be infinitesimal state transformations not arising from properties and states impossible to connect by a Hamiltonian curve. As we will see later, this is the classical analogue of the quantum superselection rules.

---

[24]This condition is: given any 3-form $\alpha \in \Omega^3(S)$, $\iota_B d\iota_B \alpha = 0$, where $\iota_B \alpha := \alpha(B \wedge \cdot) \in \Omega^1(S)$. (See N. P. Landsman. *Mathematical Topics Between Classical and Quantum Mechanics.* New York: Springer, 1998, p. 66)

[25]See ibid., Definition I.2.3.1, p. 66, and the comments following it.

[26]Cf. ibid., Proposition I.2.3.7, p.68.

The point of the last two paragraphs was to show that the Lie structure present on the algebraic side of properties is the exact analogue of the symplectic structure present on the geometric side of states. Either of them can be seen as induced by the other. Now, let us discuss the relation with the fundamental conceptual triad of Kinematics.

It is tempting to conclude that the existence of two structures on the algebra of properties is a manifestation of the two roles of properties-as-quantities and properties-as-transformations. This is enforced by the remark that the Jordan product is point-wise multiplication and therefore relies on the fact that, when evaluated on a state, the properties produce real numbers. Moreover, it can be shown that all Hamiltonian vector fields preserve the symplectic structure and are thus infinitesimal state transformations (in the specific sense given above). The following picture emerges: classical properties are defined by their numerical role, which yields the *commutative* algebra $(\mathcal{C}^\infty(S, \mathbb{R}), \cdot)$ of properties-as-quantities. To this is added a second structure, the Lie product or Poisson bracket, which allows to introduce the second, transformational role of properties. The map

$$v_- : \mathcal{C}^\infty(S, \mathbb{R}) \longrightarrow \Gamma(TS)_H$$
$$f \longmapsto v_f = \{f, \cdot\}$$

is presently seen as the technical device that captures the role of properties-as-transformations. The Lie algebra—a fortiori *non-commutative* algebra—of properties-as-transformations is then the algebra of Hamiltonian vector fields $(\Gamma(TS)_H, [\cdot, \cdot])$. As as I have explained in the short discussion following Equation II.2, this algebra is *almost* the same as the algebra $(\mathcal{C}^\infty(S, \mathbb{R}), \{\cdot, \cdot\})$: it becomes exactly the same if, instead of considering the algebra of functions over the state space, one considers functions "up

to a constant"[27]:

$$(\Gamma(TS)_H, [\cdot, \cdot]) \simeq (\mathcal{C}^\infty(S, \mathbb{R})/\{df = 0\}, \{\cdot, \cdot\}).$$

In any case, the role of properties-as-transformations is indeed completely governed by the algebraic Lie structure or by the geometric symplectic structure. Under this light, the existence of a symplectic two-form on the space of states is a necessary manifestation of the fundamental transformational role of properties in Classical Kinematics.

All in all, the double-role of properties provides us with a rather satisfactory conceptual understanding of the mathematical arena of Classical Kinematics. Let us summarize the picture in a diagram:

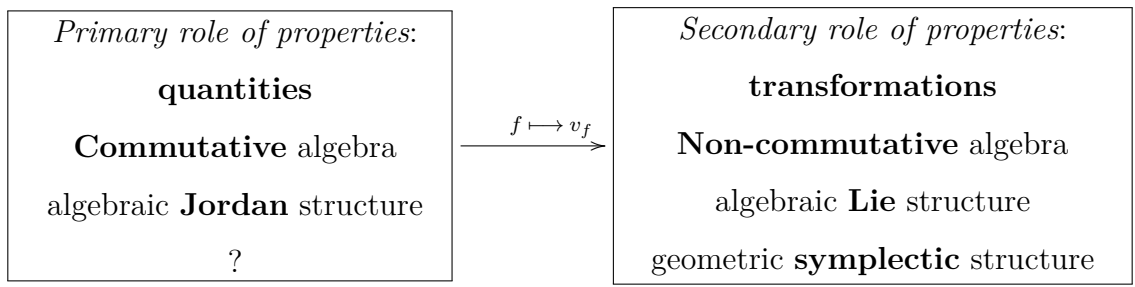

**Fig. II.2** – The role of properties in the standard formulation of Classical Kinematics

From this perspective, it becomes transparent that one cannot capture the transition from the Classical to the Quantum by a loose statement such as: "quantization is the transition from the commutative to the non-commutative"—statement unfortunately far too common in the literature. A similar claim may perhaps be pertinent, but it has to be rendered much more precise than this. Before turning to a description of Quantum Kinematics in its standard formulation, let us make some final remarks on the state-quantity-transformation interplay in the Classical context:

---

[27]Indeed, the center of the Lie algebra $(\mathcal{C}^\infty(S, \mathbb{R}), \{\cdot, \cdot\})$ is the set of *locally* constant functions, as may easily be seen from the definition of the Hamiltonian vector fields in terms of the symplectic structure. If one considers only *connected* symplectic manifolds, then the center of the Lie algebra are just the constant functions: $Z(\mathcal{C}^\infty(S, \mathbb{R}) \simeq \mathbb{R}$.

– *States and values.* Define a *valuation* on a real algebra $\mathcal{U}_\mathbb{R}$ as a function $\lambda : \mathcal{U}_\mathbb{R} \longrightarrow \mathbb{R}$ for which the functional composition principle (FUNC) holds—that is: if $f, g \in \mathcal{U}_\mathbb{R}$ are such that $g = A(f)$ for some real-valued function $A \in \mathcal{C}^\infty(\mathbb{R})$ then $\lambda[g] = A[\lambda(f)]$[28]. Any state $p \in S$ obviously defines a valuation $\lambda_p$ on the algebra of properties $\mathcal{C}^\infty(S, \mathbb{R})$ by $\lambda_p(f) = f(p)$. This means there is a consistent way of assigning a well-defined value to any physical property for any classical state. This is certainly a pedantic formulation of a trivial fact, but it becomes pertinent when compared to the Kochen-Specker theorem in Quantum Mechanics (cf. next section, page 164).

– *Quantities and transformations.* The condition of anti-symmetry of the Lie product may as well be read as a condition of *invariance*. Because of the linearity of the product, we have:

$$\big(\forall f, g \in \mathcal{C}^\infty(S, \mathbb{R}), \{f, g\} = -\{g, f\}\big) \Longleftrightarrow \big(\forall f \in \mathcal{C}^\infty(S, \mathbb{R}), \{f, f\} = 0\big).$$

The right hand side of the equivalence says that the *function $f$* is left invariant by the *infinitesimal transformation $v_f$*. Thus, the anti-symmetry of the product is equivalent to demanding any property to be invariant under the transformations it generates. In other words, defining on the space of states the equivalence relation: $p \overset{f}{\sim} q$ iff there exists a Hamiltonian curve generated by $f$ relating $p$ and $q$, we have

$$p \overset{f}{\sim} q \Longrightarrow f(p) = f(q).$$

---

[28] See for example C. Isham and J. Butterfield. "Topos Perspective on the Kochen-Specker Theorem: I. Quantum States as Generalized Valuations". In: *International Journal of Theoretical Physics* 37 (1998), pp. 2669–2733. URL: http://arxiv.org/abs/quant-ph/9803055, p. 2671. FUNC formalizes the intuition that, if $f$ is to have the value $\alpha$, then e.g. $f^2$ ought to have the value $\alpha^2$.

We therefore arrive to the following compatibility condition of the two fundamental role of properties:

> **Kinematical compatibility of transformations with quantities:** states that are distinguished by a property-as-quantity cannot be related by the associated property-as-transformation.

The converse is however not true, as is immediate by considering constant functions: states that are indistinguishable for a given property-as-quantity are not necessarily related by the property-as-transformation.

– *States and transformations.* A state $p$ is invariant under the transformations generated by the property $f$ only when $p$ is a critical point of the function.[29]

## II.1.2  Standard Quantum Kinematics

We now turn to the Quantum arena. In order to facilitate the comparison with the Classical case, the goal is to analyze the standard formulation of Quantum Kinematics, following as much as possible the analysis of the previous section. The standard formulation of Quantum Kinematics is casted into the language of Hilbert spaces, and the first thing to do is to identify again the four fundamental structures: the space of states, the Lie group of transformations, the Lie algebra of infinitesimal transformations and the algebra of properties.

However, a small cautionary remark is necessary before we proceed. Indeed, because any Hilbert space $\mathcal{H}$ is at the same time a manifold and a vector space, there is a natural isomorphism between the tangent bundle $T\mathcal{H}$ and $\mathcal{H} \times \mathcal{H}$. It is the following: given $(\phi, \psi) \in \mathcal{H} \times \mathcal{H}$, define $V_\phi \in T_\psi \mathcal{H}$ by

$$\forall f \in \mathcal{C}^\infty(\mathcal{H}, \mathbb{R}), V_\phi[f](\psi) = \frac{d}{ds} f(\psi + s\phi)\big|_{s=0}. \tag{II.3}$$

---

[29]Indeed, to say that the state $p$ is invariant under the transformations generated by the property $f$ is equivalent to demanding the Hamiltonian vector field $v_f$ to vanish at point $p$. Since, by definition, the symplectic two-form is non-degenerate and we have $\omega(v_f, \cdot) = df$, the condition $v_f\big|_p = 0$ is equivalent to $df\big|_p = 0$, which means $p$ is a critical point of the function $f$.

As a result of this, there is a constant ambiguity on the nature of the objects being handled in this formalism. In particular, it is so for operators, which can be conceived in at least two ways:

i) an operator is a map $F : \mathcal{H} \longrightarrow \mathcal{H}$ that to an element $\phi$ associates a new element $F(\phi)$. From this usual point of view, an operator is therefore a transformation of the Hilbert space onto itself.

ii) an operator is a map $F : \mathcal{H} \longrightarrow T\mathcal{H}$ that to an element $\phi$ associates the vector $V_{F(\phi)} \in T_\phi \mathcal{H}$ (defined through Equation II.3). From this point of view, an operator appears now to be a vector field on $\mathcal{H}$. Accordingly, some vector fields on $\mathcal{H}$ are oftentimes described as operators.

The subtle point one needs to be cautious about is that the two points of view are *not* equivalent, for there are expressions which make sense from the former perspective but not from the latter. For example, the composition of two operators is perfectly well defined from the point of view of operators as transformations, but is impossible to understand from the point of view of vector fields[30]. On the other hand, other expressions, such as the commutator, make sense from both perspectives. With this in mind, let us proceed.

The **space of states** is described by a (usually infinite-dimensional) Hilbert space $\mathcal{H}$. This a complex vector space *equipped with one extra-structure*: a hermitian inner product defining a norm $\| \cdot \|$ such that $(\mathcal{H}, \| \cdot \|)$ is a complete metric space. States of the system are described by rays of the space of states—that is, by one-dimensional subspaces of $\mathcal{H}$.

The **(Lie) group of global transformations** is the group $Aut(\mathcal{H}) = U(\mathcal{H})$ of unitary transformations. It is the subgroup of linear operators $A : \mathcal{H} \longrightarrow \mathcal{H}$ such that $A^{-1} = A^{*}$[31].

The **Lie algebra of infinitesimal state transformations** is the Lie algebra

---

[30]Diagrammatically, this is obvious. If $F$ is an arrow $\mathcal{H} \xrightarrow{F} \mathcal{H}$, then $F^2$ is simply the arrow 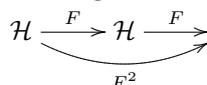 . But if $F$ is an arrow $\mathcal{H} \xrightarrow{F} T\mathcal{H}$, then $F^2$ is a mystery...

[31]Strictly speaking, $U(\mathcal{H})$ is only a *Lie* group when $\mathcal{H}$ is finite-dimensional. When this fails to be the case, $U(\mathcal{H})$ is not even a manifold because it is infinite-dimensional.

$(\mathcal{B}_{i\mathbb{R}}(\mathcal{H}), [\cdot, \cdot])$ of (bounded) *anti*-hermitian operators[32]. This is precisely a point where we take advantage of the ambiguity inherent to Hilbert spaces and describe vector fields as linear operators. Then, the Lie product $[\cdot, \cdot]$ of the Lie algebra $(\mathcal{B}_{i\mathbb{R}}(\mathcal{H}), [\cdot, \cdot])$ may be thought of as the commutator of operators, or as arising in the usual way from the composition law of the group $U(\mathcal{H})$.

Finally, the **algebra of properties** is described by a *non-associative* Jordan-Lie algebra $(\mathcal{U}_{\mathbb{R}}, \bullet, \star)$.

**Definition II.2.** A **non-associative Jordan-Lie algebra** is a real (usually infinite-dimensional) vector space *equipped with two extra-structures*: a Jordan product $\bullet$ and a Lie product $\star$ such that

i) $\bullet$ is a bilinear symmetric product,

ii) $\bullet$ is not associative: $(F \bullet G) \bullet H - F \bullet (G \bullet H) = (F \star H) \star G$ (associator rule),

iii) $\star$ is a bilinear anti-symmetric product,

iv) $\star$ satisfies the Jacobi identity: $F \star (G \star H) + G \star (H \star F) + H \star (F \star G) = 0$,

v) $F \star (G \bullet H) = (F \star G) \bullet H + G \bullet (F \star H)$[33].

A physical property is described by an element of such an algebra. Of course, quantum properties are standardly described as bounded self-adjoint operators:

$$(\mathcal{U}_{\mathbb{R}}, \bullet, \star) \simeq (\mathcal{B}_{\mathbb{R}}(\mathcal{H}), \frac{1}{2}[\cdot, \cdot]_+, \frac{i}{2}[\cdot, \cdot]).$$

---

[32]Again, for finite-dimensional Hilbert spaces, this is clear: any operator $V \in \mathcal{B}_{i\mathbb{R}}(\mathcal{H})$ defines a one-parameter group of unitary operators through exponentiation: $e^{tV} \in U(\mathcal{H}), t \in \mathbb{R}$. In the infinite-dimensional case, despite the fact that $U(\mathcal{H})$ is not a Lie group, one is still allowed to claim that the Lie algebra $\mathfrak{u}(\mathcal{H})$ associated to the group $U(\mathcal{H})$ consists of all anti-self-adjoint (not necessarily bounded) operators. This is because of Stone's theorem which shows there is a one-to-one correspondence between anti-self-adjoint operators and continuous one-parameter unitary groups. For a precise discussion of Stone's theorem, see R. Abraham, J. E. Marsden, and T. S. Ratiu. *Manifold, Tensor Analysis, and Applications*. 2nd ed. New York: Springer-Verlag, 1988, pp. 529–536.

[33]The name of this type of algebras comes from the fact that, given these axioms, $(\mathcal{U}_{\mathbb{R}}, \star)$ is a Lie algebra and $(\mathcal{U}_{\mathbb{R}}, \bullet)$ is a Jordan algebra. A real Jordan algebra $(\mathcal{U}_{\mathbb{R}}, \bullet)$ is a commutative algebra such that $F \bullet (G \bullet F^2) = (F \bullet G) \bullet F^2$ (Landsman, op. cit., Definition I.1.1.1., p. 37). Thus, although in general a Jordan algebra need not be associative, it must be power-associative. These algebras are of course called after Pascual Jordan, who introduced them in 1933 when trying to abstract the algebraic structure of quantum properties. An excellent introduction to the mathematical subject of Jordan algebras is found in K. McCrimmon. *A Taste of Jordan Algebras*. New York: Springer-Verlag, 2004.

The Jordan and Lie products are related to the composition of operators $\circ$, by means of the anti-commutator and ($i$-times) the commutator respectively:

$$F \star G := \frac{i}{2}[F, G] = \frac{i}{2}(F \circ G - G \circ F)$$
$$F \bullet G := \frac{1}{2}[F, G]_+ = \frac{1}{2}(F \circ G + G \circ F)^{[34]}.$$

Comparison with the algebra of classical properties shows these two algebras are strikingly similar. The only point where they differ is axiom ii): whereas the Jordan product of classical kinematics is associative, the quantum one is not. However, some suspicion may be raised against this comparison, for indeed the anti-commutator between two different self-adjoint operators barely ever shows up in Quantum Mechanics. One may then wonder whether the quantum Jordan product has not been artificially introduced here in order to force the analogy with the Classical case. But this suspicion fades away as soon as one realizes the intimate connection between the Jordan product and the familiar—and certainly fundamental!—operation of taking *squares*[35]. Indeed, with only the Lie product at our disposal, there would be no way of defining the square $F^2$ of a given property $F$, since $F \star F = 0$ ($\star$ is anti-symmetric). With the introduction of the Jordan product, one simply puts $F^2 := F \bullet F$. Conversely, if one supposes the square $F^2$ to be somehow a meaningful operation, then it is possible to define the Jordan product by

$$F \bullet G = \frac{1}{4}\big((F + G)^2 - (F - G)^2\big).$$

---

[34] The choice of representing properties by *bounded* self-adjoint operators is not without controversies. In particular, this restriction excludes the usual position and momentum operators, but in return, it greatly simplifies the mathematical treatment. For instance, if we remove the boundedness, it is simply false that the set of all self-adjoint operators is closed under $\frac{i}{2}[\cdot, \cdot]$.

Moreover, it is important to stress that the Lie product on bounded self-adjoint operators is *not* the commutator: the multiplication by the complex number $i$ in the definition is a necessary one. This is because the commutator of two self-adjoint operators yields an *anti*-self-adjoint operator. On the other hand, the two factors $\frac{1}{2}$ are only a convenient normalization in order to obtain the associator rule as written in axiom *ii*), but other choices are possible. For instance, another normalization is $F \star G := \frac{i}{\hbar}[F, G]$, which forces $\kappa = \frac{\hbar^2}{4}$ (cf. the next definition), but allows to write the canonical commutation relations between position and momentum operators as $P \star X = 1$.

[35] Recall, for example, the opening question of Heisenberg's seminal paper of 1925: "If instead of a classical quantity $x(t)$ we have a quantum-theoretical quantity, what quantum-theoretical quantity will appear in place of $x^2(t)$?" (see section I.1.1.a, page 21).

Therefore, if one is willing to consider squares of quantum properties, one is forced to consider the quantum Jordan product[36].

The comparison of the algebraic structure of classical and quantum properties motivates the definition of a general (not necessarily non-associative) Jordan-Lie algebra, which encapsulates both the classical and the quantum cases[37]:

**Definition II.3.** A **general Jordan-Lie algebra** is a real (possibly infinite-dimensional) vector space equipped with a Jordan product $\bullet$ and a Lie product $\star$ such that

i) $\bullet$ is a bilinear symmetric product,

ii) $\star$ is a bilinear anti-symmetric product,

iii) $\star$ satisfies the Jacobi identity: $F \star (G \star H) + G \star (H \star F) + H \star (F \star G) = 0$,

iv) $F \star (G \bullet H) = (F \star G) \bullet H + G \bullet (F \star H)$,

v) there exists $\kappa \in \mathbb{R}$ such that $(F \bullet G) \bullet H - F \bullet (G \bullet H) = \kappa (F \star H) \star G$ (associator rule).

Only the last axiom differentiates classical and quantum properties. When $\kappa = 0$, the Jordan product is associative and one gets the definition of a Poisson algebra describing classical properties (cf. Definition II.1, page 148). When $\kappa = 1$, one gets the previous definition for the algebra of quantum properties with a non-associative Jordan product. In fact, whenever $\kappa \neq 0$, one may always rescale the Lie product as to yield $\kappa = 1$. Therefore, the world of Jordan-Lie algebras is sharply divided into the sole cases of $\kappa = 0$ (corresponding to Classical Mechanics) and $\kappa = 1$ (corresponding to Quantum Mechanics). In this precise sense, one can say that the transition from classical properties to quantum properties is the transition from associativity to non-associativity—rather than from commutativity to non-commutativity—and that *the real difference between classical and quantum properties lies on the Jordan side.*

---

[36]This point of view on the Jordan product is expressed in J. C. Baez. "Division Algebras and Quantum Theory". In: *Foundations of Physics* 42.7 (2012), pp. 819–855. URL: http://arxiv.org/abs/1101.5690, p. 8.

[37]Landsman, op. cit., Definition I.1.1.2., pp. 37–38.

That this is so need not be a surprise. With the previous experience of Classical Kinematics in mind, one should expect the associations

$$\text{Lie structure} \quad \longleftrightarrow \quad \text{properties-as-transformations}$$

$$\text{(II.4)}$$

$$\text{Jordan structure} \quad \longleftrightarrow \quad \text{properties-as-quantities}$$

to be valid also in the quantum arena. And it seems reasonable to say that the most important differences between classical and quantum properties involve their behavior as quantities.

One particular way of perceiving the difference of behavior between classical and quantum properties-as-quantities is the following. Suppose we have at our disposal some 'numerical pairing' $\langle \cdot, \cdot \rangle$ which allows to assign a number $\langle f, \sigma \rangle$ to any given state $\sigma$ and property $f$, and consider the collection of numbers

$$\mathcal{N}_\sigma^f := \{\langle f, \sigma \rangle, \langle f^2, \sigma \rangle, \langle f^3, \sigma, \rangle, \ldots\}$$

where each power $f^n$ is constructed using the Jordan product (e.g., $f^3 := f \bullet f \bullet f$). The set $\mathcal{N}_\sigma^f$ may be seen as encoding the numerical role of the property $f$ with respect to the state $\sigma$. Now, one can ask: Is there more information contained in the data of the whole of $\mathcal{N}_\sigma^f$ than in the data of the single number $\langle f, \sigma \rangle$? In Classical Mechanics, the answer is negative: the numerical pairing $\langle f, \sigma \rangle$ is the evaluation of the real-valued function $f$ on the point $\sigma$ and the Jordan product is point-wise multiplication. Therefore, *by definition*, one has $\langle f^n, \sigma \rangle = (\langle f, \sigma \rangle)^n$: the whole set of numbers $\mathcal{N}_\sigma^f$ is known if one knows the first element $\langle f, \sigma \rangle$. In other words, the numerical role of classical properties is *single-layered*: it may be reduced to the data of a single number (the value of the property on the state). In Quantum Mechanics, on the other hand, the situation is very different. Take for example the numerical pairing $\langle \cdot, \cdot \rangle$ to be defined by:

$$\forall F \in \mathcal{B}_\mathbb{R}(\mathcal{H}), \forall [\phi] \in \mathbb{P}\mathcal{H}, \langle F, [\phi] \rangle := \frac{(\phi, F\phi)}{(\phi, \phi)} \qquad \text{(II.5)}$$

where $(\cdot, \cdot)$ is the Hermitian product on $\mathcal{H}$. According to the standard statistical interpretation, the number $\langle F, [\phi] \rangle$ does of course not represent the definite value of the property $F$ on the state $[\phi]$, but rather the *expected* value. Therefore, in general $\langle F^2, [\phi] \rangle \neq (\langle F, [\phi] \rangle)^2$, and the collection of numbers $\mathcal{N}^F_{[\phi]}$ is not determined by the data of one single number: the numerical role of quantum properties is in fact a complex, *multi-layered* structure[38]. Now, this existence of multiple layers in the numerical role of quantum properties simply encodes the statistical nature of Quantum Mechanics. Indeed, the standard deviation $\Delta F([\phi])$ of the property $F$ at the state $[\phi]$—which is a sensible numerical quantity only insofar as the theory is statistical: a classical property has a definite value on a state and there is nothing more to it—is exactly a measure of the non-coincidence between the first two numerical layers of the quantum property-as-quantity:

$$\Delta F([\phi]) = \left| \langle F^2, [\phi] \rangle - \left( \langle F, [\phi] \rangle \right)^2 \right|^{\frac{1}{2}}.$$

Hence, under the light of the familiar fact that Quantum Mechanics is statistical in nature whereas Classical Mechanics is not, the statement that "the real difference between the Classical and the Quantum lies on the quantitative side of properties" appears as a very natural remark.

This provides some heuristic control over the plausibility that, indeed, the algebraic Jordan structure is the one which encapsulates the differences between the two Kinematics. But the associations (II.4) remain to be checked in detail for the Quantum case. That the Lie structure encapsulates again the role of properties-as-transformations may easily be seen, for most of what has been said in the previous section concerning the algebra of classical properties can automatically be transposed in similar terms to the Quantum. In particular, the quantum analogue of the classical map $v_- : \mathcal{C}^\infty(S, \mathbb{R}) \longrightarrow \Gamma(TS)_H$, used to capture the transformational role of classical

---

[38]At this point, one could perhaps speculate that this feature is highly dependent on our choice for the numerical pairing $\langle \cdot, \cdot \rangle$, and that a modification of Equation II.5 could yield a set of numbers $\mathcal{N}^F_{[\phi]}$ that behave exactly as the classical $\mathcal{N}^f_\sigma$. In this regard, the much discussed Kochen-Specker theorem is a no-go. It establishes that, if $\dim(\mathcal{H}) > 2$, it is simply not possible to choose in the Quantum Kinematical arena a numerical pairing such that $\langle F^n, [\phi] \rangle = (\langle F, [\phi] \rangle)^n$. More precisely, it says that no valuations on $\mathcal{B}_\mathbb{R}(\mathcal{H})$ exist (cf. page 154 for the definition of a valuation). For a short discussion on the Kochen-Specker theorem see for example R. Hermens. "Quantum Mechanics, From Realism to Intuitionism". MA thesis. Radboud University Nijmegen, 2010. URL: http://arxiv.org/abs/1002.1410, pp. 32-ff.

properties, is here defined as

$$V_- : \ \mathcal{B}_\mathbb{R}(\mathcal{H}) \longrightarrow \mathcal{B}_{i\mathbb{R}}(\mathcal{H})$$
$$F \longmapsto V_F := iF. \tag{II.6}$$

In other words, given a quantum property $F$, the associated infinitesimal state transformation is simply the anti-hermitian operator obtained multiplying by $i$. In fact, this relation—between the existence of a Lie structure on the algebra of properties and the transformational role of properties in Kinematics—works even better than in Classical Kinematics, since the map $V_-$ is a canonical isomorphism of Lie algebras between $(\mathcal{B}_\mathbb{R}(\mathcal{H}), i[\cdot, \cdot])$ and $(\mathcal{B}_{i\mathbb{R}}(\mathcal{H}), [\cdot, \cdot])$.[39] This means: *to consider quantum properties solely in their role of properties-as-transformations—that is, to ignore their role of properties-as-quantities—corresponds exactly to forgetting the algebraic Jordan structure.* In other words, the map from properties to properties-as-transformations may be written

$$V_- : \ (\mathcal{B}_\mathbb{R}(\mathcal{H}), \frac{1}{2}[\cdot, \cdot]_+, \frac{i}{2}[\cdot, \cdot]) \longrightarrow (\mathcal{B}_\mathbb{R}(\mathcal{H}), \frac{i}{2}[\cdot, \cdot])$$
$$F \longmapsto F.$$

En passant, this last point shows that, whereas in Classical Kinematics there was an emphasis on properties-as-quantities and properties-as-transformations were secondary—properties were first defined as functions and only then one could associate vector fields to them—, Quantum Kinematics, at least in its standard Hilbert space formulation, presents the reverse situation: we clearly have properties-as-transformations and the reading of properties-as-quantities is more involved. This sheds light on the fact that quantum properties are easier to define through their transformational role (cf. the citation from Townsend on page 145). Thus, we have the diagram:

---

[39]One could make the choice of multiplying by $-i$ instead of $i$, defining thus the vector field $V_F \in \Gamma(T\mathcal{H})$ by $V_F\big|_\phi := -iF\phi$. With this new choice, the one-parameter group of unitary transformations associated to the property $F$ is the group of operators $\exp(-itF)$, as usual. The (slight) default is that the map $V_-$ becomes an *anti*-isomorphism of Lie algebras.

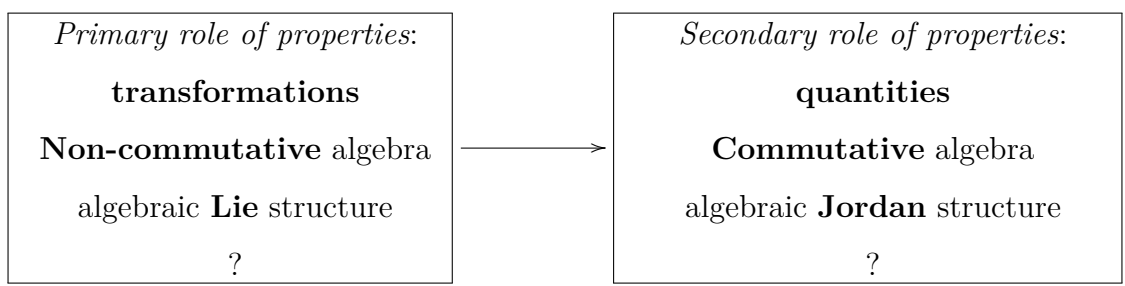

**Fig. II.3** – The role of properties in the standard formulation of Quantum Kinematics.

As a side remark, comparison of Figure II.3 and Figure II.2 (page 153) clearly shows in which way the widespread image

"classical = commutative ; quantum = non-commutative"

arises from of a wrong analogy between the two Kinematics. Indeed, instead of comparing either the full algebras of properties (with both the Jordan and Lie structures), or else the two non-commutative algebras of properties-as-transformations, the wrong characterization of the Classical/Quantum transition compares the primary algebra of classical properties with the primary algebra of quantum properties. It therefore compares classical properties-as-quantities with quantum properties-as-transformations...[40]

---

[40]This mistake was there since the very beginning of Quantum Mechanics. In his seminal paper, Heisenberg writes:

> Whereas in classical theory $x(t)y(t)$ is always equal to $y(t)x(t)$, this is not necessarily the case in quantum theory.
>
> (W. Heisenberg. "Quantum-theoretical Re-interpretation of Kinematic and Mechanical Relations". In: *Sources of Quantum Mechanics*. Ed. by B. Van der Waerden. New York: Dover Publications, Inc., 1967, pp. 261–276, p. 266)

In the same vein, one can find in the second paper from Born, Heisenberg and Jordan:

> We introduce the following basic quantum-mechanical relation: $\boldsymbol{pq} - \boldsymbol{qp} = \frac{h}{2\pi i}\boldsymbol{1}$. [...] One can see from [this equation] that in the limit $h = 0$ the new theory would converge to classical theory, as is physically required.
>
> (M. Born, W. Heisenberg, and P. Jordan. "On Quantum Mechanics II". in: *Sources of Quantum Mechanics*. Ed. by B. Van der Waerden. New York: Dover Publications, Inc., 1967, pp. 321–384, p. 327)

It is clear that they were comparing the commutator in Quantum Mechanics with point-wise multiplication in Classical Mechanics (despite the fact that, by the time of the second quoted paper, Dirac had already shown the quantum commutator should be compared to the classical Poisson bracket).

Returning to the discussion of the double role of properties in the quantum arena, it is interesting to note that two of the most notable distinguishing features of the Quantum with respect to the Classical Kinematics—namely, the existence of a condition for states to have a definite value of a property, and the existence of a compatibility condition for properties—may be reformulated as invariance conditions:

1. *Existence of a condition for states to have a definite value of a property.* As we have already commented, not all states have a definite value of a property. This is the case if and only if the variance $\widetilde{F^2} - (\tilde{F})^2$ vanishes, which happens to be so if and only if $\phi$ is an eigenvector of the self-adjoint operator: $F\phi = \tilde{F}(\phi)\phi$. In other words, the operator $F$ must leave the state $[\phi]$ invariant (since the rays $[\tilde{F}(\phi)\phi]$ and $[\phi]$ are equal). In this way, we arrive at the following statement: *a physical state $[\phi]$ has a definite value of the property $F$ if and only if it is invariant under the transformations generated by the property*[41].

2. *Existence of a compatibility condition for properties.* Two properties are said to be compatible if they are simultaneously measurable. As von Neumann showed in his book, properties are compatible if and only if their commutator vanishes[42]. But, since $i[F, \cdot]$ is the infinitesimal transformation associated to the property $F$, this is readily reformulated in terms of invariance: *two quantum properties $F$ and $G$ are compatible if and only if $F$ is invariant under the transformations generated by $G$ (or viceversa).*

The possibility of this reformulation hints to the idea that the Quantum is characterized by a particular interplay between the transformational and numerical role of properties. This idea will be become more and more precise as we advance in the analysis of the two kinematical arenas. It is also at the origin of an attempt, developed by Gabriel Catren, to construct a group-theoretical ontology of Mechanics[43]. For the moment, we will content ourselves with leaving the idea latent in the background.

---

[41] A familiar example of this is given by the fact that states with a sharply defined value of energy are called *stationary* states (since they can analogously be defined as states that do not evolve in time (i.e., are invariant under time translations)).

[42] Von Neumann, op. cit., p. 228.

[43] See for instance Catren, op. cit. or G. Catren. "On the Relation Between Gauge and Phase Symmetries". In: *Foundations of Physics* 44 (2014), pp. 1317–1335.

All of this having been said, the relation between the numerical role of quantum properties and the Jordan structure remains nonetheless obscure. In this regard, the present account of the Quantum arena is still frustrating. Moreover, we have here completely lost the interplay, which was felt so satisfactory in the Classical arena, between the algebraic picture arising from properties and the geometric picture arising from states. This motivates the attempt to change our perspective and describe Quantum Kinematics with a more geometric language, somewhat leaving ashore the standard Hilbert space formalism.

## II.2 The Quantum seen from the Classical: the geometric formulation

The comparison of the standard formulations of both Kinematics has brought out a striking structural similarity between the algebras of classical and quantum properties. They both present two products—one commutative and one anti-commutative—whose existence may be seen as a manifestation of the fundamental double role of the properties of a physical system. On the other hand, the classical and quantum descriptions of the space of states seem at first sight not to have any points in common. Ashtekar and Schilling nicely explain the situation:

> While the classical framework is *geometric and non-linear*, the quantum description is intrinsically *algebraic and linear*. Indeed, the emphasis on the underlying linearity is so strong that none of the standard textbook postulates of quantum mechanics can be stated without reference to the linear structure of $\mathcal{H}$.[44]

One could then be inclined to think that, although the non-associativity of the Jordan product has been spotted as the main difference between classical and quantum properties, the really crucial difference between Classical and Quantum Kinematics lies in the nature of the space of states. For in the popular conception of Quantum Mechanics, the linearity of the space of states is concomitant of the *superposition principle*, which in turn is often regarded as one—or perhaps *the*—fundamental feature of the theory, as Dirac asserts:

> For this purpose [of building up quantum mechanics] a new set of accurate laws of nature is required. One of the most fundamental and the most drastic of these is the *Principle of Superposition.*[45]

---

[44]A. Ashtekar and T. A. Schilling. "Geometrical Formulation of Quantum Mechanics". In: *On Einstein's Path: Essays in Honor of Engelbert Schücking*. Ed. by A. Harvey. New York: Springer, 1997, pp. 23–65. URL: http://arxiv.org/abs/gr-qc/9706069, p. 25, authors' emphasis.

With this principle in hand, the knowledge of any two different states of the physical system being considered allows one to construct a whole set of new states. In the Hilbert space picture, given the states $\phi, \psi \in \mathcal{H}$, this new collection is simply the set of all possible complex linear combinations $\varphi = a\phi + b\psi$, for $a, b \in \mathbb{C}$. At best, physicists had previously encountered superposition of *waves* and solutions to linear equations of motion, but this superposition of *states* certainly was a first, with no analogue in Classical Mechanics. Here is again Dirac insisting on this point:

> The nature of the relationships which the superposition principle requires to exist between the states of any system is of a kind that cannot be explained in terms of familiar physical concepts. One cannot in the classical sense picture a system being partly in each of two states and see the equivalence of this to the system being completely in some other state. There is an entirely new idea involved, to which one must get accustomed and in terms of which one must proceed to build up an exact mathematical theory. [...]
>
> *It is important to remember, however, that the superposition that occurs in quantum mechanics is of an essentially different nature from any occurring in the classical theory.*[46]

From this perspective, the apparently radical difference between the geometric space of classical states and the linear space of quantum states may be perceived as the natural—and almost necessary—manifestation of this "drastic" new feature of the Quantum. But in claiming so, one forgets a central point, which indicates this whole idea cannot be the end of the story: unlike in the classical arena, *in the quantum arena states are not described by points of the Hilbert space.* Rather, states are described by rays.

[45] P. A. M. Dirac. *The Principles of Quantum Mechanics.* 4th ed. Oxford: Oxford University Press, 1958, p. 4, Dirac's emphasis

[46] Ibid., pp. 12–14, Dirac's emphasis.

This sole remark suffices to raise great deal of suspicion towards Hilbert spaces. In a letter to Garrett Birkhoff, von Neumann mentions it as one motivating reason to seek a reformulation of the quantum theory in new terms:

> I would like to make a confession which may seem immoral: I do not believe absolutely in Hilbert spaces any more. [...] Because:
>
> (1) The vectors ought to represent the physical *states*, but they do it redundantly, up to a complex factor, only.[47]

Be that as it may, it is certainly a sufficient remark to realize that quantum mechanics is not as obviously linear as one initially may have thought: the "true" quantum space, in which points do represent states, is the projective Hilbert space $\mathbb{P}\mathcal{H}$, a genuine *non-linear* manifold.

The principle of superposition has certainly been a powerful idea, with a strong influence on the heuristics of the Quantum, and its link with the linearity of Hilbert spaces has been in my opinion one of the main reasons for the widespread use of the standard formalism. However, in the attempt to compare Classical and Quantum Kinematics, due care should be taken to express both Kinematics in as similar terms as possible. It becomes therefore natural to attempt a reformulation of quantum mechanics in a language resembling the classical one—that is, to forget Hilbert spaces

---

[47]Letter from von Neumann to G. Birkhoff, J. von Neumann. *John von Neumann: Selected Letters.* Ed. by M. Rédei. History of Mathematics. American Mathematical Society, 2005, p. 59, author's emphasis. In fact, this was not the unique—and not even the main—reason for von Neumann to be dissatisfied with Hilbert spaces, as is clear from the continuation of the letter:

> (2) And besides the *states* are merely a derived notion, the primitive (phenomenologically given) notion being the *qualities*, which correspond to the *linear closed subspaces*.
>
> But if we wish to generalize the lattice of all linear closed subspaces from a Euclidean space to infinitely many dimensions, then one does not obtain Hilbert space, but that configuration that Murray and I called "case $II_\infty$".

For a careful discussion of von Neumann's attitude towards Hilbert spaces, see M. Rédei. "Why John von Neumann Did Not Like The Hilbert Space Formalism of Quantum Mechanics (and What He Liked Instead)". In: *Studies In History and Philosophy of Science Part B: Studies In History and Philosophy of Modern Physics* 27.4 (1996), pp. 493–510.

and to develop the quantum theory directly in terms of the geometry of $\mathbb{P}\mathcal{H}$. In particular, this means redefining the algebra of properties—since one needs to forget about operators—and giving a new, geometrical account of the principle of superposition.

The task of this reformulation is sometimes referred to as the "*geometric or delinearization program*". Its explicit goal is to reestablish the fruitful link, witnessed in the Classical arena, between the geometry of the space of states and the algebraic structures of properties. Kibble's article *"Geometrization of Quantum Mechanics"*[48] is often cited as the initiator of the program, although many ideas and results were previously known, as one may see from the introduction of that article[49]. I regard the works of Ashtekar and Schilling on one side, and Cirelli, Manià and their collaborators on the other, as those having managed to pursue this road the furthest[50]. The former provides an excellent physical insight into what is gained from this reformulation, while the latter furnishes the most elegant mathematical description of the geometrical setting. Hughston and people surrounding him have also played an important role[51]. They all clearly express the goals of their program:

> The delinearization program, by itself, is not related in our opinion to attempts
> to construct a non linear extension of QM with operators which act non linearly
> on the Hilbert space $\mathcal{H}$. The true aim of the delinearization program is to free
> the mathematical foundations of QM from any reference to linear structure and
> to linear operators. It appears very gratifying to be aware of how naturally
> geometric concepts describe the more relevant aspects of ordinary QM, suggesting
> that the geometric approach could be very useful also in solving open problems

---

[48]T. Kibble. "Geometrization of Quantum Mechanics". In: *Communications in Mathematical Physics* 65.2 (1979), pp. 189–201.

[49]There is for example Mielnik's article *"Geometry of Quantum States"* which tries to describe physical systems in terms of the geometry determined by the transition probability structure and argues against "the old opinion that the only reasonable mathematical schemes to describe quantum phenomena [are] those related to Hilbert spaces".

[50]See in particular: T. A. Schilling. "Geometry of Quantum Mechanics". PhD thesis. The Pennsylvania State University, 1996; Ashtekar and Schilling, op. cit.; Cirelli, Gatti, and Manià, op. cit.

[51]See D. C. Brody and L. P. Hughston. "Geometric Quantum Mechanics". In: *Journal of geometry and physics* 38.1 (2001), pp. 19–53. URL: http://arxiv.org/abs/quant-ph/9906086, and references therein.

in Quantum Theories.[52]

In the same vein, Schilling writes in the introduction of his Ph.D. thesis:

> The goal of this thesis is a formulation of the postulates of standard quantum mechanics in a language which is intrinsic to the true space of states. The intent is to lay a foundation by which one may study, for example, the classical limit [...]. The desired formalism shall be valid for the generic quantum theory [...]. It should be emphasized that we are seeking a description of *ordinary* quantum mechanics; we introduce no new input, but merely acknowledge mathematical structures which are already inherent to the standard formalism. The difference is one of semantics, but a potentially useful one.
>
> The description presented here allows one to adopt a viewpoint in which the Hilbert space is a *fiducial* structure, not an essential ingredient.[53]

In regard to the questions that concern us in this chapter—namely, the link between: i) the two-fold role of properties, ii) the Jordan and Lie structures of the algebra of quantum properties, iii) the geometric structures present on the space of states—the hope is that the geometric reformulation of the quantum arena will provide us with new insights[54].

\* \* \* \* \*

The main result from which the whole geometric program springs is the fact that the projective space $\mathbb{P}\mathcal{H}$ is a Hermitian symmetric space with automorphism group $G = U(\mathcal{H})$[55]. This furnishes a completely new start to the description of the quantum kinematical arena, which strongly resembles the standard description of the classical kinematical arena of subsection II.1.1. It is the following.

---

[52]Cirelli, Gatti, and Manià, op. cit., p. 268.

[53]Schilling, op. cit., p. 4, Schilling's emphasis.

[54]It is important to clearly distinguish the 'geometric or delinearization program' from the program of 'geometric quantization' which we won't discuss here and is completely unrelated. The first aims at a reformulation of quantum mechanics which avoids Hilbert spaces. The second is geared towards an explicit construction of the quantum description of a system for which the classical description is given. But the resulting quantum description is still "intrinsically algebraic and linear", since it is based on Hilbert spaces. What is 'geometric' about geometric quantization is the means by which the Hilbert space is constructed: roughly, one starts with the symplectic manifold describing the classical system, considers a complex line bundle over it and defines the Hilbert space in terms of the sections of this bundle. The program of geometric quantization was started by Jean-Marie Souriau and Bertram Kostant. A standard reference is Woodhouse, op. cit.

[55]Cirelli, Gatti, and Manià, op. cit., section 2.

The quantum **space of states** is described by a (usually infinite-dimensional) Hermitian symmetric space $(M, G, s, \omega, J)$. This is a symmetric space $(M, G, s)$[56] *equipped with two extra-structures*: a symplectic 2-form $\omega$ and an integrable almost complex structure $J$[57] that is compatible with the symplectic structure: for any two vectors $v, w \in T_x M$, $\omega(Jv, Jw) = \omega(v, w)$. A state of the system is described by a point of the state space.

Given these geometrical structures, one can naturally define a Riemannian metric $g$ on the quantum space of states by:

$$\forall v, w \in T_x M, g(v, w) := \omega(v, Jw).$$

In fact, a Hermitian symmetric space is usually defined as a quintuple $(M, G, s, g, J)$ where $g$ is a Riemannian metric and $J$ is an invariant almost complex structure[58]. The previous equation is then perceived as the definition of the symplectic form. The important fact for us is that the quantum space of states is both a symplectic manifold and a Riemannian manifold, and has thus a very rich geometry whose meaning will be explored in the following subsections..

The **Lie group of global transformations** is the group $G$ of automorphisms of the Hermitian symmetric space. By supposition, one takes $G \simeq U(\mathcal{H})$, as would be the case for a projective Hilbert space. These transformations are diffeomorphisms that are both symplectomorphisms and isometries.

---

[56]A *symmetric G-space* is a homogeneous space $(M, G)$ together with an involutive diffeomorphism $s : M \to M$ which has an isolated fixed point. A *homogeneous G-space* is a manifold $M$ on which: *i)* $G$ acts smoothly and transitively, and *ii)* for any $x \in M$, the isotropy group $G_x$ is a Lie subgroup of $G$. (See Note 7 of S. Kobayashi and K. Nomizu. *Foundations of Differential Geometry.* Vol. 1. New York: Wiley, 1963, pp. 300-ff or, for the infinite-dimensional case Cirelli, Gatti, and Manià, op. cit., p. 3.)

[57]An *almost complex structure* is a smooth tensor field $J \in \Gamma(TM \otimes T^*M)$ such that for all $x \in M, J_x^2 = -1$.

[58]Associated to the Riemannian metric, there is a unique torsion-free metric compatible affine connection $\nabla$ (the so-called Levi-Civita connection). An almost complex structure $J$ is said to be *invariant* if $\nabla J = 0$. See for example S. Kobayashi and K. Nomizu. *Foundations of Differential Geometry.* Vol. 2. New York: Wiley, 1969, p. 259, and also Cirelli, Gatti, and Manià, op. cit., p. 6.

The **Lie algebra of infinitesimal state transformations** is the Lie algebra associated to the group of global transformations. It is hence isomorphic to the algebra $(\mathcal{B}_{i\mathbb{R}}(\mathcal{H}), [\cdot, \cdot])$ of bounded anti-self-adjoint operators, but is here interpreted as the algebra of smooth vector fields which preserve both the symplectic and Riemannian structures.

Now, the first important step in the route to a complete geometrical formulation of the quantum kinematical arena is to provide a geometrical description of the non-associative Jordan-Lie algebra of quantum properties. With the classical arena in mind, the natural strategy is to consider the collection $\mathcal{C}^\infty(\mathbb{P}\mathcal{H}, \mathbb{R})$ of smooth functions over the quantum space of states, and to define two products $\bullet$ and $\star$, induced by some of the geometrical structures of the quantum space of states, so that $(\mathcal{B}_\mathbb{R}(\mathcal{H}), \frac{1}{2}[\cdot, \cdot]_+, \frac{i}{2}[\cdot, \cdot])$ is isomorphic to $(\mathcal{C}^\infty(\mathbb{P}\mathcal{H}, \mathbb{R}), \bullet, \star)$ or to a subalgebra of it. Here, there are no surprises: the anti-commutative Lie product will be defined by means of the anti-symmetric bilinear form $\omega$, whereas the commutative Jordan product will be defined by means of the symmetric bilinear form $g$. But to see this, it is necessary to understand the link between the geometry of the projective Hilbert space and the operations of the original Hilbert space.

## II.2.1   The quantum space of states as a classical space of states

Symplectic geometry is not characteristic of Classical Mechanics. As has been hinted at and will now be explained in detail, it plays an equally important role in the Quantum arena. Any space of states, be it classical or quantum, is a symplectic manifold and the difference between both arenas is to be looked for elsewhere. In a way, this was expected: section II.1 showed us that classical and quantum properties-as-transformations behave in exactly the same way, and that the symplectic form could be perceived as the geometrical manifestation of the transformational role of properties.

All the geometrical features of the projective Hilbert space can be understood in purely group-theoretical terms, using the "well known" theory of Hermitian symmetric spaces as can be found in the book by Kobayashi and Nomizu[59]. This is the approach of Cirelli and his collaborators. However, our aim for a new geometrical characterization of quantum properties necessitates we approach the geometry of the "true" quantum space of states in a somewhat indirect fashion, passing through the geometric structures already present at the level of the Hilbert space $\mathcal{H}$. In this, we follow the approach of Ashtekar and Schilling.

As they explain, in order to clearly perceive the geometrical structures inherent to Hilbert spaces, it is best to change perspectives and consider $\mathcal{H}$ *from the point of view of real numbers rather than complex numbers.* First, one views $\mathcal{H}$ as real vector space equipped with a complex structure $J$. This simply means that the multiplication of a vector by a complex number is now considered as the result of two operations— multiplication by real numbers and action of the linear operator $J$: for $z \in \mathbb{C}$ and $\phi \in \mathcal{H}$, we have $z\phi = \text{Re}(z)\phi + \text{Im}(z)J\phi$. Second, one also decomposes the hermitian product of two vectors into its real and imaginary parts, and uses once again the canonical identification $T\mathcal{H} \simeq \mathcal{H} \times \mathcal{H}$[60] to define the tensor $\Omega \in \Gamma(T^*\mathcal{H} \otimes T^*\mathcal{H})$ by

$$\Omega(V_\phi, V_\psi) := 4\text{Im}(\langle \phi, \psi \rangle). \tag{II.7}$$

The skew-symmetry of the hermitian product $\overline{\langle \phi, \psi \rangle} = -\langle \psi, \phi \rangle$ entails the anti-symmetry of $\Omega$, which is hence a 2-form. The fact that the hermitian product is positive-definite and non-degenerate implies $\Omega$ is both closed and non-degenerate. Thus, we arrive at the following:

**Result:** By means of Equation II.7, a Hilbert space $\mathcal{H}$ may naturally be endowed with a symplectic structure $\Omega$.[61]

Note that, since by definition unitary transformations preserve the hermitian product, they automatically preserve the symplectic structure as well: $U(\mathcal{H}) \subset Symp(\mathcal{H})$.

[59]As John Baez once wrote on the n-category café blog: "'well-known', in the peculiar sense that mathematicians use this term, meaning at least ten people think it's old hat". (Entry of November 25th 2010, `https://golem.ph.utexas.edu/category/2010/11/stateobservable_duality_part_1.html`)

[60]Cf. the comment opening subsection II.1.2, page 155.

[61]Ashtekar and Schilling, op. cit., p. 6.

The whole machinery of symplectic geometry can now be deployed in the context of Hilbert spaces. In particular, to any smooth real-valued function on $\mathcal{H}$, one can associate a vector field which preserves the symplectic structure (it is the symplectic gradient, as defined on page 151) and the algebra $\mathcal{C}^\infty(\mathcal{H}, \mathbb{R})$ becomes a Poisson algebra, with Poisson bracket denoted $\{\cdot, \cdot\}_\mathcal{H}$.

In relation to what has been already discussed about the Quantum Kinematical arena, the immediate question is: does the symplectic geometry inherent to Hilbert spaces furnish a new point of view from which to understand both the Lie structure and the transformational role of quantum properties? The answer is of course positive, and to see this it suffices to consider again the map $\sim$ that, to a self-adjoint operator, associates its (unnormalized) expectation value function:

$$\sim: \ \mathcal{B}_\mathbb{R}(\mathcal{H}, \mathbb{R}) \longrightarrow \mathcal{C}^\infty(\mathcal{H}, \mathbb{R})$$
$$F \longmapsto \widetilde{F} \qquad \text{where } \widetilde{F}(\phi) := \langle \phi, F\phi \rangle.$$

The first thing to notice is that the symplectic gradient $V_{\widetilde{F}}$ of the function $\widetilde{F}$ coincides with the infinitesimal state transformation associated to the quantum property $F$ (cf. page 162). In other terms, we have:

$$V_{\widetilde{F}} = V_F = -iF. \tag{II.8}$$

Moreover, the map $\sim$ is evidently injective (if the expectation value of an operator $F$ vanishes for all vectors in $\mathcal{H}$, then $F = 0$), and is in fact a morphism of Lie algebras[62]:

$$\{\tilde{F}, \tilde{K}\}_\mathcal{H} = \frac{i}{2}\widetilde{[F, K]}. \tag{II.9}$$

We therefore get the injection

$$(\mathcal{B}_\mathbb{R}(\mathcal{H}, \mathbb{R}), \tfrac{i}{2}[\cdot, \cdot]) \lhook\joinrel\longrightarrow (\mathcal{C}^\infty(\mathcal{H}, \mathbb{R}), \{\cdot, \cdot\}_\mathcal{H})$$

Together, equations II.8 and II.9 show that, *as far as the Lie structure of quantum*

---

[62]See ibid., equation 2.6., p. 8, or Landsman, op. cit., equation I.2.38., p. 74. The difference of factors and signs corresponds to different choices for the normalization of the symplectic structure and the definition of the Poisson bracket.

*properties-as-transformations is concerned, we might as well forget the commutator of self-adjoint operators and reason in terms of the inherent symplectic structure of $\mathcal{H}$ and the Poisson bracket of expectation-value functions.*

This is certainly a very satisfactory point of view but it is not yet what we are looking for. Recall: our goal is to characterize the algebra of quantum properties in terms of the inherent geometry of the "true" quantum space of states $\mathbb{P}\mathcal{H}$. Moreover, we cannot yet forget self-adjoint operators altogether since the definition of quantum properties-as-transformations still involves them...

With the role of symplectic geometry at the level of $\mathcal{H}$ well understood, it is now easier to discuss the projective space $\mathbb{P}\mathcal{H}$. The idea is to consider the unit sphere $\mathbb{S}\mathcal{H}$—i.e. the set of all unit vectors in $\mathcal{H}$—and to use the pair of arrows

$$\mathcal{H} \xleftarrow{\quad i \quad} \mathbb{S}\mathcal{H} \xrightarrow{\quad \tau \quad} \mathbb{P}\mathcal{H} \tag{II.10}$$

to induce a geometry on $\mathbb{P}\mathcal{H}$ from the geometry of the Hilbert space. The left arrow is simply the injection saying that $\mathbb{S}\mathcal{H}$ is a submanifold of $\mathcal{H}$. The right arrow is the projection describing the unit sphere as a $U(1)$-fiber bundle over the projective Hilbert space. In other words, it describes $\mathbb{P}\mathcal{H}$ as a quotient: $\mathbb{P}\mathcal{H} \simeq \mathbb{S}\mathcal{H}/U(1)$. Then, the symplectic form on the new quantum space of states is simply the unique non-degenerate and closed 2-form $\omega \in \Omega^2(\mathbb{P}\mathcal{H})$ such that $\tau^*\omega = i^*\Omega$ (the pull-back of $\omega$ to the unit sphere coincides with the restriction to $\mathbb{S}\mathcal{H}$ of the symplectic form on $\mathcal{H}$)[63]. The induced Poisson bracket on $\mathcal{C}^\infty(\mathbb{P}\mathcal{H}, \mathbb{R})$ will be denoted by $\{\cdot, \cdot\}_{\mathbb{P}\mathcal{H}}$.

---

[63] Of course, one needs to be sure that such a 2-form does exist. There are several ways to see this is indeed the case, but it necessitates a technical machinery that goes beyond was has been explained so far. For completeness, I here shortly mention two closely related constructions involving the so-called symplectic reduction.

First, from the point of view of Hamiltonian constrained systems, one considers $\mathcal{H}$ as the initial phase space, equipped with the first class constraint $C(\phi) := \langle \phi, \phi \rangle - 1 = 0$. The constrained surface is then $\mathbb{S}\mathcal{H}$ and the reduced phase space is $\mathbb{P}\mathcal{H}$. Since the initial phase space was symplectic, one knows from the general theory of reduced phase spaces that the result is also symplectic. (See Schilling, op. cit., pp. 28–31.)

Second, from the point of view of the Marsden-Weinstein reduction, one considers the natural action of $U(1)$ on $\mathcal{H}$. This is a strongly Hamiltonian action and the momentum map $\mu : \mathcal{H} \longrightarrow u(1)^* \simeq \mathbb{R}$ is given by $\mu(\phi) = \langle \phi, \phi \rangle$. Then, $\mathbb{P}\mathcal{H} \simeq \mu^{-1}(1)/U(1)$. (See Landsman, op. cit., p. 328)

For a third point of view, using the decomposition in symplectic leaves of any Poisson space, such as $\mathcal{H}^* := \mathcal{H} \setminus \{0\}$, see ibid., p. 74.

The route towards a full geometrical description of quantum properties-as-transformations is continued by repeating the above procedure for Hilbert spaces. Indeed, consider the following map associating a real-valued function on the space of states to a self-adjoint operator:

$$\hat{} \colon \quad \mathcal{B}_{\mathbb{R}}(\mathcal{H}) \quad \longrightarrow \quad \mathcal{C}^{\infty}(\mathbb{P}\mathcal{H}, \mathbb{R}) \tag{II.11}$$

$$F \quad \longmapsto \quad \hat{F} \qquad \text{with } \hat{F}(p) = \langle \phi, F\phi \rangle \text{ and } \phi \in \tau^{-1}(p)^{\mathbf{64}}.$$

As was the case at the level of $\mathcal{H}$, we apparently have now two different ways in which to assign a smooth vector field on $\mathbb{P}\mathcal{H}$ to a self-adjoint operator $F \in \mathcal{B}_{\mathbb{R}}(\mathcal{H})$:

i) by means of the vector field $V_F$ defined on $\mathcal{H}$: define $v_F \in \Gamma(T\mathbb{P}\mathcal{H})$ by $v_F := \tau_* V_F$ (the push-forward of $V_F$ is possible since it is tangent to $\mathbb{S}\mathcal{H}$),

ii) by means of the newly defined function $\hat{F}$: define $v_{\hat{F}} \in \Gamma(T\mathbb{P}\mathcal{H})$ as the symplectic gradient of $\hat{F}$.

Happily, it just so happens that the two definitions coincide:

$$v_F = v_{\hat{F}}. \tag{II.12}$$

In other words, the symplectic flow on $\mathbb{P}\mathcal{H}$ generated by the function $\hat{F}$ coincides with the projection of the unitary flow on $\mathcal{H}$ generated by the self-adjoint operator $F^{\mathbf{65}}$.

Moreover, $\hat{}$ is again a morphism of Lie algebras[66]:

$$\{\hat{F}, \hat{K}\}_{\mathbb{P}\mathcal{H}} = \frac{i}{2} \widehat{[F, K]} \tag{II.13}$$

and we get the injection

$$(\mathcal{B}_{\mathbb{R}}(\mathcal{H}), \tfrac{i}{2}[\cdot, \cdot]) \hookrightarrow (\mathcal{C}^{\infty}(\mathbb{P}\mathcal{H}, \mathbb{R}), \{\cdot, \cdot\}_{\mathbb{P}\mathcal{H}}). \tag{II.14}$$

There is however no hope for this map to be an isomorphism, as may be seen by considering finite-dimensional Hilbert spaces: in this case, $\mathcal{B}_{\mathbb{R}}(\mathcal{H})$ is finite-dimensional whereas $\mathcal{C}^{\infty}(\mathbb{P}\mathcal{H}, \mathbb{R})$ is infinite-dimensional. The Lie algebra of self-adjoint operators is in fact a quite small subalgebra of the algebra of smooth functions!

---

[64] This definition of $\hat{F}(p)$ is clearly independent from the choice of the representative $\phi \in \tau^{-1}(p)$.

[65] Cf. Ashtekar and Schilling, op. cit., p. 12, or Landsman, op. cit., eq. I.2.45, p. 75.

[66] See Ashtekar and Schilling, op. cit., eq. 2.14, p. 12, or Landsman, op. cit., eq. I.2.42, p. 75.

We can now safely conclude: in the Quantum Kinematical arena, **the algebra of properties, with regard to its Lie structure, may be viewed as a <u>subalgebra</u> of $(\mathcal{C}^\infty(\mathbb{PH}, \mathbb{R}), \{\cdot, \cdot\}_{\mathbb{PH}})$, the algebra of smooth real-valued functions over the space of states equipped with the Poisson bracket. This algebraic Lie structure—or equivalently, the geometric symplectic structure that induces it—completely describes the role of quantum properties-as-transformations**.

This should be felt as an impressive merger of the Quantum and Classical Kinematical arenas. For if one omits the underlined word "subalgebra", this conclusion applies equally well to the Classical! Or, to put it in another way, as long as one chooses the right algebra of properties from the start, if one considers the quantum space of states *and pretends it is a classical space of states* by focusing solely on its inherent symplectic structure, one will get nonetheless the right *quantum* properties-as-transformations. The remark of this led Kibble to claim, in his seminal article on the geometry of Quantum Mechanics:

> From this point of view, the essential difference between classical and quantum mechanics lies not in the set of states (save for the infinite dimensionality) nor in the dynamic evolution [i.e, nor in properties-as-transformations], but rather in the choice of the class of observables, which is far more restricted in quantum than in classical mechanics.[67]

This restriction in the choice of the class of properties may appear mysterious. Any property is described by a real-valued function on the space of states, yet not any function describes some quantum property. Why is it so? But to ask this is to forget that, in the game of perceiving the Quantum through the looking glass of the Classical, we have been ignoring a large part of the geometry inherent to the Quantum. If we have not yet managed to provide an intrinsic characterization of the algebra of quantum properties, it is precisely because we have so far only considered the symplectic structure of the quantum space of states. However, as was quickly explained just before the beginning of subsection II.2.1, there is at least one additional geometric structure:

---

[67]Kibble, op. cit., p. 190.

a Riemannian metric $g$. If one takes it into account, the geometric definition of a kinematical property becomes possible and is beautiful in its simplicity:

> **Physical properties**: a property of a physical system is a smooth real-valued *function* on the space of states $S$ to which an infinitesimal *state transformation* can be naturally associated. That is, it is a function whose associated transformation preserves *all* the available structures present in the kinematical space of states. The set of these is denoted $\mathcal{C}^\infty(S, \mathbb{R})_\mathcal{K}$.

This definition is explicit in the work of Ashtekar and Schilling[68]. It applies equally well to Classical and Quantum Kinematics. Infinitesimal state transformations are vector fields preserving all the available kinematical structure. In the classical case, this means only to preserve the symplectic structure, and any smooth function $f$ may thus do the job, as its symplectic gradient $v_f$ automatically verifies $\mathcal{L}_{v_f}\omega = 0$. But in the quantum case, there is also the metric to preserve. Accordingly, only those functions for which the symplectic gradient is as well a Killing vector will qualify as properties. As the two authors prove, these functions exactly coincide with the functions $\widehat{F}$ that are real expectation-value maps of self-adjoint operators $F$[69]. It is important to notice that this last point only applies to the *projective* Hilbert space. Were one to insist on working at the level of $\mathcal{H}$, this geometrical characterization of properties would fail, for there are too many functions preserving both the symplectic and Riemannian structures which do not arise as expectation-value maps of operators.

The general definition of a physical property enlightens the importance of their double role. For *it is precisely this two-fold role, numerical and transformational, that serves as a definition of what a physical property is*. The standard definition of classical properties only involved their numerical role—they were defined as functions on the space of states—and did not apply to Quantum Kinematics. Conversely, the standard definition of quantum properties only involved their transformational role—they were defined as operators acting on states—and did not apply to Classical Kinematics. A posteriori, it is therefore most natural that the general definition of a physical property, be it classical or quantum, should explicitly mention properties-as-quantities and properties-as-transformations.

---

[68] Ashtekar and Schilling, op. cit., See Corollary 1, p. 14, and the comment following it.

[69] The proof of this result may also be found in Schilling, op. cit., Corollary 3.5, p. 41, or in R. Cirelli, M. Gatti, and A. Manià. "On the Nonlinear Extension of Quantum Superposition and Uncertainty Principles". In: *Journal of Geometry and Physics* 29.1 (1999), pp. 64–86, Proposition 5.2., p. 75. This result is also mentioned in N. P. Landsman. "The Infinite Unitary Group, Howe Dual Pairs, and the Quantization of Constrained Systems". In: *arXiv preprint* (1994). URL: https://arxiv.org/abs/hep-th/9411171, p. 15.

From this perspective, Kibble's enthusiastic claim appears to be partially wrong. True: "the essential difference between classical and quantum mechanics lies [...] in the choice of the class of observables". But, against Kibble, this difference is also present in the set of states: the quantum one is equipped with an additional geometric structure. Thus, the "real" difference between the two kinematical arenas lies no more in the algebraic structure of the properties than in the geometry of the states. Properties and states are the two Janus faces of Kinematics and in both may one clearly perceive the distinction[70]. We turn now to the investigation of this additional geometrical structure of the Quantum.

## II.2.2   The additional geometric structure of the Quantum

The conceptual analysis of section II.1 showed two important drawbacks of the standard formulation of Quantum Kinematics were: the difficulty to relate the algebraic structures of properties to geometric features of states, and the obscure physical meaning of the Jordan product. The geometric formulation of quantum mechanics obviously tries to address the first point and the hope is that by doing so, as a spin-off, it will also clarify the second one. In relation to this, the realization of there being *two* geometrical structures on the quantum side can only be felt as encouraging: there were *two* algebraic structures (a Jordan and a Lie product) and *two* roles for properties (numerical and transformational). Moreover, the essential difference between classical and quantum properties was found to lie on the Jordan side of properties, and it now seems to be found as well on the Riemannian side of states. Hence, it is hard not to

---

[70] Janus is the roman god of beginnings and ends and is almost systematically represented as a head with two faces looking in opposite directions. I take this image from a conference given in Paris by Klaas Landsman.

conjecture the following links:

properties-as-transformations ⟷ Lie algebra ⟷ symplectic geometry properties-as-quantities ⟷ Jordan algebra ⟷ Riemannian geometry.

Now, recall the definition of the symplectic structure at the level of the Hilbert space $\mathcal{H}$: it involved the imaginary part of the hermitian product (Equation II.7, page 172). By analogy, one may use the real part to define a second tensor $G \in \Gamma(T\mathcal{H} \otimes T\mathcal{H})$ by:

$$G(V_\phi, V_\psi) := 4\mathrm{Re}(\langle \phi, \psi \rangle). \tag{II.15}$$

This time, the skew-symmetry, positive-definiteness and non-degeneracy of the hermitian product respectively imply the symmetry, positive-definiteness and non-degeneracy of $G$. Thus, we have:

**Result:** By means of Equation II.15, a Hilbert space $\mathcal{H}$ may naturally be endowed with a Riemannian metric $G$.[71]

Again, since by definition unitary transformations preserve the hermitian product, they automatically preserve the Riemannian structure as well: $U(\mathcal{H}) \subset Isom(\mathcal{H})$. At the level of $\mathcal{H}$, it therefore becomes transparent why the vector field associated to a self-adjoint operator is a Killing vector.

At this point, we may use again diagram II.10 (page 174) to induce a Riemannian metric on the space of states. In the symplectic case, we regarded the isomorphism $\mathbb{S}\mathcal{H}/U(1) \simeq \mathbb{P}\mathcal{H}$ as the second stage of symplectic reduction and this sufficed to insure $\mathbb{P}\mathcal{H}$ was also symplectic. Instead, we now adopt towards this isomorphism a different perspective, called by Ashtekar and Schilling the "Killing reduction"[72]. It is the following: first, the restriction $i^*G$ of the metric $G$ to the unit sphere is again a metric and $\mathbb{S}\mathcal{H}$ becomes then a Riemannian manifold in its own right. Second, one regards the action of $U(1)$ on $\mathcal{H}$ as the one-parameter group of transformations generated by the vector field $V_{Id} \in \Gamma(T\mathcal{H})$ associated to the identity self-adjoint operator. By the

---

comment just above, we know that $V_{Id}$ is a Killing vector for $G$. Moreover, this vector field is tangent to $\mathbb{S}\mathcal{H}$ and is hence also a Killing vector for $i^*G$. In this way, the isomorphism $\mathbb{S}\mathcal{H}/U(1) \simeq \mathbb{P}\mathcal{H}$ describes the projective Hilbert space as the space of all trajectories of the Killing vector field $V_{Id}$. By a result of Geroch[73], we know that the resulting manifold is also Riemannian. We denote the Riemannian metric on $\mathbb{P}\mathcal{H}$ by $g$.

Hence, the Riemannian and symplectic structures of the quantum space of states $\mathbb{P}\mathcal{H}$ may be seen as arising from the decomposition of the hermitian product of $\mathcal{H}$ into its real and imaginary part. Combine to this the fact that, for self-adjoint operators, the Jordan product $\bullet = \frac{1}{2}[\cdot,\cdot]_+$ and Lie product $\star = \frac{i}{2}[\cdot,\cdot]$ may also be seen as arising from the decomposition into real and imaginary parts of the usual composition of operators[74]:

$$\text{for } A, B \in \mathcal{B}_{\mathbb{R}}(\mathcal{H}),\ A \circ B = A \bullet B - iA \star B$$

and you get a new strong hint that the algebraic Jordan structure of quantum properties must be governed by the Riemannian geometry of the states.

This conjectured Riemann-Jordan link is finally proven by defining the following commutative product on $\mathcal{C}^\infty(\mathbb{P}\mathcal{H}, \mathbb{R})$:

$$\forall f, k \in C^\infty(\mathbb{P}\mathcal{H}, \mathbb{R}),\ \ f \bullet k := g(v_f, v_k) + f \cdot k.$$

To the usual point-wise multiplication of functions $f \cdot g$, the metric adds a "Riemannian bracket" $(f, k) := g(v_f, v_k)$. Thus, in a loose sense, the presence of the Riemannian structure allows to *deform* the usual commutative and associative algebra of functions into a commutative but non-associative algebra.

Because of the similarity with the definition of the Poisson bracket, this definition of the Riemannian bracket seems most natural. There is however a sense in which it is misleading: it explicitly involves the symplectic gradient of the functions and it would therefore seem that the new commutative product depends on both the metric and the symplectic structure of the space of states. However, this is not true: as is shown by

---

[73]Appendix of R. Geroch. "A Method for Generating Solutions of Einstein's Equations". In: *Journal of Mathematical Physics* 12.6 (1971), pp. 918–924, cited in Ashtekar and Schilling, op. cit., p. 12.

[74]In the same way that for a complex number $z$, $\mathrm{Re}(z)$ and $\mathrm{Im}(z)$ are both real while $z$ is not, here $A \bullet B$ and $A \star B$ are both self-adjoint even though $A \circ B$ is not.

Schilling, it depends only on the Riemannian metric[75].

Now, as Ashtekar and Schilling show, the morphism $\widehat{\phantom{x}}\colon\ \mathcal{B}_{\mathbb{R}}(\mathcal{H}) \longrightarrow \mathcal{C}^{\infty}(\mathbb{P}\mathcal{H}, \mathbb{R})$ respects the Jordan product[76]:

$$\hat{F} \bullet \hat{K} = \frac{1}{2}\widehat{[F, K]}_{+}. \tag{II.16}$$

This equation finishes the complete geometrical characterization of the quantum algebra of properties, *for both its Lie and Jordan structures*, for we now have the following isomorphism of non-associative Jordan-Lie algebras:

$$\boxed{\left(\mathcal{B}_{\mathbb{R}}(\mathcal{H}), \frac{1}{2}[\cdot, \cdot]_{+}, \frac{i}{2}[\cdot, \cdot]\right) \simeq \left(\mathcal{C}^{\infty}(\mathbb{P}\mathcal{H}, \mathbb{R})_{\mathcal{K}}, \bullet, \{\cdot, \cdot\}_{\mathbb{P}\mathcal{H}}\right).} \tag{II.17}$$

With this geometrical reformulation of the quantum kinematical arena, it is easier to understand the sought-for link between the numerical role of properties and their Jordan structure. This may also be now translated as the question of the link between the properties-as-quantities and the geodesical structure of the space of states. As is well known, one crucial characteristic of the Quantum is its probabilistical or statistical dimension. This can be separated into two different yet related aspects:

i) *The value of a property*: given a property $f$ and a state $p$, one cannot in general assign a *definite* value of the property to that state. Instead, one associates an *expected* value $f(p)$ and an *indeterminacy* $\Delta f(p)$.

ii) *The probability of a value*: given a property $f$ and a possible value $\lambda$ of that property, to each state $p$ one can assign a probability $\Pr(p, \lambda)$ of finding $\lambda$ as result of a measurement.

A very satisfactory insight of the present geometrical reformulation is that it allows to understand these two aspects of quantum properties-as-quantities in terms of two different yet related operations on the space of states enabled by the presence of the Riemannian structure, namely:

---

[75]Schilling, op. cit., eq. 2.42, p. 24.

[76]Ashtekar and Schilling, op. cit., eq. 2.23, p. 15.

1) The definition of a *length* for tangent vectors:

$$\text{for } v \in T\mathbb{P}\mathcal{H}, \|v\|^2 := g(v, v).$$

2) The definition of a *distance* between two states:

$$\text{for } p, q \in \mathbb{P}\mathcal{H}, d_g(p, q) := \inf \Big\{ \int_\Gamma \sqrt{g(v_\Gamma(t), v_\Gamma(t))} dt \,\Big|\, \Gamma \in \text{Path}(p, q) \Big\}.$$

where $\text{Path}(p, q)$ is the set of parametrized paths between $p$ and $q$. In the same way, one can define a distance between a state $p$ and a subset of states $\Sigma \subset \mathbb{P}\mathcal{H}$, which we denote by the same symbol $d_g(p, \Sigma)$[77].

The two results which render manifest the relation between i), ii) and 1), 2) are the following. For a quantum property $f \in \mathcal{C}^\infty(\mathbb{P}\mathcal{H}, \mathbb{R})_\mathcal{K}$, a state $p \in \mathbb{P}\mathcal{H}$, $\lambda$ a possible value of the property and $\Sigma_\lambda \subset \mathbb{P}\mathcal{H}$ the set of all states having $\lambda$ as definite value of the property $f$, we have[78]

$$(\Delta f)^2(p) \;=\; \|v_f|_p\|^2 = g(v_f|_p, v_f|_p) = f \bullet f(p) - \big(f(p)\big)^2, \qquad (\text{II}.18)$$

$$\Pr(p, \lambda) \;=\; \cos^2(d_g(p, \Sigma_\lambda)). \qquad (\text{II}.19)$$

The last equation explains in which way the probabilistic features of the quantum kinematical arena are completely governed by the geodesical structure of the quantum space of states. It is useful to rewrite it in terms of two vectors $\phi$ and $\psi$ of the initial Hilbert space. It then says that, for the respective two points $[\psi], [\phi] \in \mathbb{P}\mathcal{H}$, we have

$$d_g([\psi], [\phi]) = \arccos(|\langle \psi, \phi \rangle|).$$

From this, it is clear that there exists a maximum distance between any two states, called the *diameter* of $\mathbb{P}\mathcal{H}$. Two states separated by such a maximal distance are called *antipodal*. As is transparent from the equation, antipodality at the level of $\mathbb{P}\mathcal{H}$ is the geometrical translation of orthogonality at the level of $\mathcal{H}$[79]. The further two states are from each other, the less probable is a transition between them. In other words, *the quantum transition probability is nothing else than a measure of the separation between states!*

---

[77]To do this, simply take $d_g(p, \Sigma) := \inf \big\{ d_g(p, q) \big| q \in \Sigma \big\}$.

[78]See ibid., eq. 2.26, p.15, and eq. 2.34, p. 20, or also Cirelli, Gatti, and Manià, "The Pure State Space of Quantum Mechanics as Hermitian Symmetric Space", p. 12 and p.8.

[79]See ibid., p. 8.

On the other hand, Equation II.18 is conceptually very rich and well deserves a few comments. First, it shows that the quantum indeterminacy of a property may be equally well characterized in terms of the Riemannian structure or in terms of the Jordan structure. Second, since $(\Delta f)^2 = \langle f^2 \rangle - \langle f \rangle^2$, the expression of the indeterminacy in terms of the Jordan product implies that $\langle f^2 \rangle = f \bullet f$. Specifically, if the property $\mathtt{f}$ is represented by the function $f$, then the property $\mathtt{f}^2$ is represented by the function $f \bullet f$ and not by the function $f^2 = f \cdot f$ as one could have naively thought. This clearly establishes—if there were still doubts—that the Jordan product defined through the Riemannian structure is the true commutative product of quantum properties.

Third, and most important for us, this same equation sheds a new light on the relation between the two roles of properties in Quantum Mechanics, for it expresses the beautiful and satisfactory fact that *the indeterminacy of a physical property-as-quantity is precisely a measure of how much a state is changed by the property-as-transformation*. In particular, we recover as a special case the invariance ↔ definite-valuedness relation already noticed with the standard Hilbert space formulation (cf. page 164): states with definite values of the property $f$ correspond to those being invariant under the transformations generated by $f$. Also, as noted by Ashtekar and Schilling, we rederive the geometric interpretation of the uncertainty in energy found by Anandan and Aharonov in 1990:

> [...] the uncertainty in energy $\Delta E$ for an arbitrary quantum system [...] is the magnitude of the *velocity* of the system in the projective Hilbert space. It follows that the evolution of the system in $\mathbb{P}\mathcal{H}$ completely determines $\Delta E$; no other information from $H$ is needed to determine $\Delta E$.[80]

From a conceptual perspective, this is, in my opinion, one of the most important results of the whole geometric reformulation of the quantum kinematical arena. It brings to the fore the existence of an interplay, *characteristic of the quantum*, between properties-as-quantities and properties-as-transformations:

> **Quantum interplay of quantities and transformations.** Properties-as-quantities provide a quantitative description of how the respective properties-as-transformations affect each state of the system.

---

[80]J. Anandan and Y. Aharonov. "Geometry of Quantum Evolution". In: *Physical review letters* 65.14 (1990), pp. 1697–1700, p. 1699

We had earlier seen how the transformational role of properties was consistent with its numerical role: properties-as-transformations only related states that were not separated (distinguished) by the property-as-quantity (cf. page 155). This compatibility was valid for both Classical and Quantum Kinematics. But now we see that, in the Quantum, a second compatibility condition is required, one which establishes a *reciprocal* interdependence of the two roles: the transformation respects the quantity and, in return, the quantity describes the transformation. This trait—which had already been hinted at in the discussion of the standard Hilbert space formulation of Quantum Mechanics (cf. page 164)—is completely absent in the classical realm: classical properties-as-quantities are independent of properties-as-transformations; the former do not seem to encode any information whatsoever about the latter, and it is because of this that classical uncertainties are unheard of.

In the same spirit, it is possible to provide a geometrical interpretation of Heisenberg's famous indeterminacy principle. To recall, instead of considering the indeterminacy of one single property $f$, Heisenberg's principle considers a pair of them. In the usual Hilbert space formulation, it is written: for $F, K \in \mathcal{B}_{\mathbb{R}}(\mathcal{H})$ and $\phi \in \mathcal{H}$,

$$\Delta_\phi F \Delta_\phi K \geqslant \left| \langle \frac{i}{2}[F, K] \rangle_\phi \right|.$$

That is, the product of the indeterminacies in $F$ and $K$ must be bigger than the mean value of the property $\frac{i}{2}[F, K]$. Now, using Equations II.13 (page 175), II.19 (page 182) and the definition of the Poisson bracket in terms of the symplectic form (page 150), this inequality can be rewritten as

$$\|v_{\widehat{F}}\| \, \|v_{\widehat{K}}\| \geqslant |\omega(v_{\widehat{F}}, v_{\widehat{K}})|.$$

As noted by Cirelli, Gatti and Manià, this simply expresses the uniform continuity of the symplectic form with respect to the topology induced on the tangent space by the Riemannian metric[81]. This constitutes yet another blow to the popular view that swiftly opposes the Quantum to the continuum: Heisenberg indeterminacy principle, often perceived as one of the most striking features characteristic of the Quantum, appears as a statement of *continuity* of the geometrical structures on the quantum space of states[82]!

Table II.1 below summarizes the conceptual understanding of the kinematical arenas as for now. Following Schilling, it could be tempting to say that, from the geometrical point of view, the "fundamental distinction between the classical and quantum formalisms is the presence, in quantum mechanics, of a Riemannian metric. While the symplectic structure serves exactly the same role as that of classical mechanics, the metric describes those features of quantum mechanics which do not have classical analogues."[83] This is the view found in the vast majority of works on the geometrization of quantum mechanics: the quantum would have one additional geometric structure,

---

[81]Cirelli, Gatti, and Manià, op. cit., p. 12.

[82]Of course, the view which systematically associates Quantum Mechanics with some sort of fundamental discretization of physical phenomena has been incessantly criticized from the very beginning of the theory. For example, in the seminal "three-man paper" *"Zur Quantenmechanik II"* from 1925, one can read the following:

> [...] a particularly important trait in the new theory [of Quantum Mechanics] would seem to us to consist on the way in which *both continuous and line spectra arise in it on equal footing*, i.e., as solutions of one and the same equation of motion [...]; obviously, in this theory, *any distinction between 'quantized' and 'unquantized' motion ceases to be at all meaningful*, since the theory contains no mention of a quantization condition [...].

(Born, Heisenberg, and Jordan, op. cit., pp. 322–323, my emphasis.)

In the same vein, only a few weeks after this paper, the physicist Cornelius Lanczos wrote:

> This much, however, we do believe that we are allowed to conclude [...]: that the modifications, which we must apply to our classical views in order to reach an understanding of the quantum problems, must lie in a totally different direction than could be characterized simply by the contrast between continuum and discontinuum; and that the solution of the quantum mystery should have hardly anything to do with a quantum-like re-interpretation of geometry or infinitesimal calculus.

(C. Lanczos. "Über eine feldmäßige Darstellung der neuen Quantenmechanik". In: *Zeitschrift für Physik* 35 (1926), pp. 812–830, cited in J. Mehra and H. Rechenberg. *The Historical Development of Quantum Theory. Volume 3: The Formulation of Matrix Mechanics and Its Modifications*. New York: Springer-Verlag, 1982, p. 216)

Despite this, the association Quantum–Discrete still seems to have a strong heuristic influence, specially in the research of quantum gravity.

[83]Schilling, op. cit., p. 48.

|  | **Classical Kinematics** | **Quantum Kinematics** |
|---|---|---|
| **States** | points $p$ of a symplectic manifold $(M, \omega)$ | points $p$ of a symmetric Hermitian manifold $(M, \omega, g, J)$ |
| **Properties** | $\mathcal{C}^\infty(M, \mathbb{R})_\mathcal{K}$ Smooth real-valued functions whose transformations preserve the geometric structures | $\mathcal{C}^\infty(M, \mathbb{R})_\mathcal{K}$ Smooth real-valued functions whose transformations preserve the geometric structures |
| **Geometric structures of states** | One geometric structure: <br><br> ♠ a symplectic 2-form $\omega$ | Two geometric structures: <br><br> ♠ a symplectic 2-form $\omega$ <br> ♥ *a Riemannian metric g* |
| **Algebraic structures of properties** | Two algebraic structures: <br><br> ♠ Anti-commutative Lie product $\{f, k\} = \omega(V_f, V_k)$ (induced by symplectic) <br><br> ♥ Jordan product Commutative *and associative* $f \bullet k = f \cdot k$ | Two algebraic structures: <br><br> ♠ Anti-commutative Lie product $\{f, k\} = \omega(V_f, V_k)$ (induced by symplectic) <br><br> ♥ Jordan product Commutative *but non-associative* $f \bullet k = f \cdot k + g(V_f, V_k)$ (induced by Riemannian) |
| **Roles of properties** | Two roles of properties: <br><br> ♠ Properties-as-transformations captured by Lie product <br><br> ♥ properties-as-quantities captured by Jordan product $\Delta f = 0$ *independent of transformations* | Two roles of properties: <br><br> ♠ Properties-as-transformations captured by Lie product <br><br> ♥ properties-as-quantities captured by Jordan product $\Delta f = g(v_f, v_f)$ *dependent on transformations* |

**Table II.1** – Comparison of the two kinematical arenas in their geometric formulation.

*with no analogue in the classical,* and, in a loose sense, to quantize would mean to add a Riemannian metric to the space of states.

Nonetheless, this does *not* seem to be the impression conveyed by the comparative table. By the end of section II.1, we had felt the description of the quantum kinematical arena was not quite right. The harmonious balance between the geometric structures, the algebraic structures and the double role of properties displayed in standard classical kinematics was not found in standard quantum kinematics. But at present the situation seems to have surprisingly reversed: the quantum description shines, and the praised beauty of the classical has somewhat faded away. For something seems to be *missing* in the description of classical kinematics—more precisely, *one* geometrical structure on the space of states that would induce the associative Jordan product. In fact, one gets the impression that this structure is the "Riemannian metric" $g = 0$. Indeed, setting $g$ to vanish in the quantum formulas yields the classical ones. Of course, $g = 0$ is not an actual metric, so it cannot be that simple. But this does suggest there may be yet another way of formulating the two kinematical arenas. A way in which they both exhibit the same two kinds of geometrical structures, but it just so happens that one of these structures is trivial—and hence unnoticed—in Classical Kinematics.

The search for this lost structure of the Classical constitutes my main motivation to explore the $C^*$-algebraic formulation of Mechanics. This will be investigated in the next section. However, before we turn to this, let me close the discussion of the geometric formalism by one small subsection. It lies somewhat aside the main discussion of the conceptual triad of Kinematics and may be skipped by the reader.

## II.2.3 The geometrical formulation of the superposition principle

As I commented at the beginning of section II.2, the idea of "superposition" has been perceived, since the early stages of the theory, as one of the hallmarks of quantum mechanics. Moreover, this "most fundamental and most dramatic law of nature" has been almost systematically associated to the mathematical idea of "linearity", up to the point physicists often refer to the "quantum principle of *linear* superposition". This is most explicitly stated in Dirac's *Principles of Quantum Mechanics*:

> The superposition process is a kind of additive process and implies that states can in some way be added to give new states. The states must therefore be connected with mathematical quantities of a kind which can be added together to give other quantities of the same kind. The most obvious of such quantities are vectors.[84]

But this association is also manifest in much more recent literature on the foundations of Quantum Mechanics. For example, in their attempt to present a modification of standard Quantum Mechanics that would solve the measurement problem—which is now known as the GRW model—Ghirardi, Rimini and Weber declare:

> Despite the success of quantum mechanics in accounting with striking accuracy for a vast variety of physical phenomena, this theory presents crucial conceptual difficulties, about which a lively scientific debate is still going on. Almost all the difficulties can be traced back to the problem of accounting for the behavior of macroscopic objects and for their interactions with microscopic ones, and *are strictly related to the occurrence (allowed by the theory) of linear superpositions of macroscopically distinguishable states* of a macroscopic system (a typical example being the macroscopically different pointer positions of a measuring apparatus). This very fact, i.e., that *the linearity of quantum theory unavoidably leads one to consider such superpositions*, constitutes a basic difficulty for all trials of deriving a unified description of the physical reality from microscopic to macroscopic phenomena.[85]

Yet, after discovering the geometrical reformulation of quantum mechanics, we know that the association of superposition and linearity is not inherent to Quantum Mechanics but rather emerges from a particular formulation of it. We know the de-linearization program must provide a description of the principle of superposition that does not appeal to linearity in any way. But the precise translation of this principle into geometrical ideas remains yet to be seen. A discussion of this important point is surprisingly lacking in the exposition of Ashtekar and Schilling, but the details can be found in the article by Cirelli, Gatti, and Manià (section 3) or in the article of Brody and Hughston (section 4). There is also a nice short article by Alejandro Corichi dedicated solely to this issue[86].

---

[84]P. A. M. Dirac. *The Principles of Quantum Mechanics.* 4th ed. Oxford: Oxford University Press, 1958, p. 15.

[85]G. C. Ghirardi, A. Rimini, and T. Weber. "Unified Dynamics for Microscopic and Macroscopic Systems". In: *Physical Review D* 34.2 (1986), pp. 470–491, p. 470, emphasis is mine.

In its core, the *Quantum Superposition Principle is a claim about the ability to generate new possible states from the knowledge of just a few*: given the knowledge of states $p_1$ and $p_2$, one can deduce the existence of an infinite set $S_{p_1, p_2}$ of other states which are equally accessible to the system[87]. In the standard Hilbert space formalism, superposition is described by $\mathbb{C}$-linearity: for two different states $\psi_1, \psi_2 \in \mathcal{H}$, any superposition of them can be written as $\phi = a\psi_1 + b\psi_2$, with $a, b \in \mathbb{C}$. If one only cares about the *set* of all superpositions of two states, one may then argue that the principle of superposition is captured by the canonical association of a two-dimensional complex vector space to any pair of states. In other words, it is captured by the existence of a map

$$V : \mathcal{H} \times \mathcal{H} \longrightarrow \mathrm{Hom}(\mathbb{C}^2, \mathcal{H})$$

where the linear map $V_{\psi_1, \psi_2} : \mathbb{C}^2 \to \mathcal{H}$ is an injection iff $\psi_1$ and $\psi_2$ are independent vectors.

This slight reformulation of the linear superposition paves the way to its geometrical translation. Indeed, one only needs to replace the Hilbert space by the real quantum space of states $\mathbb{P}\mathcal{H}$, and the linear space $\mathbb{C}^2$ by its projective analogue, the complex projective line $\mathbb{P}\mathbb{C}^2$. In this way, the quantum superposition principle is now

---

[86]A. Corichi. "Quantum Superposition Principle and Geometry". In: *General Relativity and Gravitation* 38.4 (2006), pp. 677–687. URL: http://arxiv.org/abs/quant-ph/0407242.

[87]In his article, Corichi argues there are two different ideas usually associated to the notion of "superposition", and that due care should be taken to distinguish them. On the one hand, there is the particular relation between states that allows one to construct new states from old ones. On the other hand, there is the important idea of quantum interferences, manifest for example in the treatment of the double-slit experiment. Therein, one separately considers the experimental setups with only *one* slit: first, only slit $S_1$ exists and this produces the probability amplitude $\psi_1$, then only slit $S_2$ exists and produces the probability amplitude $\psi_2$. The probability amplitude resulting from the actual double slit experiment is then found by *adding* $\psi_1$ and $\psi_2$. Therefore, one studies a given experimental arrangement by decomposing it into basic blocks, and then *superposing* the effects of each of these blocks. Corichi calls the first idea the *principle of superposition of states*, and corresponds to the notion of "generation of states" we have just mentioned in the main text. It is a kinematical relation valid for any quantum-mechanical system. The second idea he calls the *principle of decomposition*, and is more related to particular experimental setups than to the consideration of general physical systems.

translated into the existence of a map

$$S : \mathbb{P}\mathcal{H} \times \mathbb{P}\mathcal{H} \longrightarrow \mathrm{Hom}(\mathbb{P}\mathbb{C}^2, \mathbb{P}\mathcal{H})$$

where, for $p_1 \neq p_2$, the map $S_{p_1,p_2}$ is now a monomorphism in the category of Hermitian symmetric spaces (or, equivalently, in the category of Kähler manifolds). Since the complex projective line is in fact (isomorphic to) the Riemannian sphere—and thus, as real manifolds, one has $\mathbb{P}\mathbb{C}^2 \simeq S^2$—the geometrical reformulation of the quantum superposition of states becomes quite simple:

> **Geometrical reformulation of the superposition principle:** given a quantum system with space of states $M$ and two states $p_1, p_2 \in M$, there exists a *canonical* two-sphere $S_{p_1,p_2} \simeq S^2 \subset M$ containing them.

This two-sphere can be thought as the non-linear span generated by the two states. As noted by Cirelli, Gatti, and Manià, $S_{p_1,p_2}$ can equivalently be characterized as the smallest totally geodesic submanifold of the space of states containing $p_1$ and $p_2$[88].

One may however worry that this geometrical reformulation only deals with the set of all superpositions as a *whole*, but does not allow to describe *single* superpositions in the way it can be achieved in the Hilbert space formalism. But this is not so. For, given two states $p_1$ and $p_2$, the task of characterizing the different states which arise as superposition of these two is simply the task of defining, for the two-sphere $S_{p_1,p_2}$, a coordinate system in which the points $p_1$ and $p_2$ play a preferred role. And there is no obstruction for this to be done. For example, if one recalls that the complex projective line $\mathbb{P}\mathbb{C}^2$ is equivalently defined as the manifold obtained by adding a point at infinity to the complex plane, one sees that any point of $S_{p_1,p_2}$ may be characterized by a number $z \in \mathbb{C} \cup \{\infty\}$ such that, moreover, $z(p_1) = 0$ and $z(p_2) = \infty$[89].

---

[88]Cirelli, Gatti, and Manià, op. cit., p. 9. A totally geodesic submanifold of $M$ is a submanifold $S \subset M$ for which all geodesics in $M$ through any point $p \in S$ lie in $S$ (for small values of the parameter of the geodesics). This purely metric characterization of the set of superpositions of two states is also found in V. Cantoni. "Superposition of Physical States: a Metric Viewpoint". In: *Helvetica Physica Acta* 58 (1985), pp. 956–968, p. 961.

[89]To deal with single superpositions, Cirelli, Gatti, and Manià adopt a different strategy, based on a close examination of the geodesic structure of $\mathbb{P}\mathcal{H}$. Instead, I have here followed the simpler approach of Corichi.

Now, whereas the association of the set $S_{p_1,p_2}$ of all superpositions for a given pair of states $p_1$ and $p_2$ was canonical, the 'coordinatization'

$$z : S_{p_1,p_2} \longrightarrow \mathbb{C} \cup \{\infty\}$$
$$p \longmapsto z(p)$$

on the contrary, is not: it depends on *arbitrary* choices. Of course, the same is true in the language of Hilbert spaces, and it is interesting to recall how this characterization of single superpositions with numbers $z \in \mathbb{C} \cup \{\infty\}$ is achieved in this latter setting. Given two independent vectors $\psi_1, \psi_2 \in \mathcal{H}$, there is a canonically associated two-dimensional complex vector space $V_{\psi_1,\psi_2}$ and there is a unique way in which to write each of its elements in terms of $\psi_1, \psi_2$:

$$\forall \phi \in V_{\psi_1,\psi_2}, \ \exists!(a,b) \in \mathbb{C}^2, \ \phi = a\psi_1 + b\psi_2.$$

To describe the *states* found by superposition—that is, the relevant *rays* of the Hilbert space—one needs to arbitrarily choose one representative vector for each ray. This is achieved by an arbitrary choice of normalization (e.g., $\phi = \frac{1}{\sqrt{|a|^2 + |b|^2}}(a\psi_1 + b\psi_2)$) and of an overall phase factor (e.g., $a \in \mathbb{R}$). Only after this do we find that any state $[\phi]$ which arises as a superposition of the states $[\psi_1]$ and $[\psi_2]$ is uniquely written as

$$\phi(z) = \frac{1}{\sqrt{1 + |z|^2}}(\psi_1 + z\psi_2), \quad \text{where } z := \frac{b}{a} \in \mathbb{C} \cup \{\infty\}$$

Therefore, in this unnecessarily twisted fashion, we reach the same 'coordinatization' of the space of all superpositions as we had arrived at straightforwardly through the geometric formulation. In particular, we see that $\phi(0) = \psi_1$ and $\phi(\infty) = \psi_2$.

We have then a complete description of all the features of the Quantum Superposition Principle which entirely avoids any mention to $\mathbb{C}$-linearity and Hilbert spaces. To conclude this discussion on the geometric superposition principle, it should be noted that the two-sphere property of the quantum space of states may be perceived as one of its fundamental geometric features. Indeed, as it will be seen when discussing algebraic mechanics, it turns out that the two-sphere property is one of the very few axioms needed to characterize the pure state space of a $C^*$-algebra[90]. In this way, the principle of superposition becomes indeed one of the most basic laws of Quantum Mechanics.

## II.3 The Classical seen from the Quantum: the algebraic formulation

As became evident by the end of section II.1, the standard perspective on Kinematics fails to provide a satisfactory articulation of the Classical and the Quantum. And—it was felt—the main reason for this failure was the clumsy description of quantum states in the Hilbert space formalism. From the standard point of view of Classical Hamiltonian Mechanics—which, in order to describe a physical system, specifies *first* a space of states, and *only then* considers the algebra of functions—this was a consequence of not considering from the start the right quantum space of states. The natural strategy was hence to reformulate the Quantum directly from the projective Hilbert space $\mathbb{P}\mathcal{H}$. As we have seen in section II.2, this geometric program achieves its aims exceedingly well.

Nevertheless, from the point of view of Quantum Mechanics, this geometric strategy is certainly not the most natural one. For *the strength of the Hilbert space formalism lies in its description of the algebra of quantum properties.* Indeed, in the same way that given a differentiable manifold the set of all smooth real valued functions is a very natural real algebra to consider, given a Hilbert space it is also quite natural to study the algebra of self-adjoint operators. Thus, in the standard quantum formalism, the description of the algebra of properties is straightforward. This is to be contrasted with the situation in the geometrical formulation: one should not forget that it took almost twenty years to find the geometrical characterization of quantum properties—from Kibble's articles to the theorems of Schilling. With this remark taken to its full-blown consequences, the problematic description of quantum states of the

---

[90]See Landsman, *Mathematical Topics Between Classical and Quantum Mechanics*, pp. 105–107, and also the end of section II.3 of this thesis (page 229).

Hilbert space formalism is perceived under a very different light. The deep roots of the problem appear not to lie in the fact we had started the kinematical description with the wrong space of states, but rather: *in the fact we had insisted on starting with a space of states instead of starting with an algebra of properties*.

Accordingly, the main strategy of the algebraic approach to Mechanics is to reverse the order in which a physical system is described: specify first an abstract algebra of properties, and only then construct a space of states. It explores another possible dynamic of the State/Property couple, complementary to the one examined by the geometric program. This will furnish a reformulation of the Quantum arena *from within*, but it will accentuate the apparent "incommensurability" of the Classical/Quantum couple. To overcome it, the classical arena will also need to be rethought.

Historically, this algebraic strategy was the first serious attempt to reformulate Quantum Mechanics after von Neumann's introduction of Hilbert spaces. Already in 1934, only two years after the publication of his book, von Neumann himself started investigating this route in a joint paper with Jordan and Wigner[91]. The geometric formulation came only much later: it had to wait for the revival, in the second half of the twentieth century, of the interest for the foundations of Classical Mechanics and the progressive understanding of the importance of symplectic geometry in this setting. Thus, the algebraic approach overwhelmingly dominated the landscape of research in quantum foundations. It developed mainly into two different branches which may both be seen as originating from works published by von Neumann in 1936:

– Quantum Logic. It was launched by the joint paper of von Neumann and Garrett Birkhoff *"The Logic of Quantum Mechanics"*[92] and continued to be developed in the 1960's, principally by Josef-Maria Jauch and his student Constantin Piron in Geneva, so that this approach is sometimes referred to as the "Geneva approach

---

[91]P. Jordan, J. von Neumann, and E. P. Wigner. "On an Algebraic Generalization of the Quantum Mechanical Formalism". In: *Annals of Mathematics* 35 (1934), pp. 29–64. (Reprinted in: J. von Neumann. *Collected Works*. Ed. by A. H. Taub. Oxford: Pergamon Press, 1961, Vol. II, pp. 409–444).

[92]J. von Neumann and G. Birkhoff. "The Logic of Quantum Mechanics". In: *Annals of Mathematics* 37.4 (1936), pp. 823–843. (Reprinted in: J. von Neumann. *Collected Works*. Ed. by A. H. Taub. Oxford: Pergamon Press, 1961, Vol. IV, pp. 105–125).

to quantum mechanics"[93].

– <u>$C^*$-algebras</u>. It started with von Neumann's paper *"On an Algebraic Generalization of The Quantum Mechanical Formalism (Part I)"*[94]. A decisive contribution was that of Irving Segal's 1947 article *"Postulates for General Quantum Mechanics"*[95], which made heavy use of the mathematical formalism developped by the Russian school of Israel Gelfand (to be explained in the following section). Later, the approach was taken on by Rudolf Haag and Daniel Kastler and applied to Quantum Field Theory[96].

A nice collection of some of the most important papers in the development of both approaches can be found in Hooker's *The Logico-Algebraic Approach to Quantum Mechanics. Volume I: Historical Evolution*[97]. Here, I will concentrate only on the $C^*$-algebraic approach because it connects beautifully with the geometrical approach.

Philosophically, the algebraic strategy has oftentimes been motivated by a certain underlying *operational* view of Physics. According to this view, the fundamental concepts, upon which the physical theories are to be based, must result from a detailed analysis of the experimental procedures as they take place in an actual laboratory. In other terms, a framework is operational if "all aspects are introduced with specific reference to events to be experienced"[98]. And since physical systems are only accessible

---

[93]C. Piron. "Axiomatique quantique". In: *Helvetica Physica Acta* 37 (1964), pp. 439–468.

J.-M. Jauch. *Foundations of Quantum Mechanics*. Reading: Addison-Wesley, 1968.

J.-M. Jauch and C. Piron. "What is Quantum Logic?" In: *Quanta, Essays in Theoretical Physics, dedicated to Gregor Wentzel*. Ed. by P. Freund, C. Goebel, and Y. Nambu. Chicago: University of Chicago Press, 1970, pp. 166–181.

[94]J. von Neumann. "On an Algebraic Generalization of The Quantum Mechanical Formalism (Part I)". in: *Receuil Mathématique* 1.4 (1936), pp. 415–484. (Reprinted in: J. von Neumann. *Collected Works*. Ed. by A. H. Taub. Oxford: Pergamon Press, 1961, Vol. III, pp. 492–559).

[95]I. E. Segal. "Postulates for General Quantum Mechanics". In: *Annals of Mathematics* 48.4 (1947), pp. 930–948.

[96]R. Haag and D. Kastler. "An Algebraic Approach to Quantum Field Theory". In: *Journal of Mathematical Physics* 5.7 (1964), pp. 848–861.

R. Haag. *Local Quantum Physics – Fields, Particles, Algebras*. 2nd ed. Heidelberg: Springer-Verlag, 1996.

[97]C. Hooker. *The Logico-Algebraic Approach to Quantum Mechanics. Volume I: Historical Evolution*. Dordrecht, The Netherlands: Reidel Publishing Company, 1975.

[98]D. Aerts and S. Aerts. "Towards a General Operational and Realistic Framework for Quantum Mechanics and Relativity Theory". In: *Quo Vadis Quantum Mechanics?* Ed. by A. C. Elitzur, S. Dolev, and N. Kolenda. Berlin: Springer, 2005, pp. 153–207, p. 153.

through a series of measurements, they should be defined by the set of measurable properties—or, to adopt just once the operationalist language, by the set of *observables*. A flavor of this was certainly present in the main motivation advanced by von Neumann to develop Quantum Mechanics in terms of algebras:

> [...] the *states* are merely a derived notion, the primitive (phenomenologically given) notion being the *qualities* [...][99]

But most importantly, operationalism was explicitly endorsed by the other main contributors to the $C^*$-algebraic approach. Irving Segal opened his seminal article on algebraic quantum mechanics by claiming that his theory was "strictly operational"[100], and later wrote an article specially dedicated to defending this position[101]. In a similar spirit, Haag and Kastler decided to "base their discussion on the notions of "operations""[102] and all of their writings are full of operational arguments to motivate their choices and definitions[103]. More recently, in Strocchi's excellent introductory textbook to the $C^*$-algebraic approach to Quantum Mechanics, one finds that

> [...] it is not justified to extrapolate to the microscopic level the prejudices derived from our experience with the macroscopic world. The only guide [for establishing the mathematical description of quantum systems] must be the recourse to operational considerations [...].[104]

Beyond the question "Is it possible to start the description of a physical system by its algebra of properties?", the mathematical physicist with such philosophical motivations will attempt to answer the more ambitious question "Is it *necessary* to start the description of a physical system by its algebra of properties?". This rather extremist view has been called "Algebraic Imperialism" by Arageorgis[105] and has received serious criticisms[106], up to the point that Haag acknowledged that the "specific mathematical structure of Quantum Mechanics [...] is not so easily derivable from operational principles"[107]. But, as Laura Ruetsche rightly remarks, one need not adhere to any operationalist view whatsoever to become interested in the algebraic approach[108]. Rather, it becomes interesting to look carefully at the *details* of the mathematical structures involved to see whether they point in a particular direction of this debate. Under this light, the back-and-forth between states and properties acquires a new philosophical relevance.

---

[99]Letter from von Neumann to G. Birkhoff, J. von Neumann, *John von Neumann: Selected Letters*, p. 59, author's emphasis

[100]Segal, op. cit., p. 930.

[101]I. E. Segal. "The Mathematical Meaning of Operationalism in Quantum Mechanics". In: *The Axiomatic Method. With Special Reference to Geometry and Physics*. Proceedings of an International Symposium held at UC Berkeley, Dec. 26 1957-Jan. 4, 1958. Ed. by L. Henkin, P. Suppes, and A. Tarski. Amsterdam: North-Holland Publishing Co., 1959, pp. 341–352.

[102]Haag and Kastler, op. cit., p. 850.

[103]To give one example, in their joint article, they wrote: "We may say therefore that we have a complete theory if we are able in principle to compute such probabilities for every state and every operation when the state and the operation *are defined in terms of laboratory procedures.*" (Ibid., p. 850, my emphasis.)

[104]F. Strocchi. *An Introduction to the Mathematical Structure of Quantum Mechanics.* 2nd ed. Singapore: World Scientific, 2008, p. 42.

In my opinion, the two most important *clichés* emerging from the $C^*$-algebra formulation of Mechanics are the following:

- ⋄ *Quantum = Non-commutative.* $C^*$-algebras cover both classical and quantum systems. The first are described by commutative $C^*$-algebras, the second by non-commutative ones. Quantization is thus the passage from commutativity to non-commutativity.

- ⋄ *Quantum = Operational.* A classical system may equivalently be described by

---

[105]A. Arageorgis. "Fields, Particles, and Curvature: Foundations and Philosophical Aspects of Quantum Field Theory in Curved Space-Time". PhD thesis. University of Pittsburgh, 1995.

[106]See for example L. Ruetsche. *Interpreting Quantum Theories. The Art of the Possible.* Oxford: Oxford University Press, 2011, Section 6.4., pp. 132–143, and also S. J. Summers. "On the Stone – von Neumann Uniqueness Theorem and Its Ramifications". In: *John von Neumann and the Foundations of Quantum Physics.* Ed. by M. Rédei and M. Stöltzner. Dordrecht: Kluwer Academic Publishers, 2001, pp. 135–152.

[107]Haag, op. cit., p. 7. See however Aerts and Aerts, op. cit. for a recent attempt of building a (generalized) quantum theory from an operationalist stance.

[108]She writes: "The adamantly operationalist original axiomatizers establish the association between regions and their local algebras by the interpretive maneuver of identifying elements of $\mathcal{U}(\mathcal{O})$ with observables *measurable* by means of actions confined to the region $\mathcal{O}$. But the association between local observables $\mathcal{U}(\mathcal{O})$ and regions $\mathcal{O}$ needn't be mediated by the notion of measurement or ideologies totemizing that notion. [...] The operationalism of the original axioomatizers is one interpretative option." (Ruetsche, op. cit., pp. 104–105, author's emphasis.)
Although Ruetsche's comment is originally intended for Algebraic Quantum Field Theory, it equally well applies to the algebraic formulation of non-relativistic Quantum Mechanics.

its states or its properties. In the Quantum, this equivalence fails and a system must be defined by the algebraic structure of its properties/observables. This shows that, at the fundamental level, one must adopt an operational description of physical systems.

These are both widespread claims. They swamp all the literature discussing Quantum Mechanics—be it research articles on foundations or popular accounts of the theory. Again, a nice example of this can be found in Strocchi's book. He says:

> In this perspective, since a physical system is described in terms of measurements of its observables, one may take the point of view that a classical system is *defined* by its physical properties, i.e. by the algebraic structure of the set of its measurable quantities, which generate an abstract abelian $C^*$-algebra $\mathcal{A}$ with identity.[109]

And also:

> The deep philosophical conclusion [...] is that for the mathematical description of atomic systems *one needs an algebra of observables* which is *non-abelian*. Clearly, as always in the great discoveries, this is not a mathematical theorem and a great intuition and ingenuity was involved in Heisenberg foundations of Quantum Mechanics. To give up the abelian character of the algebra of observables may look as a very bold step, but it should be stressed that the commutativity of observables is a property of our mathematical description of classical *macroscopic* systems [...].[110]

In the following subsections, I will present the main points of the algebraic approach and discuss its relevance for the conceptual analysis of Kinematics we have here undertaken. In the course of it, we will see there are many doubts—to say the least—concerning the validity of the two mainstream claims. All of the technical material is standard and there exist many excellent expositions of it. I will mainly follow the first chapter of Landsman's enlightening *Mathematical Topics Between Classical and Quantum Mechanics*. Other important references are (from the most introductory to

---

[109]Strocchi, op. cit., p. 15, Strocchi's emphasis.

[110]Ibid., pp. 41–42.

the most advanced):

– F. Strocchi. *An Introduction to the Mathematical Structure of Quantum Mechanics.* 2nd ed. Singapore: World Scientific, 2008,

– J. Dixmier. *Les C\*-algèbres et leurs représentations.* 2nd ed. Paris: Gauthiers-Villars, 1969,

– J. M. G. Fell and R. S. Doran. *Representations of \*-Algebras, Locally Compact Groups, and Banach \*-Algebraic Bundles.* Vol. 1. San Diego: Academic press, 1988,

– E. M. Alfsen and F. W. Shultz. *State Spaces of Operator Algebras.* Boston: Birkhäuser, 2001,

– M. Takesaki. *Theory of Operator Algebras Vol. I.* New York: Springer, 2003.

## II.3.1    The grand algebraic analogy

Considering the description of classical and quantum properties of the first two sections, it may be surprising to hear that the "commutativity/non-commutativity" picture is still under consideration. As we have repeated many times, both classical and quantum properties fall under the general concept of Jordan-Lie algebras. Thus, any algebra of physical properties may be seen as composed of a commutative Jordan algebra of properties-as-quantities and a non-commutative Lie algebra of properties-as-transformations . In this setting, the distinction between classical and quantum lies in the associativity/non-associativity of the Jordan product. It would therefore seem that the commutativity/non-commutativity picture is but an antiquated miss-analogy one should better forget.

The story is however more involved than this simple account, based solely on the perspective of Jordan-Lie algebras. For other important types of algebras appear in both kinematical arenas. This can be seen as a consequence of a movement in striking analogy with the one that had launched the geometric program. Recall: the geometric structures of the quantum arena had emerged after a change in the point of view towards the space of states: switch from complex numbers to real numbers. As we will see, the powerful C\*-algebraic structures will emerge from the exact complementary

movement—change the perspective on the algebra of properties: extend from real numbers to complex numbers.

Jordan-Lie algebras are algebras defined over the field of real numbers. We now introduce two new different types of algebras defined over the field of complex numbers.

**Definition II.4.** A **\*-algebra** is a complex associative algebra with an involution. That is, it is a complex algebra $(\mathcal{U}, \circ)$ together with a real-linear map $^* : \mathcal{U} \to \mathcal{U}$ such that, for all $A, B, C \in \mathcal{U}$ and $\lambda \in \mathbb{C}$, we have

   i) $(A \circ B) \circ C = A \circ (B \circ C)$,

   ii) $A^{**} = A$,

   iii) $(A \circ B)^* = B^* \circ A^*$,

   iv) $(\lambda A)^* = \overline{\lambda} A^*$.

**Definition II.5.** A **C\*-algebra** $(\mathcal{U}, \circ, \, ^*, \| \cdot \|)$ is a \*-algebra together with a norm such that

   i) $(\mathcal{U}, \| \cdot \|)$ is a complex Banach space,

   ii) for all $A, B \in \mathcal{U}$, $\|A \circ B\| \leqslant \|A\| \|B\|$,

   iii) for all $A \in \mathcal{U}$, $\|A^* A\| = \|A\|^2$.

A \*-algebra or a C\*-algebra is said to be *commutative* if the associative product $\circ$ is commutative: $\forall A, B \in \mathcal{U}, \ A \circ B = B \circ A$.

The canonical example from quantum mechanics which motivates these definitions is the algebra $(\mathcal{B}(\mathcal{H}), \circ, \dagger, \| \cdot \|)$ of *bounded* linear operators on a Hilbert space $\mathcal{H}$, with the usual composition of operators as the associative product, and the action of taking adjoints as involution[111]. Of course, this is a non-commutative C\*-algebra. The canonical example of a commutative \*-algebra is the algebra $\mathcal{C}(M, \mathbb{C})$ of continuous complex-valued functions over a topological space $M$, with complex conjugation as involution and the usual point-wise multiplication as the associative product[112].

---

[111]The fact of considering only bounded operators is crucial for two reasons. First, it is necessary to insure that the map $\dagger : A \mapsto A^\dagger$ is an involution and thus that $\mathcal{B}(\mathcal{H})$ is a \*-algebra. Indeed, the identity $A^{\dagger\dagger} = A$ is in general not true for unbounded operators of an infinite-dimensional Hilbert space. Second, the boundedness condition is also necessary to define the norm that turns it into a $C^*$-algebra. Recall the definition of the norm for a bounded operator $A \in \mathcal{B}(\mathcal{H})$ (see Landsman, op. cit., p. 39):
$$\|A\| := \sup \{ \|A\psi\| \, \big| \, \psi \in \mathbb{S}\mathcal{H} \}.$$

In fact, the famous Gelfand-Naimark theorem shows that any $C^*$-algebra falls into one of these two fundamental examples. First, given *any* $C^*$-algebra $\mathcal{U}$, there exists a Hilbert space $\mathcal{H}$ such that $\mathcal{U}$ is isomorphic to a norm-closed *-subalgebra of $\mathcal{B}(\mathcal{H})$[113]. Second, given a *commutative* $C^*$-algebra $\mathcal{U}$, there exists a locally compact topological space $X$ such that $\mathcal{U} \simeq \mathcal{C}_0(X, \mathbb{C})$[114]. Thus, *in the context of $C^*$-algebras*, and only in this context, the following associations can be made:

$$\textbf{commutative algebra} \longleftrightarrow \textbf{functions over a topological space} \qquad \text{(II.20a)}$$

$$\textbf{non-commutative algebra} \longleftrightarrow \textbf{operators over a Hilbert space.} \qquad \text{(II.20b)}$$

This means that any commutative $C^*$-algebra can be thought as an algebra of functions over some topological space, whereas any non-commutative $C^*$-algebra can be thought as an algebra of bounded operators on some Hilbert space.

Now, as we saw in section II.1, the standard description of Kinematics suggests we should also make the associations

$$\textbf{classical properties} \longleftrightarrow \textbf{functions over a differentiable manifold} \qquad \text{(II.21a)}$$

$$\textbf{quantum properties} \longleftrightarrow \textbf{operators over a Hilbert space} \qquad \text{(II.21b)}$$

which, combined with II.20a and II.20b, seems to lead to the common-place picture

$$\textbf{classical physical systems} = \textbf{commutative } C^*\text{-algebra} \qquad \text{(II.22a)}$$

$$\textbf{quantum physical systems} = \textbf{non-commutative } C^*\text{-algebra.} \qquad \text{(II.22b)}$$

Nevertheless, this train of thought, swiftly leading to such a conclusion, should be regarded with suspicion. There are two main reasons for this: on the one hand, as the geometric reformulation of Mechanics has clearly shown, quantum properties may also be described by functions over a differentiable manifold. Thus, it is not at all clear in which way the Function/Operator couple can be used to distinguish the Classical from the Quantum. On the other hand—and this has been perhaps the most important point throughout—the truly relevant object in the description of physical properties is their abstract mathematical structure. That is, our attention should focus more on the collection of algebraic operations attached to the set of properties than in the particular nature of the elements of this set. Classical properties may be described by the set $\mathcal{C}^\infty(M, \mathbb{R})$, but it is a completely different thing whether one considers this set equipped with point-wise multiplication or with the Poisson bracket. Quantum properties may be described by the set $\mathcal{B}_\mathbb{R}(\mathcal{H})$, but it is a completely different situation whether one considers this set equipped with the commutator or with the anti-commutator.

---

[112]To turn this into a commutative $C^*$-algebra, one needs either to require the topological space $M$ to be compact, or else to restrict attention to the algebra $\mathcal{C}_0(M, \mathbb{C})$ of continuous complex-valued functions that vanish at infinity. Recall the definition of the norm of a continuous function $f$ on a *compact* space $M$:
$$\|f\| := \sup\{|f(p)| \, \big| \, p \in M\}.$$

[113]See ibid., Theorem I.1.1.8., p. 40, or Strocchi, op. cit., Theorem 2.3.1., p. 47.

[114]Landsman, op. cit., Theorem I.1.2.3, p. 42.

Therefore, to have a proper understanding of the (in)validity of "equations" (II.22a) and (II.22b), one needs to understand the precise mathematical relation between these newly defined complex algebras (*-algebras and $C^*$-algebras) and the real algebras we had already encountered (Jordan, Jordan-Lie and Poisson algebras). The example of bounded operators, which form a non-commutative $C^*$-algebra, and bounded self-adjoint operators, which form a non-associative Jordan-Lie algebra, indicates these two types of algebras should indeed be closely related.

With $C^*$-algebras it is the first time we explicitly consider the existence of a norm on the algebra of physical properties. Thus, it is best to start by also introducing a norm on their real counterparts.

**Definition II.6.** A **Jordan-Lie-Banach algebra** (or JLB-algebra for short)[115] is a real Jordan-Lie algebra $(\mathcal{U}_\mathbb{R}, \bullet, \star)$ equipped with a norm $\| \cdot \|$ such that

i) $(\mathcal{U}_\mathbb{R}, \| \cdot \|)$ is a Banach space,

ii) for all $A, B \in \mathcal{U}_\mathbb{R}$, we have $\|A \bullet B\| \leqslant \|A\|\|B\|$ and $\|A\|^2 \leqslant \|A^2 + B^2\|$.

Analogously, one gets the concept of a **Jordan-Banach algebra** (or JB-algebra for short) by norming a real Jordan algebra $(\mathcal{U}_\mathbb{R}, \bullet)$[116].

---

[115]Ibid., Definition I.1.14., p. 38.

[116]For the definition of a Jordan algebra, see footnote 33 (page 157).

Now, given the natural definition of morphisms between JLB-algebras and between $C^*$-algebras[117], the relation between all these algebras is captured in the following two facts:

**Theorem II.1.** *There is an equivalence between the category $\mathcal{JLB}$ of Jordan-Lie-Banach algebras and the category $\mathcal{CStar}$ of $C^*$-algebras. Moreover, this restricts to an equivalence between the full subcategories $a\mathcal{JLB}$ of associative JLB-algebras and $c\mathcal{CStar}$ of commutative $C^*$-algebras.*[118]

**Theorem II.2.** *The category $a\mathcal{JLB}$ is equivalent to the category $a\mathcal{JB}$ of associative Jordan-Banach algebras*[119].

Let us comment on these results. Theorem II.1 states that, although JLB-algebras and $C^*$-algebras are seemingly different types of algebras, they are in fact the *same*: one can switch from one point of view to the other without any loss of information. It is useful to see how this back-and-forth can be performed:

1. *From $C^*$-algebras to JLB-algebras:* given a $C^*$-algebra $(\mathcal{U}, \circ, {}^*, \|\cdot\|)$, consider the set $\mathcal{U}_{\mathbb{R}}$ of all self-adjoint elements in $\mathcal{U}$. Then, equipped with the two operations

$$A \bullet B := \frac{1}{2}(A \circ B + B \circ A) \ \text{ and } \ A \star B := \frac{i}{2}(A \circ B - B \circ A),$$

   $(\mathcal{U}_{\mathbb{R}}, \bullet, \star, \|\cdot\|)$ is a real JLB-algebra.

2. *From JLB-algebras to $C^*$-algebras:* conversely, given a JLB-algebra $(\mathcal{U}_{\mathbb{R}}, \bullet, \star, \|\cdot\|)$, consider its complexification $(\mathcal{U}_{\mathbb{R}})_{\mathbb{C}}$. For any $A, B \in \mathcal{U}_{\mathbb{R}}$, equip $(\mathcal{U}_{\mathbb{R}})_{\mathbb{C}}$ with the

---

[117]A morphism of JLB-algebras is a continuous linear map $\phi : \mathcal{U}_{\mathbb{R}} \longrightarrow \mathcal{B}_{\mathbb{R}}$ respecting both the Jordan and Lie structures: for all $A, B \in \mathcal{U}_{\mathbb{R}}$, $\phi(A \bullet B) = \phi(A) \bullet \phi(B)$ and $\phi(A \star B) = \phi(A) \star \phi(B)$ (cf ibid., Definition I.1.1.3., p. 38). On the other hand, a morphism of $C^*$-algebras is a linear map $\phi : \mathcal{U} \longrightarrow \mathcal{B}$ respecting both the associative product and the involution: for all $A, B \in \mathcal{U}$, $\phi(A \circ B) = \phi(A) \circ \phi(B)$ and $\phi(A^*) = \phi(A)^*$ (cf ibid., Definition I.1.1.7., p. 40).

Note that, for $C^*$-algebras, one does not need to require the morphisms to be continuous, as this is automatically satisfied (cf. J. Dixmier. *Les $C^*$-algèbres et leurs représentations.* 2nd ed. Paris: Gauthiers-Villars, 1969, p. 7).

[118]The main content of the theorem can be found in Landsman, op. cit., Theorem I.1.1.9., p. 40, but the result is not stated in terms of categories. The proof of the equivalence of categories is sketched in Nlab: http://ncatlab.org/nlab/show/Jordan-Lie-Banach+algebra.

[119]Ibid., p. 38.

operations

$$A \circ B := \; A \bullet B - iA \star B \; \text{ and } \; (A + iB)^* := \; A - iB$$

and with the norm $\|C\|^2 = \|C \circ C^*\|$, for any $C \in (\mathcal{U}_{\mathbb{R}})_{\mathbb{C}}$. Then, $\big((\mathcal{U}_{\mathbb{R}})_{\mathbb{C}}, \circ, {}^*, \|\cdot\|\big)$ is a $C^*$-algebra[120].

Thus, the movement from complex algebras to real algebras is performed by *restricting* to the real subset and *splitting* the product $\circ$ into its real and imaginary parts. The converse movement is achieved by *extending* the real algebra and *unifying* the two algebraic structures into one single product. These two constructions are compatible in the sense that $\big((\mathcal{U}_{\mathbb{R}})_{\mathbb{C}}\big)_{\mathbb{R}} \simeq \mathcal{U}_{\mathbb{R}}$ and $(\mathcal{U}_{\mathbb{R}})_{\mathbb{C}} \simeq \mathcal{U}$. Thus, any JLB-algebra may be seen as the real part of a $C^*$-algebra[121].

Moreover, Theorem II.1 shows that associativity at the level of JLB-algebras is equivalent to commutativity at the level of $C^*$-algebras. In the presence of this equivalence, one could be tempted to see the confirmation of the validity, *in the context of $C^*$-algebras*, of the commutativity/non-commutativity picture of the Classical/Quantum couple. Classical properties are described by associative Jordan-Lie algebras (i.e. by Poisson algebras); quantum properties are described by non-associative Jordan-Lie algebras. Thus—it would seem—by the above theorem, this is equivalent to equations II.22a and II.22b (page 200).

Yet, one must be careful to distinguish Jordan-Lie algebras from Jordan-Lie-*Banach* algebras. For the introduction of a norm is not an innocent move. Theorem II.1 reduces the question of the relation between $C^*$-algebras and Jordan-Lie algebras to the investigation of which Jordan-Lie algebras can be turned into JLB-algebras.

---

[120]Here, the associative product is first defined for elements of $\mathcal{U}_{\mathbb{R}}$ and then extended by $\mathbb{C}$-linearity to any element of $\mathcal{U}$. One may alternatively define it by the quite obscure formula

$$(A + iB) \circ (C + iD) := (A \bullet C + B \star C + A \star D - B \bullet D) + i(B \bullet C + A \bullet D + B \star D - A \star C)$$

whose only merit is to define the product $\circ$ directly on the whole complex algebra. See for example
http://ncatlab.org/nlab/show/Jordan-Lie-Banach+algebra.

[121]En passant, this finally explains the notation we have adopted from the beginning for abstract Jordan-Lie algebras.

Theorem II.2 partially answers this later question. It states that a JLB-algebra is associative if and only if its Lie structure vanishes. In other words, non-trivial Poisson algebras (e.g. the algebra of smooth functions over a symplectic manifold) cannot in general be normed and therefore do not fall under the theory of JLB-algebras. By the same token, it appears that the theory of $C^*$-algebras is unable to capture precisely those algebraic structures which describe classical properties! Quite to the opposite of what could have been initially thought, *the combination of the above two theorems definitively <u>invalidates</u> the idea that quantization is the transition from commutativity to non-commutativity*. For to equate classical systems with commutative $C^*$-algebras would amount to boldly ignoring the symplectic structure of Classical Kinematics...

Despite the failure of $C^*$-algebras to encompass all of Classical and Quantum Kinematics, there is no doubt of the fruitfulness of this algebraic approach to Quantum Mechanics. While the machinery of $C^*$-algebras may be unable to perceive the Poisson structure of Classical Kinematics, it nonetheless may be used to have new insights on the meaning of the algebraic Jordan structure, as will be seen in the next section. Moreover, this algebraic approach has already provided another interesting characterization of the distinction between classical and quantum properties. Before, the world of Jordan-Lie algebras was sharply divided into two groups, depending on the value of $\kappa$ in the associator rule $(F \bullet G) \bullet H - F \bullet (G \bullet H) = \kappa (F \star H) \star G$[122]. Either $\kappa = 0$ and you were in the Classical arena dealing with Poisson algebras; or $\kappa = 1$ and you were in the Quantum arena dealing with non-associative Jordan-Lie algebras. But the conceptual meaning of this change, from associativity to non-associativity, was not transparent. Now, with the introduction of *-algebras, we see that non-associative Jordan-Lie algebras precisely correspond to those Jordan-Lie algebras arising as the real part of complex *-algebras. Thus, instead of using the associator rule as a means to classify Jordan-Lie algebras, we use the relation with the category *Star* of *-algebras. Either the Jordan-Lie algebra stems from a *-algebra, in which case you are dealing with an algebra of quantum properties; or the Jordan-Lie algebra cannot be complexified into a *-algebra, in which case you are dealing with an algebra of classical

---

[122]Cf Definition II.3 (page 159) and the paragraph following it.

properties[123].

This new characterization of the distinction between quantum and classical properties serves well our purpose of understanding the two-fold role of physical properties in Mechanics. For it shows that, in the quantum arena, when passing from real numbers to complex numbers, the Jordan and Lie structures appear as two faces of one single structure. This *unification* of the two algebraic structures is impossible in the classical case and thus characterizes the Quantum.

> **Quantum unification of Jordan and Lie.** The Jordan and Lie structures defined on the set of quantum properties may always be seen as originating from one single structure: they are the real and imaginary part of the associative product of a $C^*$-algebra.

Since the Jordan and Lie structures respectively govern the numerical and transformational roles of physical properties, this is the algebraic reformulation of the quantum interplay of quantities and transformations (cf. page 184).

The relation between the different types of algebras used to describe classical and quantum properties is summarized in Figure II.4 below.

## II.3.2 States and representations of algebras

Recall: in the algebraic formulation of Kinematics, the starting point of the mathematical description of a physical system should be an abstract algebra, which is intended to describe the properties of the system. The point of the last section was to clarify the precise algebras that must be considered for this purpose in both kinematical arenas: in the classical case, one should start from a real Poisson algebra; in the quantum case, one has the choice between real JLB-algebras and complex $C^*$-algebras. From these, the goal is now to define states and investigate the structure of the space of states. In the following subsections, I present and comment the main definitions and results of the algebraic formulation of Quantum Kinematics. Then, we will turn to the Classical arena.

---

[123]One may wonder what is found by complexifying the real algebra of classical properties. Given a Poisson algebra $(\mathcal{U}_\mathbb{R}, \cdot, \{\cdot, \cdot\})$, one can analogously consider its complexification $(\mathcal{U}_\mathbb{R})_\mathbb{C}$ and define the operations: $A \circ B := A \cdot B - i\{A, B\}$ and $(A + iB)^* := A - iB$, where $A, B \in \mathcal{U}_\mathbb{R}$. The problem is that the product $\circ$ is no longer associative: $((\mathcal{U}_\mathbb{R})_\mathbb{C}, \circ, {}^*)$ is a non-associative complex algebra with involution. This is a mathematical structure of a much less studied type than *-algebras. Moreover, if one considers a non-associative algebra with involution $(\mathcal{U}, \circ, {}^*)$, its real part will in general fail to be a Poisson algebra. Thus, the equivalence between the complex and real points of view on the algebra of physical properties is lost in the classical arena.

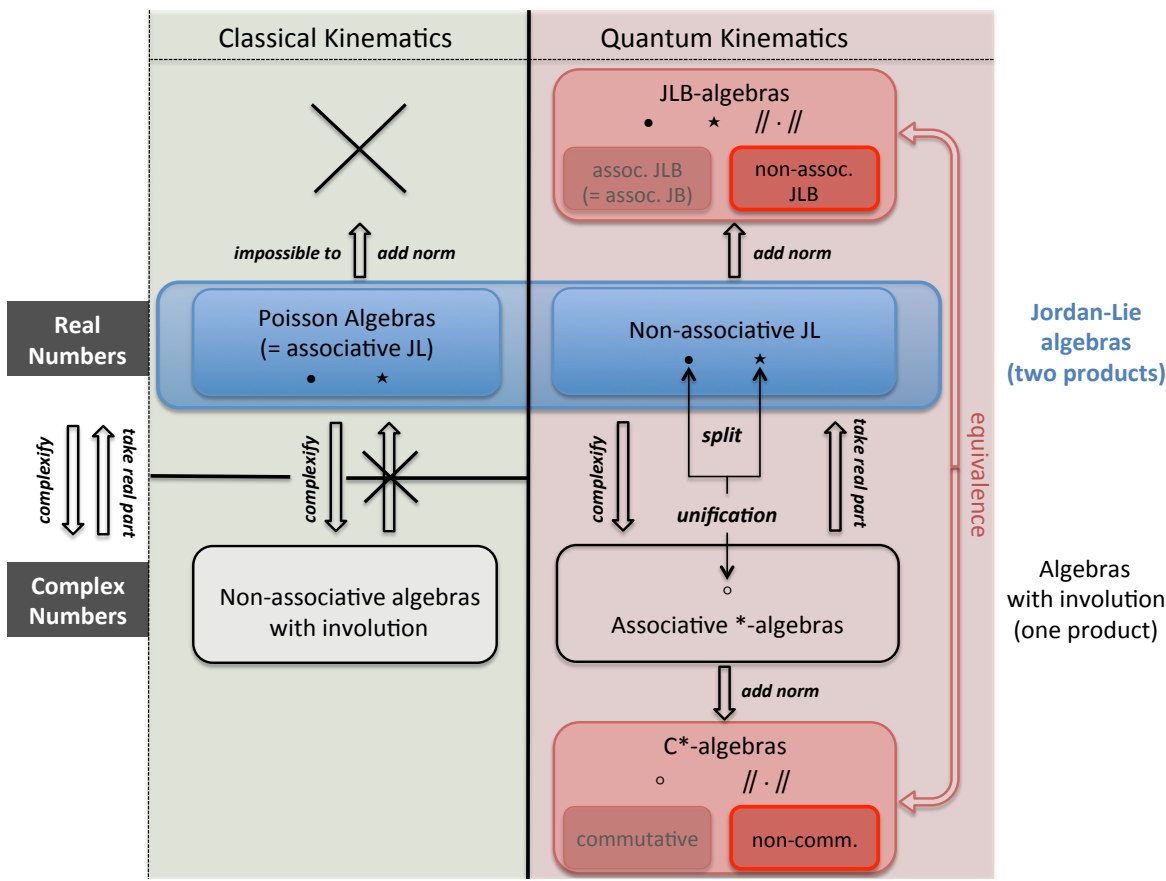

**Fig. II.4** – **The grand algebraic analogy.** Jordan-Lie algebras are the only type of algebras able to cover both classical and quantum kinematics. These are real algebras equipped with two products. By complexification, the quantum side may be equivalently be stated in terms of $C^*$-algebras. In this movement, the two products are unified into one single associative product. This product is commutative if and only if the Lie structure trivially vanishes. Thus, commutative $C^*$-algebras correspond to no physical system whatsoever and quantum systems are described by non-commutative $C^*$-algebras or, equivalently, by non-associative Jordan-Lie-Banach algebras.

The fundamental relation that allows to define the notion of "state" from that of "property" is the *numerical pairing* between the two: from a property $f$ and a state $\rho$, one should be able to produce a number $\langle f, \rho \rangle$. In the geometric approach, the primitive concept was that of state and the property $f$ was defined by the collection of these numbers for all different states. In other words, properties were defined as functions over the space of states and the numerical pairing was rather denoted as $f(\rho)$ (read: '$f$ of $\rho$'). In the algebraic approach, this is simply turned around: the property is the primitive concept and states are defined by the collection of the pairings $\langle f, \rho \rangle$ for all different properties. Accordingly, the numerical pairing will now be denoted by $\rho(f)$ (read: '$\rho$ of $f$').

**Definition II.7.** Given an abstract unital[124] $C^*$-algebra $\mathcal{U}$, a **state** $\rho$ is a normalized positive linear functional over $\mathcal{U}$[125]. This means:

   i) it is a linear map $\rho : \mathcal{U} \longrightarrow \mathbb{C}$,

   ii) for all $A \in \mathcal{U}_{\mathbb{R}}^+$, $\rho(A) \geqslant 0$, where $\mathcal{U}_{\mathbb{R}}^+ := \{ B^*B \, \big| \, B \in \mathcal{U} \}$ (positivity),

   iii) $\rho(\mathbb{I}) = 1$ (normalization).

The definition may equivalently be stated in terms of JLB algebras[126]. The set of states is denoted by $\mathcal{S}(\mathcal{U})$.

One can immediately remark that, so defined, the space of states is *not* a linear space—fact we had already insisted upon in the geometric formulation. Indeed, as linear functionals, one can consider complex linear combinations of the two given states $\rho$ and $\sigma$. But, because of the positivity and normalization conditions, the resulting functional $a\rho + b\sigma$ will in general fail to be a state, unless $a, b \in \mathbb{R}^+$ and $a + b = 1$. Whenever these two conditions are met, then the combination $a\rho + b\sigma$ of the two states is again a state. In other words, the space of states fails to be a complex linear space but is, instead, a *convex* set. This important fact motivates the following definition:

**Definition II.8.** A **pure state** is a state lying on the boundary of $\mathcal{S}(\mathcal{U})$—that is, it is a state that cannot be written as a weighted sum of two different states. The space of pure states is denoted by $\mathcal{P}(\mathcal{U})$. By definition, we have $\mathcal{P}(\mathcal{U}) = \partial\mathcal{S}(\mathcal{U})$. A state that is not pure is called a **mixed state**.

---

[124]The *unit* $\mathbb{I}$ of a $C^*$-algebra, if it exists, is the neutral element of the associative product: for all $A \in \mathcal{U}$, $A \circ \mathbb{I} = \mathbb{I} \circ A = A$. Any non-unital $C^*$-algebra can be turned into a unital $C^*$-algebra in a canonical fashion (in categorical terms, this means that the forgetful functor from the category $\mathcal{CStar}_1$ of unital $C^*$-algebras to the category $\mathcal{CStar}$ has a (left) adjoint). The same holds for JLB-algebras, where the unit is the neutral element of the Jordan product. Thus the restriction to unital algebras is of no consequence and will be often tacitly assumed. (Cf. ibid., Proposition I.2.1., p. 41.)

[125]Alfsen and Shultz, op. cit., p. 50.

[126]In which case one regards a state as a linear map $\rho : \mathcal{U}_{\mathbb{R}} \longrightarrow \mathbb{R}$ and $\mathcal{U}_{\mathbb{R}}^+ := \{ A^2 \, \big| \, A \in \mathcal{U}_{\mathbb{R}} \}$.

As usual, it is insightful to consider the two main examples of $C^*$-algebras. First, if one takes $\mathcal{U} = \mathcal{B}_0(\mathcal{H})$ (the non-commutative $C^*$-algebra of compact operators), then $\mathcal{S}(\mathcal{U})$ is the set of all density matrices and the pure state space is $\mathcal{P}(\mathcal{U}) = \mathbb{P}\mathcal{H}$[127]: given an element $[\psi] \in \mathbb{P}\mathcal{H}$, the corresponding state $\rho_\psi$ is defined by

$$\text{for all } A \in \mathcal{B}_0(\mathcal{H}), \, \rho_\psi(A) := \frac{\langle \psi, A\psi \rangle}{\langle \psi, \psi \rangle}$$

which obviously does not depend on the choice of the representative $\psi \in \mathcal{H}$ of $[\psi] \in \mathbb{P}\mathcal{H}$. Second, if one takes $\mathcal{U} = C_0(X, \mathbb{C})$ (the commutative $C^*$-algebra of compact functions), then $\mathcal{S}(\mathcal{U})$ is the set of all probability measures on $X$, and the pure state space is $\mathcal{P}(\mathcal{U}) \simeq X$[128]. Thus, the algebraic definition of pure states exactly corresponds to the usual "states" we have been handling in this chapter, whereas mixed states correspond to the notion of state of classical or quantum *statistical* physics.

Another crucial notion in this algebraic formulation of Kinematics is that of a representation.

**Definition II.9.** A **representation** of a $C^*$-algebra $\mathcal{U}$ (on a Hilbert space $\mathcal{H}$) is a morphism of $C^*$-algebras $\pi : \mathcal{U} \longrightarrow \mathcal{B}(\mathcal{H})$. Analogously, a representation of a JLB-algebra $\mathcal{U}_\mathbb{R}$ is a morphism of JLB-algebras $\pi : \mathcal{U}_\mathbb{R} \longrightarrow \mathcal{B}_\mathbb{R}(\mathcal{H})$.

Of course, a representation of a $C^*$-algebra induces a representation of the associated JLB-algebra of self-adjoint elements. A representation is **non-degenerate** if 0 is the only vector belonging to the kernel of all the representatives:

$$\text{(for all } A \in \mathcal{U}, \pi(A)\psi = 0) \Longrightarrow \psi = 0.$$

A representation on $\mathcal{H}$ is **cyclic** if there exists a vector $\psi \in \mathcal{H}$ such that $\pi(\mathcal{U})\psi$ is dense in $\mathcal{H}$[129]. In other words, $\psi$ is a cyclic vector for the representation $\pi$ if the smallest closed subspace containing $\psi$ which is invariant under all $\pi(\mathcal{U})$ is the whole Hilbert space $\mathcal{H}$. A closely related notion is that of an **irreducible** representation:

---

[127] Landsman, op. cit., Corollary I.1.1.6., p. 57 and Proposition I.2.1.2., p. 61.

[128] Ibid., p. 55 and Proposition I.2.1.4., p. 61.

[129] A subspace $\mathcal{K}$ is said to be *dense in $\mathcal{H}$* if, for every $\psi \in \mathcal{H}$, there exists a sequence of elements of $\mathcal{K}$ which converges to $\psi$ in the norm of $\mathcal{H}$.

this is a representation $\pi$ for which there are no closed subspaces of $\mathcal{H}$ invariant under all $\pi(\mathcal{U})$ (other than the trivial ones: the whole space $\mathcal{H}$ and the 0 vector). Hence, a representation is irreducible if—and, in fact, only if—*any* non-zero vector in $\mathcal{H}$ is cyclic[130]. Finally, two representations $\pi_1$ and $\pi_2$ are **equivalent** (denoted $\pi_1 \sim \pi_2$) if there exists an isomorphism $U : \mathcal{B}(\mathcal{H}_1) \to \mathcal{B}(\mathcal{H}_2)$ such that the following diagram commutes:

$$
\begin{array}{ccc}
 & \mathcal{U} & \\
{\scriptstyle \pi_1}\swarrow & & \searrow{\scriptstyle \pi_2} \\
\mathcal{B}(\mathcal{H}_1) & \xrightarrow[U]{\sim} & \mathcal{B}(\mathcal{H}_2).
\end{array}
$$

From the abstract point of view that is ours, two equivalent representations are considered to be equal[131].

In fact, the notions of "states" and "representations of the algebra of properties" are closely related. Indeed, we have the two following 'movements':

♠ *Representations allow to define states.* Given a $C^*$-algebra $\mathcal{U}$ and a representation $\pi$ of it on a Hilbert space $\mathcal{H}$, any non-zero vector in $\mathcal{H}$ allows to define a state $\rho_\psi$ by

$$
\text{for all } A \in \mathcal{U}, \ \rho_\psi(A) := \frac{\langle \psi, \pi(A)\psi \rangle}{\langle \psi, \psi \rangle}.
$$

States of $\mathcal{U}$ arising in this way are called **vector states**. For non-degenerate representations, it is clear that two vector states are equal if and only if the vectors defining them are collinear.

♥ *States allow to define representations.* The fact that the information contained in a representation of a $C^*$-algebra is enough to build some states should not surprise. Much less evident is the converse statement. This is known as the GNS construction (Gelfand-Naimark-Segal) and is certainly a fundamental result in the theory of $C^*$-algebras. Given a state $\rho$, there exists a triple $(\mathcal{H}_\rho, \pi_\rho, \psi_\rho)$ such that $\pi_\rho : \mathcal{U} \to \mathcal{B}(\mathcal{H}_\rho)$ is a cyclic representation of $\mathcal{U}$ and $\psi_\rho$ is a normalized cyclic vector such that, for all $A \in \mathcal{U}$, $\rho(A) = \langle \psi_\rho, \pi_\rho(A)\psi_\rho \rangle$[132].

---

[130] Ibid., Proposition I.2.2.2, p. 63). For representations of JLB-algebras, it is this last property which defines the notion of irreducibility (N. P. Landsman. "Classical and Quantum Representation Theory". In: *arXiv preprint* (1994). URL: http://arxiv.org/abs/hep-th/9411172, Definition 6, p. 23).

[131] In other words, in the category $\mathcal{R}ep(\mathcal{U})$ of all representations of the algebra $\mathcal{U}$, equivalence is the pertinent notion of isomorphism.

Through the GNS construction, the *algebraic* problem of studying representations and the *geometric* problem of studying states become entangled. Much of the strength and conceptual importance of the GNS construction comes from the unveiling of this entanglement. This can be very well perceived by investigating the back-and-forth between states and representations established by the two complementary movements ♠ and ♥:

1. First, does any state arise from some representation of the algebra of properties $\mathcal{U}$ through ♠? *Answer:* Yes. This is clear from the GNS construction, since for any $\rho \in \mathcal{S}(\mathcal{U})$ we have $\rho(A) = \langle \psi_\rho, \pi_\rho(A)\psi_\rho \rangle$ for all $A \in \mathcal{U}$. In other words, *all states are vector states.*

2. Conversely, does any representation arise from a state through ♥? *Answer:* No, since GNS representations are necessarily cyclic whereas a general representation of $\mathcal{U}$ need not be so. However, any non-degenerate representation is a direct sum of cyclic representations[133]. Therefore, the problem of classifying all non-degenerate representations may be reduced to the problem of classifying all cyclic representations: they constitute, so to speak, the basic building blocks. The

---

[132]Roughly, the construction is as follows. First, one considers the set $\mathcal{N}_\rho := \{A \in \mathcal{U} \,|\, \rho(A^*A) = 0\} = \{B \in \mathcal{U}_\mathbb{R}^+ \,|\, \rho(B) = 0\}$. This may be thought as the subset of positive properties which are *invisible* to the state $\rho$, in the sense that, for any physical property $C$ and any "invisible" property $B$, the two numbers $\rho(C + B)$ and $\rho(C)$ are equal. Thus, from the point of view of the state $\rho$ it makes better sense to consider physical properties only "up to invisible properties". This is done by defining the quotient $\mathcal{U}/\mathcal{N}_\rho$. The Hilbert space is then the closure of this quotient $\mathcal{H}_\rho := \overline{\mathcal{U}/\mathcal{N}_\rho}$. The representation is defined simply by projecting left multiplication to the quotient: for $A, B \in \mathcal{U}$, we have $\pi_\rho(A)[p(B)] := p(A \circ B)$, where $p : \mathcal{U} \twoheadrightarrow \mathcal{N}_\rho$ is the projection. Finally, the cyclic vector $\psi_\rho$ is defined as the projection of the unit of the $C^*$-algebra : $\psi_\rho := p(\mathbb{I})$. For a detailed description of the GNS construction, see Alfsen and Shultz, op. cit., pp. 51–53. The original construction (which is casted in terms of normed *-rings and does not mention yet the notion of "state") is found in I. Gelfand and M. Naimark. "On the Imbedding of Normed Rings Into the Ring of Operators in Hilbert Space". In: *Matematicheskii Sbornik* 12 (1943), pp. 197–213, pp. 204-ff. The construction as it is now known, in terms of $C^*$-algebras and states, was first described in I. E. Segal. "Irreducible Representations of Operator Algebras". In: *Bulletin of the American Mathematical Society* 61 (1947), pp. 69–105, pp. 77–78.

[133]Landsman, *Mathematical Topics Between Classical and Quantum Mechanics*, Proposition I.1.5.2, p. 53.

appropriate question is thus:

$2^{\text{bis}}$. Does any *cyclic* representation arise from a state through ♥? *Answer:* Yes. Given a cyclic representation $\pi(\mathcal{U})$ on $\mathcal{H}$ and a cyclic vector $\psi \in \mathcal{H}$, the GNS-representation associated to the vector state $\rho_\psi$ is equivalent to $\pi(\mathcal{U})$[134].

In the light of this, the algebraic classification problem can be solved by studying the geometry of the space of states. More precisely, if one denotes by $\mathcal{C}ycl(\mathcal{U})$ the set of all equivalence classes of cyclic representations (the description of which is the main goal of the algebraic problem just mentioned), and one defines on the space of states the equivalence relation

$$(\rho \sim \sigma) \Longleftrightarrow (\pi_\rho \text{ is equivalent to } \pi_\sigma)$$

then, by the points 1 and $2^{\text{bis}}$ above, we have

$$\mathcal{C}ycl(\mathcal{U}) \simeq \mathcal{S}(\mathcal{U})/\sim . \tag{II.23}$$

The problem is now to understand which states are equivalent. In particular, one may wonder what happens if one applies twice the GNS construction. Consider the following sequence (cf. diagram below): i) start with a state $\rho$; ii) construct the associated GNS representation $(\mathcal{H}_\rho, \pi_\rho, \psi_\rho)$; iii) consider a vector $\psi \neq \psi_\rho \in \mathcal{H}_\rho$ and the associated vector state $\rho_\psi$; iv) apply again the GNS construction to define the representation $(\mathcal{H}_{\rho_\psi}, \pi_{\rho_\psi}, \psi_{\rho_\psi})$.

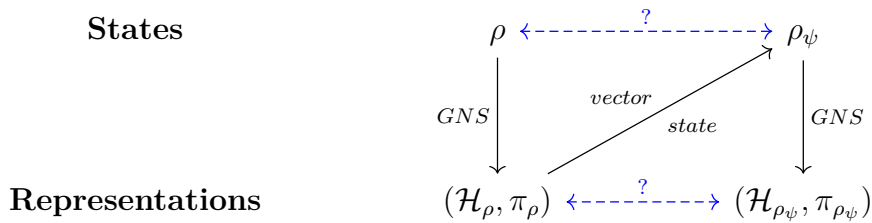

As we have already commented, in general the states $\rho$ and $\rho_\psi$ will differ: only by choosing $\psi$ collinear to $\psi_\rho$, will the two states be equal. Moreover, it is clear that if $\psi \in \mathcal{H}_\rho$ is not itself a cyclic vector for $\pi_\rho$, the representations $\pi_{\rho_\psi}$ and $\pi_\rho$ cannot be equivalent[135]. But one may still hope for the two states to be equivalent whenever $\psi$ is a cyclic vector for $\pi_\rho$. This would be a nice feature, furnishing some kind of stability of the GNS construction: given a representation and a vector, one would like the GNS construction to produce the same initial representation. And this is indeed the case[136]:

$$\rho \sim \rho_\psi \Longleftrightarrow (\psi \text{ is cyclic for } \pi_\rho).$$

---

[134]Ibid., Proposition I.1.5.5, p. 54.

Of particular interest are then those representations for which *any* vector of the carrier Hilbert space is cyclic. These are the only representations not containing the seeds of any other representation. They are completely stable in the sense that, from any such representation $\pi$, the GNS construction will inevitably lead back to the starting point. Symbolically, this may be written as

$$\pi \text{ stable } \Longleftrightarrow \heartsuit \circ \spadesuit(\pi) \sim \pi.$$

It turns out this stability property characterizes both irreducible representations and pure states[137]. Thus, we have the following triangle of equivalent notions:

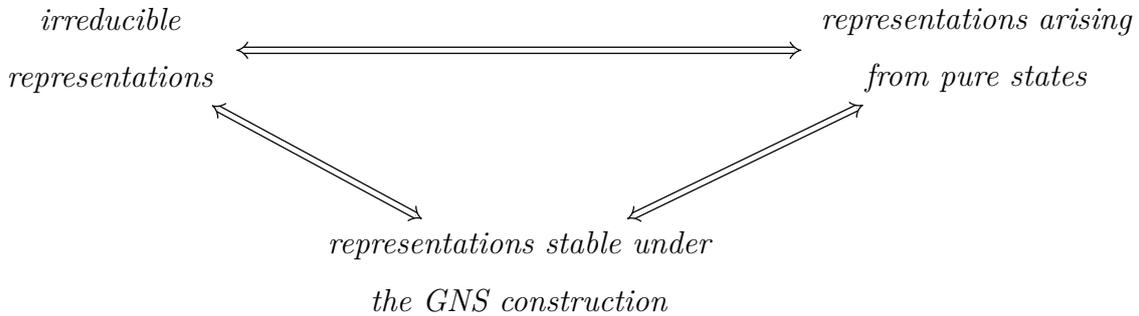

This triangle allows one to understand much of the geometry of the pure state space. First, it provides a connection between geometry and algebra analogous to Equation II.23, but at the level of pure states this time:

$$\widehat{\mathcal{U}} := \mathit{Irrep}(\mathcal{U}) \simeq \mathcal{P}(\mathcal{U})/\sim . \tag{II.24}$$

---

[135]Indeed, there are two ways to perceive $\rho_\psi$: as a vector state defined by $\psi$ or, after the second GNS construction, as a vector state defined by $\psi_{\rho_\psi}$. In other words, supposing both vectors to be unit vectors, we have, for all $A \in \mathcal{U}$, $\langle \psi, \pi_\rho(A)\psi \rangle = \langle \psi_{\rho_\psi}, \pi_{\rho_\psi}(A)\psi_{\rho_\psi} \rangle$. If $\pi$ and $\pi_\rho$ are equivalent, since $\psi_{\rho_\psi}$ is cyclic, then $\psi$ must also be.

[136]Ibid., Proposition I.1.5.5., p. 54.

[137]Ibid., Proposition I.2.2.2, p. 63 and Theorem I.2.2.3, p. 64.

The equivalence classes of pure states are labelled by irreducible representations of the algebra of properties. Moreover, an irreducible representation $\pi_\alpha : \mathcal{U} \longrightarrow \mathcal{B}(\mathcal{H}_\alpha)$ allows to define many different equivalent states. In fact, given a state $\rho$, any other state equivalent to it will arise as a vector state of the Hilbert space $\mathcal{H}_\rho$ constructed by the GNS construction. Thus, there is a bijection between the set of states equivalent to $\rho$ and $\mathbb{P}\mathcal{H}_\rho$. This yields the following result[138]:

**Theorem II.3.** *Given an abstract $C^*$-algebra $\mathcal{U}$, its space of pure states is described as the disjoint union*

$$\mathcal{P}(\mathcal{U}) = \bigsqcup_{\alpha \in \widehat{\mathcal{U}}} \mathbb{P}\mathcal{H}_\alpha.$$

Through this theorem, the algebraic formulation of Quantum Kinematics succeeds in making contact with the geometrical approach. It shows the extent to which the use of $C^*$-algebras as starting point in the definition of physical systems is a generalization of the previous two formulations. For only when $\widehat{\mathcal{U}} = \{*\}$ will one recover the description of the quantum space of states as a projective Hilbert space (or as a Hermitian symmetric space). In particular, it is so when $\mathcal{U} = \mathcal{B}(\mathcal{H})$ (the unique irreducible representation being then the defining one) or when $\mathcal{U}$ is the Weyl $C^*$-algebra $\mathcal{A}_W$ (the uniqueness of the irreducible representation being proven by the Stone-von Neumann theorem)[139]. As Landsman points out, these are the most common examples of non-relativistic quantum systems and, because of this, one may have the impression that the $C^*$-algebraic formalism truly constitutes a generalization only when considering infinite-dimensional systems—that is, quantum fields. But, as we will see in the next chapter, this is far from true, as there are many important examples of $C^*$-algebras admitting inequivalent irreducible representations which are pertinent in the non-relativistic context[140].

---

[138] Ibid., Theorem I.2.5.4, pp. 72–73.

[139] The precise definition of the Weyl $C^*$-algebra may be found in Strocchi, op. cit., pp. 60–61. More generally, whenever $\mathcal{U}$ is a simple $C^*$-algebra (in the sense it admits no non-trivial ideals), the dual space $\widehat{\mathcal{U}}$ will be reduced to one single point.

[140] See N. P. Landsman. "Quantization and Superselection Sectors I. Transformation Group $C^*$-algebras". In: *Rev. Math. Phys* 2 (1990), pp. 45–72.

It is therefore important to truly consider the quantum space of pure states in its wholeness, instead of considering separately each "leaf" $\mathbb{P}\mathcal{H}_\alpha$ as had been done so far. This means to rethink in this slightly more general setting the results of the geometric approach (section II.2). In particular, we must investigate whether it is still possible to reconstruct the algebra of properties from the space of states.

## II.3.3    Reconstructing the algebra of properties from the pure state space

At the risk of becoming repetitive, let me recall once again the conceptual issues at stake in this task of studying the oscillations between states and properties. There is first the obvious question of the logical relation between these two fundamental notions of Kinematics: are properties/observables logically prior to states? By definition of what operationalism is, any such approach to Physics seems forced to answer positively to this question: the entire description of a physical system *must* be based upon experimental facts—and hence the notion of 'state' must be constructed from the measurable quantities (cf. page 195). Now, the algebraic formulation has succeeded in showing that the algebraic structure of properties *can* indeed be the starting point of the descriptions of quantum systems. This is certainly an encouraging result for operationalism. But, needless to say, there is a big step between possibility and necessity, between what can be and what must be. The success of the algebraic approach only shows operationalism is a technically viable philosophical position. If one is seeking for mathematical reasons of preferring it over other philosophical perspectives, then one should attempt to prove that "states" cannot be perceived as the primitive notion. This means attempting to prove that one cannot reconstruct the $C^*$-algebra of properties from the geometry of the space of pure states. If this reconstruction turns out to be possible in the general case—as it was for $\mathcal{U}_\mathbb{R} = \mathcal{B}_\mathbb{R}(\mathcal{H})$—then it would appear that the mathematical formalism of Mechanics is neutral regarding the logical hierarchy between states and properties.

Second, there is the question of articulating the numerical and transformational roles of properties. A recurrent point in the analysis of the different formalisms in which to cast the kinematical arenas has been the quantum compatibility between both roles. This was first hinted at in the standard Hilbert space formalism (page 164) and progressively became more precise with the geometric formulation (page 184) and the $C^*$-algebras (page 205). Thus, it remains to see how—if at all—this important trait manifests itself at the level of the quantum space of pure states.

As I have already mentioned, in the setting of $C^*$-algebras, the cornerstone result around which one must reflect in order to understand the precise interplay between the algebraic and geometric aspects is the following:

**Theorem II.4** (Commutative Gelfand-Naimark). *For any commutative $C^*$-algebra $\mathcal{U}$, there is a canonical isomorphism between $\mathcal{U}$ and $\mathcal{C}_0(\mathcal{P}(\mathcal{U}), \mathbb{C})$, given by the* ***Gelfand transform****:*

$$\widehat{\ } : \mathcal{U} \longrightarrow \mathcal{C}_0(\mathcal{P}(\mathcal{U}), \mathbb{C})$$
$$A \longmapsto \widehat{A} \qquad where \ \widehat{A}(\rho) := \rho(A).$$

*The space of pure states $\mathcal{P}(\mathcal{U})$ is a locally compact Hausdorff space. Moreover, it is compact if and only if $\mathcal{U}$ is unital*[141].

Thus, commutative $C^*$-algebras may effectively be reconstructed from states. This is a very well-known and much commented result from 1943[142]. Now, even better, the equivalence between the algebraic and spatial points of view can be stated in the most precise mathematical manner if one restricts attention to unital $C^*$-algebras and uses the language of category theory:

**Theorem II.5.** *The categories $c\mathcal{C}Star_1$ of unital commutative $C^*$-algebras and $\mathcal{C}pt$ of compact Hausdorff spaces are dual to each other. That is, $c\mathcal{C}Star_1$ and $\mathcal{C}pt^{op}$ are equivalent.*[143]

---

[141] Idem, *Mathematical Topics Between Classical and Quantum Mechanics*, Definition I.2.1.6 and Theorem I.2.1.7, p. 62.

[142] Gelfand and Naimark, op. cit.

[143] The equivalence is given by the functors $\mathcal{P} : c\mathcal{C}Star_1 \longrightarrow \mathcal{C}pt^{op}$ (which associates to a $C^*$-algebra $\mathcal{U}$ its pure state space $\mathcal{P}(\mathcal{U})$) and $\mathcal{C} : \mathcal{C}pt \longrightarrow c\mathcal{C}Star_1^{op}$ (which associates to a space $X$ the algebra of complex valued continuous functions $\mathcal{C}(X)$). The Gelfand transform is then $\widehat{\ } = \mathcal{C} \circ \mathcal{P}$ and the theorem says this functor is naturally isomorphic to the identity functor $1_{c\mathcal{C}Star_1}$. For the equivalence stated

This is sometimes called the *Gelfand duality* (or Gelfand-Naimark duality)[144]. It is a very suggestive theorem since it shows that many properties of compact spaces (in fact, any construction or theorem expressed in the language of categories) can be immediately reformulated in terms of unital commutative $C^*$-algebras (and vice-versa)[145] . This is certainly one of the main reasons of the mathematical interest for this type of algebras and it explains why the general study of $C^*$-algebras is often called "non-commutative topology". Moreover, this theorem is oftentimes regarded as the result that truly launched the whole algebraic approach[146].

Notwithstanding this, our primary interest lies, of course, in non-commutative $C^*$-algebras (recall Figure II.4, page 206) and we must thus study the possibility of reconstructing the algebra also in this case. Here, it is important to stress the existence of many different spaces from which one could expect this reconstruction to be possible.

---

in terms of these two functors, see for example J. M. Gracia-Bondía, J. C. Várilly, and H. Figueroa. *Elements of Noncommutative Geometry.* Boston: Birkhäuser, 2011, pp. 9-10.

[144]As a side remark—and for completeness—let us note that this duality does not hold if one considers non-unital commutative $C^*$-algebras (or locally compact Hausdorff spaces): the category *cCstar* is not equivalent to *LocCpt$^{op}$*. Indeed, the commutative Gelfand-Naimark theorem asserts that the functor $\mathcal{C}_0$ : *LocCpt* $\longrightarrow$ *cCStar$^{op}$* (which associates to a space $X$ the algebra of complex valued continuous functions $\mathcal{C}(X)$ vanishing at infinity) is essentially surjective. However, this functor fails to be full. Despite this, the procedure of unitalization of a commutative $C^*$-algebra (which may be seen as a functor $(\cdot)^+$ : *cCStar* $\longrightarrow$ *cCStar$_1$*) does admit an analogue on the topological side: it is the Alexandroff one-point compactification of locally compact Hausdorff spaces (which may be seen as a functor $(\cdot)_+$ : *LocCpt* $\longrightarrow$ *Cpt*). If one denotes by $X_+$ the resulting compactified space, one has $C_0(X)^+ \simeq C(X_+)$. For details, see I. Dell'Ambrosio. "Categories of $C^*$-algebras". Lecture Notes. URL: http://math.univ-lille1.fr/~dellambr/exercise_C_algebras.pdf.

[145]For a nice dictionary between topological concepts and their algebraic translation, see N. E. Wegge-Olsen. *K-theory and $C^*$-algebras: a Friendly Approach.* New York: Oxford University Press, 1993, p. 24.

[146]Here is Strocchi commenting on it: "From the point of view of general philosophy, the picture emerging from the Gelfand theory of abelian $C^*$-algebras has far reaching consequences and it leads to a rather drastic change of perspective. In the standard description of a physical system the geometry comes first: one first specifies the coordinate space, (more generally a manifold or a Hausdorff topological space), which yields the geometrical description of the system, and *then* one considers the abelian algebra of continuous functions on that space. By the Gelfand theory the relation can be completely reversed: one may start from the abstract $C^*$-algebra, which in the physical applications may be the abstract characterization of the observables, in the sense it encodes the relations between the physical quantities of the system, and then one reconstructs the Hausdorff space such that the given $C^*$-algebra can be seen as the $C^*$-algebra of continuous functions on it. In this perspective, one may say that the algebra comes first, the geometry comes later. The total equivalence between the two points of view indicates a purely algebraic approach to geometry [...]." (Strocchi, op. cit., p. 15, author's emphasis.)

Indeed, given a $C^*$-algebra $\mathcal{U}$, one can consider[147]:

- the space of pure states $\mathcal{P}(\mathcal{U})$,

- the space $\widehat{\mathcal{U}}$ of equivalence classes of irreducible representations,

- the space $\Omega(\mathcal{U}) = \mathrm{Hom}_{\mathcal{C}Star}(\mathcal{U}, \mathbb{C})$ of all non-zero $C^*$-morphisms from $\mathcal{U}$ to $\mathbb{C}$,

- the space $\mathcal{M}(\mathcal{U})$ of maximal ideals of $\mathcal{U}$,

- the space $\mathrm{Prim}(\mathcal{U})$ of primitive ideals of $\mathcal{U}$[148].

These are all a priori different yet related spaces. For example, elements of $\Omega(\mathcal{U})$ are particular instances of elements of $\widehat{\mathcal{U}}$ (they are one-dimensional irreducible representations), and are also particular elements of $\mathcal{P}(\mathcal{U})$ (they are pure states which are also multiplicative). Thus, we have $\widehat{\mathcal{U}} \hookleftarrow \Omega(\mathcal{U}) \hookrightarrow \mathcal{P}(\mathcal{U})$ but it is clear that, in the general case, these injections are not bijections.

However, it just so happens that in the commutative case this plethora of different spaces is invisible, for all these notions of space coincide:

$$\text{if } \mathcal{U} \in \mathrm{ob}(\textit{cCstar}), \text{ then } \mathcal{P}(\mathcal{U}) \simeq \widehat{\mathcal{U}} \simeq \Omega(\mathcal{U}) \simeq \mathcal{M}(\mathcal{U}) \simeq \mathrm{Prim}(\mathcal{U})^{[149]}.$$

The space is then called the (Gelfand) **spectrum** of the commutative $C^*$-algebra and is denoted $\mathrm{Spec}(\mathcal{U})$[150]. The points of the spectrum are sometimes called the *characters*

---

[147]Here, I just give the definition of these spaces as *sets*. The complete definition should also mention the particular topology defined on each of these sets, but to do so would imply an excessively technical digression for the purpose at hand.

[148]An ideal is said to be *primitive* if it is the kernel of an irreducible representation of $\mathcal{U}$. See for example Alfsen and Shultz, op. cit., Definition 5.20., p. 208.

[149]The bijection between $\widehat{\mathcal{U}}$ and $\Omega(\mathcal{U})$ is proven by Schur's lemma (Landsman, op. cit., Proposition I.2.2.2., p. 63). Since all irreducible representations are one dimensional, then for $\alpha \in \widehat{\mathcal{U}}$, $\mathbb{P}\mathcal{H}_\alpha$ is reduced to a point and it becomes clear from Theorem II.3 (page 213) that $\mathcal{P}(\mathcal{U}) \simeq \widehat{\mathcal{U}}$. Finally, the proof of the bijection between $\Omega(\mathcal{U})$ and $\mathcal{M}(\mathcal{U})$ is found in M. Takesaki. *Theory of Operator Algebras Vol. I*. New York: Springer, 2003, Proposition 3.8, p. 15.

[150]The terminological choice is of course not innocent, as this new notion of spectrum generalizes the usual notion of spectrum of a linear operator. Recall: for a linear operator $A$, its spectrum is the set $\mathrm{Sp}(A) = \{z \in \mathbb{C} \,|\, (A - z\mathbb{I}) \text{ is not invertible}\}$. This definition works for any element of a unital (non-commutative) $C^*$-algebra $\mathcal{U}$. The result which establishes the link between both notions of spectrum is the following: given $A \in \mathcal{U}$, consider the commutative $C^*$-algebra generated by $A$ and $\mathbb{I}$, denoted $C^*(A)$. Then, $\mathrm{Spec}(C^*(A)) \simeq \mathrm{Sp}(A)$ (Landsman, op. cit., Theorem I.1.2.4.2.).

For a more extensive investigation on the notion of spectrum (in particular, on the link with the physicists' use of the word "spectrum" for light and atoms), see the interesting article (in french) P. Cartier. "Notion de spectre". In: *Première école d'été : Histoire conceptuelle des mathématiques - Dualité Algèbre-Géométrie*. Maison des Sciences de l'Homme. Universidade de Brasilia, 2008, pp. 232–of $\mathcal{U}$. The commutative Gelfand-Naimark theorem shows there is the same information in the commutative algebra $\mathcal{U}$ or in its spectrum $\mathrm{Spec}(\mathcal{U})$.

When turning to the non-commutative case, the spatial degeneracy splits and the question arises of which space to choose in order to extend the Gelfand duality[151]. In the context of our investigations, it seems we must choose the space of pure states as hypothetical starting point, but it is important to keep in mind that this is not the only mathematically sound possibility[152]. This is the route followed by Landsman: he presented an explicit reconstruction of the algebra of properties from the space of pure states in his article *"Poisson Spaces With a Transition Probability"*[153].

The steps of Landsman's construction are best understood when compared to the work of Schilling and Ashtekar. As already noted, most of the tasks of the geometric program may be seen as particular instances of the general problems arising in the $C^*$-algebraic approach. Indeed, with the Gelfand theory at hand, we can now recognize that the key map considered by Ashtekar and Schilling, which allowed to transform self-adjoint operators into real-valued functions over the projective space (Equation II.11, page 175), is nothing but the Gelfand transform for the JLB-algebra $\mathcal{U}_{\mathbb{R}} = \mathcal{B}_{\mathbb{R}}(\mathcal{H})$ (in which case $\mathcal{P}(\mathcal{U}_{\mathbb{R}}) = \mathbb{P}\mathcal{H}$). In their case, the map was found to be injective (Equation II.14, page 175) and this fact showed there was indeed hope of reconstructing the

---

242. URL: http://semioweb.msh-paris.fr/f2ds/docs/dualite_2008/dualite_doc_final_2008.pdf.

[151]There is also the terminological question of whether one should keep using the word "spectrum" for non-commutative $C^*$-algebras. The most common decisions seem to be either: i) to reserve the notion of "spectrum" only for commutative $C^*$-algebras (e.g. Dixmier, Landsman, Takesaki); ii) to define the spectrum of a general $C^*$-algebra as $\mathrm{Spec}(\mathcal{U}) := \widehat{\mathcal{U}}$ equipped with the so-called Jacobson topology (e.g., Alfsen and Shultz, p. 210 and also Fell and Doran, p. 556).

[152]For example, Akemann has shown how to reconstruct the $C^*$-algebra from the space of maximal ideals (C. Akemann. "A Gelfand Representation Theory for C*-algebras". In: *Pacific Journal of Mathematics* 39.1 (1971), pp. 1–11).

[153]N. P. Landsman. "Poisson Spaces With a Transition Probability". In: *Review of Mathematical Physics* 9.1 (1997), pp. 29–57. URL: http://arxiv.org/abs/quant-ph/9603005. His reconstruction may also be found in the third section ("From Pure States to Observables") of the first chapter of his book *Mathematical Topics Between Classical and Quantum Mechanics*. As he explains, many ideas are motivated by the work of Alfsen, Hanche-Olsen and Shultz, who characterized those compact convex sets arising as space of states of a $C^*$-algebra (E. M. Alfsen, H. Hanche-Olsen, and F. W. Shultz. "State Spaces of $C^*$-algebras". In: *Acta Mathematica* 144 (1980), pp. 267–305 and F. W. Shultz. "Pure States as Dual Objects for $C^*$-algebras". In: *Communications in Mathematical Physics* 82 (1982), pp. 497–509).

algebra of properties as functions over the space of states. In fact, the result is more general[154]:

**Proposition II.6.** *Let $\mathcal{U}_{\mathbb{R}}$ be any JLB-algebra and denote by $\widehat{\mathcal{U}_{\mathbb{R}}}$ its image through the Gelfand transform $\widehat{\phantom{.}}$: $\mathcal{U}_{\mathbb{R}} \longrightarrow \mathcal{C}(\mathcal{P}(\mathcal{U}_{\mathbb{R}}), \mathbb{R})$ (defined as before by $\widehat{A}(\rho) := \rho(A)$ for $A \in \mathcal{U}_{\mathbb{R}}$ and $\rho \in \mathcal{P}(\mathcal{U}_{\mathbb{R}})$). Then, as partially ordered Banach spaces, we have*

$$\mathcal{U}_{\mathbb{R}} \simeq \widehat{\mathcal{U}_{\mathbb{R}}} \subset \mathcal{C}(\mathcal{P}(\mathcal{U}_{\mathbb{R}}), \mathbb{R}).$$

In other words, whenever the Gelfand transform is considered as defined only on the real algebra $\mathcal{U}_{\mathbb{R}}$, it will always be an injection. As Landsman stresses, the theorem fails if the Gelfand transform is extended to the whole $C^*$-algebra. From this point of view, it is thus better to work in the category of real JLB-algebras, the extension to their complex counterparts being useful only for commutative $C^*$-algebras.

Therefore, we know that, even in the general situation, the algebra of physical observables lies somewhere inside the set of all real-valued functions over the space of pure states. The task remains then to characterize the JLB-algebra $\widehat{\mathcal{U}_{\mathbb{R}}}$ inside $\mathcal{C}(\mathcal{P}(\mathcal{U}_{\mathbb{R}}), \mathbb{R})$. The general solution provided by Landsman will mimic in every aspect the definition of physical properties found by Ashtekar and Schilling for the particular case of $\mathcal{U}_{\mathbb{R}} = \mathcal{B}_{\mathbb{R}}(\mathcal{H})$: in order to characterize $\widehat{\mathcal{U}_{\mathbb{R}}}$ inside $\mathcal{C}(\mathcal{P}(\mathcal{U}_{\mathbb{R}}), \mathbb{R})$, we will need to first bring out the structures naturally present in any pure state space. $\widehat{\mathcal{U}_{\mathbb{R}}}$ will then appear to be the subset of functions respecting those additional structures[155].

Since the space of pure states $\mathcal{P}(\mathcal{U})$ can always be described as a disjoint union of projective Hilbert spaces, the relevant mathematical structures need to be somehow generalizations of the symplectic and Riemannian structures present in each leaf $\mathbb{P}\mathcal{H}_{\alpha}$[156]. In his article, Landsman proposes the following two structures:

---

[154]Landsman, *Mathematical Topics Between Classical and Quantum Mechanics*, Theorem I.2.1.7., p. 62.

[155]Let me be clear: I am *not* claiming that these results of Landsman's work were influenced by the work of Schilling. My claim is simply that, given the particular order of exposition *I have chosen* for this chapter, we can conceptually relate both works and use Schilling's to help *us* understand the core of Landsman's reconstruction.

[156]A disjoint union of symplectic manifolds is not necessarily a symplectic manifold: it may not even be a manifold!

**Definition II.10.** A **Poisson space**[157] is a Hausdorff topological space $\mathcal{P}$ together with a collection $S_\alpha$ of symplectic manifolds, as well as continuous injections $\iota_\alpha : S_\alpha \hookrightarrow \mathcal{P}$, such that

$$\mathcal{P} = \bigsqcup_\alpha \iota_\alpha(S_\alpha).$$

**Definition II.11.** A symmetric **transition probability space**[158] is a set $\mathcal{P}$ equipped with a function $\mathrm{Pr} : \mathcal{P} \times \mathcal{P} \longrightarrow [0, 1]$ such that for all $\rho, \sigma \in \mathcal{P}$

i) $\mathrm{Pr}(\rho, \sigma) = 1 \iff \rho = \sigma$,

ii) $\mathrm{Pr}(\rho, \sigma) = \mathrm{Pr}(\sigma, \rho)$ (i.e. $\mathrm{Pr}$ is symmetric).

The function $\mathrm{Pr}$ is called a *transition probability*[159].

The notion of a Poisson space coined by Landsman should not come as a surprise. First, it is clear that pure state spaces are indeed Poisson spaces: the notion is almost hand made in order to cover them. Moreover, it sounds reasonable to say they are a generalization of the notion of a symplectic manifold. In fact, the definition is strongly motivated by the important result that any Poisson manifold can be written as a disjoint union of symplectic manifolds[160].

Less transparent is the fact that the transition probability structure is the correct generalization in the present context of the Riemannian structure found on the projective Hilbert spaces. Recall the two main functions of the Riemannian metric $g$ on $\mathbb{P}\mathcal{H}$: first, it allowed to define a distance $d_g(p, q)$ between two states $p$ and $q$; second, it enabled to construct the Jordan product between two properties. In turn, $d_g(p, q)$ was used to define the crucial transition probability function $\Pr(p,q) := \cos^2(d_g(p,q))$. In the present context, however, this definition does not suffice, for it only allows to compute the transition probabilities between equivalent states—that is, between pure states belonging to the same leaf $\mathbb{P}\mathcal{H}_\alpha$. From the point of view of Riemannian geometry, the question of the transition probability between inequivalent states appears to be non-sensical, and it is so because the distance between two points belonging to different leaves of the pure state space cannot be defined. Yet, there are strong indications these transition probabilities should be defined. Indeed, there is the following alternative characterization of inequivalent pure states due to Hepp:

> Two pure states $\rho$ and $\sigma$ of a $C^*$-algebra $\mathcal{U}$ are inequivalent if and only if, for each representation $\pi(\mathcal{U})$ on a Hilbert space $\mathcal{H}$ containing unit vectors $\psi$ and $\varphi$ such that $\rho_\psi = \rho$ and $\rho_\varphi = \sigma$, one has $\langle \psi, \pi(A)\varphi \rangle = 0$ for all $A \in \mathcal{U}$.[161]

---

[157]Ibid., Definition I.2.6.2, p. 76. This notion was introduced for the first time by Landsman in "Poisson Spaces With a Transition Probability", p. 38. His definition also includes a linear subspace $\mathcal{U}_\mathbb{R}(\mathcal{P}) \subset \mathcal{C}_L^\infty(\mathcal{P}, \mathbb{R})$ which separates points and is closed under the Poisson bracket: $\{f, g\}(\iota_\alpha(q)) = \{\iota_\alpha^* f, \iota_\alpha^* g\}_\alpha(q)$, where $q \in S_\alpha$. I nonetheless find the inclusion of this subspace slightly unnatural at this point. This subspace $\mathcal{U}_\mathbb{R}(\mathcal{P})$ will only become important when defining the key notion of a Poisson space with transition probability (cf. Definition II.12 and footnote 164, page 222).

[158]Landsman, *Mathematical Topics Between Classical and Quantum Mechanics*, Definition I.2.7.1, pp. 80–81.

[159]This concept was introduced for the first time in 1937 by von Neumann in a series of lectures delivered at Pennsylvania State College. The manuscript was only published in 1981, after von Neumann's death (J. von Neumann. *Continuous Geometries with a Transition Probability*. Vol. 252. American Mathematical Society, 1981).

[160]This is the so-called "symplectic decomposition of a Poisson manifold" (see Landsman, op. cit., Theorem I.2.4.7, p. 71). For the definition of a Poisson manifold, see page 151.

In the light of this, one should extend the definition of the transition probabilities as follows:

$$\Pr(\rho,\sigma) = \begin{cases} \cos^2(d_{g_\alpha}(\rho,\sigma)) & \text{if } \rho,\sigma \in \mathbb{P}\mathcal{H}_\alpha \\ 0 & \text{if } \rho \nsim \sigma. \end{cases} \tag{II.25}$$

In this way, the pure state space of a $C^*$-algebra is equipped with a transition probability and becomes a symmetric transition probability space. In fact, as Landsman explains, Mielnik has shown that the boundary $\partial K$ of any compact convex set $K$ (such as $\mathcal{P}(\mathcal{U}) = \partial\mathcal{S}(\mathcal{U})$) may naturally be equipped with a transition probability, and it can be proven that Mielnik's transition probability coincides with the one just defined[162]. This shows that the function $\Pr$ is indeed an intrinsic object attached to the pure state space. Therefore, as long as states are concerned, one could attempt to ignore the Riemannian structure and place the transition probability structure as the central concept. The problem of course is to know whether the Jordan product can also be defined solely in terms of the transition probability stucture.

---

[161] K. Hepp. "Quantum Theory of Measurement and Macroscopic Observables". In: *Helvetica Physica Acta* 45 (1972), pp. 237–248, Lemma 1, p. 240 (cited in N. P. Landsman. "Between Classical and Quantum". In: *Philosophy of Physics (Handbook of the Philosophy of Science) 2 volume set.* Ed. by J. Butterfield and J. Earman. Vol. 1. Amsterdam: North-Holland Publishing Co., 2007, pp. 417–554. URL: http://arxiv.org/abs/quant-ph/0506082, p. 502).

[162] For the details, see idem, "Poisson Spaces With a Transition Probability", pp. 33-ff.

Now, recall Ashtekar and Schilling's geometric characterization of the algebra of properties (page 177). Associated to the symplectic structure $\omega$ was the set of functions $\mathcal{C}^\infty(S, \mathbb{R})_\omega$ preserving it. Similarly, to the Riemannian metric $g$ one associated the set $\mathcal{C}^\infty(S, \mathbb{R})_g$. Then, the algebra of properties was simply found to be

$$\mathcal{C}^\infty(S, \mathbb{R})_\mathcal{K} := \mathcal{C}^\infty(S, \mathbb{R})_\omega \cap \mathcal{C}^\infty(S, \mathbb{R})_g.$$

This idea may be immediately transposed to the general situation. One considers the function space $\mathcal{C}_{Prob}(\mathcal{P}, \mathbb{R})$ intrinsically related to a transition probability space and the function space $\mathcal{C}^\infty_{Pois}(\mathcal{P}, \mathbb{R})$ intrinsically associated to a Poisson space[163]. Then, for a space that is both a Poisson space and a transition probability space, one defines:

$$\mathcal{U}_\mathbb{R}(\mathcal{P}) = \mathcal{C}^\infty_{Pois}(\mathcal{P}, \mathbb{R}) \cap \mathcal{C}_{Prob}(\mathcal{P}) \tag{II.26}$$

At this point, it is not clear what the structure of this algebra of functions is. It nonetheless allows to define the key concept and state the two main theorems of Landsman's construction:

**Definition II.12.** A **Poisson space with a transition probability**[164] is a set that is both a transition probability space and a Poisson space and for which:

   i) $\mathcal{U}_\mathbb{R}(\mathcal{P})$ separates points,

   ii) $\mathcal{U}_\mathbb{R}(\mathcal{P})$ is closed under the Poisson bracket,

   iii) the Hamiltonian flow defined by each element of $\mathcal{U}_\mathbb{R}(\mathcal{P})$ preserves the transition
        probabilities (unitarity condition).

---

[163]These function spaces are defined as follows. $\mathcal{C}^\infty_{Pois}(\mathcal{P}, \mathbb{R})$ is the set of all $f \in \mathcal{C}(\mathcal{P}, \mathbb{R})$ such that their restrictions to any $S_\alpha$ is smooth: $\iota^*_\alpha f \in \mathcal{C}^\infty(S_\alpha, \mathbb{R})$. (Idem, *Mathematical Topics Between Classical and Quantum Mechanics*, Definition I.2.6.2.3, p. 76.)

On the other hand, the definition of $\mathcal{C}_{Prob}(\mathcal{P})$ is more involved. One considers first the functions $\mathrm{Pr}_\rho : \mathcal{P} \to \mathbb{R}$ such that $\mathrm{Pr}_\rho(\sigma) := \mathrm{Pr}(\rho, \sigma)$, and defines $\mathcal{C}^{00}_{Prob}(\mathcal{P})$ as the real vector space generated by these functions. Then $\mathcal{C}_{Prob}(\mathcal{P}) := \left( \overline{\mathcal{C}^{00}_{Prob}(\mathcal{P})} \right)^{**}$. The reason why this is the function space intrinsically associated to a transition probability space is not clear to me. This is however explicitly stated by Landsman on repeated occasions. (Ibid., Definition I.3.1.1, p. 84.)

[164]Since my notion of Poisson space differs from Landsman's, this definition is different from the one found in Landsman's book (Definition I.3.1.4, p. 86). However, my notion of Poisson space with a transition probability should coincide with Landsman's.

**Theorem II.7.** *The pure state space of a $C^*$-algebra is a Poisson space with transition probability*[165].

**Theorem II.8.** *Let $\mathcal{U}$ be a $C^*$-algebra, $\mathcal{U}_{\mathbb{R}}$ be the JLB-algebra of self-adjoint elements and $\mathcal{P}(\mathcal{U})$ be the space of pure states of $\mathcal{U}$. Then, we have the isomorphism of JLB-algebras*[166]

$$\mathcal{U}_{\mathbb{R}}\big(\mathcal{P}(\mathcal{U})\big) \simeq \mathcal{U}_{\mathbb{R}}.$$

Theorem II.7 recognizes the kind of spaces under which fall the pure state spaces of a $C^*$-algebra. Theorem II.8 is two-fold. First, it implicitly says that the algebra $\mathcal{U}_{\mathbb{R}}(\mathcal{P})$ intrinsically associated to a Poisson space with a transition probability may be endowed with a Jordan product $\bullet$ and a norm in such a way that $\mathcal{U}_{\mathbb{R}}(\mathcal{P})$ becomes a JLB-algebra. It turns out that this Jordan product is defined solely in terms of the transition probability, as was needed[167]. Moreover, the unitarity condition is a compatibility condition recognizing the fact that the two fundamental structures of the pure state spaces are not independent from each other. This is the analogue, at the level of states, of the Leibniz rule—which relates, at the level of properties, the otherwise independent Jordan and Lie structures[168].

Second, in the same way that the Gelfand-Naimark theorem shows that any JLB-algebra may be realized as a certain subalgebra of bounded self-adjoint operators on a Hilbert space $\mathcal{H}$, this theorem shows that any JLB-algebra may equally well be realized as a certain subalgebra of real-valued functions over some topological space. Thus, the association of non-commutativity to operators (II.20b, page 200) is by no means a necessary one. More importantly, it appears that any $C^*$-algebra, be it commutative or not, can be recovered from its space of pure states. This establishes, for the quantum kinematical arena, the *complete equivalence between the point of view of states and*

---

[165]Ibid., Theorem I.3.1.5, p. 86.

[166]Ibid., Theorem I.3.2.1 (combined with equations (3.2) and (3.6)), pp. 85–88.

[167]The explicit construction is found in ibid., Section I.3.3., pp. 88–90. It uses the fact that any element $F \in \mathcal{U}_{\mathbb{R}}(\mathcal{P})$ can be uniquely written as a linear combination of functions of the type $\mathrm{Pr}_\rho$ (spectral resolution). In turn, this allows to define the square of a property $F^2$ and subsequently the Jordan product by the formula $F \bullet G = \frac{1}{4}\big((F+G)^2 - (F-G)^2\big)$.

[168]Ibid., Section I.3.4., pp. 90–92. See also the definition of a Poisson algebra and the comment following it (page 148).

*the point of view of properties.* This is summarized in the following diagram[169] (Figure II.5), which replaces and generalizes the one emerging from the geometric program of Ashtekar, Schilling, Cirelli, etc (page 178).

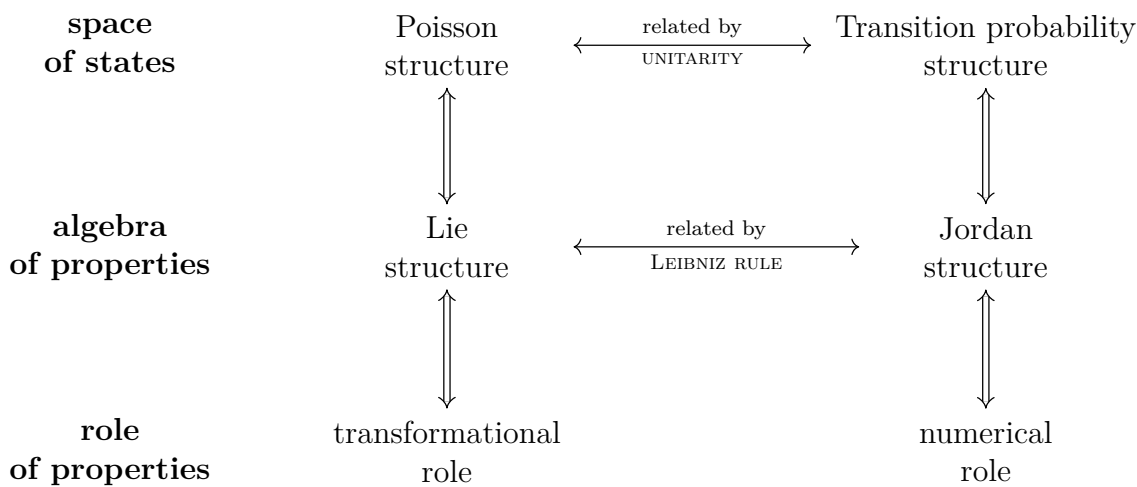

**Fig. II.5** – The interplay between the geometry of states
and the algebra of properties in Quantum Kinematics.

## II.3.4   New look into the Classical and characterization of the Quantum

As the last three subsections show, through the $C^*$-algebraic approach we reach a transparent understanding of the conceptual structure of Quantum Kinematics. The main three highlights were:

1. The realization that the quantum compatibility between the numerical and transformational role of properties could be characterized by the possibility of unifying the two real algebraic structures of properties into a single complex one (subsection II.3.1).

2. The unveiling of the close relation between states of a physical system and representations of the algebra of properties (subsection II.3.2).

---

[169]See also ibid., Table 1, p. 8.

3. The proof of the complete equivalence between the definition of physical systems based upon properties or based upon states, and the establishment of a precise dictionary between the geometric and algebraic structures (subsection II.3.3).

However, since Poisson algebras escape from the theory of $C^*$-algebras, we have focused on the discussion of Quantum Kinematics, somehow losing grasp on Classical Kinematics and forgetting the task of comparing the two arenas. Recall the drawback of $C^*$-algebras in regard to the Classical: the only Poisson algebras arising as the real part of a $C^*$-algebra are those whose Poisson bracket trivially vanishes (cf. Figure II.4, page 206). Hence, from their point of view, the Lie structure of classical properties is invisible. This phenomenon can also be understood geometrically. If one considers a commutative $C^*$-algebra $\mathcal{U}$, all its irreducible representations are necessarily one-dimensional. The decomposition

$$\mathcal{P}(\mathcal{U}) = \bigsqcup_{\alpha \in \widehat{\mathcal{U}}} \mathbb{P}\mathcal{H}_\alpha$$

of the pure state space into a disjoint union of symplectic manifolds still holds, but becomes now a trivial decomposition: each projective Hilbert space is reduced to a point and the decomposition simply says that the space of pure states is the disjoint union of its points, considered as symplectic 0-dimensional manifolds. Thus, what remains unexplained from the point of view of $C^*$-algebras is the fact that, beyond the symplectic structure of each leaf $\mathbb{P}\mathcal{H}_\alpha$, in Classical Mechanics there is also a "transversal" symplectic structure tying together the different leaves.

Yet, $C^*$-algebras do perceive the Jordan product of Classical Mechanics and the insights from Landsman's construction may be used in order to adopt a new look at the geometric origin of this structure. Recall: in the geometrical formulation of Classical Kinematics, it had appeared that the presence of a Jordan product for classical properties—which in this case is simply point-wise multiplication—was not mirrored by the existence of some particular structure on the classical space of states (cf. Table II.1, page 186). This lack of a structure on the classical space of states which would be the analogue of the Riemannian metric found on the quantum space was one of the main drawbacks of the whole geometric approach to Kinematics (as presented in section II.2). But we now know that the quantum Jordan product may equally be thought as stemming from a transition probability. It is then natural to attempt to see the classical Jordan product as arising from this same structure.

In the case of a commutative $C^*$-algebra, Equation II.25 (page 221) reduces to:

$$\Pr(\rho, \sigma) = \delta_{\rho,\sigma} = \begin{cases} 1 & \text{if } \rho = \sigma \\ 0 & \text{if } \rho \neq \sigma \end{cases} \tag{II.27}$$

and indeed the Jordan product defined by this transition probability coincides with point-wise multiplication[170]. This is a completely trivial structure, adding no further information to a topological space[171]. Because of this, it had been (rightly) disregarded in Classical Kinematics. Nonetheless, the presence of this trivial transition probability in the classical kinematical arena becomes interesting when compared to the Quantum. One realizes that, although $C^*$-algebras are unable to cover both quantum and classical systems, the type of spaces emerging from this approach—that is, Poisson spaces with a transition probability—do encompass classical and quantum spaces of states: both pure state spaces of non-commutative $C^*$-algebras and symplectic manifolds equipped with the trivial transition probability (II.27) satisfy the axioms of Definition II.12 (page 222).

Therefore, it becomes natural to compare the two kinematical arenas in the common geometric language of Poisson spaces with a transition probability. As it will turn out, this language manages to capture with unmatched clarity the conceptual difference between Classical and Quantum Kinematics. The key lies in comparing the way in which the two geometrical structures interact with each other. In the quantum case, the unitarity condition imposes a very strong constraint to the Poisson structure: given the space $\mathbb{P}\mathcal{H}_\alpha$ and the transition probabilities (II.25, page 221), the requirement of unitarity uniquely determines the symplectic structure[172]. In turn, in the classical

---

[170] Cf. the comment following Proposition 3 in idem, "Poisson Spaces With a Transition Probability", p. 44.

[171] In the sense that any topological space may be seen as a transition probability space equipped with the trivial transition probability.

[172] Idem, *Mathematical Topics Between Classical and Quantum Mechanics*, Theorem I.3.8.2, p. 103. Therein, the Poisson bracket is determined *up to a multiplicative constant*. However, if one further imposes that the associator rule of the Jordan-Lie algebra be given by $(f \bullet g) \bullet h - f \bullet (g \bullet h) = \{\{f, h\}, g\}$

case, the exact opposite happens: since the transition probabilities are trivial, *any* symplectic structure whatsoever will automatically be unitary. Thus, although unitarity is present in both kinematical arenas, the effect of this compatibility condition is radically different in the two situations. Whereas in the Quantum unitarity closely ties together the two fundamental geometric structures of the space of states, in the Classical unitarity imposes no restriction and the Poisson structure remains completely independent from the transition probability structure. Either the two geometric structures go hand in hand, or they do not discuss with each other[173].

This last point can be rendered more precise if one considers the following two equivalence relations on the space of states:

– Equivalence defined by the Poisson structure: two states $\rho$ and $\sigma$ are said to be *transformationally equivalent* if they can be connected by a piecewise smooth Hamiltonian curve. We denote this equivalence by $\rho \underset{T}{\sim} \sigma$ and the equivalence classes under this relation are called the *symplectic leaves* of the space of states[174].

– Equivalence defined by the transition probability structure: two states $\rho$ and $\sigma$ are said to be *numerically equivalent* if they belong to the same *sector*[175]. We denote this equivalence by $\rho \underset{N}{\sim} \sigma$ and the equivalence classes under this relation are by definition the *sectors* of the space of states.

These two different equivalence relations may be seen as two different notions of *connectedness* of the space of states. 'Transformational equivalence' is connectedness from the point of view of properties-as-transformations: two states $\rho$ and $\sigma$

---

(instead of $(f \bullet g) \bullet h - f \bullet (g \bullet h) = \kappa\{\{f, h\}, g\}$ as Landsman does), then the determination becomes unique.

[173] As Landsman points out, this feature of the Quantum could have been already noticed in section II.2. Indeed, given the natural symplectic form on $\mathbb{P}\mathcal{H}$, the Riemannian metric $g$ is completely fixed (up to a constant) by the demand that it be invariant under the Hamiltonian flows generated by the functions $\widetilde{F} \in \mathcal{C}^\infty(\mathbb{P}\mathcal{H}, \mathbb{R})$. (Idem, "Poisson Spaces With a Transition Probability", p. 47)

[174] The terminological decision of calling this equivalence relation "transformational" is mine. For the rest, cf. idem, *Mathematical Topics Between Classical and Quantum Mechanics*, definition I.2.4.3., p. 70.

[175] Given a transition probability space $(\mathcal{P}, \mathrm{Pr})$, two subsets $\mathcal{S}_1$ and $\mathcal{S}_2$ are said to be *orthogonal* if, for any $\rho \in \mathcal{S}_1$ and any $\sigma \in \mathcal{S}_2$, $\mathrm{Pr}(\rho, \sigma) = 0$. A subset $\mathcal{S} \subset \mathcal{P}$ is said to be a *component* if $\mathcal{S}$ and $\mathcal{P} \setminus \mathcal{S}$ are orthogonal. Finally, a *sector* is a component which does not have any non-trivial components (cf. ibid., Definition I.2.7.2., p. 80).

are transformationally inequivalent ($\rho \underset{T}{\nsim} \sigma$) if and only if it is impossible to connect them by a physical transformation—that is, if and only if it is impossible to find $g_f \in \mathrm{Aut}(\mathcal{P})$ generated by a physical property $f$ such that $g_f(\rho) = \sigma$. In other words, a symplectic leaf is by definition a transformationally connected component of the space of states. In the same way, 'numerical equivalence' is connectedness from the point of view of transitions: two states are numerically inequivalent ($\rho \underset{N}{\nsim} \sigma$) if and only if it is impossible to find a collection of intermediate states $\chi_1, \ldots, \chi_n$ such that the chain of transitions or 'transitional path' $\rho \to \chi_1 \to \ldots \to \chi_n \to \sigma$ has a non-vanishing probability—that is, if and only if for any choice $\chi_1, \ldots, \chi_n \in \mathcal{P}$, one has $\mathrm{Pr}(\rho, \chi_1)\mathrm{Pr}(\chi_1, \chi_2) \ldots \mathrm{Pr}(\chi_{n-1}, \chi_n)\mathrm{Pr}(\chi_n, \sigma) = 0$. Thus, a sector is a transitionally connected component of the space of states.

Therefore, each of the two geometric structures of the space of states produces a certain 'image' of this space. In Classical Kinematics, where one considers as space of states $\mathcal{P}_{cl}$ a symplectic manifold with transition probabilities $\mathrm{Pr}(\sigma, \rho) = \delta_{\sigma, \rho}$, the two images are at odds from each other: from the point of view of the Poisson structure, the space of states is completely connected (any two states are transformationally equivalent), whereas from the point of view of the transition probability structure the space of states is completely disconnected (no two different states are numerically equivalent). In other words, we have

$$* = (\mathcal{P}_{cl}/\underset{T}{\sim}) \neq (\mathcal{P}_{cl}/\underset{N}{\sim}) = \mathcal{P}_{cl}.$$

On the other hand, in Quantum Kinematics the hand-in-hand of the two geometric structures is captured in the fact the two images coincide. That is, we have the following result[176]:

---

**Quantum compatibility of Poisson and transition probability structures.**
On the quantum space of states $\mathcal{P}_{qu}$, the notions of transformational equivalence and numerical equivalence coincide: $\rho \underset{T}{\sim} \sigma \Longleftrightarrow \rho \underset{N}{\sim} \sigma$. In other words,

$$(\mathcal{P}_{qu}/\underset{T}{\sim}) = (\mathcal{P}_{qu}/\underset{N}{\sim})$$

---

[176]That this holds for the pure state space of any $C^*$-algebra should be clear from Theorem II.3 (page 213).

In the light of the whole analysis we have undertaken in this chapter, the existence of such a quantum compatibility condition should not come as a surprise. After all, this statement, expressed in the intrinsic language of the space of states as Poisson space with a transition probability, is the analogue of the quantum unification of the Jordan and Lie structure (page 205), or of the quantum interplay between the numerical and transformational roles of properties (page 184).

The surprising fact, that was perhaps difficult to foresee in the previous formulations, is that this compatibility between properties-as-transformations and properties-as-quantities constitutes precisely the core of the difference between Classical Mechanics and Quantum Mechanics. Indeed, given a Poisson space with a transition probability $\mathcal{P}$, Landsman has provided the following axiomatic characterization of when a space is a quantum space of states[177]:

**Theorem II.9** (Characterization of the quantum space of states)**.** *A uniform Poisson space with a transition probability $\mathcal{P}$ is the pure state space of a finite-dimensional $C^*$-algebra if:*

    *QM 1) The sectors and the symplectic leaves of $\mathcal{P}$ coincide,*

    *QM 2) $\mathcal{P}$ has the two-sphere property.[178]*

Axiom QM 2) encodes the quantum superposition principle (cf. subsection II.2.3, page 192); axiom QM 1) encodes the quantum compatibility between properties-as-quantities and properties-as-transformations. These two may be seen as the real fundamental differences between the Classical and the Quantum. The former has been stressed since the birth of Quantum Mechanics (cf. Dirac's quote on page 166). The latter seems to have been the blind spot on the conceptual analysis of Quantum Kinematics.

---

[177]Ibid., Theorem I.3.9.2., p. 105 and Corollary I.3.9.3., p. 106.

[178]Two more technical axioms are necessary in the case of infinite-dimensional $C^*$-algebras.

## II.4   Conclusion

The key remark that launched our analysis of the Classical and the Quantum Kinematical arenas was the realization of the two-fold role of physical properties with respect to states. To distinguish them, I introduced the terminology of properties-as-quantities and properties-as-transformations. From that moment on, the conceptual discussion of Kinematics became subject to the requirement of building *articulations* for three couples of fundamental concepts: Classical/Quantum, State/Property and Quantity/Transformation[179]. Out of the three, the first pair is the most slippery one and the main goal concerning it has been simply to get a hold on it: to find all-embracing languages in which it is possible to cast both theories and *formulate their distinction.* In other words, regarding the Classical/Quantum couple, we have just been searching for perspectives from which to take a *static* picture with the two poles clearly distinguished. Thus, at this stage of the investigation, we have not yet been concerned with the possible transitions between these poles—"quantization" and, in the opposite direction, "classicalisation" (as Brody and Hughston propose to name it[180]). On the contrary, regarding the other two couples, the point of interest has lied in their *dynamics.* For the State/Property couple, it has been the movement of oscillation—Is it possible to freely transit from states to properties, and from properties to states, or is there some kind of priority of one pole over the other?—and, as we

---

[179]Here, the word "articulation" is meant in the precise sense found in the work of the French philosopher Gilles Châtelet. He says:

> Articulation does not claim to reconcile two contrasts $A$ and $-A$; it gets round their confrontation. [...T]o articulate is always to allow oneself a new envelopment, to discover a material that is more *ductile* than that of the sides. An articulation does not link together two contents or two separate segments which preexisted it; it grasps the very emergence of these sides from an indifference point.
>
> In its fork, the articulation carries the product and productivity. It always participates in the liberation of a dimension. [...] It is indeed the articulation that makes it possible to situate oneself beyond all opposition, and therefore to overcome all opposition. For it is a matter neither of saving the old dualisms (subject/object, form/content, etc) nor of letting one self be submerged in the confusion of some 'primordial soup'. A suitable articulation no doubt allows a positive integration of all the forces imprisoned by contrasts, but it is always accompanied by the birth of a singularity.

(G. Châtelet. *Figuring Space: Philosophy, Mathematics and Physics.* Trans. by R. Shore and M. Zagha. Dordrecht, The Netherlands: Springer Science & Business Media, 2000, p. 94) have seen, this question is intimately linked with the articulation of the couple Geometry/Algebra in Mathematics. For the Quantity/Transformation, the question has been to understand the precise interplay between these two fundamental roles of physical properties. Therefore, from the State/Property and Quantity/Transformation couples sprang several questions tending to order the conceptual analysis of the Classical and Quantum kinematical arenas (see Figure II.6 below).

[180]Brody and Hughston, op. cit., p. 2.

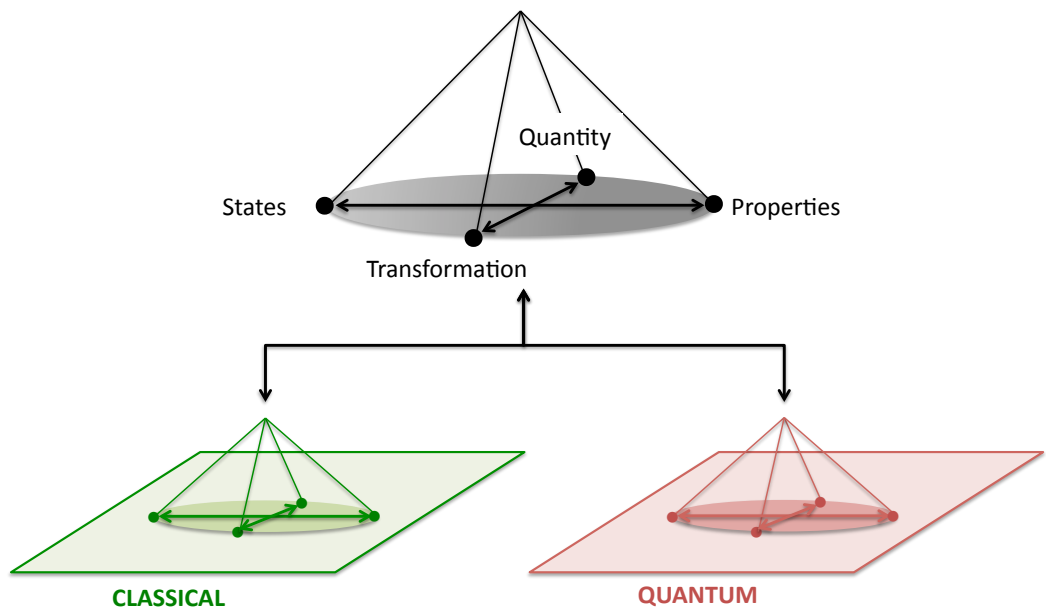

**Fig. II.6** – **The conceptual tensions of the analysis.** The dynamical articulations of the couples State/Property and Quantity/Transformation lead to a static picture of the Classical/Quantum couple.

In the last three sections, we have performed this analysis from the perspective of three different mathematical formulations and it seems we have finally reached a satisfactory global picture of the situation, which helps us in clarifying the conceptual difference between the Classical and the Quantum. It is the following (cf. also Figure II.7, page 234).

Common to both Kinematical arenas is the fact that the full description of physical properties is the conjunction of properties-as-quantities and properties-as-transformations. At the algebraic level of properties, this two-fold role gets translated into the existence of two real structures: a Jordan product which governs the numerical role, and a Lie product which governs the transformational role. Accordingly, the language of Jordan-Lie algebras is the common algebraic language which covers both Classical and Quantum Kinematics. The geometrical level of states mirrors the algebraic level in every respect: herein, the two-fold role manifests itself by the presence of two geometric structures—a probability structure and a Poisson structure (which respectively stem from the Jordan and Lie product and from which the Jordan and Lie product can be defined)—and the common geometric language is that of uniform Poisson spaces with a transition probability. One often restricts attention to the simpler case where the Poisson space has only one symplectic leaf. Then, the Poisson structure is equivalent to a symplectic 2-form and the non-trivial transition probability structure of the Quantum may be perceived as arising from a Riemannian metric (the transition probability is the distance between two points). In this way, one recovers the geometric formulation of Classical and Quantum Kinematics in terms of symplectic manifolds and Hermitian symmetric spaces.

With the use of either Jordan-Lie algebras or Poisson spaces with a transition probability, one may sharply characterize the difference between the two Kinematics. At the algebraic level, the difference lies in the associativity/non-associativity of the Jordan product, whereas at the geometric level it lies in the triviality/non-triviality of the probability structure. In other words, from the restricted point of view of the symplectic/Lie structure, the Classical and the Quantum are indistinguishable. At the conceptual level, this means that the real difference between Classical and Quantum lies in the numerical role of properties.

The numerical role of a property $f$ vis-à-vis a given state $\sigma$ is captured by the string of numbers

$$\mathcal{N}_\sigma^f = \{f(\sigma), f^2(\sigma), f^3(\sigma), \ldots\}$$

found by repeated use of the Jordan product. In Classical Mechanics, the Jordan product is point-wise multiplication so, *by definition*, we have $f^n(\sigma) := \big(f(\sigma)\big)^n$. Thus, we see that the Jordan product of classical properties-as-quantities is precisely defined in such a way that there is no more information in the data of the whole $\mathcal{N}_\sigma^f$ than in the first term of the string. Accordingly, one may reduce the numerical role of classical properties to the datum of the sole number $f(\sigma)$. This is to be contrasted with the situation in Quantum Kinematics. Therein, the non-associativity of the Jordan product (or equivalently the non-triviality of the transition probability structure) entails that there is no general relation between the numbers of $\mathcal{N}_\sigma^f$ and one cannot reduce the information contained in $\mathcal{N}_\sigma^f$ to the data of some of the numbers in the string. In this sense, the complete description of a quantum property-as-quantity is achieved by a complex, multi-layered structure, whereas a classical property-as-quantity is single-layered. But the existence of multiple numerical layers in the numerical role of quantum properties simply encodes the statistical nature of Quantum Mechanics. Therefore, the statement that "the real difference between the Classical and the Quantum lies on the numerical side of properties" appears as a very natural remark.

We still need to describe the articulation between the two roles of properties. There is a first compatibility condition which holds in the two kinematical arenas. At the geometrical level, this is captured by unitarity: the Hamiltonian flow of any physical property preserves the transition probabilities. At the algebraic level, the kinematical compatibility becomes the Leibniz rule: properties-as-transformations act as derivations on properties-as-quantities. At the conceptual level, this simply means that the transformational role of properties respects their numerical role. On top of this, Quantum Kinematics exhibits a second compatibility condition which ensures the complete consistency between the two roles of physical properties. Algebraically, this is seen in the unification of the two real structures into a single complex one (which in turn allows to reformulate Quantum Kinematics in terms of $C^*$-algebras instead of non-associative JL-algebras); geometrically, it is expressed in the coincidence of the natural foliations of the pure state space produced by the two geometric structures. Conceptually, this highlights the fact that, in Quantum Kinematics, the Quantity must describe the Transformation (e.g. the indeterminacy of the property-as-quantity $\Delta f(\sigma)$ describes the change of the state by the property-as-transformation).

This may be turned around: *given the two-fold role of properties in Kinematics, the demand that the two roles be consistent with each other may be seen as the <u>defining trait of the Quantum</u>.* This Quantum compatibility condition forces the numerical role to be multi-layered, the Jordan product to be non-associative and the transition probability to be non-trivial.

Figure II.7 below attempts to summarize the situation we have reached.

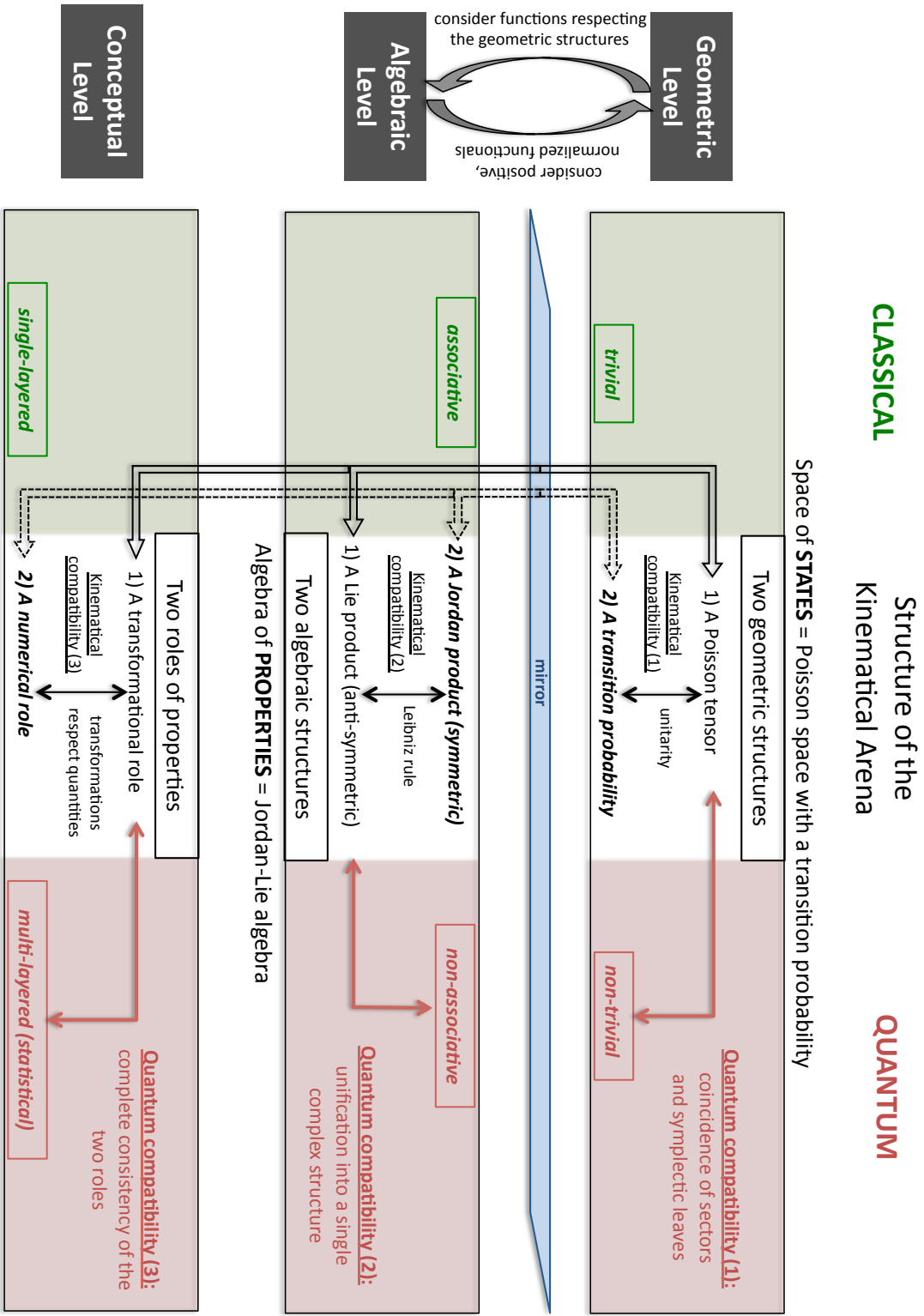

Fig. II.7 – The structure of the kinematical arenas.

# Chapter III

# Constructing the Mathematical Description of a Physical System

The preceding chapter was devoted to the study, from several perspectives, of the fundamental mathematical structures underlying the formalisms of Classical and Quantum Kinematics. But these abstract mathematical structures—symplectic manifolds, Poisson algebras, Hilbert spaces, $C^*$-algebras, etc.—furnish only the general theoretical tools used to describe the Kinematics of any given physical system. They constitute, so to speak, the bare canvas supporting the paintings of Classical and Quantum systems. Just as the artist, who first needs to carefully study the raw material he has decided to work with, and to become familiar with the constraints and possibilities it offers, so we had to get acquainted with these homogeneous Kinematical arenas. Accordingly, our aim in Chapter II was to understand at this very broad level the subtle interplay between the two major concepts of Kinematics: states and properties. Now, the time has come to embark in the actual process of constructing the mathematical description of a physical system. This means starting our "chase for individuation" and considering different techniques used in mathematical physics to break the homogeneity of the Classical and Quantum Kinematical arenas in order to introduce discernibility into the picture.

# III.1    The general strategy for introducing discernibility

So far, the problem has been presented from the perspective of states: given a connected symplectic manifold $S$ (respectively, a projective Hilbert space $\mathbb{PH}$), the action $Aut(S) \circlearrowleft S$ (resp. $Aut(\mathbb{PH}) \circlearrowleft \mathbb{PH}$) is transitive. This entails that, considered as elements of abstract mathematical structures, the points of the state space are qualitatively indiscernible individuals. But, if we endorse the descriptive perspective on the mathematical description of a physical system (Chapter I, page 15), we expect all the physical information to be encoded in the mathematical structure itself. In particular, we expect the different states of the system to be individuated without the need for an arbitrary coordinate frame. Therefore, the consideration of bare symplectic manifolds or projective Hilbert spaces does not suffice, and we need to look for ways of *enriching* these initial structures, thus endowing them with enough descriptive power to individuate each point of the state space. Now—it is important to remark this—the problem could have been equally well analyzed from the point of view of properties: if the set of all properties of a physical system is to be described by a mathematical structure $\mathcal{U}_{\mathbb{R}}$, and if one expects to be able to talk about *the* representative $f \in \mathcal{U}_{\mathbb{R}}$ of *the* physical property $\mathtt{f}$, then there must be an intrinsic structural way of individuating $f$ among the various elements of $\mathcal{U}_{\mathbb{R}}$.

Of course, the problem of individuating states and the problem of individuating properties are just two sides of the same problem. If one prefers to develop the theory from the point of view of states, the starting point is the abstract space of states $\mathcal{P}$ and the set of properties is then conceived as a particular set of functions on this space: $\mathcal{U}_{\mathbb{R}} \simeq \mathcal{C}^{\infty}(\mathcal{P}, \mathbb{R})_{\mathcal{K}}$. Two states $x$ and $y$ are qualitatively indiscernible (denoted by $x \sim y$) if they are related by an automorphism of $\mathcal{P}$, and one may use this to define indiscernible properties by the following requirement:

$$f \sim g \iff \forall x \in \mathcal{P}, \exists \phi \in Aut(\mathcal{P}) \text{ such that } g(x) = f(\phi(x)). \tag{III.1}$$

Conversely, if one prefers to develop the theory from the point of view of properties,

the starting point is the abstract algebra of properties $\mathcal{U}_\mathbb{R}$ and states are conceived as a particular kind of linear functionals on this algebra. In this case, two properties $f$ and $g$ are qualitatively indiscernible (denoted by $f \approx g$) if they are related by an automorphism of $\mathcal{U}_\mathbb{R}$, and indiscernible states are defined by the analogue of (III.1):

$$x \approx y \iff \forall f \in \mathcal{U}_\mathbb{R}, \exists \Phi \in Aut(\mathcal{U}_\mathbb{R}) \text{ such that } y(f) = x(\Phi(g)). \qquad \text{(III.2)}$$

These two approaches to discernibility coincide, in the sense that $x \sim y \Leftrightarrow x \approx y$ and $f \sim g \Leftrightarrow f \approx g$. Indeed, any automorphism of the space of states $\mathcal{P}$ induces an automorphism of the algebra of properties $\mathcal{U}_\mathbb{R}$, and vice-versa[1].

(III.1) and (III.2) show that the ability to individuate all states is equivalent to the ability to individuate all properties: if it is possible to individuate any state, (III.1) becomes $f \sim g \Leftrightarrow f = g$. However, one must be careful with the fact that the introduction of some degree of individuation within the abstract structure of properties does not necessarily entail an introduction of some degree of individuation within the space of states. In other words, the homogeneity of the space of states does not imply the homogeneity of the algebra of properties. Indeed, there exist qualitatively discernible properties even when any two states are qualitatively indiscernible (to see this, it suffices to consider two properties $f, g \in \mathcal{U}_\mathbb{R}$ whose spectra do not coincide)[2].

---

[1] More precisely, on the classical side one has a canonical isomorphism between $Aut(S)$ and $Aut(\mathcal{C}^\infty(S, \mathbb{R}))$: given any morphism of Poisson algebras $L_{alg} : \mathcal{C}^\infty(S_1, \mathbb{R}) \to \mathcal{C}^\infty(S_2, \mathbb{R})$, there exists a unique morphism of symplectic manifolds $L_{geo} : S_2 \to S_1$ such that $L_{alg} = L_{geo}^*$ (see N. P. Landsman. *Mathematical Topics Between Classical and Quantum Mechanics*. New York: Springer, 1998, corollary I.2.6.5, p. 77).

On the quantum side, one has $Aut(\mathcal{B}_\mathbb{R}(\mathcal{H})) \subset Aut(\mathbb{P}\mathcal{H})$. The difference between automorphisms of the quantum space of states and automorphisms of the quantum algebra of properties lies in the possibility of considering anti-unitary operators on $\mathcal{H}$ (which are anti-linear and hence do not belong to the $C^*$-algebra of bounded *linear* operators). Put differently, $Aut(\mathbb{P}\mathcal{H})$ is canonically isomorphic to the group of automorphisms *and anti-automorphisms* of $\mathcal{B}_\mathbb{R}(\mathcal{H})$. Indeed, as we will see later in greater detail, Wigner proved that any automorphism of $\mathbb{P}\mathcal{H}$ is induced by a unitary or anti-unitary operator on $\mathcal{H}$. On the other hand, Kaplansky proved in 1952 that, for von Neumann algebras which are type I factors (as is the case of $\mathcal{B}(\mathcal{H})$), all *-automorphisms are inner: for any $\alpha \in Aut(\mathcal{B}(\mathcal{H}))$, there exists a unitary operator $U$ such that $\alpha(A) = UAU^*$, for any $A \in \mathcal{B}(\mathcal{H})$. In other words, $Aut(\mathcal{B}(\mathcal{H})) \simeq U(\mathcal{H})/U(1)$ (I. Kaplansky. "Algebras of Type I". in: *Annals of Mathematics* 56.3 (1952), pp. 460–472, Theorem 3, p. 470).

[2] According to equation (III.1), two properties $f$ and $g$ are qualitatively indiscernible if there exists an automorphism $\phi$ of the space of states such that $f$ is the pull-back of $g$ by $\phi$. But the pull-back of a function has the same spectrum as the initial function. Hence, properties whose spectra do not coincide are necessarily discernible.

This need to go beyond the Kinematical arena studied in Chapter II in order to describe a physical system can therefore be perceived from both the geometric and algebraic perspectives. And indeed, one finds in the literature of both fields statements where the problem is touched upon. A clear example of this is Brody and Hughston's article *"Geometric Quantum Mechanics"*, where the two authors write:

> The specification of a physical system implies further geometrical structure on the state space [than the data of a projective Hilbert space]. Indeed, the point of view we suggest is that *all* the relevant physical details of a quantum system can be represented by additional projective geometrical features.[3]

At the other side of the spectrum, there is for instance the abstract algebraic work of Irving Segal:

> The set of all self-adjoint elements of an *abstract $C^*$*-algebra forms then a physical system [...].
>
> The complete description of a physical system involves however not only the statement of the mathematical character of the algebra of bounded observables, *but also a labelling of the observables, a kind of physical-mathematical dictionary.* This is clearly visible e.g. in the fact that in elementary quantum mechanics it is assumed that the bounded observables consist of all bounded hermitian operators on a countably-dimensional Hilbert space, irrespective of the number of degrees of freedom of the system.
>
> Now there is evidently no mathematical labelling scheme that will be applicable to a perfectly general $C^*$-algebra of observables. However, *the physically relevant $C^*$-algebras all involve implicitly or explicitly a labelling scheme whose mathematical structure is of essential importance* in the theory. [...] The treatment of these labelling matters involves additional elements of mathematical structure [...].[4]

---

[3] D. C. Brody and L. P. Hughston. "Geometric Quantum Mechanics". In: *Journal of geometry and physics* 38.1 (2001), pp. 19–53. URL: http://arxiv.org/abs/quant-ph/9906086, p. 25, authors' emphasis. Their insistence on representing "all the relevant physical details" can also be seen as a perfect illustration of what I have called the 'descriptive perspective'.

[4] I. E. Segal. "Mathematical Problems of Relativistic Physics". In: *Proceedings of the Summer Conference, Boulder, Colorado*. Ed. by M. Kac. American Mathematical Society, 1960, pp. 8–9, my emphasis.

The latter passage seems particularly enlightening to understand our goal. The construction of the physical-mathematical dictionary, the making explicit of the "labelling scheme" which allows to identify the physical properties and/or states is precisely what we are after. But the important point stressed here by Segal is that the labelling of properties/states cannot be just the subjective and ethereal move of fixing, once and for all, a *choice* of names for the elements of the mathematical structures. Rather, if the labelling is to have some definite, unambiguous *meaning*, then it must inescapably be governed by some *additional mathematical structures* yet to be considered[5].

Let me shortly comment a simple and concrete situation to illustrate the point: the mathematical description of a non-relativistic quantum particle in one-dimensional space. We consider two operators $J$ and $K$ on a Hilbert space $\mathcal{H}$ of which we know only that they obey the algebraic relation

$$i[J, K] = \mathbb{I},$$

and which we of course intend to represent position and linear momentum. Then, the question is: which operator should we pick to represent linear momentum and which should we pick to represent position? Finding an answer would amount to finding a [...] we have been allowing observables such as $S_u$ to 'represent' spin in the $u$-direction, but what precisely *is* this relationship of 'representation'? How may the connection between formalism and physical fact be made, or understood? [...]

---

[5]The idea that the meaning of a labelling scheme must be printed in the formalism itself is also highlighted in Michael Dickson's review of the philosophical problems arising in non-relativistic Quantum Mechanics. Therein, he devotes one section to discuss "the issue of how the formalism of quantum theory gets empirical content" and poses the following question:

It is crucial to understand that the issue here is not about how to engineer a spin-measuring device, for example. Rather, **it is about what it *means* to 'have' spin-up in the $u$-direction (for example) and how this meaning is captured in the formalism**.

[...] What, in other words, is the relationship between the elements of the mathematical formalism that we have described and physical matters of fact? And finally, **why do we pick one map (POVM) rather than another to represent some given physical quantity?**

(M. Dickson. "Non-relativistic Quantum Mechanics". In: *Philosophy of Physics (Handbook of the Philosophy of Science) 2 volume set*. Ed. by J. Butterfield and J. Earman. Vol. 1. Amsterdam: North-Holland Publishing Co., 2007, pp. 275–415. URL: http://philsci-archive.pitt.edu/3321/, pp. 327-328. The italics are from the author, the bold type is mine.)

labelling scheme. Now, since position $Q$ and momentum $P$ should verify

$$i[P, Q] = \mathbb{I}$$

a plausible choice would be $P = J$ (meaning: "the physical property 'momentum' is represented by the operator $J$") and $Q = K$. But, with the information at our disposal, this choice is completely arbitrary: one could equally well decide at this stage that $P = -K$ and $Q = J$, or that $P = J + K$ and $Q = K$, etc. This is the type of situation to be avoided if one adheres—as Segal, Brody and Hughston seem to do in the above quotes—to the descriptive perspective on mathematical definitions of physical systems. The existence of several different choices on the physical interpretation of the abstract mathematical elements is felt as the indication that the description is still incomplete: the definition of position and momentum must involve further mathematical structures than the sole commutation relation, and the goal becomes to reveal what these structures are.

All attempts to elucidate the labelling scheme at work in Classical and Quantum Kinematics follow the same strategy: roughly, one introduces into the kinematical arena new abstract mathematical structures which clothe the bare initial kinematical structures, and thus partially break the homogeneity. To be more precise, we need to first distinguish between three sorts of structures:

– First, there are what I call the *fundamental kinematical structures*. These are the abstract mathematical structures that constitute the *starting point* in the description of the Classical and Quantum kinematical arenas. For example, from the algebraic perspective of properties, these could be an abstract Poisson algebra for Classical Kinematics and an abstract non-commutative $C^*$-algebra for Quantum Kinematics.

– Moreover, there are what I call *internal structures*. These are structures explicitly *built from*—and therefore, by definition, *intrinsically related to*—the fundamental kinematical structures. In the previous chapter, we encountered many of those: given an abstract symplectic manifold $(S, \omega)$, it is for instance possible to construct the Poisson algebra $\left(\mathcal{C}^\infty(S, \mathbb{R}), \cdot, \{\cdot, \cdot\}\right)$, the group $Aut(S)$, the Lie algebra $\Gamma(TS)_H$ of Hamiltonian vector fields, etc.

– Finally, there are what I call *external! structures*. Contrary to the previous case, these are abstract structures with *a priori no relation* to the fundamental kinematical structures.

Now, because of the difference in the type of relation they bear to the kinematical arena, internal and external structures will have quite different roles in the mathematics of Kinematics. The consideration of structures of the first sort allows to gain insight about the constitution of the kinematical arena[6]. The whole previous chapter can be seen as an example of this, but a simpler example is the fact that the group of automorphisms captures the degree of discernibility within an abstract structure. Therefore, the typical questions arising in this context will involve the amount of information about the initial structure encoded in a second structure. The question may take the form of a *reconstruction problem*—e.g., is it possible to recover the symplectic manifold $(S, \omega)$ from the Poisson algebra $\left(\mathcal{C}^\infty(S, \mathbb{R}), \cdot, \{\cdot, \cdot\}\right)$?—or, when the reconstruction is impossible, of a *loss of information problem*—e.g., which information of the symplectic manifold is invisible when considering its group of automorphisms? Crucial for our purposes is the remark that internal structures are unable to break the homogeneity of the kinematical arena, since they do not introduce any new information into the picture. At best, they allow to capture an intrinsic trait of the fundamental structure which could have been spotted by other means (for example, the degree of discernibility within $(S, \omega)$ could have been studied using the Poisson algebra of functions and definition (III.2, page 237), instead of using the action of the group of automorphisms).

Therefore, the addition of further elements of mathematical structure will necessarily involve the consideration of external structures. Since these have a priori no relation to the fundamental kinematical structures, *the additional information will lie precisely in the specification of a relation between an internal structure and an external one.* If we denote by $\mathcal{K}$ the fundamental kinematical structure (a symplectic manifold, a Jordan-Lie algebra, etc.) and by $\mathcal{E}$ a certain external structure, the new abstract data that we wish to use to describe a physical system will then be a triple $(\mathcal{K}, \mathcal{E}, \rho)$ where $\rho$ establishes a relation between $\mathcal{K}$ and $\mathcal{E}$.

---

[6]This should resonate with the words of Weyl: "[...] what has indeed become a guiding principle in modern mathematics is this lesson: *Whenever you have to do with a structure-endowed entity* $\Sigma$ *try to determine its group of automorphisms*, the group of those element-wise transformations which leave all structural relations undisturbed. You can expect to gain a deep insight into the constitution of $\Sigma$ in this way" (H. Weyl. *Symmetry.* Princeton: Princeton University Press, 1952 (reprinted in 1989), p. 144, author's emphasis).

At this point, it just remains to clarify this key notion of *relation*. This is most naturally achieved by recasting the whole discussion in the language of category theory. The general strategy appears then to be as follows:

i) Start with two categories $\mathcal{Ext}$ and $\mathcal{Kin}$. An object $\mathcal{E}$ of $\mathcal{Ext}$ is an external abstract structure, and an object $\mathcal{K}$ of $\mathcal{Kin}$ is the fundamental kinematical structure. For example, $\mathcal{Kin}$ would be the category $\mathcal{PoissMan}$ of Poisson manifolds or the category $\mathcal{JL}$ of Jordan-Lie algebras, and $\mathcal{Ext}$ could be $\mathcal{Grp}$ (the category of groups).

ii) Internal structures correspond to the images of the objects of $\mathcal{Kin}$ by various functors $\mathcal{F} : \mathcal{Kin} \longrightarrow \mathcal{D}$. Precisely, it would be the functoriality of the assignment $\mathcal{F}$ that would convey a definite meaning to the notion of an "intrinsic construction". In this way, the problems about the possible reconstructions or losses of information would translate into questions about whether the functor $\mathcal{F}$ admits an adjoint or, better, whether it establishes an equivalence of categories.

iii) The choice of a relation between a fundamental kinematical structure $\mathcal{K}$ and an external structure $\mathcal{E}$ is now given by the choice of two functors $\mathcal{F}_1 : \mathcal{Kin} \longrightarrow \mathcal{D}$ and $\mathcal{F}_2 : \mathcal{Ext} \longrightarrow \mathcal{D}$ and a morphism $\rho : \mathcal{F}_2(\mathcal{E}) \longrightarrow \mathcal{F}_1(\mathcal{K})$. Accordingly, $\rho$ is more often called a *representation* of $\mathcal{E}$ in $\mathcal{K}$.[7].

It thus appears that, in our chase for individuation, the central problem will be to study the possible *transits* between internal and external structures. In particular, this general representational strategy for adding new elements of mathematical structure into the kinematical arena raises three main questions:

---

[7]Despite its elegance, this categorical account of the general strategy strategy has one major caveat. The problem is that there are many constructions that one would clearly would like to qualify as 'intrinsic' but nonetheless fail to be functorial. The most compelling example is the assignment which associates to a given structure its group of automorphisms: in general, a morphism $X \longrightarrow Y$ in a certain category $\mathcal{C}$ does not induce a morphism of groups between $\mathrm{Aut}_{\mathcal{C}}(X)$ and $\mathrm{Aut}_{\mathcal{C}}(Y)$. Sufficient conditions on the category $\mathcal{C}$ for the automorphism assignment to be a (contravariant) functor are found in M. Linckelmann. "Alperin's weight conjecture in terms of equivariant Bredon cohomology". In: *Mathematische Zeitschrift* 250.3 (2005), pp. 495–513, Proposition 2.2.

Notwithstanding this, I will keep its main idea: that the key objects are morphisms, in a certain category $\mathcal{D}$, between objects built out of an abstract external structure and objects built out of the fundamental kinematical structure.

a) *Choice of the type of external structures.* Which type of external structures are relevant to the mathematical description of a physical system?

b) *The representation problem.* For a fixed type of external structures, there will be in general many different ways of representing them in the kinematical arena: there will be different natural candidates for the category $\mathcal{C}$ and different possible choices for the object $\mathcal{C}(\mathcal{K})$ within this category. How do these different representations compare and is there a privileged choice?

c) *The individuation problem.* For a fixed representation of an external structure, has the homogeneity of the kinematical arena been completely broken? In other terms, does a given triple $(\mathcal{K}, \mathcal{E}, \rho)$ satisfy the requirement of individuation?

In what follows we shall consider in turn the representation and individuation problems for one particular type of structures which have played a crucial role in the foundations of Mechanics: groups.

## III.2 Introducing discernibility through groups (1): the representation problem

It comes as no surprise that the first type of external structures we will consider are (Lie) groups, and their infinitesimal version Lie algebras. Their paramount importance in the development of Classical and Quantum Mechanics is beyond doubt and a striking evidence of this is the fact that the first book ever written on the foundations of Quantum Mechanics was Weyl's *Quantenmechanik und Gruppentheorie* of 1928. Since then, the crucial role of groups in both Classical and Quantum Mechanics has been underlined almost systematically. Thus, I feel there is no need to justify my choice of studying group-theoretical techniques in Kinematics.

However, before we plunge into a detailed analysis of the formalism related to groups, I should make a preliminary comment, for there is an important point in which my motivation for such a study greatly differs from the traditional point of view on the role of groups in Physics. A simple and clear account of this usual conception of groups may be found—again—in Dickson's excellent review of the philosophical problems arising in Quantum Mechanics. He writes:

> There is a traditional account of one way that groups have been related to empirical content. Take any group $G$, and consider its action on a set $S$. If two elements of $S$ are connected by an element of $G$, then call them "equivalent". One can readily verify that $G$ thus partitions $S$ into equivalence classes, and we can say, then, that $G$ is a group of symmetries on $S$, in the sense that the elements of $S$ connected by an element of $G$ are in some important sense 'the same'.[8]

Thus, because of their relation to symmetries, groups are most often perceived as a means to introduce a certain notion of *sameness* into the space of states of a physical system. From the general lines of our discourse, it should be clear that this is *not* the way *external* groups should be considered here. In our case, the necessity of considering groups stems from the requirement of breaking the homogeneity of the kinematical arena. The present situation is therefore the exact opposite of the one described by Dickson: given the abstract data of e.g. a symplectic manifold $S$, all its elements are a priori 'the same' and we are interested in finding "some important sense in which they would be 'different'". And we are hoping groups will indeed furnish such a sense. Thus, we are here trying to *perceive groups as a means to introduce a certain notion of **difference***—in other words, as a means to define labels of properties and to introduce discernibility among states[9].

Luckily, despite this important conceptual difference between the present approach and the usual one, all group-theoretical techniques remain useful. Indeed, whether one is interested in implementing the notion of symmetry in Mechanics (as were Weyl and Wigner) or trying to break the homogeneity of the kinematical arena by means of groups (as we are), the technical question it leads to remains the same: given an abstract group $G$, what does it mean to introduce it or represent it in the homogeneous arena?

There is, in mathematics, a general notion of group representation:

---

[8]Dickson, op. cit., p. 328.

[9]Again, it is important to insist on the fact that the groups we are here considering in order to introduce discernibility are *external* groups, rather than internal (cf. page 241).

**Definition III.1.** Given an abstract group $G$ and an object $C$ of a certain category $\mathcal{C}$, a **representation of** $G$ **on** $C$ is a morphism of groups $L : G \longrightarrow Aut_{\mathcal{C}}(C)$.[10]

In this sense, we see that a representation of an abstract group on $C$ is almost the same as the choice of a privileged subgroup among all automorphisms of the abstract object. Here, this subgroup is simply the set $L(G) \subset Aut_{\mathcal{C}}(C)$ of representatives of elements of $G$[11].

Whenever it happens that both $G$ and $Aut_{\mathcal{C}}(C)$ are Lie groups, a representation of $G$ on $C$ allows to define also a representation on $C$ of the infinitesimal version of the group, the Lie algebra $\mathfrak{g}$. This is achieved through the functor $\textbf{\textit{Lie}}$ from the category of Lie groups to the category of Lie algebras. The induced representation of $\mathfrak{g}$ on $C$ is simply the morphism $\textbf{\textit{Lie}}(L) : \mathfrak{g} \longrightarrow \textbf{\textit{Lie}}(Aut_{\mathcal{C}}(C))$. More generally, a representation of $\mathfrak{g}$ on $C$ is a morphism of Lie algebras

$$\rho : \mathfrak{g} \longrightarrow \textbf{\textit{Lie}}(Aut_{\mathcal{C}}(C)).$$

Those $\mathfrak{g}$-representations that are of the form $\rho = \textbf{\textit{Lie}}(L)$ for some $G$-representation $L$ are called *integrable*.

With these notions at hand, it would thus seem that the technical manner in which groups and Lie algebras are introduced into Kinematics is transparent. However, as we will now see, the situation is in fact more involved.

## III.2.1  In Classical Kinematics

We first specialize the above discussion to the context of Classical Kinematics. Therein, the starting point can be taken to be either a symplectic manifold $(S, \omega)$ (geometric point of view, emphasis on states) or its Poisson algebra of real-valued functions $\left(\mathcal{C}^\infty(S, \mathbb{R}), \cdot, \{\cdot, \cdot\}\right)$ (algebraic point of view, emphasis on properties). As we saw in the last chapter, one may choose any of the two points of view, since they are equivalent. For simplicity, we will often adopt the geometric point of view.

---

[10]See, for example, S. Lang. *Algebra*. 3rd ed. New York: Springer GTM, 2002, p. 54.

[11]The remark that any group may be seen as a category (with only one object and only invertible arrows) provides a categorical reformulation of the general notion of group representation. Given an abstract group $G$, a representation of it is simply any functor from $G$. Accordingly, one sometimes finds that a general functor $\mathcal{F} : \mathcal{C} \to \mathcal{D}$ is called a representation of the category $\mathcal{C}$ in the catgeory $\mathcal{D}$. However, this functorial point of view on representations seems to me less natural when working with representations of other algebraic structures which cannot so easily be viewed as special categories (such as Lie algebras, Poisson algebras or $C^*$-algebras).

In this case, $Aut_C(C)$ becomes $Symp(S)$, the group of all symplectomorphisms of the space of states $S$[12], and the general notion of group representation yields what is usually called a **symplectic or Poisson $G$-action on** $S$[13]—that is, a morphism of groups

$$G \xrightarrow{\;\;L\;\;} Aut(S)$$[14].

The infinitesimal analogue of these group actions is immediate. A *Poisson $\mathfrak{g}$-action* on $S$ is a Lie algebra morphism

$$\mathfrak{g} \xrightarrow{\;\;\rho\;\;} \Gamma(TS)_\omega$$

where $\Gamma(TS)_\omega$ is the Lie algebra of vector fields preserving the symplectic structure[15].

---

[12]Hereafter, I will take a morphism of symplectic manifolds $\phi: S \longrightarrow S'$ to be a morphism of differentiable manifolds such that the pull-back $\phi^*: \mathcal{C}^\infty(S', \mathbb{R}) \longrightarrow \mathcal{C}^\infty(S', \mathbb{R})$ is a morphism of Poisson algebras. This choice is by no means undisputed: Alan Weinstein has been suggesting for some time that the correct "symplectic category" to consider for Classical Mechanics should rather be defined in such a way to include Lagrangian correspondences as morphisms. For the detailed reasons pushing him to do so, see A. Weinstein. "Symplectic Categories". In: *Proceedings of Geometry Summer School, Lisbon*. 2009. URL: https://arxiv.org/pdf/0911.4133v1.pdf.

[13]One can also sometimes find the term "canonical action" (for example in J. E. Marsden and T. S. Ratiu. *Introduction to Mechanics and Symmetry. A Basic Exposition of Classical Mechanical Systems.* 2nd ed. New York: Springer, 1999), which is related to the use of "canonical" in the expression "canonical commutation relations". "Canonical" having such a different meaning in mathematics (as in "canonically isomorphic"), I will try to avoid the use of the word in its first sense.

[14]From the algebraic point of view, one would have rather considered a morphism $L_{alg}: G \longrightarrow Aut\left(\mathcal{C}^\infty(S, \mathbb{R})\right)$. But since $Aut(S)$ and $Aut\left(\mathcal{C}^\infty(S, \mathbb{R})\right)$ are canonically isomorphic the algebraic and geometric notions of classical $G$-representation coincide. I will refrain from systematically presenting the two points of view, as this would considerably slow down the reading, and will only perform this oscillation when there is an insight to gain in doing so.

[15]There are some technical subtleties I am omitting here, which are related to the ambiguity of the expression "the Lie algebra of vector fields". If one considers the Lie algebra of vector fields $\Gamma(TS)$ with the usual commutator, a *left* $G$-action induces a map $\rho: \mathfrak{g} \to (\Gamma(TS), [\cdot, \cdot])$ which is in fact a *anti*-morphism of Lie algebras. In other words, when integrable, a morphism of Lie algebras $\rho: \mathfrak{g} \to (\Gamma(TS), [\cdot, \cdot])$ integrates to a *right* $G$-action (which is a group *anti*-morphism $R: G \to Aut(S)$). Of course, it suffices to consider the Lie algebra $(\Gamma(TS), -[\cdot, \cdot])$ in order to make the statement in the main text rigorously true: a *left* $G$-action induces a morphism of Lie algebras $\rho: \mathfrak{g} \to (\Gamma(TS), -[\cdot, \cdot])$. Because one wants to perceive $\rho$ as $(Lie)(L)$, one sometimes says that "the Lie algebra of the group of diffeomorphisms is the Lie algebra of vector fields with *minus* the commutator", but this is again not rigorous since neither $Diff(S)$ nor $Symp(S)$ are Lie groups. (See for example D. Alekseevsky and P. W. Michor. "Differential Geometry of $\mathfrak{g}$-manifolds". In: *Differential Geometry and its Applications* 5.4 (1995), pp. 371–403. URL: http://arxiv.org/abs/math/9309214.)

Now, it would be wrong to think that Poisson actions are necessarily the central notion by means of which Lie groups and Lie algebras are introduced into Classical Kinematics. In practice, groups are introduced through a more specific type of actions, namely (strongly) Hamiltonian actions. These are usually defined as follows[16].

**Definition III.2.** A Poisson $G$-action on a symplectic manifold $(S, \omega)$, with associated Poisson $\mathfrak{g}$-action $\rho : \mathfrak{g} \to \Gamma(TS)_\omega$, is said to be **Hamiltonian** if, for any $X \in \mathfrak{g}$, the 1-form $\omega(\rho(X), \cdot)$ is exact.[17]

The condition that $\omega(\rho(X), \cdot)$ be exact implies the possibility of constructing a linear map called the **co-momentum map**, defined by

$$\hat{J} : \mathfrak{g} \longrightarrow \mathcal{C}^\infty(S, \mathbb{R})$$
$$X \longmapsto \hat{J}(X) \qquad \text{where } d\hat{J}(X) := \omega(\rho(X), \cdot).$$

From this, one can construct a second map, called the **momentum map**, defined by:

$$J : S \longrightarrow \mathfrak{g}^*$$
$$x \longmapsto J(x) \qquad \text{where } J(x)[X] := \hat{J}(X)(x).$$

**Definition III.3.** A Hamiltonian $G$-action $L : G \to Aut(S)$ is said to be **strongly Hamiltonian** if the momentum map is Co-equivariant—that is, if, for every $g \in G$, the following diagram commutes:

$$
\begin{array}{ccc}
S & \xrightarrow{\ J\ } & \mathfrak{g}^* \\
{\scriptstyle L(g)}\downarrow & & \downarrow{\scriptstyle Co(g)} \\
S & \xrightarrow{\ J\ } & \mathfrak{g}^*
\end{array}
$$

---

[16] Again, all the definitions and technical details that follow are standard and may be found in several textbooks. In my opinion, the best place to learn about group actions in Classical Mechanics is Marsden and Ratiu's *Introduction to Mechanics and Symmetry. A Basic Exposition of Classical Mechanical Systems.* They spend several chapters discussing Hamiltonian actions and provide an extensive list of examples. For a conceptual understanding of the general situation, the few pages of Landsman's *Mathematical Topics Between Classical and Quantum Mechanics* on the subject are particularly enlightening (pp. 178-191). Finally, Iglesias-Zemmour's *Symétries et moment* was also an important reading for my understanding of this topic.

[17] P. Iglesias-Zemmour. *Symétries et moment.* Paris: Hermann, Éditeurs des Sciences et des Arts, 2000, pp. 101–102. Note that, for any Poisson $G$-action, the 1-form $\omega(\rho(X), \cdot)$ is necessarily closed. Indeed, since by definition $\rho(X)$ preserves the symplectic structure, we have $\mathcal{L}_{\rho(X)}\omega = 0 = \iota_{\rho(X)}d\omega + d(\omega(\rho(X), \cdot)) = d(\omega(\rho(X), \cdot))$.

where $Co : G \to Aut(\mathfrak{g}^*)$ denotes the co-adjoint action[18].

**Definition III.4.** A Hamiltonian $\mathfrak{g}$-action is said to be **strongly Hamiltonian**[19] if the co-momentum map $\widehat{J}$ is a morphism of Lie algebras—that is, if, for every $X, Y \in \mathfrak{g}$, we have

$$\widehat{J}([X,Y]) = \{\widehat{J}(X), \widehat{J}(Y)\}_S.$$

In this latter case, one sometimes says that the momentum map is 'infinitesimally equivariant'. This is because the Co-equivariance of the momentum map implies its infinitesimal equivariance (in other words, if $L : G \to Aut(S)$ is strongly Hamiltonian, then so is the associated $\mathfrak{g}$-action $dL$). The converse is true only if $G$ is connected[20].

In its modern form, the momentum map was independently introduced around 1965 by the American mathematician Bertram Kostant and the French Jean-Marie Souriau, although, with the wisdom of hindsight, a version of it can already be found in the work of Sophus Lie (1890)[21]. This concept has become a notion of the uttermost importance for the foundations of Classical Kinematics. Marsden and Ratiu describe

---

[18] The co-adjoint action of $G$ on $\mathfrak{g}^*$ is usually defined in terms of the adjoint action of $G$ on $\mathfrak{g}$ by: $\forall \theta \in \mathfrak{g}^*, \forall X \in \mathfrak{g}, (Co(g)\theta)(X) = \theta(Ad(g^{-1})X)$. In turn, the adjoint action is defined by: if $X \in \mathfrak{g}$ is the tangent vector at the identity to the parametrized curve $\gamma$, then $Ad(g)X$ is defined as the tangent vector at the identity to the parametrized curve $g\gamma g^{-1}$. In other terms,

$$Ad(g)X := \frac{d}{dt}(g\gamma g^{-1})(t)\big|_{t=0} \quad \text{where } \frac{d\gamma(t)}{dt}\big|_{t=0} = X \text{ and } \gamma(0) = e.$$

See Landsman, op. cit., p. 184.

[19] A remark on terminology, for there are slight variations from one reference to another with respect to the various notions I have just introduced. First, there is the harmless variation between 'momentum' and 'moment' map (or mapping). This is explained from the fact that the terminology was introduced *in French* by Jean-Marie Souriau in his article *"Quantification géométrique. Applications"*. Therein, he used the word "moment" because it generalized the notion of angular momentum (in French: "*moment* angulaire"). Despite this, the first usages of this notion in English kept the French word (e.g. Marsden and Weinstein's *"Reduction of Symplectic Manifolds With Symmetry"* in 1974, and also the English translation of Souriau's book). Nowadays, most people use "momentum" but notable exceptions are Guillemin and Sternberg (who use "moment", perhaps because it also generalizes the notion of moment of inertia (in French: "moment d'inertie")) and Woit (who uses "momentum map" to refer to what I have called co-momentum).

One needs to be a little bit more careful with the notion of "Hamiltonian action". I am here using the terminology of Landsman, which also agrees with that of Iglesias-Zemmour. However, in many textbooks (e.g., Abraham and Marsden, Marsden and Ratiu, Puta), 'Hamiltonian actions' refer to what I call 'strongly Hamiltonian actions'...

[20] Marsden and Ratiu, op. cit., Theorem 12.3.2, p. 402.

[21] For more extensive references of the original papers dealing with the momentum map, see ibid., pp. 369–370.

it as a "rich concept that is ubiquitous in the modern developments of geometric mechanics" and that "has led to surprising insights into many areas of mechanics and geometry"[22]. Even stronger, Souriau turns the existence of a momentum map into one of the four fundamental principles of non-relativistic symplectic mechanics[23].

### III.2.1.a  Importance of (strongly) Hamiltonian actions

The question for us is: *Why?* Why do Hamiltonian actions and the momentum mapping play such an important role in the foundations of Classical Mechanics? Why should we consider classical state spaces endowed with an action of $G$ *and* a momentum map, instead of simply considering general Poisson actions?

An often cited motivation for introducing the momentum map is its relation to Noether's theorem. Given a Hamiltonian action $G \circlearrowright S$ with momentum map $J$, and a property $h \in \mathcal{C}^\infty(S, \mathbb{R})$ which is $G$-invariant, then any property of the form $\widehat{J}(X) \in \mathcal{C}^\infty(S, \mathbb{R})$ (for $X \in \mathfrak{g}$) is constant along the flow of the Hamiltonian vector field $v_h$[24]. This result is the geometric reformulation of Noether's first theorem. Indeed, if one thinks of $h$ as the Hamiltonian of the system, the above statement is saying that whenever $G$ is a symmetry group of the Hamiltonian, the functions $\widehat{J}(X)$ are conserved quantities. The co-momentum map appears thus as a very powerful tool to build conserved quantities for a system with symmetries.

But this clearly cannot be the whole story. The relation to Noether's theorem only succeeds in explaining why Hamiltonian actions are something valuable in the light of a particular quest, but not why they appear as a vault upon which rests the general theory. In other words, it only allows to understand why they are *convenient*, but not why they are *fundamental*[25]. Contrary to the utilitarian approach in which the concept becomes meaningful a posteriori, through the usefulness of its applications, we are looking for a perspective presenting Hamiltonian and strongly Hamiltonian actions as something *natural* to consider in the first place—that is, a perspective from which one could almost anticipate the concept before its introduction[26].

---

[22]Ibid., p. 365.

[23]J.-M. Souriau. *Structure of Dynamical Systems. A Symplectic View of Physics*. Trans. by C. Cushman-de Vries. Boston: Birkhäuser, 1997, p. 155. The other three principles are: i) that the space of motions be a connected symplectic manifold; ii) that the space of motions of a composite system of independently evolving parts be the symplectic direct product of the spaces of motions of each part; iii) that for an isolated system, the space of motions be endowed with a Poisson action of the Galileo group. Recall that, for Souriau, the fundamental symplectic manifold appearing in Classical Mechanics should not be viewed as the space of instantaneous states but rather as the space of states extended in time (that is, the space of motions).

[24]See Landsman, op. cit., Proposition I.1.2.2.

A resolute attempt to answer the question in these terms is found in Gabriel Catren's article *"On the Relation Between Gauge and Phase Symmetries"*. Therein, it is argued that the existence of a momentum map in the classical space of states should be perceived as a footprint left by Quantum Mechanics inside Classical Mechanics[27]. It nonetheless seems to me that one should at least try to understand the central importance of Hamiltonian actions without any mention of the Quantum, thus attempting to consider the momentum map as an entirely classical notion.

A step in the sought-for direction is to take seriously the representation problem alluded to at the end of the introduction to this chapter. Indeed, Definition III.1 (page 245) recognizes the fact that, from any mathematical structure whatsoever, it is possible to define at least one group: the automorphism group. But besides this general representational strategy, there exist other equally sound possibilities which are attached to the particularities of Classical Kinematics. These alternatives stem from two sources: the ability to define several different Lie algebras from a symplectic manifold, and the ability to construct both a Poisson space and a Poisson algebra from a given Lie group.

---

[25]Moreover, there are many other reasons why the momentum map is useful. As Marsden and Ratiu say, "this concept is more than a mathematical reformulation of a concept that simply describes the well-known Noether theorem" (Marsden and Ratiu, loc. cit.). Important examples of other applications are the construction of new symplectic manifolds out of Hamiltonian actions by means of the so-called 'Marsden-Weinstein reduction' (Landsman, op. cit., section IV.1.5.) and the classification of transitive symplectic actions by Kostant's coadjoint orbit covering theorem (Marsden and Ratiu, op. cit., Theorem 14.4.5., p. 465).

[26]The *naturalness* of mathematical concepts is here an essential point, and a philosophical clarification of this idea would certainly constitute a valuable work which is largely overdue in the philosophy of mathematics. Very few authors seem to tackle this question. David Corfield spends a small section of his book *Towards a Philosophy of Real Mathematics* trying to elucidate this issue ("The Conceptual and the Natural", pp. 223-230). More recently, Luca San Mauro and Giorgio Venturi have published an article solely dedicated to the notion of naturalness (L. San Mauro and G. Venturi. "Naturalness in Mathematics". In: *From Logic to Practice.* Ed. by G. Lolli, M. Panza, and G. Venturi. Springer, 2015, pp. 277–313), but further work remains to be done. Let me just briefly comment that I do not regard 'naturalness' as an intrinsic property of a concept, but rather as a property of the *place* a concept occupies *within an expository discourse.* The general idea would be that the naturalness of a concept is intimately linked to the notion of *continuity* (of the process of developing a theory) or of *inevitability* (of the introduction of the concept). Thus, a concept would appear as unnatural when its introduction into the exposition constitutes a moment of rupture which could not have been foreseen.

[27]Indeed, Catren's approach emphasizes the idea that $\mathfrak{g}^*$ "encodes the unitary representation theory of [the group] $G$" (p. 1321). This is inspired by Kirillov's orbit method, which establishes for certain Lie groups (e.g., abelian, nilpotent) a correspondence between the Co-adjoint orbits $\mathfrak{g}^*/G$ and the unitary irreducible representations of the group. Hence, the presence of a momentum map in Classical Mechanics can only be fully understood from the vantage viewpoint of Quantum Mechanics and its Hilbert space formulation.

Let us review these various notions of representation on the classical arena. First, given a symplectic manifold $(S, \omega)$, there are at least three different Lie algebras one can construct:

i) $\Gamma(TS)_\omega$ (vector fields preserving the symplectic structure),

ii) $\Gamma(TS)_H$ (Hamiltonian vector fields),

iii) $\left(\mathcal{C}^\infty(S, \mathbb{R}), \{\cdot, \cdot\}\right)$ (smooth real-valued functions where one forgets point-wise multiplication and keeps only the Poisson bracket).

These are related by the following diagram of Lie algebras:

$$\mathbb{R} \longhookrightarrow \mathcal{C}^\infty(S, \mathbb{R}) \overset{v_-}{\longtwoheadrightarrow} \Gamma(TS)_H \overset{\iota}{\longhookrightarrow} \Gamma(TS)_\omega \qquad \text{(III.3)}$$

where the first two arrows form a short exact sequence (the image of one arrow is the kernel of the next one) and capture the fact that $\Gamma(TS)_H \simeq \mathcal{C}^\infty(S, \mathbb{R})/\mathbb{R}$ (properties-as-transformations are properties 'up to a constant'). The existence of this triple of internal Lie algebras furnishes three possible ways of representing Lie algebras in the Classical arena—just consider morphisms from $\mathfrak{g}$ to any of $\Gamma(TS)_\omega$, $\Gamma(TS)_H$ or $\mathcal{C}^\infty(S, \mathbb{R})$. In fact, these three notions of representation exactly coincide with the three different notions of $\mathfrak{g}$-actions on $S$ we have already discussed:

– a Poisson $\mathfrak{g}$-action is a morphism of Lie algebras $\mathfrak{g} \overset{\rho}{\longrightarrow} \Gamma(TS)_\omega$,

– a Hamiltonian $\mathfrak{g}$-action is a morphism of Lie algebras $\mathfrak{g} \overset{\rho_H}{\longrightarrow} \Gamma(TS)_H$[28],

---

[28]Indeed, the requirement that the 1-form $\omega(\rho_H(X), \cdot)$ be exact (cf. Definition III.2, page 247) means there exists $f \in \mathcal{C}^\infty(S, \mathbb{R})$ such that $\omega(\rho_H(X), \cdot) = df$. In other terms, $\rho_H(X)$ is a Hamiltonian vector field.

– a strongly Hamiltonian $\mathfrak{g}$-action is a morphism of Lie algebras  $\mathfrak{g} \xrightarrow{\widehat{J}} \mathcal{C}^\infty(S, \mathbb{R})$[29].

Second, given a Lie algebra $\mathfrak{g}$, it is possible to show that its topological dual $\mathfrak{g}^*$ is a Poisson manifold (or, equivalently, that $\mathcal{C}^\infty(\mathfrak{g}^*, \mathbb{R})$ is a Poisson algebra). This important result is summarized in the following

**Theorem III.1.** *Consider the injection*  $\mathfrak{g} \xhookrightarrow{\iota} \mathcal{C}^\infty(\mathfrak{g}^*, \mathbb{R})$  *provided by the canonical identification of $\mathfrak{g}$ with $\mathfrak{g}^{**}$. Then, there exists a unique Poisson structure on $\mathfrak{g}^*$ such that the above map is an injection of Lie algebras:* $\{\widetilde{X}, \widetilde{Y}\}_{\mathfrak{g}^*} := \widetilde{[X, Y]}$, *where $X, Y \in \mathfrak{g}$ and $\widetilde{X} := \iota(X)$.*[30]

This then seems to furnish two additional strategies for representing the Lie algebra $\mathfrak{g}$ in the classical arena: one can either consider **representations of $\mathcal{C}^\infty(\mathfrak{g}^*, \mathbb{R})$ on S** (that is, morphisms of Poisson algebras  $\mathcal{C}^\infty(\mathfrak{g}^*, \mathbb{R}) \xrightarrow{J^*} \mathcal{C}^\infty(S, \mathbb{R})$)  or **realizations of $\mathfrak{g}^*$ on S** (that is, morphisms of Poisson manifolds  $S \xrightarrow{J} \mathfrak{g}^*$)[31]. Nonetheless, we have the following equivalence[32]:

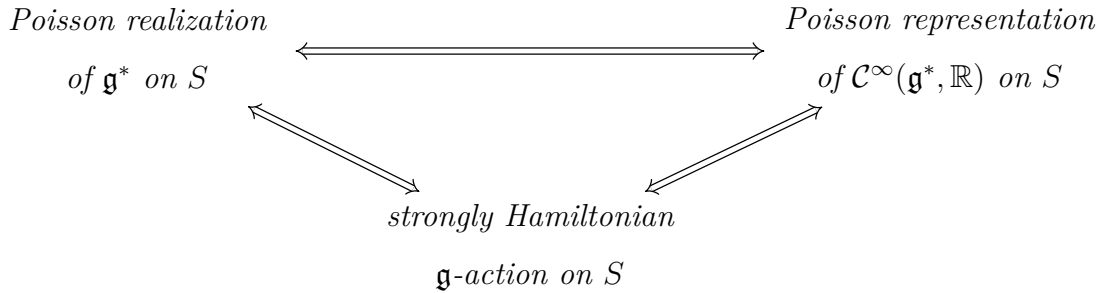

---

[29]More precisely, this last arrow is the unique infinitesimally equivariant co-momentum map associated to the strongly Hamiltonian action. The latter is the morphism $\rho$ induced by the co-momentum map through the diagram

$$\mathfrak{g} \xdashrightarrow{\widehat{J}} \mathcal{C}^\infty(S, \mathbb{R}) \xrightarrow{v_-} \Gamma(TS)_H \xhookrightarrow{\iota} \Gamma(TS)_\omega.$$

[30]Cf Landsman, op. cit., p. 179, Definition III.1.1.1. and the comment following it. Most usually, the Poisson structure on $\mathfrak{g}^*$ is defined explicitly by: $\{f, g\}_{\mathfrak{g}^*}(\theta) = \theta([df|_\theta, dg|_\theta])$, where $f, g \in \mathcal{C}^\infty(\mathfrak{g}^*, \mathbb{R})$, $\theta \in \mathfrak{g}^*$ and where one uses the identifications $T^*_\theta \mathfrak{g}^* \simeq \mathfrak{g}^{**} \simeq \mathfrak{g}$.

[31]See ibid., Definition I.2.6.1., p. 76, and also N. P. Landsman. "Lie Groupoids and Lie Algebroids in Physics and Noncommutative Geometry". In: *Journal of Geometry and Physics* 56.1 (2006), pp. 24–54. URL: http://arxiv.org/abs/math-ph/0506024, p. 41.

[32]By 'equivalence', I mean here the existence of a canonical bijection between these three sets of arrows. For the upper equivalence, see idem, *Mathematical Topics Between Classical and Quantum Mechanics*, Corollary I.2.6.5, p. 77. For the lower right equivalence, see ibid., Theorem III. 1.1.7., p. 181, or Marsden and Ratiu, op. cit., pp. 403–405 and in particular Remark 1, p. 405.

The first of these equivalences we already knew: a Poisson realization $S \xrightarrow{J} \mathfrak{g}^*$ induces a Poisson representation by taking the pullback $J^*$, and every Poisson representation is the pullback of some Poisson realization. The second equivalence stems from two facts. First, the co-momentum map $\widehat{J}$ is a morphism of Lie algebras if and only if the momentum map $J$ is a morphism of Poisson manifolds[33]. Moreover—and this is the crucial point—a Poisson representation $\mathcal{C}^\infty(\mathfrak{g}^*, \mathbb{R}) \xrightarrow{J^*} \mathcal{C}^\infty(S, \mathbb{R})$ induces a Poisson $\mathfrak{g}$-action through the diagram (where all arrows are morphisms of Lie algebras)

$$\mathfrak{g} \xhookrightarrow{\iota} \mathcal{C}^\infty(\mathfrak{g}^*, \mathbb{R}) \xrightarrow{J^*} \mathcal{C}^\infty(S, \mathbb{R}) \xrightarrow{v_-} \Gamma(TS)_H \xhookrightarrow{\iota} \Gamma(TS)_\omega.$$

$$\rho$$

In other words, as soon as one is given a Poisson realization $S \xrightarrow{J} \mathfrak{g}^*$ and considers its pull-back $\mathcal{C}^\infty(\mathfrak{g}^*, \mathbb{R}) \xrightarrow{J^*} \mathcal{C}^\infty(S, \mathbb{R})$, one can generate a diagram defining a strongly Hamiltonian $\mathfrak{g}$-action: infinitesimally equivariant momentum maps are just a particular instance of the more general concept of Poisson realizations, and the co-momentum map $\widehat{J}$ is simply the map $\widehat{J} := J^* \circ \iota$.

Hence, *strongly Hamiltonian $\mathfrak{g}$-actions are precisely those $\mathfrak{g}$-actions which are induced by a Poisson representation/realization*. Through this change of emphasis—from the notion of action to the notion of Poisson representation—the problem of studying all possible strongly Hamiltonian actions of $\mathfrak{g}$ is no longer understood as a problem focusing on a *restricted* class of $\mathfrak{g}$-actions; rather, it is the problem of studying *all the possible* representations of the Poisson algebra $\mathcal{C}^\infty(\mathfrak{g}^*, \mathbb{R})$.

All in all, the classical representation problem (for Lie groups and Lie algebras) boils down to the comparison of the three different types of $\mathfrak{g}$-actions. The situation is summarized in Figure III.1 below.

The first point highlighted by this figure is how little groups are actually involved in these constructions. Both the notions of Poisson representation and Poisson realization only appeal to the infinitesimal information of the group, and the same holds for the notions of Hamiltonian and strongly Hamiltonian action. Because of this, it is better to leave groups and the question of integrability of $\mathfrak{g}$-actions somewhat aside and focus mainly on the three central lines of the diagram.

---

[33]Landsman, op. cit., Proposition III.1.1.2, p. 179, or Marsden and Ratiu, op. cit., Theorem 12.4.1, p. 403.

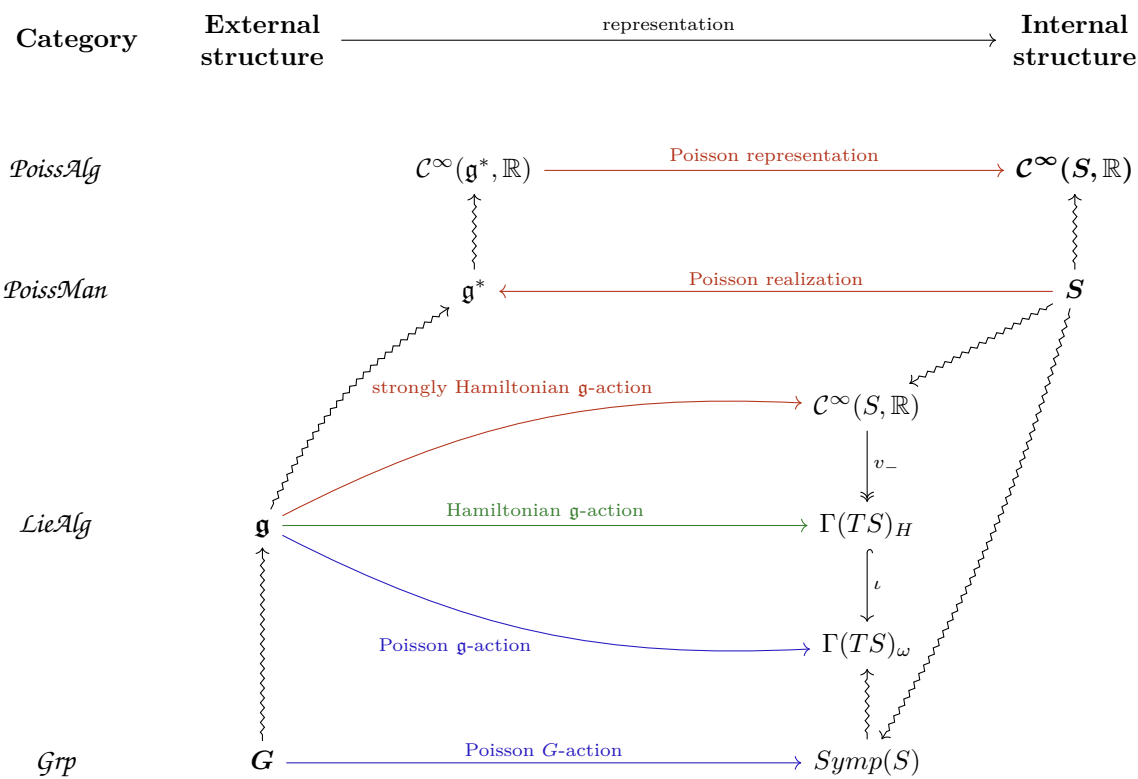

**Fig. III.1** – **The Classical representation problem for Lie groups and Lie algebras (or how to introduce them into Classical Kinematics).**
On the top right and bottom left of the 'diagram' are the initial external and internal structures. Squiggle arrows represent intrinsic constructions of new structures (some of which, but not all, being functorial), while all other arrows represent morphisms in the appropriate category. The coloured arrows represent the various a priori different transits between the external and the internal structures.

So far, this more systematic exploration of the several possible transits between the external and the internal has allowed us to understand why both Hamiltonian and strongly Hamiltonian actions appear as natural objects to consider in our attempt to construct the mathematical characterization of a classical system using Lie groups and Lie algebras. Presently, they are perceived as a representational strategy that stands on the same footing as Poisson actions. But this is not enough. If we wish to provide a rationale for Souriau's fourth axiom of non-relativistic symplectic mechanics, we need to explain why we should *not* consider general Poisson actions. In other terms, we need to explain why we should consider (strongly) Hamiltonian actions as the *preferred* representational strategy.

Here, the conceptual framework developed in the preceding chapter offers us a clear answer. As we have just seen, the introduction of an abstract external Lie algebra into the Classical arena will, in all representational strategies, lead to a distinguished set of infinitesimal state transformations (it is the image of $\mathfrak{g}$ by the action $\rho$). However, if these distinguished transformations are to be useful in the definition of the properties of the classical system being described, if they are to provide a labelling scheme for physical properties, then it is essential that the chosen representational strategy reflects the fundamental conceptual triad which constitutes the core of Kinematics (Figure II.1, page 146). In other words, it is essential to be able to interpret the represented transformations as *transformations generated by properties*. This means $\mathfrak{g}$ should be represented, not by general transformations, but rather by properties-as-transformations. And this point is precisely what distinguishes Hamiltonian actions from general Poisson actions.

### III.2.1.b  Hamiltonian vs. strongly Hamiltonian actions

Both Hamiltonian and strongly Hamiltonian actions respect the transformational role of properties, as they both allow to associate to each element of the external Lie algebra $\mathfrak{g}$ a property-as-transformation. Despite this, there also seems to be a clear advantage of the latter type of actions: they represent the external infinitesimal transformations directly as properties. In this way, *strongly Hamiltonian actions appear as those $\mathfrak{g}$-actions which respect both the transformational <u>and numerical</u> role of physical properties*. In the light of this, it would seem that one should also discard the use of Hamiltonian actions and restrict attention solely to strongly Hamiltonian actions.

However, as we will now see, the difference between Hamiltonian and strongly Hamiltonian actions is in fact not that important. To have a better grasp of this difference, it is useful to compare the diagrams defining these two types of actions.

Recall: strongly Hamiltonian actions are defined by the diagram

$$\mathfrak{g} \xrightarrow{\widehat{J}} \mathcal{C}^\infty(S, \mathbb{R}) \xrightarrow{v_-} \Gamma(TS)_H \xleftarrow{\iota} \Gamma(TS)_\omega \qquad \text{(III.4)}$$

$$\xrightarrow[\rho_H]{}$$

where all arrows are morphisms of Lie algebras, while Hamiltonian actions are defined by the diagram

$$\mathfrak{g} \dashrightarrow{\widehat{J}} \mathcal{C}^\infty(S, \mathbb{R}) \xrightarrow{v_-} \Gamma(TS)_H \xleftarrow{\iota} \Gamma(TS)_\omega \qquad \text{(III.5)}$$

$$\xrightarrow[\rho_H]{}$$

where now the co-momentum map $\widehat{J}$ is only a morphism of *vector spaces* (equivalently, the momentum map $J : S \to \mathfrak{g}^*$ is only a morphism of differentiable manifolds, instead of being a Poisson realization; or the pull-back $J^* : \mathcal{C}^\infty(\mathfrak{g}^*, \mathbb{R}) \to \mathcal{C}^\infty(S, \mathbb{R})$ is only a morphism of commutative algebras instead of being a Poisson representation) and the remaining arrows are morphisms of Lie algebras.

Diagrams (III.4) and (III.5) may look very similar. Nonetheless, they present an essential conceptual difference, which lies in the possible *oscillations* between the morphisms $\widehat{J}$ and $\rho_H$. For strongly Hamiltonian actions, one can either: *start* from the data $\widehat{J}$ and *deduce* the data $\rho_H$ (one defines it as $\rho_H := v_- \circ \widehat{J}$), or start from the $\mathfrak{g}$-action and deduce the map $\widehat{J}$ (the equivariant co-momentum map is *uniquely* defined for a strongly Hamiltonian action). This freedom of circulation breaks down for Hamiltonian actions: given a smooth map $S \xrightarrow{J} \mathfrak{g}^*$, the induced map $v_- \circ \widehat{J}$ is no longer guaranteed to be a morphism of Lie algebras. Moreover, given a Hamiltonian action, one cannot associate to it one single smooth map $S \xrightarrow{J} \mathfrak{g}^*$ but rather a *class* of such maps (the momentum map is *not* uniquely defined). In other words, whereas strongly Hamiltonian actions of $\mathfrak{g}$ on $S$ and Poisson realizations of $\mathfrak{g}^*$ on $S$ are equivalent (one can freely circulate between the two), the notion of Hamiltonian action is prior to the non-equivariant momentum map (one must start from the data $\mathfrak{g} \xrightarrow{\rho_H} \Gamma(TS)_H$).

Having said this, let us describe the *obstructions* for a Hamiltonian $\mathfrak{g}$-action to be strongly Hamiltonian[34]. We have the following:

**Theorem III.2.** *The second cohomology group of $\mathfrak{g}$ with values in $\mathbb{R}$ governs the obstruction to have strongly Hamiltonian actions. In other terms, any Hamiltonian $\mathfrak{g}$-action is strongly Hamiltonian if and only if $H^2(\mathfrak{g}, \mathbb{R}) = 0$.*[35]

Once the locus of the obstruction is properly pointed out, the immediate question becomes to investigate strategies to *bypass* such an obstruction. In this case, there are

$$\Gamma : \mathfrak{g} \times \mathfrak{g} \longrightarrow \mathbb{R}$$
$$(X, Y) \longmapsto \Gamma(X, Y) := \widehat{J}([X, Y]) - \{\widehat{J}(X), \widehat{J}(Y)\}_S.$$

---

[34]In order not to be drowned by the technical details in what is about to come, I have decided to develop part of the formalism in the footnotes. The general picture should be understandable without the reading of these. A precise exposition is found in Landsman, op. cit., Section I.1.1, pp. 178–183.

[35]Given a diagram of the form (III.5), the default of $\rho_H$ to be strongly Hamiltonian is closely related to the default of some associated co-momentum map $\widehat{J}$ to be a morphism of Lie algebras, which in turn is captured by the map

(Strictly speaking, the definition shows $\Gamma(X, Y)$ to be an element of $\mathcal{C}^\infty(S, \mathbb{R})$. Nonetheless, the fact that both $\rho$ and $v_-$ are morphisms of Lie algebras enforces the basic identity $v_{\widehat{J}([X,Y])} = v_{\{\widehat{J}(X), \widehat{J}(Y)\}_S}$, which in turn implies, if $S$ is connected, that, in fact, $\Gamma(X, Y)$ is just a real number.)

Because of the anti-symmetry and Jacobi identity of both the Lie product $[\cdot, \cdot]$ on $\mathfrak{g}$ and the Poisson bracket $\{\cdot, \cdot\}$ on S, the bilinear map $\Gamma$ satisfies two similar properties:

i) $\Gamma(X, Y) = -\Gamma(Y, X)$,

ii) $\Gamma(X, [Y, Z]) = \Gamma([X, Y], Z) + \Gamma(Y, [X, Z])$.

A bilinear function on $\mathfrak{g} \times \mathfrak{g}$ satisfying these two properties is called a *2-cocycle* on $\mathfrak{g}$ with values in $\mathbb{R}$. The set of all such 2-cocycles, denoted by $Z^2(\mathfrak{g}, \mathbb{R})$, captures therefore the obstruction to the (infinitesimal) equivariance of the co-momentum map $\widehat{J}$.

However, because the (co-)momentum map is not uniquely determined by the Hamiltonian action (each $\widehat{J}(X)$ is defined only up to a constant), the obstruction of a *given* (co-)momentum map to be equivariant is not quite the same as the obstruction of the $\mathfrak{g}$-action to be strongly Hamiltonian. Indeed, if one considers $\widehat{J}_2 := \widehat{J} - \alpha$, where $\alpha \in \mathfrak{g}^*$, we have:

$$\widehat{J}_2([X, Y]) - \{\widehat{J}_2(X), \widehat{J}_2(Y)\}_S = \widehat{J}([X, Y]) - \{\widehat{J}(X), \widehat{J}(Y)\}_S - \alpha([X, Y])$$
$$= \Gamma(X, Y) - \alpha([X, Y]).$$

A 2-cocycle for which there exists $\alpha \in \mathfrak{g}^*$ such that $\Gamma(X, Y) = \alpha([X, Y])$ is said to be *trivial* and their set is denoted by $B^2(\mathfrak{g}, \mathbb{R})$. The last equation shows that, whenever $\Gamma$ is trivial, one can effectively overcome the obstruction and find an equivariant momentum map by a simple redefinition (whence the adjective "trivial").

To build an object which does not depend on the arbitrary choice of a momentum map—and is therefore *intrinsically* related to the Hamiltonian action under consideration—one should identify 2-cocycles whose difference is trivial but not necessarily zero. Thus, one considers the quotient

$$H^2(\mathfrak{g}, \mathbb{R}) := Z^2(\mathfrak{g}, \mathbb{R})/B^2(\mathfrak{g}, \mathbb{R})$$

called the *second cohomology group* of $\mathfrak{g}$.

two possible strategies, each of them shedding a new light onto the notion of 'Hamiltonian actions'. They both have in common the fundamental idea of *transforming the negativity of the obstruction into a positive fact*: instead of considering an obstruction as a default to meet some "nice" conditions (negativity), one views them as enabling the possibility of constructing new structures (positivity):

1. *Absorb the obstruction by modifying the Poisson algebra* $\mathcal{C}^\infty(\mathfrak{g}^*, \mathbb{R})$. Instead of considering the dual of the momentum map as an arrow

$$\left(\mathcal{C}^\infty(\mathfrak{g}^*, \mathbb{R}), \{\cdot, \cdot\}_{\mathfrak{g}^*}\right) \xrightarrow{\ J^*\ } \left(\mathcal{C}^\infty(S, \mathbb{R}), \{\cdot, \cdot\}_S\right)$$

   which fails to preserve the Poisson structures, one can define a new Poisson structure on $\mathfrak{g}^*$ so that the arrow $J^*$ appears as a morphism of Poisson algebras

$$\left(\mathcal{C}^\infty(\mathfrak{g}^*, \mathbb{R}), \{\cdot, \cdot\}_{\mathfrak{g}^*}^\Gamma\right) \xrightarrow{\ J^*\ } \left(\mathcal{C}^\infty(S, \mathbb{R}), \{\cdot, \cdot\}_S\right)^{[36]}.$$

   From this perspective, we have the equivalence of notions[37]

$$\begin{array}{ccc} \textit{Hamiltonian } \mathfrak{g}\textit{-action} & & \textit{Poisson representation} \\ \textit{with cocycle } \Gamma & \Longleftrightarrow & \textit{of the algebra } \mathcal{C}_\Gamma^\infty(\mathfrak{g}^*, \mathbb{R}) \end{array}$$

2. *Absorb the obstruction by modifying the initial Lie algebra* $\mathfrak{g}$. Instead of viewing the second cohomology group $H^2(\mathfrak{g}, \mathbb{R})$ as the locus of obstructions to strongly Hamiltonian $\mathfrak{g}$-actions, one views it as the classifier of *central extensions* of $\mathfrak{g}$. Given a Hamiltonian $\mathfrak{g}$-action with cocycle $\Gamma$, this allows to construct a new Lie

---

[36]This is achieved in the following way. Define the modified Poisson structure on $\mathfrak{g}^*$ by:

$$\{\widetilde{X}, \widetilde{Y}\}_{\mathfrak{g}^*}^\Gamma := \{\widetilde{X}, \widetilde{Y}\}_{\mathfrak{g}^*} - \Gamma(X, Y)$$

for any $X, Y \in \mathfrak{g}$. In this way, the equation

$$J^*\left(\{\widetilde{X}, \widetilde{Y}\}_{\mathfrak{g}^*}\right) - \{J^*(\widetilde{X}), J^*(\widetilde{Y})\}_S = \Gamma(X, Y)$$

may be rewritten as

$$J^*\left(\{\widetilde{X}, \widetilde{Y}\}_{\mathfrak{g}^*}^\Gamma\right) - \{J^*(\widetilde{X}), J^*(\widetilde{Y})\}_S = 0.$$

This shows that the dual of the momentum map is indeed a representation on $S$ of the Poisson algebra $\mathcal{C}_\Gamma^\infty(\mathfrak{g}^*, \mathbb{R})$.

[37]Ibid., Theorem III.1.1.7, p. 181.

algebra $\mathfrak{g}_\Gamma$ which has a strongly Hamiltonian action on $S$[38]. From this perspective, we have the equivalence of notions[39]

$$\begin{array}{ccc} \textit{Hamiltonian } \mathfrak{g}\textit{-action} & & \textit{strongly Hamiltonian } \mathfrak{g}_\Gamma\textit{-action} \\ \textit{with cocycle } \Gamma & \Longleftrightarrow & \textit{in which the center acts trivially} \end{array}$$

The first move shows Hamiltonian actions to fall under the theory of Poisson representations/realizations, as it was the case for strongly Hamiltonian actions. Through the second move, we see that, lurking behind a Hamiltonian action of some Lie algebra, there is always a strongly Hamiltonian action of another closely related Lie algebra. In this way, the problem of studying all Hamiltonian actions of some *fixed* Lie algebra may be restated as the problem of studying strongly Hamiltonian actions of a certain *range* of Lie algebras (the central extensions of $\mathfrak{g}$). The result is summarized in Figure III.2.

---

[38] A central extension $\mathfrak{g}_\Gamma$ of $\mathfrak{g}$ is a short exact sequence $0 \to \mathbb{R} \to \mathfrak{g}_\Gamma \to \mathfrak{g} \to 0$ such that, in addition, for any $T \in \mathbb{R}$ and $X \in \mathfrak{g}$, $[X, T]_{\mathfrak{g}_\Gamma} = 0$. This implies $\mathfrak{g}_\Gamma = \mathfrak{g} \oplus \mathbb{R}$ as vector spaces and also the existence of $\Gamma : \mathfrak{g} \times \mathfrak{g} \to \mathbb{R}$ such that

$$[X, Y]_{\mathfrak{g}_\Gamma} = [X, Y]_{\mathfrak{g}} + \Gamma(X, Y)T.$$

Again, from the anti-symmetry and Jacobi identity for $[\cdot, \cdot]_{\mathfrak{g}_\Gamma}$, it follows that $\Gamma$ is a 2-cocycle. En passant, notice that, as a Lie algebra, $\mathcal{C}^\infty(S, \mathbb{R})$ is a central extension of the Lie algebra of Hamiltonian vector fields $\Gamma(TS)_H$.

Two central extensions $\mathfrak{g}_\Gamma$ and $\mathfrak{g}_\Lambda$ are said to be equivalent if there exists an isomorphism of Lie algebras $\mathfrak{g}_\Gamma \longleftrightarrow \mathfrak{g}_\Lambda$ such that the following diagram commutes

$$\begin{array}{ccc} & \mathfrak{g}_\Gamma & \\ & \nearrow \quad \searrow & \\ 0 \longrightarrow \mathbb{R} & \big\updownarrow & \mathfrak{g} \longrightarrow 0 \\ & \searrow \quad \nearrow & \\ & \mathfrak{g}_\Lambda & \end{array}$$

This is the case if and only if the 2-cocycles $\Gamma$ and $\Lambda$ belong to the same cohomology class. There is thus a bijection between the cohomology group $H^2(\mathfrak{g}, \mathbb{R})$ and the set of equivalence classes of central extensions of $\mathfrak{g}$. (For an excellent exposition of this, see G. M. Tuynman and W. Wiegerinck. "Central Extensions and Physics". In: *Journal of Geometry and Physics* 4.2 (1987), pp. 207–258.)

Now, the construction of a strongly Hamiltonian action of a central extension of $\mathfrak{g}$ from a Hamiltonian action $\mathfrak{g} \xrightarrow{\rho_H} \Gamma(TS)_H$, may be captured in the following commutative diagram:

$$\begin{array}{ccc} \Gamma(TS)_H & \Longleftarrow \mathcal{C}^\infty(S, \mathbb{R}) \longleftarrow & \mathbb{R} \\ \big\uparrow & \big\uparrow & \big\uparrow \\ \mathfrak{g} & \longleftarrow \quad \mathfrak{g}_\Gamma \longleftarrow & \mathbb{R} \end{array}$$

Given the Hamiltonian $\mathfrak{g}$-action (in green), one constructs the central extension and its strongly Hamiltonian action (in red) by pull-back.

[39] Landsman, op. cit., Proposition III.1.1.9, p. 182.

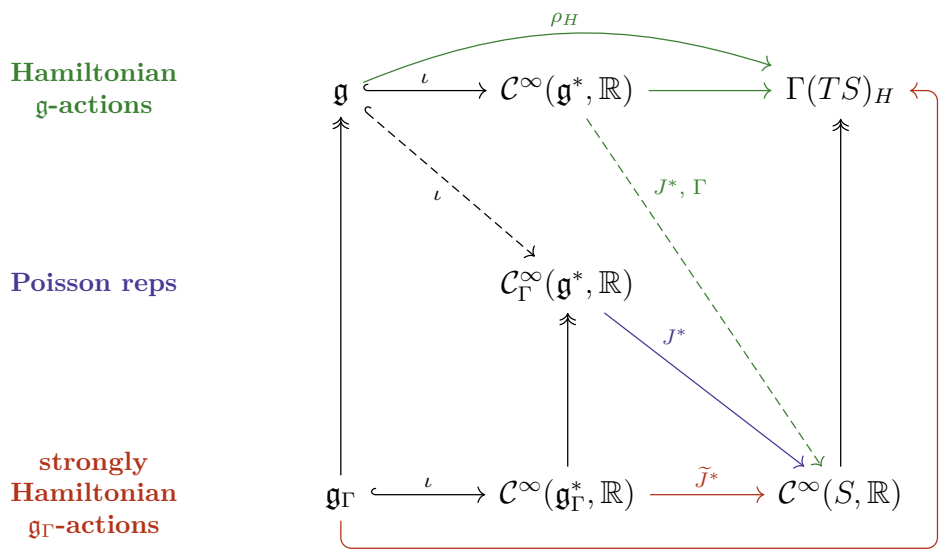

**Fig. III.2** – **Bypassing the obstruction of Hamiltonian actions.**
All diagrams are commutative and dashed arrows are the only ones which are not morphisms of Lie algebras. The bypassing of the obstruction can be visualized by the shifting of Γ as we move downwards throughout the diagram: from being a subscript characterizing the failure of an arrow to be a Lie algebra morphism (top line), it becomes a subscript characterizing a modification of the Poisson algebra (middle line), and finally a subscript of a modification of the initial Lie algebra (bottom line).[40]

---

[40]Let us pause for a moment, and adopt a more distant point of view on the local phenomenon we are discussing in order to connect it with more general discussions on the methodologies of contemporary mathematics. In his book *Synthetic Philosophy of Contemporary Mathematics*, Fernando Zalamea tries to pinpoint some of the "minimal characteristics" of contemporary mathematics which any philosophical approach of the subject should try to capture. Among them, there is the requirement of "presenting a full and faithful vision of mathematical practice, **particularly sensitive to a pendular weaving between transferences and obstructions, and between smoothings and residues**". He writes:

> [...] mathematical practice turns out to be much closer to a vision that genuinely and persistently seeks to detect, between minimal contexts of adequation, *both transferences and obstructions alike*. The notions of obstruction and residue are fundamental here, since the incessant survey of obstructions, and the reconstruction of entire maps of mathematics on the basis of certain residues attached to those obstructions, is part and parcel of both mathematical inventiveness and its subsequent demonstrative regulation. Now, the obstructions and residues acquire meaning only locally, with respect to certain contexts of adequation [...].

> (F. Zalamea. *Synthetic Philosophy of Contemporary Mathematics*. Trans. by Z. L. Fraser. Urbanomic/Sequence Press, 2012, pp. 127–128, author's empasis)

This passage fits particularly well with our present situation: we study *transferences* (from external structures to internal structures) and find *obstructions* to them. The *residue* attached to the obstruction is here the 2-cocycle. By including them in new *contexts of adequation* (central extensions of Lie algebras), the residues are *reinterpreted* and allow to *invent* new structures which *smooth out* the initial obstructions.

$* * * * *$

Given an external Lie algebra $\mathfrak{g}$, there are three ways of introducing it into the Classical Kinematical arena, which stem from the possibility of constructing three different Lie algebras out of the homogeneous symplectic manifold $(S, \omega)$. First, one can construct the Lie algebra $\Gamma(TS)_\omega$ of vector fields preserving the symplectic structure, and then consider morphisms $\mathfrak{g} \to \Gamma(TS)_\omega$. This yields the notion of Poisson $\mathfrak{g}$-actions. But these fail to take into account that physical transformations are generated by properties. Second, one can construct the Lie algebra $\Gamma(TS)_H$ of Hamiltonian vector fields, and then consider morphisms $\mathfrak{g} \to \Gamma(TS)_H$, called Hamiltonian $\mathfrak{g}$-actions. These now take into account the transformational role of properties but ignore their numerical role. Finally, one can use $\mathcal{C}^\infty(S, \mathbb{R})$ seen as a Lie algebra to consider morphisms $\mathfrak{g} \to \mathcal{C}^\infty(S, \mathbb{R})$. These define strongly Hamiltonian $\mathfrak{g}$-actions and they take into account the fundamental double role of physical properties. This is best seen from the fact they involve the full Jordan-Lie structure of the algebra of properties, as strongly Hamiltonian actions are equivalent to Poisson representations of $\mathcal{C}^\infty(\mathfrak{g}^*, \mathbb{R})$. Therefore, from the systematic consideration of all the several different ways of representing Lie algebras into the classical arena and the requirement of respecting the two-fold role of properties in Kinematics, we arrive at the following conclusion:

> As far as Lie groups and Lie algebras are concerned, strongly Hamiltonian actions should be the central objects through which to build the description of a classical system.

In fact, this will also include Hamiltonian actions since if $\mathfrak{g}$ is the central extension of some other Lie algebra $\mathfrak{h}$, then, by studying all possible strongly Hamiltonian $\mathfrak{g}$-actions one is automatically studying, among other things, all possible Hamiltonian $\mathfrak{h}$-actions. In other words, morally, Hamiltonian actions may be considered as induced actions of particular *subalgebras* of the initial external Lie algebra, the action of which is strongly Hamiltonian.

Therefore, the essential distinction, as Souriau insisted, is that between Poisson and Hamiltonian actions[41].

## III.2.2  In Quantum Kinematics

Let us now turn to the Quantum. Applied to the standard Hilbert space formalism, the general definition of group representation (cf. Definition III.1, page 245) yields the notion of *unitary representations*. These are morphisms of groups

$$G \xrightarrow{\ U\ } U(\mathcal{H})$$

where $U(\mathcal{H})$ is the group of all unitary operators on $\mathcal{H}$[42].

---

[41]As a side remark, I should mention the mathematical treatment of the obstruction of a Poisson action to be Hamiltonian. This allows to perceive how (un)restrictive this condition actually is. The main result is the following:

**Theorem III.3.** *The first cohomology group of $\mathfrak{g}$ with values in the first de Rham cohomology group of $S$ governs the obstruction to have Hamiltonian actions. In other words, any Poisson action is automatically Hamiltonian if and only if $H^1(\mathfrak{g}, \mathbb{R}) \otimes H^1_{dR}(S, \mathbb{R}) = 0$.*

That de Rham cohomology captures part of the obstruction for Hamiltonian actions is not surprising: the difference between Poisson and Hamiltonian $\mathfrak{g}$-actions is that, while the first represent $\mathfrak{g}$ as symplectic vector fields, the second represents it as Hamiltonian vector fields. In turn, if $\xi$ is a symplectic vector field, we have $d\omega(\xi, \cdot) = 0$ (it is a closed 1-form), whereas if $\xi$ is a Hamiltonian vector field, we have $\omega(\xi, \cdot) = df$ (it is an exact 1-form).

In particular, the theorem shows that there exist no Poisson actions which are not Hamiltonian whenever $H^1_{dR}(S, \mathbb{R}) = 0$ (this is the case of projective Hilbert spaces) or $[\mathfrak{g}, \mathfrak{g}] = \mathfrak{g}$ (this is the case of the Poincaré group).

For the details, see Iglesias-Zemmour, op. cit., p. 103 or Marsden and Ratiu, op. cit., pp. 370–371.

[42]It is tempting to write $U(\mathcal{H}) = Aut_{\mathcal{Hilb}}(\mathcal{H})$—that is, to view unitary operators on $\mathcal{H}$ as automorphisms of $\mathcal{H}$ in the category of Hilbert spaces where morphisms $\mathcal{H}_1 \to \mathcal{H}_2$ are isometries. However, this is not the category of Hilbert spaces usually considered. Instead, one chooses as morphisms continuous linear maps, and in this case the group of automorphisms of $\mathcal{H}$ is not $U(\mathcal{H})$. For an excellent exposition of the reasons leading to such a choice for the category $\mathcal{Hilb}$, see J. C. Baez. "Quantum Quandaries: a Category-Theoretic Perspective". In: *The Structural Foundations of Quantum Gravity*. Ed. by S. French, D. Rickles, and J. Saatsi. New York: Oxford University Press, 2006. URL: http://arxiv.org/abs/quant-ph/0404040. Notwithstanding this, I will sometimes write $Aut(\mathcal{H})$ for the group of unitaries in order to stress the similarities with Classical Kinematics.

Associated to a unitary *G*-representation is its infinitesimal version, called a *quantum* $\mathfrak{g}$-*representation*. This is a morphism of Lie algebras[43]

$$\rho : \mathfrak{g} \longrightarrow \big(\mathcal{B}_{i\mathbb{R}}(\mathcal{H}), \tfrac{1}{2}[\cdot, \cdot]\big).$$

Now, from the perspective of Hilbert spaces, one could naively think that unitary *G*-representations $G \xrightarrow{\ U\ } U(\mathcal{H})$ are the quantum analogue of Poisson *G*-actions $G \xrightarrow{\ L\ } Aut(S)$. However, we know from the preceding chapter that Hilbert spaces should not be viewed as one of the main mathematical structures of the quantum kinematical arena. Instead, the starting point should be either the projective Hilbert space $\mathbb{P}\mathcal{H}$ (geometric point of view, emphasis on states) or the JLB-algebra $\mathcal{B}_{\mathbb{R}}(\mathcal{H})$ (algebraic point of view, emphasis on properties)[44]. Therefore, the natural sort of morphisms to consider in the Quantum, which are analoguous to Poisson *G*-actions, are in fact *ray representations*:

$$G \xrightarrow{\ L\ } Aut(\mathbb{P}\mathcal{H}),$$

where $Aut(\mathbb{P}\mathcal{H})$ is the group of continuous maps of $\mathbb{P}\mathcal{H}$ into itself which preserve both the symplectic and Riemannian structures (cf. section II.2)[45].

Given this, an obvious question arises: If ray representations are the quantum analogue of Poisson actions in Classical Kinematics, which are the quantum analogues of Hamiltonian and strongly Hamiltonian actions?

---

[43]A word of caution however. The relation between *G*-representations and $\mathfrak{g}$-representations in the quantum case is quite delicate. There is first the problem that, when $\mathcal{H}$ is infinite-dimensional, $U(\mathcal{H})$ is not a Lie group (it is not a manifold). We have already mentioned this problem in chapter II and how to deal with it using Stone's theorem (cf. footnote 32, page 157). Morever, a unitary *G*-representation is usually not smooth over all of $\mathcal{H}$, fact which renders more difficult to define an associated $\mathfrak{g}$-action. This is nonetheless possible in a dense subset of $\mathcal{H}$ called the *essential G-smooth part of $\mathcal{H}$*. See, for example, Marsden and Ratiu, op. cit., pp. 322–323.

[44]In the first point of view, one then views $\mathcal{B}_{\mathbb{R}}(\mathcal{H})$ as the algebra of functions $\mathcal{C}^{\infty}(\mathbb{P}\mathcal{H}, \mathbb{R})_{\mathcal{K}}$. In the second point of view, one views $\mathbb{P}\mathcal{H}$ as the space of pure states $\mathcal{P}(\mathcal{B}_{\mathbb{R}}(\mathcal{H}))$.

[45]This of course was stressed from the very beginning by Weyl and Wigner. Cf., for example, this well-known passage of Weyl: "In quantum theory the representations take place in system space [in our terminology: space of states of the system]; but this is to be considered as a ray rather than a vector space, for a pure state is represented by a ray rather than a vector" (H. Weyl. *The Theory of Groups & Quantum Mechanics.* Trans. by H. Robertson. New York: Dover Publications, Inc., 1931, pp. 180–181).

### III.2.2.a   The quantum analogue of (strongly) Hamiltonian actions

To answer this, it helps to start by establishing the quantum representation problem, analogue of Figure III.1. Recall: the classical representation problem had stemmed from the existence of the following diagram of internal Lie algebras (cf. III.3, page 251):

$$\mathbb{R} \longhookrightarrow \mathcal{C}^\infty(S, \mathbb{R}) \xrightarrow{\ v_-\ } \Gamma(TS)_H \longhookrightarrow \Gamma(TS)_\omega.$$

It turns out that, in the Quantum, a similar diagram of internal *groups* exists:

$$U(1) \longhookrightarrow U(\mathcal{H}) \xrightarrow{\ p\ } Aut_U(\mathbb{P}\mathcal{H}) \longhookrightarrow Aut(\mathbb{P}\mathcal{H}). \tag{III.6}$$

Let us explain this diagram. First, a unitary operator $U \in U(\mathcal{H})$ induces a ray transformation $\mathtt{U} \in Aut(\mathbb{P}\mathcal{H})$ by $\mathtt{U}[\varphi] := [U\varphi]$, where $\varphi \in \mathcal{H}$ and $[\varphi] \in \mathbb{P}\mathcal{H}$. The set of all ray transformations which are induced by unitary operators on $\mathcal{H}$ is denoted $Aut_U(\mathbb{P}\mathcal{H})$. Now, two unitary operators differing only by a complex number will define the same ray transformation. Hence, elements of $Aut_U(\mathbb{P}\mathcal{H})$ are in fact what Bargmann calls "unitary operator rays"[46]—that is, elements of the quotient group $U(\mathcal{H})/U(1)$. In other words, we have the isomorphism $Aut_U(\mathbb{P}\mathcal{H}) \simeq U(\mathcal{H})/U(1)$. The left hand side of the diagram is then fairly obvious: it states that $U(1)$ is a subgroup of $U(\mathcal{H})$ and, since it is a normal subgroup, the quotient $U(\mathcal{H})/U(1)$ is itself a group.

Second—this is the more difficult part—we need to understand whether or not any particular ray transformation may be seen as stemming from some unitary operator on $\mathcal{H}$. In other words, we ask whether the injection $U(\mathcal{H})/U(1) \longhookrightarrow Aut(\mathbb{P}\mathcal{H})$ is also a surjection. This question was settled in 1931 by Wigner in his book *Group Theory and Its Applications to the Quantum Mechanics of Atomic Spectra*[47]. The proof of Wigner's theorem is also found in S. Weinberg. *The Quantum Theory of Fields*. Vol. 1. New York: Cambridge University Press, 1996, pp. 91–96.

---

[46]V. Bargmann. "On Unitary Ray Representations of Continuous Groups". In: *Annals of Mathematics* 59.1 (1954), pp. 1–46.

[47]For a clean and simple mathematical exposition of this, I nonetheless refer the reader to Bargmann's article *"On Unitary Ray Representations of Continuous Groups"*.

**Theorem** (WIGNER). *Consider the group $\widetilde{U}(\mathcal{H})$ of all unitary and anti-unitary operators on $\mathcal{H}$. Then, one has the isomorphism of groups:*

$$Aut(\mathbb{P}\mathcal{H}) \simeq \widetilde{U}(\mathcal{H})/U(1)^{[48]}.$$

In other words, to get all possible automorphisms of the quantum space of states, one needs to consider both unitary and anti-unitary operators. This may come as a surprise, for anti-unitary operators are scarcely ever used in non-relativistic Quantum Mechanics: the only relevant anti-unitary operator seems to be the time reversal operator[49] [50].

In any case, following exactly the same procedure used in Classical Kinematics, diagram (III.6) leads to the consideration of three possible strategies for representing external groups in the Quantum arena.

**Definition III.5.** For a given abstract group G,

– a **ray representation** is a morphism of groups $\;G \xrightarrow{\;L\;} Aut(\mathbb{P}\mathcal{H}),$

– a **projective ray representation** is a morphism of groups $\;G \xrightarrow{\;\mathtt{U}\;} Aut_U(\mathbb{P}\mathcal{H}),$

– a **unitary representation** is a morphism of groups $\;G \xrightarrow{\;U\;} U(\mathcal{H}).$

---

[48] $U$ is an anti-unitary operator on $\mathcal{H}$ if $U$ is anti-linear—that is, for $\varphi, \psi \in \mathcal{H}$ and $a, b \in \mathbb{C}$, one has $U(a\varphi + b\psi) = \bar{a}U\varphi + \bar{b}U\psi$—and preserves the hermitian product ($\langle U\varphi, U\psi \rangle = \langle \varphi, \psi \rangle$).

Wigner's theorem is usually stated as: given a one-to-one transformation of the projective Hilbert space onto itself which preserves the transition probabilities, there exists a unique (up to a factor of modulus 1) unitary or anti-unitary operator on $\mathcal{H}$ which extends this ray transformation.

[49] It is quite easy to be convinced of the fact that the time reversal operator $\Theta$ must be anti-unitary. Indeed, if $\boldsymbol{x}$ and $\boldsymbol{p}$ are the position and linear momentum operators, one should have $\Theta\boldsymbol{x}\Theta^\dagger = \boldsymbol{x}$ and $\Theta\boldsymbol{p}\Theta^\dagger = -\boldsymbol{p}$. Applying this to the commutation relations $[\boldsymbol{x}, \boldsymbol{p}] = i$ leads to

$$\Theta i \Theta^\dagger = \Theta[\boldsymbol{x}, \boldsymbol{p}]\Theta^\dagger = -[\boldsymbol{x}, \boldsymbol{p}] = -i.$$

Hence $\Theta$ must be anti-linear.

[50] The dissymmetry between unitary and anti-unitary operators in Quantum Mechanics seems to originate in the following fact: the square of a unitary *or anti-unitary* operator is a unitary operator. Because of this, all elements of a Lie group $G$ which are connected to the identity are necessarily represented by unitary operators. To prove this, one uses the followings two facts (cf. ibid., p. 2):

1. There exists a neighborhood $\mathcal{N}$ of the identity such that: *i)* every group element in $\mathcal{N}$ is the square of some element, and *ii)* every group element connected to the identity can be written as a finite product of elements in $\mathcal{N}$.

2. The square of a unitary or anti-unitary ray operator is a unitary ray operator.

Therefore, for *connected* Lie groups, one can safely ignore anti-unitary operators.

Comparison of diagrams (III.3) and (III.6) suggests quite a different analogy from the naive one proposed by focusing on the Hilbert space formalism: the morphisms of groups $G \xrightarrow{\mathbb{U}} Aut_U(\mathbb{P}\mathcal{H})$ and $G \xrightarrow{U} U(\mathcal{H})$ should be thought as the quantum analogue of the morphisms of Lie algebras $\mathfrak{g} \xrightarrow{\rho_H} \Gamma(TS)_H$ and $\mathfrak{g} \xrightarrow{\widehat{J}} \mathcal{C}^\infty(S, \mathbb{R})$ respectively. Put differently, this approach hints at the following analogy:

| **Classical Kinematics** | **Quantum Kinematics** |
| --- | --- |
| Hamiltonian $\mathfrak{g}$-actions | projective ray $G$-representations |
| strongly Hamiltonian $\mathfrak{g}$-actions | unitary $G$-representations |
| co-momentum map $\mathfrak{g} \xrightarrow{\widehat{J}} \mathcal{C}^\infty(S, \mathbb{R})$ | unitary map $G \xrightarrow{U} U(\mathcal{H})$ |

**Table III.1** – The quantum analogue of (strongly) Hamiltonian actions

Now, since strongly Hamiltonian $\mathfrak{g}$-actions may equivalently be seen as representations of a classical algebra of properties constructed out of $\mathfrak{g}$ (cf. page 252), one may wonder if it is also possible to perceive unitary $G$-representations as being induced by representations of a certain quantum algebra of properties constructed out of $G$. The answer is positive, for besides the usual Hilbert space approach to unitary representations, there exist two other different yet equivalent points of view:

– from the *perspective of $C^*$-algebras*: to any group $G$, one can associate a $C^*$-algebra, called the *group $C^*$-algebra* and denoted by $C^*(G)$[51]. Then, a unitary

---

[51]It is defined as follows. For $f, g \in \mathcal{C}^\infty(G, \mathbb{C})$, respectively define the *convolution* and *involution* products by

$$f * g(x) := \int\limits_G dy f(xy^{-1}) g(y) \; ; \; f^*(x) := \overline{f(x^{-1})}.$$

This turns $\mathcal{C}^\infty(G, \mathbb{C})$ into a *-algebra. Then, define $C^*(G)$ as the norm closure of this convolution algebra in the norm $\|f\| := \sup_\pi \|\pi(f)\|$, where $\pi : \mathcal{C}^\infty(G, \mathbb{C}) \to \mathcal{B}(\mathcal{H})$ is a bounded non-degenerate representation of $G$ on $\mathcal{H}$ is equivalent to a non-degenerate representation of $C^*(G)$ on $\mathcal{H}$—that is, to a morphism of $C^*$-algebras

$$C^*(G) \xrightarrow{\ \pi\ } \mathcal{B}(\mathcal{H}).$$

In particular, irreducible unitary representations of $G$ correspond to irreducible representations of the group $C^*$-algebra[52].

– from the *perspective of JLB-algebras*: this is a trivial reformulation of the previous point of view. One defines the Jordan-Lie-Banach algebra $JL(G) := C^*(G)_{\mathbb{R}}$ and uses the fact that any morphism of $C^*$-algebras restricts to a morphism of JLB-algebras of the self-adjoint parts (and viceversa). Thus, a unitary (irreducible) representation of $G$ is equivalent to an (irreducible) representation of $JL(G)$ on $\mathcal{H}$—that is, to a morphism of JLB-algebras[53]:

$$JL(G) \xrightarrow{\ \pi\ } \mathcal{B}_{\mathbb{R}}(\mathcal{H}).$$

Hence, we see that, indeed, unitary $G$-representations arise from taking into account the double role of quantum properties. The quantum representation problem, analogue of Figure III.1, may then be depicted by the diagram below (Figure III.3).

In the spirit of the comment made immediately after Figure III.1, an important point should be remarked here: the proposed analogy foreshadows *a surprising merger between the physical process of quantization—from the Classical to the Quantum—and the mathematical process of integration—from the Local to the Global*. Whereas the Poisson algebra $\mathcal{C}^\infty(\mathfrak{g}^*, \mathbb{R})$ associated to a Lie group $G$ depends only on the infinitesimal information $\mathfrak{g}$, the group JLB-algebra $JL(G)$ does encode global information:

---

morphism of associative *-algebras. The group $C^*$-algebra is commutative if and only if the group is abelian. (Landsman, op. cit., Definition III.1.7.4, p. 204.)

[52] See ibid., Corollary III.1.7.5, p. 204. In this way, the unitary dual $\widehat{G}$ of any group may be described as the set of symplectic leaves of the pure state space of $C^*(G)$: $\widehat{G} = \mathcal{P}(C^*(G))/\sim$. In particular, if $G$ is abelian, $\widehat{G} = \mathcal{P}(C^*(G))$, or, what amounts to the same, $C^*(G) = \mathcal{C}_0(\widehat{G}, \mathbb{C})$.

[53] This is a point I continue to stress: all mentions of $C^*$-algebras in Quantum Mechanics can systematically be replaced by an analogous statement in terms of *real* JLB-algebras. By the same token, one can develop the whole of Quantum Mechanics without using complex numbers. This is a point Landsman repeats many times (for example, in *"Classical and Quantum Representation Theory"*, pp. 2 and 17).

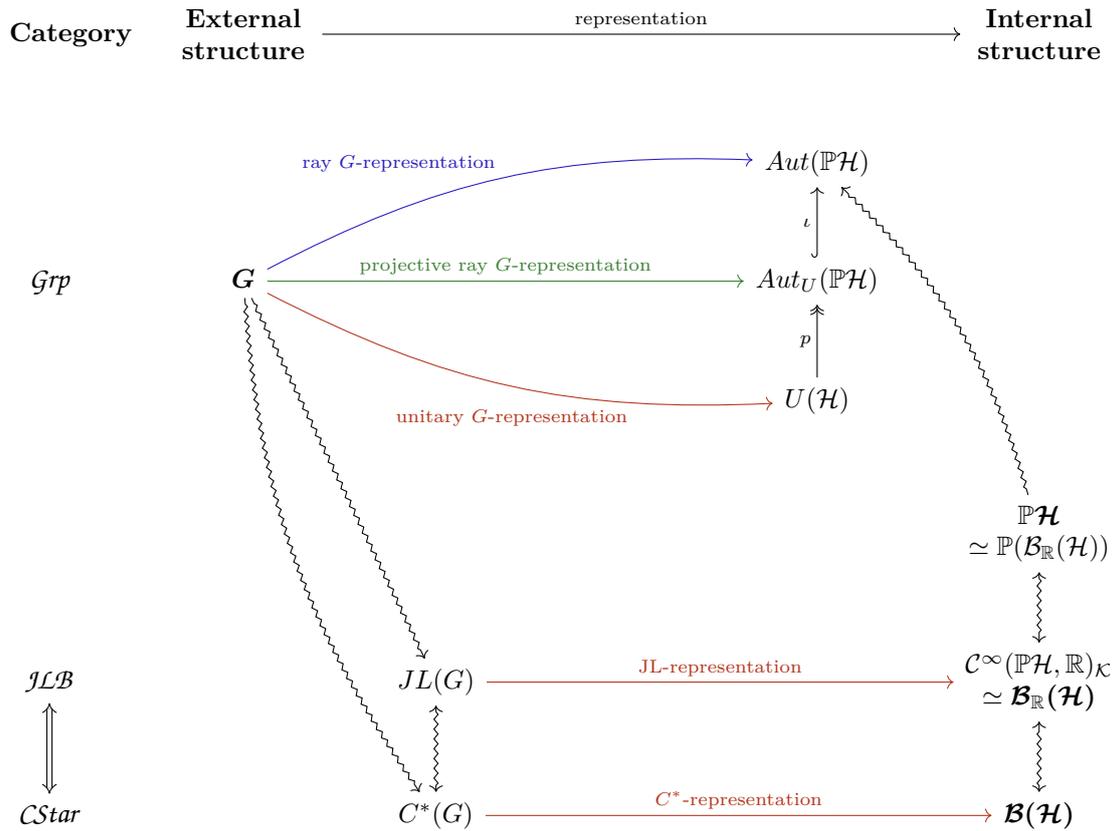

**Fig. III.3 – The Quantum representation problem for Lie groups**.
On the top left and bottom right of the 'diagram' are the initial external and internal structures. Squiggle arrows represent intrinsic constructions of new structures (some of which, but not all, being functorial), while all other arrows represent morphisms in the appropriate category. The coloured arrows represent the various a priori different transits between the external and the internal structures.

two non-isomorphic groups with same Lie algebra will have non-isomorphic group algebras. In fact, as we will progressively unveil during the remainder of the chapter, this seems to capture a deep difference between Classical and Quantum Kinematics: the former is attached to infinitesimal transformations, while the latter is attached to global transformations.

### III.2.2.b  Pursuing the analogy: unitary vs. projective representations

A first strategy for exploring this analogy is to analyze the quantum concepts through the looking glass of symplectic geometry. Recall that, from the geometric point of view, the quantum space of states may be perceived as a classical space of states with extra structure (section II.2). In particular, the quantum space of states is a symplectic manifold and it is always enlightening, for the comparison of both Kinematics, to treat it as if it were a classical space of states.

Consider then a unitary representation $G \xrightarrow{\ U\ } U(\mathcal{H})$. It induces a ray representation through the following diagram of groups

$$G \xrightarrow{\ U\ } U(\mathcal{H}) \xrightarrow{\ p\ } Aut_U(\mathbb{P}\mathcal{H}) \xhookrightarrow{\ \iota\ } Aut(\mathbb{P}\mathcal{H}).$$
$$\underset{\mathtt{U}}{\underbrace{\phantom{\hspace{6cm}}}}$$

In order to distinguish the map $U$ from the map $\mathtt{U}$, let us call the latter a ***unitary ray representation***. Since $Aut(\mathbb{P}\mathcal{H}) \hookrightarrow Symp(\mathbb{P}\mathcal{H})$, $\mathtt{U}$ is a Poisson $G$-action. The question, of course, is which kind of Poisson action it is, and the answer is the expected one: it is a strongly Hamiltonian $G$-action!

The idea of the proof is quite simple: given the unitary representation $U$, the associated infinitesimal version is a map $dU : \mathfrak{g} \longrightarrow \mathcal{B}_{\mathbb{R}}(\mathcal{H})$. But we also have $\mathcal{B}_{\mathbb{R}}(\mathcal{H}) \simeq \mathcal{C}^{\infty}(\mathbb{P}\mathcal{H}, \mathbb{R})_{\mathcal{K}}$ (cf. Equation II.17). Therefore, a unitary representation allows to generate the diagram of Lie algebras

$$\mathfrak{g} \xrightarrow{\ dU\ } \mathcal{B}_{\mathbb{R}}(\mathcal{H}) \simeq \mathcal{C}^{\infty}(\mathbb{P}\mathcal{H}, \mathbb{R})_{\mathcal{K}} \subset \mathcal{C}^{\infty}(\mathbb{P}\mathcal{H}, \mathbb{R}) \xrightarrow{\ v_-\ } \Gamma(T\mathbb{P}\mathcal{H})_H$$
$$\underset{\rho_H}{\underbrace{\phantom{\hspace{8cm}}}}$$

which corresponds exactly to the definition of a strongly Hamiltonian $\mathfrak{g}$-action on $\mathbb{P}\mathcal{H}$ (cf. page 251). This shows that, when perceived from the perspective of symplectic geometry, unitary representations are strongly Hamiltonian actions. Moreover, it clarifies the relation between the co-momentum map in Classical Kinematics and the unitary representation in Quantum Kinematics. Indeed, the co-momentum map of the unitary ray representation $\mathtt{U}$ is simply the derivative of the unitary map: $\widehat{J} = dU$[54].

In this way, this symplectic approach allows to relate, in a very transparent fashion, the central group-theoretical notion of Classical Kinematics (strongly Hamiltonian actions) to unitary representations in Quantum Kinematics. In fact, we have the following characterization of unitary ray representations among all strongly Hamiltonian

---

[54]Because of the technical difficulties alluded to in footnote 15 (page 246), this does not constitute a rigorous proof. A careful derivation is found in Marsden and Ratiu, op. cit., pp. 376–377 and 394–395. In the same vein, it is possible to prove that projective ray representations yield Hamiltonian actions.

actions on $\mathbb{P}\mathcal{H}$:

**Proposition III.4.** *A strongly Hamiltonian $\mathfrak{g}$-action on $\mathbb{P}\mathcal{H}$ stems from a unitary ray $G$-representation if and only if*

  *i)  it acts by infinitesimal isometries (Killing vectors),*

 *ii)  it is integrable.*

Condition *i)* appeals to the non-trivial structure distinguishing the Quantum arena from the Classical one. Condition *ii)* suggests a new difference between the Classical and the Quantum, which was not perceived at the level of the homogeneous arenas.

This notwithstanding, a clearer manifestation that Table III.1 is indeed the correct analogy is to study more closely the difference between unitary $G$-representations and projective ray representations, while comparing it with the relation between strongly Hamiltonian and Hamiltonian $\mathfrak{g}$-actions explored in the previous section. As we will now see, the analysis is strikingly similar to that conducted in subsection III.2.1.b.

Recall: a unitary representation $G \xrightarrow{\;U\;} U(\mathcal{H})$ induces a ray representation through the following diagram of groups

$$G \xrightarrow{\;U\;} U(\mathcal{H}) \xrightarrow{\;p\;} Aut_U(\mathbb{P}\mathcal{H}) \xleftarrow{\;\iota\;} Aut(\mathbb{P}\mathcal{H}). \qquad (\text{III.7})$$

This should be compared with diagram (III.4, page 256) defining strongly Hamiltonian actions: the morphism of groups $U$ inducing the unitary ray representation $\mathtt{U}$ is the quantum analogue of the classical infinitesimally equivariant co-momentum map inducing the strongly Hamiltonian action $\rho_H$.

On the other hand, given a projective ray representation $\mathtt{U}$ of a Lie group, it is always possible to construct a map $U : G \longrightarrow U(\mathcal{H})$ by picking a representative for each unitary operator ray $\mathtt{U}(g)$. However, in general $U$ will fail to be a morphism of groups (and hence a unitary $G$-representation). Indeed, the relation $\mathtt{U}(g)\mathtt{U}(h) = \mathtt{U}(gh)$ implies only $U(g)U(h) = m(g,h)U(gh)$, where $m(g,h) \in U(1)$. A map $U$ verifying this condition is called a *projective representation* of $G$ on $\mathcal{H}$. Projective ray representations are thus defined by the existence of the following diagram:

$$G \dashrightarrow^{U} U(\mathcal{H}) \xrightarrow{\ p\ } Aut_U(\mathcal{H}) \xhookrightarrow{\ \iota\ } Aut(\mathbb{P}\mathcal{H}). \qquad \text{(III.8)}$$

This is the quantum analogue of diagram (III.5, page 256) defining Hamiltonian actions, and all the comments below it equally apply here.

Indeed, there are two key differences between diagrams (III.7) and (III.8). The first obvious one—already mentioned—is that $U$ fails to be a morphism of Lie groups for general projective ray representations (in the same way that the co-momentum map fails to be a morphism of Lie algebras for general Hamiltonian actions). The second difference lies in the possible transits between the morphisms $U$ and $\mathtt{U}$. For unitary ray representations, one can either: *start* from the data $U$ and *deduce* the data $\mathtt{U}$ (one defines it as $\mathtt{U} := p \circ U$), or start from the unitary ray representation and deduce the morphism $U$ (the unitary representation on $\mathcal{H}$ is *uniquely* defined for a unitary ray representation). This freedom of circulation breaks down for projective ray representations: given a map $G \xrightarrow{U} U(\mathcal{H})$, the induced map $p \circ U$ is no longer guaranteed to be a morphism of Lie groups. Moreover, given a ray representation, one cannot associate to it one single map $G \xrightarrow{U} U(\mathcal{H})$ but rather a *class* of such maps (the representative $U(g)$ of $\mathtt{U}(g)$ is *not* uniquely defined). In other words, whereas unitary ray representations of $G$ on $\mathbb{P}\mathcal{H}$ and unitary representations of $G$ on $\mathcal{H}$ are equivalent (one can freely circulate between the two), the notion of ray representation is prior to the "non-equivariant" map $U$ (one must start from the data $G \xrightarrow{\mathtt{U}} Aut(\mathbb{P}\mathcal{H})$)[55].

With this at hand, we can now describe the obstruction for projective ray representations to be unitary[56]:

**Theorem III.5.** *The second cohomology group of $G$ with values in $U(1)$ governs the obstruction to have unitary ray representations. In other terms, any projective ray representation is unitary if and only if $H^2(G, U(1)) = 0$.*[57]

---

[55]Notice how these last ten lines are, literally, a copy-paste of the paragraph following diagram (III.5) on page 256.

[56]Again, I will develop part of the formalism in the footnotes in order to stress the general lines of the investigation. All the mathematical details may be found in Landsman, op. cit., pp. 187–197.

This is to be compared with Theorem III.2 (page 257). As in the classical case, there are two possible strategies for bypassing this obstruction, each of them enforcing the analogy between unitary (respectively projective) ray representations on the Quantum side and strongly Hamiltonian (respectively Hamiltonian) actions on the Classical side:

1. *Absorb the obstruction by modifying the group $C^*$-algebra $C^*(G)$.* From the algebraic point of view, a projective representation of $G$ on $\mathcal{H}$ is seen as arising from a map

$$C^*(G) \xrightarrow{\ \pi\ } \mathcal{B}(\mathcal{H})$$

which fails to be a morphism of $C^*$-algebras. Instead, one can modify the convolution and involution products and define a new $C^*$-algebra in such a way that

---

[57]See ibid., Proposition III.1.5.2., p. 197. Given a diagram of the form (III.8), the failure of $U$ to be a morphism of Lie groups is measured by the function

$$m : G \times G \longrightarrow U(1)$$
$$(g, h) \longmapsto m(g, h) := U(g)U(h)U(gh)^{-1}$$

Because of the associativity of composition and the existence of an identity element in both $G$ and $U(\mathcal{H})/U(1)$, the function $m$ satisfies the properties
  i)  $m(g, h)m(gh, k) = m(g, hk)m(h, k)$ for all $g, h, k \in G$,
  ii) $m(e, g) = m(g, e) = 1$ for all $g \in G$.
A function $m : G \times G \longrightarrow U(1)$ satisfying these two properties is called a *multiplier*. The set of all multipliers is denoted by $Z^2(G, U(1))$.

But the map $U$ is not uniquely determined by the projective ray representation (each $U(g)$ is defined up to an element of $U(1)$), and thus $Z^2(G, U(1))$ still does not capture the obstruction to have unitary ray representations. Indeed, consider a different choice: $U'(g) = b(g)U(g)$ with $b(g) \in U(1)$. Then, we have

$$m'(g, h) = U'(g)U'(h)U'(gh)^{-1} = \frac{b(g)b(h)}{b(gh)}m(g, h)$$

Two multipliers satisfying this equation are said to be *equivalent*. A multiplier $m$ for which there exists $b : G \longrightarrow U(1)$ such that $m(g, h) = \frac{b(gh)}{b(g)b(h)}$ is said to be *trivial* and their set is denoted $B^2(G, U(1))$. The last equation shows that, whenever $m$ is trivial, one can redefine the map $U$ associated to the ray representation $\mathtt{U}$ in such a way that $U$ is a unitary representation.

Therefore, the object intrinsically associated to a given projective representation—that is, an element of

$$H^2(G, U(1)) := Z^2(G, U(1))/B^2(G, U(1)).$$

This is called the *second cohomology group* of $G$ with values in $U(1)$.

the map $\pi$ appears as a morphism of $C^*$-algebras[58]

$$C^*(G, m) \xrightarrow{\ \pi\ } \mathcal{B}(\mathcal{H}).$$

From this perspective, we have the equivalence of notions[59]

$$
\begin{array}{ccc}
\textit{projective ray} & & \\
\textit{G-representation} & \Longleftrightarrow & \textit{C}^*\textit{-algebra} \\
\textit{with multiplier m} & & \textit{C}^*(G, m)
\end{array}
\Longleftrightarrow
\begin{array}{c}
\textit{JLB-algebra} \\
\textit{representation of} \\
\textit{JL}(G, m)
\end{array}
$$

2. *Absorb the obstruction by modifying the initial Lie group $G$.* One reinterprets the locus of the obstruction—that is, the second cohomology group $H^2(G, U(1))$—as the *classifier of central extensions* of $G$ by $U(1)$. Given a projective representation of $G$ on $\mathcal{H}$, this allows to construct a new Lie group $G_m$ which has a unitary representation on $\mathcal{H}$[60]. From this perspective, we have the equivalence

---

[58]For the *twisted* group $C^*$-algebra $C^*(G, m)$, the convolution and involution products are defined as follows (see ibid., p. 202):

$$f * g(x) := \int_G dy\, m(xy^{-1}, y) f(xy^{-1}) g(y),$$
$$f^*(x) := \overline{m(x, x^{-1}) f(x^{-1})}.$$

For two equivalent multipliers $m$ and $m'$, the associated twisted group $C^*$-algebras are isomorphic.

[59]Ibid., Corollary III.1.7.5, p. 204.

[60]A central extension $G_m$ of $G$ by $U(1)$ is a short exact sequence $e \to U(1) \to G_m \to G \to e$ such that, in addition, $U(1)$ is contained in the center of $G_m$. This implies $G_m$ is a $U(1)$-principal bundle over $G$.

Two central extensions $G_m$ and $G_n$ are said to be equivalent if there exists an isomorphism of Lie groups $G_m \longleftrightarrow G_n$ such that the following diagram commutes

$$
e \longrightarrow U(1)
\begin{array}{c}
\nearrow\ G_m\ \searrow \\
\downarrow \\
\searrow\ G_n\ \nearrow
\end{array}
G \longrightarrow e
$$

It can be shown that there is a one-to-one correspondence between equivalence classes of $U(1)$-central extensions of $G$ and the set $H^2(G, U(1))$ of equivalence classes of multipliers on $G$ (see Tuynman and Wiegerinck, op. cit., proposition 3.4, p. 6).

Now, the construction of a unitary representation of a central extension of $G$ from a projective ray representation $G \xrightarrow{\mathrm{u}} Aut_U(\mathbb{P}\mathcal{H})$, may be captured in the following commutative diagram:

$$
\begin{array}{ccccc}
Aut_U(\mathbb{P}\mathcal{H}) & \longleftarrow & U(\mathcal{H}) & \longleftrightarrow & U(1) \\
\uparrow & & \uparrow & & \uparrow \\
G & \longleftarrow & G_m & \longleftrightarrow & U(1)
\end{array}
$$

of notions[61]

<div style="text-align:center">

*projective ray representation*        *unitary $G_m$-representation in*
*of G with multiplier m* $\Longleftrightarrow$ *which the center acts trivially*

</div>

Both strategies show that projective ray representations fall under the representation theory of JLB-algebras. Moreover, through the second move, the study of all projective ray representations of a certain group $G$ can be transformed into the problem of studying all unitary ray representations of $G$ and its $U(1)$-central extensions, in exact analogy with the translation of the question of studying all Hamiltonian $\mathfrak{g}$-actions into the question of studying all strongly Hamiltonian actions of $\mathfrak{g}$ and its $\mathbb{R}$-central extensions.

<div style="text-align:center">

\* \* \* \* \*

</div>

We have then arrived at the following two proposals:

**Classical System 2.** A classical system is characterized by a strongly Hamiltonian $\mathfrak{g}$-action—that is, by the data of a triple $S_{\mathfrak{g}} := (S, \mathfrak{g}, \widehat{J})$, where $S$ is an abstract symplectic manifold, $\mathfrak{g}$ is an abstract Lie algebra and $\widehat{J} : \mathfrak{g} \longrightarrow \mathcal{C}^{\infty}(S, \mathbb{R})$ is a morphism of Lie algebras called the (infinitesimally equivariant) co-momentum map.

**Quantum System 2.** A quantum system is characterized by a unitary (ray) $G$-representation—that is, by the data of a triple $\mathcal{H}_G := (\mathcal{H}, G, U)$, where $\mathcal{H}$ is an abstract Hilbert space, $G$ is an abstract (Lie) group and $U : G \longrightarrow U(\mathcal{H})$ is a morphism of groups.

As it was the case for the first naive definitions proposed (cf. p. 139), we now need to assess whether these more sophisticated versions may or may not be regarded as acceptable candidates for the mathematical description of a physical system. In particular, we need to understand whether or not these new abstract structures meet the requirement of individuation (page 17).

---

Given the projective ray representation (in green), one constructs the central extension and its strongly Hamiltonian action (in red) by pull-back. (Compare this discussion with footnote 38, page 259.)

[61]Landsman, op. cit., Proposition III.1.5.1.

# III.3 Introducing discernibility through groups (2): the individuation problem

As we concluded at the end of the first chapter, to investigate the amount of individuation within an abstract structure $\mathcal{S}$ one should study the action of the automorphism group $Aut(\mathcal{S}) \circlearrowright \mathcal{S}$ (cf. page 135). Therefore, our first task is to determine the group of automorphisms for the structures $\mathcal{H}_G$ and $S_{\mathfrak{g}}$. This turns out to be a delicate matter, as we will presently see.

## III.3.1 Identity and the group of automorphisms

### III.3.1.a The quantum case

Let us start by discussing the situation in Quantum Kinematics. Perhaps, the naive expectation would be the following: by passing from a bare abstract Hilbert space $\mathcal{H}$ to a unitary $G$-representation $\mathcal{H}_G$, we reduce the group of automorphisms from $U(\mathcal{H})$ to $U(G)$. Put differently, one could imagine that endowing an abstract Hilbert space $\mathcal{H}$ with a morphism $G \xrightarrow{\;U\;} U(\mathcal{H})$ is a way of selecting, among all available automorphisms $(U(\mathcal{H}))$, those that are to be considered 'physically meaningful' $(U(G))$. Since the problem with the quantum kinematical arena was, essentially, that the group of automorphisms was too big—$Aut(\mathcal{H}) \circlearrowright \mathbb{P}\mathcal{H}$ is a transitive action—, this reduction from $U(\mathcal{H})$ to $U(G)$ should then represent a step forward in dealing with the problematic homogeneity of the space of states[62].

A minute of reflexion shows however that the identity $Aut(\mathcal{H}_G) = U(G)$ cannot be true in general. To see this, simply consider the trivial representation, for which $U(G) = Id$. This should be tantamount to not introducing the group at all, and thus should have no impact on the homogeneity of the quantum space of states. Yet, according to the above identity, by endowing an abstract Hilbert space with such a $G$-representation, we would mysteriously pass from a completely homogeneous structure

---

[62]This was at least *my* initial expectation, as is manifest in my article *"The Mathematical Description of a Generic Physical System"*, p. 346. And it certainly was quite a naive thought...

where no elements can be individuated to a rigid structure where all elements can be individuated: indeed, if $Aut(\mathcal{H}_G) = U(G)$ were true, we would have $Aut(\mathcal{H}_G) = Id$ for the trivial representation.

This only shows that we must proceed with due care in the determination of the group of automorphisms of the structure $\mathcal{H}_G$, which, in turn, should follow from reflecting on the appropriate notion of morphism for this kind of structures. What is then the good definition of the "category of unitary representations"? The following definition is often found in the literature:

**Definition III.6.** The **category $\mathcal{URep}(G)$ of $G$-unitary representations** has $G$-unitary representations $(\mathcal{H}, G, U)$ as objects and $G$-equivariant linear maps as morphisms. In other words, a morphism $(\mathcal{H}, G, U) \overset{\phi}{\longrightarrow} (\mathcal{H}', G, U')$ is given by a linear map $\mathcal{H} \overset{\phi}{\longrightarrow} \mathcal{H}'$ such that, for all $g \in G$, the following diagram commutes:

$$
\begin{array}{ccc}
\mathcal{H} & \overset{\phi}{\longrightarrow} & \mathcal{H}' \\
{\scriptstyle U(g)}\downarrow & & \downarrow{\scriptstyle U'(g)} \\
\mathcal{H} & \underset{\phi}{\longrightarrow} & \mathcal{H}'
\end{array}
$$

Such a map is usually called an *intertwiner* between the unitary representations $U$ and $U'$[63].

Given this definition, we see that an isomorphism of unitary $G$-representations corresponds to the usual notion of equivalence. Therefore, conceived as an object of the category $\mathcal{URep}(G)$, the group of automorphisms of the structure $\mathcal{H}_G$—sometimes called the group of "symmetries of the unitary representation"[64]—is

$$Aut_{\mathcal{URep}(G)}(\mathcal{H}_G) = \{G\text{-equivariant unitary operators of } \mathcal{H}\}. \tag{III.9}$$

---

[63]For this definition (in the case of linear representations) see for example C. Procesi. *Lie Groups: An Approach through Invariants and Representations.* New York: Springer, 2007, p. 12.

That this is a natural category to consider is further supported if one recalls that groups may themselves be seen as categories (with only one object and all arrows being isomorphisms). Indeed, from this perspective a unitary $G$-representation is a functor from $G$ to $\mathcal{Hilb}$ and the obvious choice for the category $\mathcal{URep}(G)$ should then be the category of functors $\mathcal{Hilb}^G$ where morphisms are natural transformations between functors. But the definition of an intertwiner is precisely that of a natural transformation between the functors $U$ and $U'$ and hence both definitions for the category $\mathcal{URep}(G)$ coincide.

[64]Ibid., p. 3.

This may be rephrased by remarking that an automorphism of the structure $\mathcal{H}_G$ is simply a unitary operator on $\mathcal{H}$ which commutes with any $U(g)$. Thus, we can also write:

$$Aut_{\mathcal{U}\mathcal{R}ep(G)}(\mathcal{H}_G) = \text{ centralizer of } U(G) \text{ in } U(\mathcal{H})^{[65]}.$$

Let us explore some consequences of this result to evaluate its soundness. First, we see that if $G \xrightarrow{U} U(\mathcal{H})$ is the trivial representation, then $Aut_{\mathcal{U}\mathcal{R}ep(G)}(\mathcal{H}_G) = U(\mathcal{H})$ and the homogeneity of the quantum space of states has not been broken whatsoever, as it should be. Second, if $U$ is an *irreducible* representation, the automorphism group gets severely reduced: by Schur's lemma we have $Aut_{\mathcal{U}\mathcal{R}ep(G)}(\mathcal{H}_G) = U(1)$, which in turn projects into the trivial group when passing to the projective space $\mathbb{P}\mathcal{H}^{[66]}$. Therefore, if Definition III.6 is correct, by considering irreducible unitary $G$-representations on $\mathcal{H}$ instead of considering bare Hilbert spaces, we manage to pass from a *transitive* action $Aut(\mathcal{H}) \circlearrowleft \mathbb{P}\mathcal{H}$ to a *trivial* action $Aut(\mathcal{H}_G) \circlearrowleft \mathbb{P}\mathcal{H}$. In this case, all states of the physical system described by the structure $\mathcal{H}_G$ are stable under the action of the automorphism group and reveal themselves to be qualitatively discernible individuals, as we so wish.

There is a certain evident appeal in the fact that *irreducible* representations appear as precisely those quantum structures satisfying the requirement of individuation. The idea that quantum systems should be described by these particular structures cannot but resonate with "Wigner's definition" of quantum elementary particles as irreducible unitary representations of the Poincaré group[67]. Unfortunately, there is—yet again—a

---

[65] Given a group $G$ and a *subset* $S \subset G$, the centralizer of $S$ in $G$ is defined as $Z_G(S) := \{ g \in G | \forall s \in S, gs = sg \}$. The centralizer is necessarily a group.

[66] **Schur's Lemma:** Let $G$ be a Lie group and consider an intertwiner $\phi$ between two irreducible unitary representations $U$ and $U'$. Then either $\phi = 0$ or $\phi$ is an equivalence.

From this one concludes that any non-vanishing intertwiner of an irreducible representation with itself must be proportional to the identity map. The proof of both results may be found in any book on group representation theory. See for example A. W. Knapp. *Lie Groups Beyond an Introduction.* 2nd ed. Boston: Birkhäuser, 2002, p. 240.

[67] Although Wigner did not explicitly define elementary particles in this way, it is attributed to him. This is discussed in some detail by Yuval Ne'eman and Shlomo Sternberg. They write:

Ever since the fundamental paper of Wigner on the irreducible representations of the Poincaré group, it has been a (perhaps implicit) definition in physics that an elementary particle "is" an irreducible representation of the group, $G$, of "symmetries of nature". (Y. Ne'eman and S. Sternberg. "Internal Supersymmetry and Superconnections". In: *Symplectic Geometry and Mathematical Physics: Actes du colloque en l'honneur de Jean-Marie Souriau.* Ed. by P. Donato et al. Boston: Birkhäuser, 1991, pp. 326–354, p. 327.)

serious problem with this idea.

Indeed, were it to be true, we would have at our disposal two drastically different manners in which to conceive an abstract projective Hilbert space $\mathbb{P}\mathcal{H}_n$. If, on the one hand, one decides to consider an $n$-dimensional Hilbert space as an object of the category $\mathcal{Hilb}$, then $\mathbb{P}\mathcal{H}_n$ would appear as a homogeneous structure: the automorphism group $U(\mathcal{H}_n) \simeq U(n)$ acts transitively on $\mathbb{P}\mathcal{H}_n$. On the other hand, one could use the fact that $U(\mathcal{H}_n)$ is represented on $\mathcal{H}_n$ (the representation being the tautological triple $(\mathcal{H}_n, U(\mathcal{H}_n), Id)$) in order to perceive $\mathcal{H}_n$ as an object of the category $\mathcal{URep}(U(\mathcal{H}_n))$. In this case, since the representation is obviously irreducible, its automorphism group $Aut_{\mathcal{URep}(U(\mathcal{H}_n))}(\mathcal{H}_n)$ is simply $U(1)$ and now $\mathcal{H}_n$ would show to be a rigid structure whose automorphism group acts trivially on $\mathbb{P}\mathcal{H}_n$. It would therefore appear that the characteristics of the projective Hilbert space depend crucially on the *choice* of the category to which it belongs. And this would be an embarrassing situation: if the homogeneity of $\mathbb{P}\mathcal{H}$ is not an intrinsic feature of the structure but depends on the arbitrary choice of a point of view—should we see $U(\mathcal{H}) \circlearrowright \mathcal{H}$ as the action of the automorphism group on the structure defining the group, or rather as the irreducible representation of the group $U(\mathcal{H})$ on $\mathcal{H}$?—then, the fundamental problem leading our whole investigation—namely, the chase for individuation in the mathematical structures of Quantum Kinematics—suddenly disappears into thin air...

It seems that the only way out of this problem is to question the validity of Equation III.9 and therefore of Definition III.6. To see what has gone wrong and why $\mathcal{URep}(G)$ is not the correct category to consider for our inquiries, let us ask the following question: when should we consider two abstract unitary representations $U, U' : G \longrightarrow U(\mathcal{H})$ to be equal? The usual 'material' set theoretical criterion of identity would be:

$$U' = U \iff \forall g \in G, U'(g) = U(g)$$

$$\iff \begin{array}{ccc} G & \xrightarrow{\;U\;} & U(\mathcal{H}) \\ \Big\| & & \Big\| \\ G & \xrightarrow[U']{} & U(\mathcal{H}) \end{array} \quad \text{commutes.} \tag{III.10}$$

But we know from section I.2 that this is not the correct answer when dealing with *abstract* Hilbert spaces. In this case, we need to take into account that the appropriate notion of identity is isomorphism—this is the *definition* of an abstract structure[68]. Thus, we get the more sophisticated answer:

$$U' \equiv U \iff \exists \phi \in U(\mathcal{H}) \text{ such that } \forall g \in G, U'(g) = \phi U(g)\phi^{-1}$$

$$\iff \exists \Phi \in Inn(U(\mathcal{H})) \text{ such that } \quad \begin{array}{ccc} G & \xrightarrow{U} & U(\mathcal{H}) \\ \Big\| & & \wr \Big\downarrow_{\Phi} \\ G & \xrightarrow[U']{} & U(\mathcal{H}) \end{array} \quad \text{commutes}[69]. \tag{III.11}$$

This of course is the notion of equivalence of unitary $G$-representations, which was also the notion of isomorphism in the category $\mathcal{URep}(G)$. In this way, we see that this latter category is intimately related to the choice of (III.11) as the proper criterion of identity for unitary $G$-representations.

However, when dealing with an abstract unitary $G$-representation on $\mathcal{H}$, we should take into account not only the fact that the Hilbert space is abstract but also that *the group $G$ is itself an abstract structure.* And, as comparison of the diagrams (III.10) and (III.11) clearly shows, neither of the two proposed criteria of identity includes this. Indeed, in order to incorporate the abstract nature of the Hilbert space $\mathcal{H}$, we passed from the criterion of identity (III.10) to the criterion (III.11) by replacing the equality $U(\mathcal{H}) = U(\mathcal{H})$ by the isomorphism $U(\mathcal{H}) \simeq U(\mathcal{H})$. Now, the same must be done with the equality $G = G$. Therefore, it appears that the correct answer to the question "When are two unitary representations $U, U' : G \longrightarrow U(\mathcal{H})$ equal?" is then:

---

[68] Recall in particular Makkai's claim that "isomorphism is the real equality in Abstract Mathematics" (see page 88).

[69] Given a group $G$, any element $g_0$ of the group gives rise to an automorphism $\phi_{g_0} \in Aut(G)$ through conjugation $\left(\forall g \in G, \ \phi_{g_0}(g) := g_0 g(g_0)^{-1}\right)$. Any such automorphism is called an *inner automorphism* of the group. The set of all inner automorphisms forms a subgroup of $Aut(G)$ denoted by $Inn(G)$. Given $\phi \in U(\mathcal{H})$, I will denote by $\Phi \in Inn(U\mathcal{H}))$ the associated inner automorphism of $U(\mathcal{H})$.

**Criterion of identity for unitary representations.**  Given two unitary representations $U$ and $U'$ of an abstract group $G$ on an abstract Hilbert space $\mathcal{H}$, we have

$$U' \simeq U \iff \exists (\phi, \alpha) \in U(\mathcal{H}) \times Aut(G) \text{ such that } \forall g \in G, U'(\alpha(g)) = \phi U(g)\phi^{-1}$$

$$\iff \exists (\Phi, \alpha) \in Inn(U(\mathcal{H})) \times Aut(G) \text{ such that } \begin{array}{ccc} G & \xrightarrow{\ U\ } & U(\mathcal{H}) \\ {\scriptstyle \alpha}\big\downarrow{\scriptstyle \wr} & & {\scriptstyle \wr}\big\downarrow{\scriptstyle \Phi} \\ G & \xrightarrow[\ U'\ ]{} & U(\mathcal{H}) \end{array} \text{ commutes.}^{70}$$

(III.12)

Only in this way do we take into account the fact that both the Hilbert space $\mathcal{H}$ and the external group $G$ used to describe a quantum system are abstract structures[71].

---

[70]Since this is not the usual criterion of identity used by mathematicians when working with group representations, let me give an explicit example of two unitary representations which are *not* equivalent but nonetheless should be, in my view, considered as essentially the same. Consider the following:
– $G$ is the abelian abstract group $\{1, a, b, b^{-1}\}$ with the multiplication rules:

$$a^2 = 1,\ b^2 = (b^{-1})^2 = a,$$

– $U : G \longrightarrow U(1)$ is the one-dimensional unitary representation defined by

$$U(1) = Id,\ U(a) = -Id,\ U(b) = iId \text{ and } U(b^{-1}) = -i\,Id,$$

– $\alpha$ is the only possible non-trivial automorphism of $G$, namely that which exchanges $b$ with $b^{-1}$ while leaving 1 and $a$ fixed.
Then, the new unitary representation $U' = U \circ \alpha$ is simply the complex conjugate of $U$, which is *not* equivalent to $U$. (Indeed, suppose there exists a linear map $\phi : \mathbb{C} \longrightarrow \mathbb{C}$ such that $\forall g \in G,\ U(\alpha(g)) = \phi U(g)\phi^{-1}$. Then, in particular, we would have, $\forall z \in \mathbb{C},\ -iz = U'(b)z = \phi(U(b)\phi^{-1}(z)) = \phi(i\phi^{-1}(z)) = iz$, which is impossible.)
Therefore, from the standard point of view of representation theory, $U$ and $U'$ will be considered as different representations. Yet, from an abstract structuralist point of view, the elements $b$ and $b^{-1}$ are indiscernible. The group $G$ is isomorphic to the group of complex numbers $\{1, -1, i, -i\}$ with the usual multiplication, and the representations $U$ and $U'$ differ only in our arbitrary choice to equate $b$ with $i$ or $-i$. They should then be considered as two different descriptions/presentations/coordinatizations of the *same* situation.
I thank Christine Cachot for finding this example.

[71]It is enlightening to recast this discussion on the identity of unitary representations in the categorical language described in footnote 63 (page 276). From this perspective, the question is when to consider the two functors $U$ and $U'$ equal. The first criterion of identity corresponds to considering that two functors $F_1, F_2 : \mathcal{C} \longrightarrow \mathcal{D}$ are equal whenever they coincide on objects and arrows:

$$F_1 = F_2 \iff \forall C, C' \in Ob(\mathcal{C}), F_1(C) = F_2(C) \text{ and } F_1(\mathrm{Hom}_{\mathcal{C}}(C, C')) = F_2(\mathrm{Hom}_{\mathcal{C}}(C, C')).$$

But in category theory there is no sense of talking about equality of *objects*: the strongest claim one can make about the identity on objects should be isomorphism. Therefore, in the above criterion $F_1(C) = F_2(C)$ should be replaced by $F_1(C) \simeq F_2(C)$, and one should then demand the existence of a natural bijection between the sets $F_1(\mathrm{Hom}_{\mathcal{C}}(C, C'))$ and $F_2(\mathrm{Hom}_{\mathcal{C}}(C, C'))$. In other words, one considers two functors to be equal whenever there exists a natural isomorphism between them. Contrary to the naive criterion of identity, which considers the collection of functors $\mathcal{D}^{\mathcal{C}}$ to form a set, this second criterion recognizes the fact that $\mathcal{D}^{\mathcal{C}}$ is a category.

This motivates to look for a new category of unitary representations in which the notion of isomorphism coincides with the criterion of identity (III.12). This is achieved by rather considering a category where the group $G$ is not fixed:

**Definition III.7.** The category $\mathcal{URep}$ of unitary representations has unitary representations $(\mathcal{H}, G, U)$ as objects. A morphism $(\mathcal{H}, G, U) \xrightarrow{(\phi, \alpha)} (\mathcal{H}', G', U')$ is given by a linear map $\mathcal{H} \xrightarrow{\phi} \mathcal{H}'$ *together with a morphism of groups* $G \xrightarrow{\alpha} G'$ such that, for all $g \in G$, the following diagram commutes[72]:

$$
\begin{array}{ccc}
\mathcal{H} & \xrightarrow{\phi} & \mathcal{H}' \\
{\scriptstyle U(g)}\downarrow & & \downarrow{\scriptstyle U'(\alpha(g))} \\
\mathcal{H} & \xrightarrow{\phi} & \mathcal{H}'
\end{array}
$$

With this new definition, the group of automorphisms of the structure $\mathcal{H}_G$ changes into:

$$
Aut_{\mathcal{URep}}(\mathcal{H}_G) = \left\{ (\phi, \alpha) \in U(\mathcal{H}) \times Aut(G) \;\middle|\; \begin{array}{ccc} G & \xrightarrow{U} & U(\mathcal{H}) \\ {\scriptstyle \alpha}\downarrow\wr & & \wr\downarrow{\scriptstyle \Phi} \\ G & \xrightarrow{U} & U(\mathcal{H}) \end{array} \text{ commutes} \right\}. \quad \text{(III.13)}
$$

Let us again explore this result and compare it with the previous proposal (Equation III.9, page 276). First, if the unitary $G$-representation is the trivial one, then $Aut_{\mathcal{URep}}(\mathcal{H}_G) \simeq U(\mathcal{H}) \times Aut(G)$. This is the expected result: considering the trivial representation $(\mathcal{H}, G, U)$ should be equivalent to considering the pair of *independent*

---

However, this second criterion of identity corresponds to the criterion of Equation III.11, which we have also rejected. Now, the problem is most clearly perceived. In a given category, contrary to what happens with identity on objects, identity on *morphisms* (between two given objects) is primitive: it is a matter of fact whether two arrows $f, g : C \longrightarrow C'$ are or are not equal. Put differently, in category theory, the collection $Hom_C(C, C')$ is described as a *set*. Therefore, by describing $G$ as a category—and thus describing the elements as arrows of the category—we lose the automorphisms of the group and forget the idea that elements of the group might be indiscernible...

[72]In fact, this definition is a particular case of the so-called *Grothendieck construction* (I thank Zhen Lin for pointing this to me). Given a functor $F : \mathcal{C} \longrightarrow \mathcal{Cat}$, the Grothendieck construction defines the category $\mathbf{\Gamma(F)}$, where objects are pairs $(A, x) \in Ob(\mathcal{C}) \times Ob(F(A))$ and morphisms $f : (A, x) \longrightarrow (A', x')$ are pairs $(f_0, f_1)$ where $f_0 \in Hom_C(A, A')$ and $f_1 \in Hom_{F(A')}(F(f_0)x, x')$.

The category $\mathcal{URep}$ is the Grothendieck construction for the functor $F : \mathcal{Grp}^{op} \longrightarrow \mathcal{Cat}$ which associates to a group $G$ the category $\mathcal{URep}(G)$.

In the case where the functor $F$ lands in sets (considered as discrete categories), the category $\mathbf{\Gamma(F)}$ is called the *category of points* of the functor $F$ and is denoted by $\int F$. This special case is discussed in S. Mac Lane and I. Moerdijk. *Sheaves in Geometry and Logic: A First Introduction to Topos Theory.* New York: Springer-Verlag, 1992, pp. 41–44. For the general definition, see the Wikipedia entry "Grothendieck construction".

structures $(\mathcal{H}, G)$. Therefore, the group $Aut_{\mathcal{URep}}(\mathcal{H}_G)$ should be the product of the two automorphism groups. In particular, we see again that, by considering the trivial representation, one does not break the homogeneity of the space of states.

Second—and this is the crucial difference between (III.9) and (III.13)—there is now, for *any* unitary $G$-representation $\mathcal{H}_G$, a canonical morphism of groups

$$\pi_Q : G \longrightarrow Aut_{\mathcal{URep}}(\mathcal{H}_G). \qquad (\text{III.14})$$

Indeed, the canonical map is simply the product $\pi_Q := U \times Inn$, where $G \xrightarrow{\ U\ } U(\mathcal{H})$ is the unitary map and $G \xrightarrow{\ Inn\ } Aut(G)$ is the map of inner automorphisms (cf. footnote 69, page 279)[73]. In other words, in this new category, elements of the group $G$ invariably give rise to (non-trivial) automorphisms of the structure $\mathcal{H}_G$. Hence, the existence of this canonical map may be perceived as a sophisticated version of the initial naive expectation that the elements of the group $G$ should be the automorphisms of the structure $\mathcal{H}_G$ (cf. page 275).

In this way, we see that even for an irreducible representation the group of automorphisms of $\mathcal{H}_G$ will in general act non trivially on $\mathbb{P}\mathcal{H}$. In particular, we recover the fact that for an abstract Hilbert space, even when seen as equipped with the irreducible representation of $U(\mathcal{H})$, its group of automorphisms acts transitively on $\mathbb{P}\mathcal{H}$ (since $U(\mathcal{H}) \subset Aut_{\mathcal{URep}}(\mathcal{H}_{U(\mathcal{H})})$).

### III.3.1.b  The classical case

Having dealt with the group of automorphisms of $\mathcal{H}_G$ in quite some detail, it is now a simpler task to transpose the discussion over to the Classical arena. As we have learned, the problem is better dealt with if one addresses first the related issues of determining: i) a criterion of identity for abstract strongly Hamiltonian $\mathfrak{g}$-actions, and ii) the correct category *sHam* of strongly Hamiltonian actions associated to the chosen criterion of identity.

---

[73]More explicitly, given an element $g_0 \in G$, we have $\pi_Q(g_0) := \big(U(g_0), \phi_{g_0}\big) \in U(\mathcal{H}) \times Aut(G)$ and the condition $U\big(\alpha(g)\big) = \phi U(g)\phi^{-1}$ (Equation III.13) becomes the trivial equation $U\big(g_0 g(g_0)^{-1}\big) = U(g_0)U(g)U(g_0)^{-1}$.

Following the case with unitary $G$-representations, there are again three possible levels of identity one may consider. Two strongly Hamiltonian actions $S_{\mathfrak{g}} := (S, \mathfrak{g}, \widehat{J})$ and $S'_{\mathfrak{g}'} := (S', \mathfrak{g}', \widehat{J}')$ are said to be:

i) *identical* (denoted $S_{\mathfrak{g}} = S'_{\mathfrak{g}'}$) if

$$\begin{cases} S = S' \\ \mathfrak{g} = \mathfrak{g}' \\ \forall X \in \mathfrak{g}, \ \widehat{J}(X) = \widehat{J}'(X), \end{cases} \tag{III.15}$$

ii) *equivalent* (denoted $S_{\mathfrak{g}} \equiv S'_{\mathfrak{g}'}$) if

$$\begin{cases} S \simeq S' \\ \mathfrak{g} = \mathfrak{g}' \\ \exists \phi \in \mathrm{Iso}(S, S') \text{ such that } \begin{array}{c} S \\ \phi \downarrow \wr \quad \searrow^{J} \\ S' \xrightarrow{J'} \mathfrak{g}^* \end{array} \text{ commutes}^{74}, \end{cases} \tag{III.16}$$

iii) *isomorphic* (denoted $S_{\mathfrak{g}} \simeq S'_{\mathfrak{g}'}$) if

$$\begin{cases} S \simeq S' \\ \mathfrak{g} \simeq \mathfrak{g}' \\ \exists (\phi, \alpha) \in \mathrm{Iso}(S, S') \times \mathrm{Iso}(\mathfrak{g}, \mathfrak{g}') \text{ such that } \begin{array}{ccc} S & \xrightarrow{J} & \mathfrak{g}^* \\ \phi \downarrow \wr & & \wr \uparrow \alpha^* \\ S' & \xrightarrow{J'} & (\mathfrak{g}')^* \end{array} \text{ commutes}^{75}. \end{cases} \tag{III.17}$$

---

[74]Instead of using the momentum maps, this third condition may also be written in terms of the co-momentum maps $\widehat{J}$ and $\widehat{J}'$ as

$$\exists \phi \in \mathrm{Iso}(S, S') \text{ such that } \begin{array}{c} \mathcal{C}^\infty(S, \mathbb{R}) \\ \mathfrak{g} \ \ \nearrow^{\widehat{J}} \quad \wr \downarrow \phi^* \\ \searrow_{\widehat{J}'} \ \mathcal{C}^\infty(S', \mathbb{R}) \end{array} \text{ commutes.}$$

[75]Similarly, in terms of the co-momentum map, this third condition becomes:

$$\exists (\phi, \alpha) \in \mathrm{Iso}(S, S') \times \mathrm{Iso}(\mathfrak{g}, \mathfrak{g}') \text{ such that } \begin{array}{ccc} \mathfrak{g} & \xrightarrow{\widehat{J}} & \mathcal{C}^\infty(S, \mathbb{R}) \\ \alpha \downarrow \wr & & \wr \uparrow \phi^* \\ \mathfrak{g}' & \xrightarrow{\widehat{J}'} & \mathcal{C}^\infty(S', \mathbb{R}) \end{array} \text{ commutes.}$$

These criteria are the three analogues of (III.10), (III.11) and (III.12) of the previous section. As noted, the first corresponds to the material set-theoretical identity based on extensionality. Only the third one is a sensible criterion of identity for abstract strongly Hamiltonian actions, where both the symplectic manifold $S$ and the Lie algebra $\mathfrak{g}$ are conceived abstractly. The associated category of strongly Hamiltonian actions is defined as follows:

**Definition III.8.** The category *sHam* of strongly Hamiltonian Lie algebra actions has strongly Hamiltonian actions $(S, \mathfrak{g}, \widehat{J})$ as objects. A morphism $(S, \mathfrak{g}, \widehat{J}) \xrightarrow{(\phi, \alpha)} (S', \mathfrak{g}', \widehat{J}')$ is given by a morphism of symplectic manifolds $S \xrightarrow{\phi} S'$ (i.e., such that the pullback $\mathcal{C}^\infty(S', \mathbb{R}) \xrightarrow{\phi^*} \mathcal{C}^\infty(S, \mathbb{R})$ is a morphism of Poisson algebras) together with a morphism of Lie algebras $\mathfrak{g} \xrightarrow{\alpha} \mathfrak{g}'$ such that the following diagram commutes

$$
\begin{array}{ccc}
S & \xrightarrow{\ J\ } & \mathfrak{g}^* \\
{\scriptstyle \phi}\downarrow & & \uparrow{\scriptstyle \alpha^*} \\
S' & \xrightarrow{\ J'\ } & (\mathfrak{g}')^*
\end{array}
$$

Finally, given this definition or the criterion of identity (III.17), the group of automorphisms of the structure $S_\mathfrak{g}$ shows to be

$$
Aut_{sHam}(S_\mathfrak{g}) = \left\{ (\phi, \alpha) \in Aut(S) \times Aut(\mathfrak{g}) \,\middle|\, 
\begin{array}{ccc}
S & \xrightarrow{\ J\ } & \mathfrak{g}^* \\
{\scriptstyle \phi}\downarrow{\scriptstyle \wr} & & {\scriptstyle \wr}\uparrow{\scriptstyle \alpha^*} \\
S & \xrightarrow{\ J\ } & \mathfrak{g}^*
\end{array}
\text{ commutes} \right\}. \quad \text{(III.18)}
$$

Again, one can readily see that, if the strongly Hamiltonian $\mathfrak{g}$-action $\mathfrak{g} \xrightarrow{\rho} \Gamma(TS)_\omega$ integrates into a group action $G \xrightarrow{L} Aut(S)$, then the group elements give rise to automorphisms of the abstract structure $S_\mathfrak{g}$. In other words, as it was the case for abstract unitary representations, there exists, for any strongly Hamiltonian $G$-action, a canonical morphism of groups

$$
\pi_C : G \longrightarrow Aut(S_\mathfrak{g}). \quad \text{(III.19)}
$$

It is given by $\pi_C := L \times Ad$ where $G \xrightarrow{Ad} Aut(S)$ is the adjoint action (cf. footnote 18, page 248): with this choice, the commutativity of the diagram in (III.18) for each $\pi_C(g)$ amounts precisely to the definition of a strongly Hamiltonian $G$-action[76].

## III.3.2   The group-theoretical labelling scheme

The complete solution to the individuation problem (see page 243) for the abstract structures $S_{\mathfrak{g}}$ and $\mathcal{H}_G$ may be seen as involving the two following aspects:

i) From the perspective of states, to understand which are the smallest subsets of the space of states one can possibly expect to individuate. As we have already discussed, this means to determine the orbits of the actions $Aut_{\mathcal{URep}}(\mathcal{H}_G) \circlearrowleft \mathbb{P}\mathcal{H}$ and $Aut_{\mathcal{SHam}}(S_{\mathfrak{g}}) \circlearrowleft S$.

ii) From the perspective of properties, to understand how does one construct specific structural properties which allow to effectively define a labelling scheme that distinguishes states.

Because of the existence of the two canonical morphisms $G \xrightarrow{\pi_Q} Aut_{\mathcal{URep}}(\mathcal{H}_G)$ (III.14, page 282) and $G \xrightarrow{\pi_C} Aut_{\mathcal{SHam}}(S_{\mathfrak{g}})$ (III.19, page 284), these two aspects are related to the study of invariants of the group $G$.

### III.3.2.a   Quantum properties as labels of irreducible representations

In this respect, the most standard problem in linear representation theory is the breaking of a unitary representation into its irreducible components. Recall: a unitary representation $U : G \longrightarrow U(\mathcal{H})$ is called *reducible* if it is possible to find invariant linear subspaces $(V_i)_{i \in I}$ such that the total Hilbert space may be written as a direct sum of these : $\mathcal{H} = \bigoplus_{i \in I} V_i$. Moreover, it is called *completely reducible* if the invariant subspaces $V_i$ are the smallest possible (i.e., they contain no non-trivial invariant subspaces). Said differently, a unitary $G$-representation $U$ is completely reducible whenever it is possible

---

[76] If $\alpha = Ad(g)$ for some element $g \in G$, then $\alpha^* = Co(g^{-1})$ and the commutative diagram in (III.18) writes

$$\begin{array}{ccc} S & \xrightarrow{\;J\;} & \mathfrak{g}^* \\ {\scriptstyle L(g)}\big\downarrow & & \big\downarrow{\scriptstyle Co(g)} \\ S & \xrightarrow[\;J\;]{} & \mathfrak{g}^* \end{array}$$

which is the Co-equivariance condition of the momentum map (Definition III.3, page 247).

to write it as a direct sum of irreducible representations: $U = \bigoplus_{i \in \widehat{G}} U_i$. For compact Lie groups, any unitary representation is completely reducible[77], but in general this is not the case.

The problem of breaking a unitary group representation into its irreducible components undisputedly plays a crucial role in the description of quantum systems. Because of the clear similarity of this problem with that of breaking the space of states into the orbits of $Aut_{\mathcal{URep}}(\mathcal{H}_G)$—the first seeks to write $\mathcal{H}$ as a direct sum of the smallest possible invariant linear subspaces, while the second seeks to write $\mathbb{P}\mathcal{H}$ as a union of the smallest possible invariant subspaces—, one could be led to think that by finding the decomposition of a unitary representation one solves problem i) above. However, the similarity is misleading and the two problems are better kept apart: given a unitary representation $G \xrightarrow{U} U(\mathcal{H})$ and the associated induced action $G \xrightarrow{\mathsf{U}} Aut(\mathbb{P}\mathcal{H})$, the *irreducibility* of the former does not entail the *transitivity* of the latter[78].

In fact, the role of irreducible representations is better understood when approached in relation to properties rather than states. Since the decomposition $U = \bigoplus_{i \in \widehat{G}} U_i$ is invariant under the equivalence relation, belonging to a particular irreducible component of the abstract structure $\mathcal{H}_G = (\mathcal{H}, G, U)$ is a structural property which can be used to distinguish some states of the system. Thus, in the group-theoretical approach to the kinematical description of quantum systems, the mathematical problem of finding a parametrization of the unitary dual $\widehat{G}$ becomes the main road for building a labelling scheme. In this way, it appears as quite natural a phenomenon for quantum numbers to be "indices characterizing representations of groups" as was pointed out by Hermann Weyl[79].

---

[77]See for instance Knapp, loc. cit.

[78]This is easily seen by dimensional considerations: given a Hilbert space of dimension $d$, the projective space $\mathbb{P}\mathcal{H}$, as a *real* manifold, has dimension $2(d-1)$. Now, while most Lie groups admit irreducible unitary representations of any finite dimension, transitive actions must be at most of the dimension of the Lie group. In other words, if $2d > \dim G + 1$, then it is impossible for the action $G \circlearrowleft \mathbb{P}\mathcal{H}$ to be transitive. The converse statement is however true: if the action $G \xrightarrow{\mathsf{U}} Aut(\mathbb{P}\mathcal{H})$ is transitive, then the irreducible representation from which it stems is irreducible.

[79]Weyl, op. cit., p. xxi.

For our investigation, the important point is to understand the amount of discernibility introduced into the quantum space of states through this group-representational labelling scheme. In general, it is clear that, by simply indexing the various irreducible representations of the abstract group $G$, one does not succeed in individuating most states. There are two main reasons for this to be so. First, there is the *dimensionality* of a given unitary irreducible representation: by singling out an irreducible component $V_i$ of the structure $\mathcal{H}_G$, one designates a submanifold of the quantum space of states which is $(2\dim V_i - 2)$–dimensional. Hence, unless $V_i$ is one-dimensional, one does not individuate in this way a point of the quantum space of states. An important exception to this drawback is the case of *abelian* groups, for which any irreducible unitary representation is necessarily one-dimensional. Abelian groups play hence an important role in the labelling of states. But even in this latter case there is a second source of indistinguishability: the eventual *multiplicity* of an irreducible component. In the decomposition $U = \bigoplus_{i \in \widehat{G}} U_i$, nothing prevents the same irreducible representation from appearing twice (or more), and, in order to clearly show this possibility, the decomposition is better written as

$$U = \bigoplus_{i \in \widehat{G}} m_i U_i, \text{ where } m_i \in \mathbb{N}. \tag{III.20}$$

In such cases where $m_i \notin \{0, 1\}$, there is no sense in which one can talk about *the* linear subspace $V_i \subset \mathcal{H}$ that supports the irreducible representation $U_i$.

These issues are well perceived when looking at the paradigmatic example of the group-representational labelling scheme in Quantum Mechanics—namely, the treatment of spin states in relation to the abstract group $SU(2)$. As it turns out, there is exactly one unique irreducible representation of this group for any given dimension $d \in \mathbb{N}^{*}$[80]. Hence, the unitary dual $\widehat{SU(2)}$ is parametrized by a single variable with discrete integer values, which in physics is taken to be $s := \frac{d-1}{2}$. Only for the case $s = 0$ is this number enough to designate a state, since the representation is then one-dimensional. For the remaining cases, more work needs to be done to individuate a state. In the spirit of the representational labelling scheme, one considers a maximal torus of $SU(2)$—that is, a connected abelian subgroup $T_X \subset SU(2)$ that has the property of not being a subgroup of any other abelian subgroup of $SU(2)$—and decomposes each irreducible representation of the whole group into irreducible representations of the subgroup (which are also called '*weights*'). More specifically, one here takes $T_X \simeq U(1)$, so that $\widehat{T_X} \simeq \mathbb{Z}$. In this way, a second quantum number is constructed and the initial abstract unitary $SU(2)$-representation may be described as

$$U = \bigoplus_{s \in \widehat{SU(2)}} \bigoplus_{m \in \widehat{U(1)}} m_s U_{s,m}. \tag{III.21}$$

---

[80]Cf. M. R. Sepanski. *Compact Lie Groups*. New York: Springer, 2007, Theorem 3.32 p. 68.

$U_{s,m}$ is no other than the usual spin state $|s, m\rangle$: it denotes the *unique* one-dimensional irreducible representation of $U(1)$ labelled by $m$ which is found inside the *unique* $(2s + 1)$-dimensional irreducible representation of $SU(2)$.

Because the above involves the *arbitrary choice* of a subgroup of $G$, it could be felt that this last construction does not comply with the abstract structuralist methodology followed so far. And the objection would be valid if the definition of the quantum number $m$ relied on the ability to distinguish the specific maximal torus $T_X$ from all others. Indeed, whenever the group $G$ is conceived abstractly, any two maximal tori $T_X$ and $T_Y$ are structurally indiscernible: it is always possible to find an automorphism $\alpha_g \in Inn(G)$ such that $\alpha_g(T_X) = T_Y$[81]. But the decomposition (III.21) is in fact *independent* of the choice of the subgroup $T_X$: all maximal tori of a compact Lie group are isomorphic and the same weights will appear for any maximal torus[82]. Hence, the definition of this second quantum property does not involve any specificity of some particular maximal torus and can be considered a structural property of the abstract structure $\mathcal{H}_G$.

In the best possible scenario, where all multiplicities appearing in (III.21) are either 0 or 1, we therefore see how the labelling scheme succeeds in individuating a 'basis' of states—that is, in constructing a complete set of commuting observables. This is certainly the best situation which can be hoped for in the group-representational approach to the kinematical description of a quantum system. However, it does not automatically mean that the requirement of individuation is satisfied. Even in this case, even when it is possible to individuate a family of states from which one can generate the whole quantum space of states, there remain qualitatively indiscernible states. That this is so may be counter-intuitive from the point of view of Hilbert spaces: given a basis $\mathcal{B}$ of $\mathcal{H}$, any element $\phi \in \mathcal{H}$ can be written in a *unique* way as a linear combination of elements of $\mathcal{B}$. This implies that, if a basis of vectors can be individuated, then all elements of the Hilbert space can be as well. But—let us not forget—*a state is described by a ray*, not by an element of the Hilbert space, and the group-representational labelling scheme only achieves at best the individuation of a ray. In this case, the amount of discernibility that is introduced into the quantum space of states is therefore more difficult to grasp.

---

[81] Ibid., Corollary 5.10.b), p. 101.

[82] T. Bröcker and T. tom Dieck. *Representations of Compact Lie Groups.* 1st ed. New York: Springer, 1985, p. 184.

The simplest example of this phenomenon is the spin-$\frac{1}{2}$ quantum system, described by the fundamental representation of $SU(2)$ whose weight decomposition is

$$U_{\frac{1}{2}} = U_{\frac{1}{2},\frac{1}{2}} \oplus U_{\frac{1}{2},-\frac{1}{2}}.$$

The choice of a maximal torus $T_X$—or, equivalently, of a component of the spin $S_X$—allows to individuate two states: $|\frac{1}{2}, \frac{1}{2}\rangle$ ('spin up') and $|\frac{1}{2}, -\frac{1}{2}\rangle$ ('spin down'), and any other state results from a superposition of these two. Geometrically, the quantum space of states $\mathcal{P}$ may be depicted as a sphere with two antipodal points pinned (Figure III.4)[83]. In this picture, it becomes clear that not all states have been individuated: given the two poles, the intrinsic metric structure of the space of states only allows to distinguish the different 'parallels' of the sphere but any two states lying in the same 'parallel' are qualitatively indiscernible[84]. To further increase the amount of discernibility and finally individuate all states, it would be necessary to consider several

---

[83]Recall that, as real manifolds, one has $\mathbb{PC}^2 \simeq S^2$. Moreover, states that are orthogonal in the Hilbert space formulation are antipodal in the geometric formulation.

[84]For $\alpha \in [0, \frac{\pi}{2}]$, the parallel $P_\alpha$ is defined as the subset of states lying at distance $\alpha$ from the state $|\frac{1}{2}, \frac{1}{2}\rangle$ (or, equivalently, at distance $\frac{\pi}{2} - \alpha$ from the state $|\frac{1}{2}, -\frac{1}{2}\rangle$):

$$P_\alpha := \{p \in \mathcal{P} \mid d_g(p, |\frac{1}{2}, \frac{1}{2}\rangle) = \alpha\} = \{p \in \mathcal{P} \mid d_g(p, |\frac{1}{2}, -\frac{1}{2}\rangle) = \frac{\pi}{2} - \alpha\}.$$

$P_\alpha$ can also be defined in more familiar form in terms of the probabilities of the measurement outcome for the property $S_X$ as the subset of states such that $\Pr(S_X = \frac{\hbar}{2}) = 1 - \Pr(S_X = -\frac{\hbar}{2}) = \cos^2(\alpha)$.

different components of the spin. But this involves the consideration of various different maximal tori and the ability to qualitatively distinguish them, which, as have we just discussed, is not possible for an abstract compact Lie group.

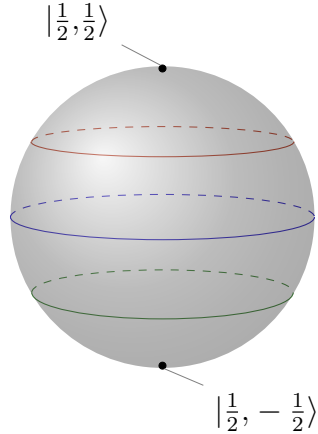

**Fig. III.4** – **The space of states of the spin-$\frac{1}{2}$ system.**
The individuation of two antipodal states through the group-representational labelling scheme only allows to distinguish the different 'parallels' (such as the green, blue and red ones), but not the states within them.

### III.3.2.b   Classical properties as labels of transitive actions

When Weyl introduced the idea that quantum numbers could be seen as labels characterizing irreducible representations, he seemed to be highlighting a characteristic novelty of the Quantum. Yet, inspired by this group-representational labelling scheme, one can attempt to build well-defined classical properties in a similar way. Given the abstract structure $\mathcal{H}_G = (\mathcal{H}, G, U)$, the key in the above construction was to use the morphism  $G \xrightarrow{\ U\ } U(\mathcal{H})$  in order to produce a decomposition of the quantum space of states which was also an invariant of the equivalence class $[U]$. In the classical case, where the abstract structure is now $S_{\mathfrak{g}} = (S, \mathfrak{g}, \widehat{J})$, one can use the momentum map to write an analogous decomposition into qualitatively discernible subsets of states.

Given the map  $S \xrightarrow{\ J\ } \mathfrak{g}^*$,  the most natural partition of $S$ to consider is the one defined by the equivalence relation

$$p \sim q \Longleftrightarrow J(p) =_{\mathfrak{g}^*} J(q), \tag{III.22}$$

where $p$ and $q$ are points in $S$. One can then write $S = \bigsqcup_{\theta \in \mathfrak{g}^*} J^{-1}(\theta)$. However, this decomposition does not immediately yield the desired result, for two subsets $J^{-1}(\theta)$ and $J^{-1}(\xi)$ are not necessarily qualitatively discernible. Indeed, this is the case only if the elements $\xi$ and $\theta$ are themselves qualitatively discernible elements of $\mathfrak{g}^*$: if there exists $\alpha \in Aut(\mathfrak{g})$ such that $\alpha^*(\theta) = \xi$ (in which case $\theta$ and $\xi$ are qualitatively *indiscernible*), then there also exists $\phi \in Symp(S)$ such that $\phi\big(J^{-1}(\theta)\big) = J^{-1}(\xi)$ and $(\phi, \alpha) \in Aut_{\mathscr{Ham}}(S_{\mathfrak{g}})$[85]. Thus, to achieve a decomposition in the spirit of (III.20), instead of considering the pre-images under the momentum map of the elements of the abstract structure $\mathfrak{g}^*$, one should rather consider the pre-images *of the smallest subsets of $\mathfrak{g}^*$ which can be individuated*. If $\mathcal{O}_l \subset \mathfrak{g}^*$ is any such subset, then its pre-image $S_l := J^{-1}(\mathcal{O}_l) \subset S$ will be also be a qualitatively discernible subset of the space of states.

The study of the amount of discernibility introduced into the classical space of states by endowing it with a strongly Hamiltonian $\mathfrak{g}$-action involves therefore a study of the amount of discernibility intrinsic to the abstract Poisson manifold $\mathfrak{g}^*$. Luckily, the latter is easily determined, for it is simply given by the *symplectic decomposition* of $\mathfrak{g}^*$. Any finite-dimensional Poisson manifold $P$ can be written as a disjoint union of symplectic manifolds, called the *symplectic leaves* of $P$, which can be viewed as maximal sets of points that are connected by a piecewise smooth Hamiltonian curve[86]. In particular, this means that for any two points of a symplectic leave there exists an automorphism of $P$ relating them. Hence, a symplectic leave of $\mathfrak{g}^*$ is precisely a maximal set of points that are qualitatively indiscernible.

In this way, we arrive at the following scheme: given the structure $S_{\mathfrak{g}} = (S, \mathfrak{g}, \widehat{J})$, use the intrinsic symplectic decomposition of the abstract Poisson manifold $\mathfrak{g}^*$

$$\mathfrak{g}^* = \bigsqcup_{l \in \mathfrak{g}^*/Aut(\mathfrak{g}^*)} \mathcal{O}_l$$

---

[85] Take $\alpha = Ad(g)$ for some $g \in G$. We know that $\big(L(g), Ad(g)\big) \in Aut_{\mathscr{Ham}}(S_{\mathfrak{g}})$ (see footnote 76 page 285). Moreover, because of the equivariance of the momentum map , for any $p \in J^{-1}(\theta)$ and any $q \in J^{-1}(\xi)$ we have $J(L(g)p) = \xi$ and $J\big(L(g^{-1})q\big) = \theta$. Hence, $L(g)\big(J^{-1}(\theta)\big) = J^{-1}(\xi)$.

[86] See Landsman, op. cit., Definition I.2.4.3 and Theorem I.2.4.7, pp. 70-71 (and also herein Chapter II, page 227).

to induce a similar decomposition of the classical space of states:

$$S = \bigsqcup_{l \in \mathfrak{g}^*/Aut(\mathfrak{g}^*)} S_l \tag{III.23}$$

where $S_l$ is the pre-image under the momentum map of the symplectic leave $\mathcal{O}_l$. In other words, this new partition of the space states arises from considering the equivalence relation

$$p \approx q \Longleftrightarrow [J(p)] =_{\mathfrak{g}^*/Aut(\mathfrak{g}^*)} [J(q)] \tag{III.24}$$

instead of the equivalence (III.22) taken initially. The subsets $S_l$ thus obtained have no reason to be symplectic submanifolds of the space of states[87]. Now, the label $l$, which runs over the set of symplectic leaves of $\mathfrak{g}^*$, can effectively be used to distinguish states of the classical system described by $S_{\mathfrak{g}}$. In this way, one achieves the structural construction of a classical number that is an 'index characterizing a symplectic leaf of $\mathfrak{g}^*$'.

This is indeed similar to Weyl's insight on quantum numbers, but perhaps not quite as much as expected. By following the analogy of Table III.1 (page 266), one would have surely conjectured the classical analogue of Weyl's 'quantum numbers as indices characterizing representations of groups'—if any—to be: 'classical numbers as indices characterizing actions of Lie algebras'. Nonetheless, with a little more work, it is possible to reformulate the above construction in exactly those terms.

The first step in this direction is to find the classical analogue of *irreducible* unitary representations. Recall: in the Quantum $\longleftrightarrow$ Classical comparison, unitary $G$-representations correspond to strongly Hamiltonian $\mathfrak{g}$-actions, in particular because the former are equivalent to representations of the JLB-algebra $JL(G)$ while the latter are equivalent to representations of the Poisson algebra $\mathcal{C}^\infty(\mathfrak{g}^*, \mathbb{R})$ (see pages 267 and 252). Moreover, *irreducible* unitary group representations are equivalent to irreducible representations of $JL(G)$. Therefore, the classical analogue of irreducible $G$-representations are those $\mathfrak{g}$-actions induced by irreducible Poisson representations of $\mathcal{C}^\infty(\mathfrak{g}^*, \mathbb{R})$. With no surprise, these correspond to *transitive* strongly Hamiltonian $\mathfrak{g}$-actions[88].

---

[87] However, it is not difficult to construct new symplectic manifolds from the subsets $S_l$ (although these will not be submanifolds of the space of states). Because $S \xrightarrow{J} \mathfrak{g}^*$ is a Poisson map, the entire $\mathfrak{g}$-orbit through an arbitrary point $p \in S_l$ will be contained in $S_l$. For any subspace $S_l$, one can therefore consider its space of $\mathfrak{g}$-orbits, denoted $S_l/\mathfrak{g}$. If the $\mathfrak{g}$-action is 'sufficiently well-behaved', the space $S_l/\mathfrak{g}$ is a manifold, in which case it can be proven to be a symplectic manifold. This construction is called the *Marsden-Weinstein symplectic reduction* and was introduced in their joint article J. E. Marsden and A. Weinstein. "Reduction of Symplectic Manifolds With Symmetry". In: *Reports on Mathematical Physics* 5.1 (1974), pp. 121–130.

---

[88]A Poisson representation $\mathcal{C}^\infty(\mathfrak{g}^*, \mathbb{R}) \xrightarrow{\;J^*\;} \mathcal{C}^\infty(S, \mathbb{R})$ is said to be *irreducible* if, for any $p \in S$, the composite map $\mathcal{C}^\infty(\mathfrak{g}^*, \mathbb{R}) \xrightarrow{\;J^*\;} \mathcal{C}^\infty(S, \mathbb{R}) \xrightarrow{\;v_-\;} \Gamma(TS)_H \xrightarrow{\;ev_p\;} T_p S$ is surjective. This means that all the infinitesimal transformations at any given point of $S$ can be generated by elements of $\mathcal{C}^\infty(\mathfrak{g}^*, \mathbb{R})$ (see Landsman, op. cit., Definition I.2.6.6 p. 78).

On the other hand, a $\mathfrak{g}$-action $\mathfrak{g} \xrightarrow{\;\rho\;} \Gamma(TS)$ is said to be *transitive* if, for any $p \in S$, the composite map $\mathfrak{g} \xrightarrow{\;\rho\;} \Gamma(TS) \xrightarrow{\;ev_p\;} T_p S$ is surjective (see Alekseevsky and Michor, op. cit., p. 6).

I have not been able to find in the mathematical literature a proof of the equivalence "irreducible representations of $\mathcal{C}^\infty(\mathfrak{g}^*, \mathbb{R})$" $\Longleftrightarrow$ "transitive strongly Hamiltonian $\mathfrak{g}$-actions" to which I could refer the reader. So, for completeness, here it is:

Consider the following commutative diagram

$$
\begin{array}{ccccc}
 & & \overset{\rho|_p}{\overbrace{\phantom{aaaaaaaaaaaaaaaaaaaa}}} & & \\
\mathfrak{g} & \xrightarrow{\;\;\rho\;\;} & \Gamma(TS)_H & \xrightarrow{\;\;ev_p\;\;} & T_p S \\
\iota \downarrow & & \uparrow v_- & & \\
\mathcal{C}^\infty(\mathfrak{g}^*, \mathbb{R}) & \xrightarrow{\;\;J^*\;\;} & \mathcal{C}^\infty(S, \mathbb{R}) & & \\
 & & \underset{J^*|_p}{\underbrace{\phantom{aaaaaaaaaaaaaaaaaaaa}}} & &
\end{array}
$$

The goal is to prove: $\rho|_p$ surjective $\Longleftrightarrow J^*|_p$ surjective.

$\Longrightarrow$ : It is a particular case of the fact that, if $f \circ g$ is surjective, then so is $f$.

$\Longleftarrow$ : We wish to show that for all $v_p \in T_p S$ there exists $X \in \mathfrak{g}$ such that $\rho|_p(X) = v_p$. In other words, we want $\omega(v_p, \cdot) = \omega(\rho|_p(X), \cdot)$.

By definition, we have $\omega(\rho(X), \cdot) = dJ^*(\tilde{X})$, where $\tilde{X} \equiv i(X)$. Moreover, $J^*(\tilde{X})(p) = J(p)(X)$ where $J : S \longrightarrow \mathfrak{g}^*$ is the momentum map. Hence, $dJ^*(\tilde{X}) = d(\langle J, X \rangle) = \langle dJ, X \rangle$.

Let $\{X_1, \ldots, X_n\}$ be a basis of $\mathfrak{g}$, and $\{\theta_1, \ldots, \theta_n\}$ the associated dual basis. With this choice, one can write $J = J_k \theta_k$ where $J_k : S \longrightarrow \mathbb{R}$, and therefore $dJ = dJ_k \theta_k$ where $dJ_k \in \Omega^1(S, \mathbb{R})$. Then, for $X = x^k X_k \in \mathfrak{g}$, we have

$$\omega(\rho(X), \cdot) = dJ_k x^k.$$

On the other hand, since by hypothesis $J^*|_p$ is surjective, there must exist $f \in C^\infty(\mathfrak{g}^*, \mathbb{R})$ such that

$$\omega(v_p, \cdot) = \omega(J^*|_p(f), \cdot) = dJ^*(f)\Big|_p$$

But $J^*(f) = f \circ J$ and hence $dJ^*(f)\Big|_p = df\Big|_{J(p)}\left(dJ\Big|_p\right)$. Therefore,

$$\omega(v_p, \cdot) = df\Big|_{J(p)}\left(dJ_k\Big|_p \theta_k\right) = dJ_k\Big|_p \cdot df\Big|_{J(p)}\left(\theta_k\right).$$

Finally, with the definitions $x^k := df\Big|_{J(p)}\left(\theta_k\right)$ and $X := x^k X_k$ (so that $x^k \in \mathbb{R}$ and $X \in \mathfrak{g}$), the last equation may be rewritten as

$$\omega(v_p, \cdot) = dJ_k\Big|_p x^k = \omega(\rho|_p(X), \cdot). \qquad \square$$

Second, one uses the following key result due to Kostant:

**Theorem III.6** (Kostant's Coadjoint Orbit Covering Theorem). *Consider a transitive strongly Hamiltonian action* $\mathfrak{g} \circlearrowleft S$. *Then, $S$ must be symplectomorphic to a symplectic leaf of $\mathfrak{g}^*$ or to a covering space of one.*[89]

Put differently, Kostant's theorem shows that the symplectic leaves of $\mathfrak{g}^*$ classify all the possible transitive strongly Hamiltonian $\mathfrak{g}$-actions. Under this light, the classical decomposition (III.23) appears to be somewhat analogous to the quantum decomposition (III.20), in the sense that it describes which transitive $\mathfrak{g}$-actions show up in the structure $S_{\mathfrak{g}}$[90]. In addition, the classical label $l$ receives a new interpretation and we can now claim:

> In the same way that the values of quantum properties can be seen as labels characterizing irreducible unitary representations of groups, the values of classical properties can be seen as labels characterizing transitive strongly Hamiltonian actions of Lie algebras.

Let us close this discussion by exhibiting the example of the group $SU(2)$, as we did in the previous section. In the present approach for introducing individuality

---

[89] The theorem is more commonly stated in terms of left transitive strongly Hamiltonian G-actions and Co-adjoint orbits (hence the name). See for reference Marsden and Ratiu, op. cit., p. 463.

The reason why I have chosen this 'unconventional' formulation is because it involves no group action whatsoever—it only involves $\mathfrak{g}^*$. By the same token, it emphasizes that only the infinitesimal information is relevant. It is taken from Landsman, op. cit., Corollary III.1.4.7, p. 195 (combined with Theorem III.1.4.4).

[90] Notwithstanding this, (III.23) is *not* the complete analogue of (III.20). What is missing here is the equivalent of the multiplicities $m_i$ appearing in (III.20), which would describe the number of times a given transitive action appears in the initial action $\mathfrak{g} \circlearrowleft S$. Victor Guillemin and Shlomo Sternberg have done some remarkable work in trying to determine the classical analogue of these multiplicities. The kernel of their idea is that the multiplicities $m_i$ are intimately related to the symplectic manifolds $S_l/\mathfrak{g}$ of the Marsden-Weinstein symplectic reduction (cf. footnote 87). This is captured in their famous conjecture that 'quantization commutes with reduction', which has subsequently played an important role in the mathematical foundations of quantization. The conceptual implications of their ideas have been explored in much detail by Gabriel Catren. See V. Guillemin and S. Sternberg. "Geometric Quantization and Multiplicities of Group Representations". In: *Inventiones Mathematicae* 67 (1982), pp. 515–538, G. Catren. "On the Relation Between Gauge and Phase Symmetries". In: *Foundations of Physics* 44 (2014), pp. 1317–1335 and also N. P. Landsman. "Quantum Mechanics and Representation Theory: the New Synthesis". In: *Acta Applicandae Mathematica* 81.1 (2004), pp. 167–189.

The main reason why I do not discuss more closely these ideas is mainly because I unfortunately still do not master the mathematical technique of these works well enough to see the forest and not the trees.

into Classical Kinematics, only the infinitesimal information of the group is relevant. Hence, it does not matter whether one considers $SU(2)$ or $SO(3)$ since they have the same Lie algebra: $\mathfrak{su}(2) \simeq \mathfrak{so}(3)$. As a vector space, one has $\mathfrak{so}(3)^* \simeq \mathbb{R}^3$, but as a *Poisson manifold* it is better thought as the disjoint union of 3-dimensional spheres of all possible radii[91]:

$$\mathfrak{so}(3)^* = \bigsqcup_{r \in \mathbb{R}^+} S_r^2$$

Therefore, transitive strongly Hamiltonian actions of $\mathfrak{so}(3)$ are classified by one continuous parameter which can take any positive real value. The space of states of the classical system described by $(S, \mathfrak{so}(3), \widehat{J})$ can be written as the disjoint union of the pre-images of these spheres and the resulting subsets of states, which we denote by $S_r$[92], will be qualitatively individuated by the value of the radius $r$. The classical property that, to any state $p \in S$, associates the value $r$ of the subset $S_r$ to which it belongs, is the magnitude of angular momentum. This explains the fact that in Classical Mechanics the possible values of angular momentum form a continuous set, contrary to the case of Quantum Mechanics where they form a discrete set. In general, the subsets $S_r$ will contain more than one state, so this construction will again not suffice to individuate single states and further techniques need to be considered.

## III.4  Conclusion

As pointed out by several authors, the description of a physical system cannot involve only the fundamental kinematical structures discussed in Chapter II. Because these are homogeneous abstract structures, the description must involve "further geometrical structure on the space of states" (Brody and Hughston) that allow to define a "labelling scheme" (Segal) and answer the question of how to specify the particular element of the algebra of properties which is to "represent a given physical quantity" (Dickson). This problem applies likewise to the Classical and Quantum formalisms and the mechanisms used to address it are also extremely similar in both cases. In this chapter, the goal was to discuss some of the underlying mathematical techniques used to specify physical properties and label states, introducing in this way a certain amount of discernibility into the Classical and Quantum arenas. The form of exposition has been chosen such that it highlights the resemblances of both Kinematics (see Table III.2, page 298).

---

[91]By '3-dimensional sphere' I mean of course spheres which can be embedded in $\mathbb{R}^3$. Hence, a 3-dimensional sphere is, in fact, a 2-dimensional manifold. In local angle coordinates $\theta$ and $\varphi$ (polar angle and azimuthal angle), the symplectic 2-form of the sphere $S_r^2$ of radius $r$ writes $\omega = r \sin(\theta) d\theta \wedge d\varphi$.

[92]Here the notation becomes tricky: $S_r^2 \subset \mathfrak{so}(3)^*$ and $S_r = J^{-1}(S_r^2) \subset S$ are not the same object!

As we have explained in Section III.1, the general strategy for endowing the kinematical description with further structure is what could be called a 'representational approach': one considers external structures $\mathcal{E}$ which have no a priori relation to the fundamental kinematical structures $\mathcal{K}$, and then looks for possible morphisms $\rho$ between them. In the case of groups, the central objects in Quantum and Classical Kinematics are respectively unitary representations—that is, group morphisms $G \xrightarrow{U} U(\mathcal{H})$—and strongly Hamiltonian actions—that is, Lie algebra morphisms $\mathfrak{g} \xrightarrow{\hat{J}} \mathcal{C}^\infty(S, \mathbb{R})$. But these are not the only possible choices. In fact, they are not even the most natural ones to consider from the point of view of states: indeed, the first guess would have been to consider morphisms from the external abstract group to the automorphisms of the space of states—in other words, to consider Poisson actions in Classical Kinematics and ray representations in Quantum Kinematics. One is hence confronted with the question of why strongly Hamiltonian actions and unitary representations play such a prominent role in the construction of the kinematical descriptions (*representation problem*).

I propose that this is intimately related to the fundamental two-fold role of physical properties highlighted in Chapter II. While both Poisson actions and strongly Hamiltonian actions allow to distinguish a specific subset of state transformations, only the latter allow to perceive these transformations as being generated by physical properties. With hindsight, this can even be considered to be the *definition* of strongly Hamiltonian actions: it is a morphism that takes abstract infinitesimal transformations and represents them unambiguously as classical properties-as-transformations. Exactly the same remark applies to unitary representations, which can be seen as morphisms taking abstract transformations and representing them unambiguously as quantum properties-as-transformations. This last point is best seen with the remark that a unitary representation is the same as a strongly Hamiltonian action on $\mathbb{P}\mathcal{H}$ which, moreover, is isometric (i.e. respects the additional geometric structure of the Quantum, see Section II.2.2). In sum, morphisms $G \xrightarrow{U} U(\mathcal{H})$ and $\mathfrak{g} \xrightarrow{\widehat{J}} \mathcal{C}^\infty(S, \mathbb{R})$ are the cornerstones in this approach to the kinematical descriptions because they respect the roles of properties-as-transformations and properties-as-quantities.

With the objects $S_\mathfrak{g} = (S, \mathfrak{g}, \widehat{J})$ and $\mathcal{H}_G = (\mathcal{H}, G, U)$ taken now as initial data, the question becomes to understand in what manner this move allows to effectively break the homogeneity of the Classical and Quantum arenas (*individuation problem*). Since these structures ought to be considered abstractly, it is necessary to first establish their correct criterion of identity. I claim that neither the usual set-theoretic identity, based on extensionality, nor the widely used notion of equivalence of group representations are acceptable candidates, for they both fail to take into account that both the fundamental kinematical structure *and* the external group are abstract. I then propose a novel criterion of identity, which applies in general to any type of morphisms between abstract objects: given any two such morphisms $\mathcal{E} \xrightarrow{\rho} \mathcal{K}$ and $\mathcal{E}' \xrightarrow{\rho'} \mathcal{K}$ in the category $\mathcal{C}$, one should ask

$$\rho =_\mathcal{C} \rho' \iff \exists (\alpha, \phi) \in \mathrm{Iso}(\mathcal{E}, \mathcal{E}') \times \mathrm{Iso}(\mathcal{K}, \mathcal{K}') \text{ such that } \begin{array}{ccc} \mathcal{E} & \xrightarrow{\rho} & \mathcal{K} \\ \alpha \downarrow \wr & & \wr \downarrow \phi \\ \mathcal{E}' & \xrightarrow{\rho'} & \mathcal{K}' \end{array}.$$

Only with this choice, do the elements of the abstract group $G$ give rise to automorphisms of the structures $\mathcal{H}_G$ and $S_\mathfrak{g}$. In this way, the individuation problem relates to the study of invariant subspaces. The decomposition of a unitary representation into irreducible components appears then as an intrinsic feature of the structure $\mathcal{H}_G$, insofar as it is an invariant of the criterion of identity, and the labels of these components allow to define a well-specified physical property. In Classical Kinematics, the symplectic decomposition of the Poisson manifold $\mathfrak{g}^*$ allows an analogous construction, but the labels refer now to transitive strongly Hamiltonian actions.

| | **Classical Kinematics** | **Quantum Kinematics** |
|---|---|---|
| **Abstract structure** | Strongly Hamiltonian $\mathfrak{g}$-action $S_{\mathfrak{g}} = (S, \mathfrak{g}, \widehat{J})$ | Unitary G-representation $H_G = (\mathcal{H}, G, U)$ |
| **Central object** | co-momentum map $\mathfrak{g} \xrightarrow{\widehat{J}} \mathcal{C}^\infty(S, \mathbb{R})$ (morphism of Lie algebras) | unitary map $G \xrightarrow{U} U(\mathcal{H})$ (morphism of groups) |
| **Respect of properties' two-fold role** | co-momentum map equivalent to Poisson representation $\mathcal{C}^\infty(\mathfrak{g}^*, \mathbb{R}) \xrightarrow{J^*} \mathcal{C}^\infty(S, \mathbb{R})$ | unitary map equivalent to JLB-representation $JL(G) \xrightarrow{U} \mathcal{B}_{\mathbb{R}}(\mathcal{H})$ |
| **Criterion of identity** | $\begin{array}{ccc} \mathfrak{g} & \xrightarrow{\widehat{J}} & \mathcal{C}^\infty(S, \mathbb{R}) \\ \alpha \downarrow \wr & & \wr \uparrow \phi^* \\ \mathfrak{g}' & \xrightarrow{\widehat{J'}} & \mathcal{C}^\infty(S', \mathbb{R}) \end{array}$ | $\begin{array}{ccc} G & \xrightarrow{U} & U(\mathcal{H}) \\ \alpha \downarrow \wr & & \wr \downarrow \Phi \\ G' & \xrightarrow{U'} & U(\mathcal{H}') \end{array}$ |
| **Properties as** | labels characterizing *transitive* strongly Hamiltonian $\mathfrak{g}$-actions | labels characterizing *irreducible* unitary $G$-representations |
| **Representat° problem** |  |  |
| **Obstruction** | $H^2(\mathfrak{g}, \mathbb{R})$ | $H^2(G, U(1))$ |
| **Absorption** | central extensions of $\mathfrak{g}$ by $\mathbb{R}$ | central extensions of $G$ by $U(1)$ |

**Table III.2** – Classical and Quantum Kinematics compared from a group-theoretical perspective.

# Conclusion

The idea of a *Chase for Individuation* was launched by the remark of a conceptual tension between a certain 'abstract way' of conceiving mathematical structures, which shows up mainly in mathematical physics, and some general features of the theoretical discourse used in the handling of Mechanics. For indeed, when the basic mathematical structures involved in Classical and Quantum Kinematics are conceived abstractly an unambiguous designation of their elements becomes problematic. Given an abstract Hilbert space or an abstract symplectic manifold, it is impossible to find a "conceptual fixation of points [...] that would enable one to reconstruct any point when it has been lost"[93]. This failure to single out elements of the structures which ought to describe the space of states or the algebra of observables of a physical system evidently clashes with the practice. The theoretical discourse is plagued with expressions referring to specific properties or states: *the* energy function $H$, the norm $p^2$ of *the* classical linear momentum, *the* angular momentum operator $L^2$, *the* spin-up state $|\frac{1}{2}, \frac{1}{2}\rangle_z$, etc. In fact, little could be said in Mechanics without this capacity to designate particular elements of the structures involved. But in the abstract approach, how is one to understand which, among all elements, is *the* mathematical representative $f$ of the physical property $\mathtt{f}$ being considered? Since there is no question of appealing to a demonstrative act—as if the elements were standing in front of our eyes and one could declare: 'take *this* element right here'—, the only reasonable stance is to consider a central task of mathematical physics to propose mechanisms of designation which allow to convey a precise meaning to those referential expressions while at the same time remaining faithful to the abstract method.

---

[93]H. Weyl. *Philosophy of Mathematics and Natural Science.* Trans. by O. Helmer. Princeton: Princeton University Press, 1949, p. 75.

This is the central problem of this thesis but the conceptual setting in which it becomes meaningful requires a great deal of clarifications. In particular, because it relies so heavily on the abstract way of conceiving Hilbert spaces, symplectic manifolds and the like, the first chapter was entirely devoted to elucidating this conception and justifying its importance for Kinematics. In this spirit, the example of the equivalence between Schrödinger's Wave Mechanics and Göttingen's Matrix Mechanics was used to indicate that below a certain level of specification the details of the mathematical construction have no relevance for the Physics and can therefore be omitted. Von Neumann showed that the development of both theories relied only on the Hilbert space structure of $L^2(\mathbb{R})$ and $l^2(\mathbb{N})$ and proposed to consider *abstract* Hilbert spaces as the *starting point* of the mathematical formulation of Quantum Mechanics. Although he did not explain in which precise way an abstract Hilbert space $\mathcal{H}$ was to differ from a particular Hilbert space, the crucial point in his methodology was to consider $\mathcal{H}$ as an independent and autonomous entity: von Neumann's precept was indeed to avoid any coordinatization and 'work directly with the abstract entity *itself*'. To arrive at the conclusion that the physical content of a theory is better grasped when one avoids superfluous technical specifications and works abstractly, I could have very well used other examples in the practice of theoretical physics. I think in particular of the description of spacetime by *abstract* Riemannian manifolds in General Relativity. This is certainly a situation worth discussing but it involves the quicksands of general covariance and diffeomorphism invariance, whose clarification would have taken us too far apart from Mechanics, and I decided to leave that exploration for future work[94].

The next step was to clarify von Neumann's core distinction between particular and abstract entities. Abstraction is a widely discussed topic in the philosophy of mathematics, but it is often viewed as a process starting from the consideration of entities of a certain kind and leading to the consideration of entities of a *new* kind.

---

[94] The classic reference for this subject is J. D. Norton. "General Covariance and the Foundations of General Relativity: Eight Decades of Dispute". In: *Reports on Progress in Physics* 56 (1993), pp. 791–858. For a discussion of this in the context of Abstract Mathematics (in the sense of Makkai), see M. Shulman. "Homotopy Type Theory: A Synthetic Approach to Higher Equalities". In: *arXiv preprint* (2016). URL: http://arxiv.org/abs/1601.05035.

With the notable exception of Marquis' work, the problem of understanding the particular/abstract distinction for entities of the *same* kind seems to be better treated in the context of mathematical structuralism. Therein, the position which insists in considering abstract structures as autonomous entities rather than as convenient linguistic tools for expressing generalizations over particulars is called *ante rem* or *sui generis* structuralism. Nonetheless, instead of analyzing this form of structuralism from the outset, I decided in my exposition to focus first on abstraction and only afterwards on structuralism. In this way, one clearly perceives which challenges to the abstract conception of structures are inherited from the general problem of abstraction and which are specific to structuralism. The upshot of the first discussion was that the handling of different well-defined *levels* of abstraction is achieved through a complex *hierarchy of identities*, where isomorphisms play a crucial role. With the consideration of a new criterion of identity, properties which appeared to be invariant at one level become intrinsic properties at the new level, and it is this feature that gives autonomy to the newly defined abstract entity. On the other hand, it is characteristic of the mathematical method of structuralism to always consider entities which are *sets* endowed with *relations*. An important challenge in this setting is then to understand the nature of the *elements* of an abstract structure, and in particular the means allowing to differentiate them. Through the analysis of the so-called 'problem of identity of structural indiscernibles', we arrived at the conclusion that abstract structures are better conceived as structured types endowed with an ungrounded primitive typed identity. This in turn allowed us to finally give a precise meaning to the *Chase for Individuation*: to ask for an 'unambiguous designation of the elements of the abstract structure $\mathcal{S}$' means to be able to reconstruct the primitive typed identity of $\mathcal{S}$ by means of structural properties.

The second chapter started initially as a small branch in the unfolding of the *Chase for Individuation*, but it eventually grew to the point of becoming a second trunk of this work. In my view, it contains the most important idea of the thesis—which is borrowed from the work of Gabriel Catren—namely: that the general relation between functions and transformations enabled by the presence of a symplectic structure on the space of states is to be viewed as a *constitutive* ingredient in the definition of *both* classical and quantum properties. Any physical property certainly has a quantitative dimension. But it also possesses a transformational one, and the physical interpretation of a given quantity should not be separated from the study of the transformations it generates. The second chapter thematized this twofold role and investigated the precise articulation between the two dimensions. Under the focus of this question, many features of the mathematical formalisms of Classical and Quantum Kinematics shone with a different light, and we reached a new understanding of some of the crucial differences between the two theories. Indeed, the claim was that the Quantum exhibits a compatibility between properties-as-quantities and properties-as-transformations which is lacking in the Classical.

In the standard formulation of both Kinematics, one could already remark that Jordan-Lie algebras—*i.e.*, algebras endowed with *two* products, one commutative (Jordan) and one anti-commutative (Lie)—offer a common language in which to describe the algebras of both classical and quantum properties. Here, the sole difference between the two lies in the associativity (Classical) or non-associativity (Quantum) of the Jordan product. Yet, a conceptual reading of what this means was not obvious. In this respect, the geometrical reformulation of Quantum Mechanics in terms of Hermitian symmetric spaces—which have, in particular, a symplectic and a Riemannian structure—offered some remarkable insights. Above all, there was the rigorous result that, indeed, *physical properties are defined by their twofold role*: in both theories, physical properties are described precisely by those *functions* over the space of states whose associated Hamiltonian vector field is an infinitesimal state *transformation*. This is trivial in Classical Kinematics but it also manages to characterize self-adjoint operators in Quantum Kinematics. Moreover, there was the surprising observation that the numerical values of quantum properties-as-quantities describe the behavior of quantum properties-as-transformations. More precisely, the indeterminacy $\Delta f$ of a property $f$ equals the norm of the Hamiltonian vector field associated to it.

This appeared as the first indication of a distinctive trait of the Quantum realm in the way the two roles of properties are articulated, but in order to reach a better understanding of this phenomenon it was necessary to examine as well the algebraic formulation of Quantum Mechanics in terms of $C^*$-algebras or real JLB-algebras. By studying the transitions from the algebra of properties to the space of states and vice versa, we finally hit upon the common geometric language describing both the classical and quantum spaces of states. This was the language of uniform Poisson spaces with a transition probability. The two geometric structures of the space of states mirror the two algebraic structures of the algebra of properties and the whole is a manifestation of the two roles of properties. In my view, the key result in this formulation was Landsman's characterization of quantum space states as those uniform Poisson spaces with a transition probability for which the symplectic leaves coincide with the sectors of the transition probability function. I interpreted as showing that the Quantum distinguishes itself from the Classical by the following consistency condition between the twofold role of properties: the discernibility introduced into the space of states by properties-as-transformations must not differ from the discernibility induced by properties-as-quantities.

After this long analysis of the mathematical structures constituting the arenas for Classical and Quantum Kinematics, the third and final chapter returned to the *Chase for Individuation* proper. In it we showed how many of the mathematical developments in the dealing of groups in Classical and Quantum Mechanics could indeed be understood as constituting a program for introducing discernibility into the homogeneous kinematical structures. By considering abstract morphisms of Lie groups and Lie algebras which respect the twofold role of physical properties, this program succeeds in unambiguously designating certain properties. In Classical Kinematics, the intrinsic symplectic decomposition of the abstract Poisson manifold $\mathfrak{g}^*$ allows to view classical properties as labels characterizing transitive strongly Hamiltonian actions. In Quantum Kinematics, on the other hand, the decomposition of an abstract unitary representation into irreducible components allows to view quantum properties as labels characterizing irreducible unitary representations. In this abstract constructive approach to the kinematical description of physical system, groups therefore appeared as a fundamental tool to introduce a notion of *difference* into the space of states and the algebra of properties. This does not contradict in any way the traditional view on groups as describing symmetries, since the states related by the group action are among those that this method of differentiation does not succeed in distinguishing. However, the 'sameness' of that which is related by a symmetry is not to be viewed as a consequence of the introduction of the group but rather as remnant of the underlying homogeneity of the fundamental kinematical structures.

Now, the consideration of groups is not the only strategy for introducing discernibility into the Classical and Quantum kinematical arenas. The *Chase for Individuation* offers a perspective from which one could attempt to understand many other developments in the mathematical foundations of Mechanics. An important example is the theory of *systems of imprimitivity* developed by the american mathematician George Mackey in the 1950's, which can be viewed as an elucidation of the mathematical structures lurking behind the labelling scheme that differentiates linear momentum and position. Instead of taking groups as the additional external structure to consider, this theory explores representations of *groupoids*. Therein, a quantum system is described by a unitary representation of an action Lie groupoid $G \ltimes E$, whereas a classical system is described by a Poisson representation of an action Lie algebroid $\mathfrak{g} \ltimes E$[95]. Again, one perceives the idea, already hinted at in the group-theoretical approach, that the transition from Classical to Quantum Kinematics involves a phenomenon of *integration* from infinitesimal to global transformations. This is an interesting path worth further investigation.

---

[95]This groupoid-theoretical view on systems of imprimitivity is presented in N. P. Landsman. "Lie Groupoids and Lie Algebroids in Physics and Noncommutative Geometry". In: *Journal of Geometry and Physics* 56.1 (2006), pp. 24–54. URL: http://arxiv.org/abs/math-ph/0506024.

# References

Abraham, R. and J. E. Marsden. *Foundations of Mechanics*. 2nd ed. Redwood City: Addison-Wesley Publishing Company, 1978 (cit. on pp. 14, 140, 147).

Abraham, R., J. E. Marsden, and T. S. Ratiu. *Manifold, Tensor Analysis, and Applications*. 2nd ed. New York: Springer-Verlag, 1988 (cit. on p. 157).

Adams, R. M. "Primitive Thisness and Primitive Identity". In: *The Journal of Philosophy* (1979), pp. 5–26 (cit. on p. 106).

Aerts, D. and S. Aerts. "Towards a General Operational and Realistic Framework for Quantum Mechanics and Relativity Theory". In: *Quo Vadis Quantum Mechanics?* Ed. by A. C. Elitzur, S. Dolev, and N. Kolenda. Berlin: Springer, 2005, pp. 153–207 (cit. on pp. 194, 196).

Akemann, C. "A Gelfand Representation Theory for C*-algebras". In: *Pacific Journal of Mathematics* 39.1 (1971), pp. 1–11 (cit. on p. 218).

Alekseevsky, D. and P. W. Michor. "Differential Geometry of 𝔤-manifolds". In: *Differential Geometry and its Applications* 5.4 (1995), pp. 371–403. URL: http://arxiv.org/abs/math/9309214 (cit. on pp. 246, 293).

Alfsen, E. M., H. Hanche-Olsen, and F. W. Shultz. "State Spaces of *C\**-algebras". In: *Acta Mathematica* 144 (1980), pp. 267–305 (cit. on p. 218).

Alfsen, E. M. and F. W. Shultz. *State Spaces of Operator Algebras*. Boston: Birkhäuser, 2001 (cit. on pp. 146, 198, 207, 210, 217, 218).

Anandan, J. and Y. Aharonov. "Geometry of Quantum Evolution". In: *Physical review letters* 65.14 (1990), pp. 1697–1700 (cit. on p. 183).

Angelelli, I. "Frege and Abstraction". In: *Philosophia Naturalis* 21 (1984), pp. 453–471 (cit. on p. 62).

Angelelli, I. "Adventures of Abstraction". In: *Poznarí Studies in the Philosophy of the Sciences and the Humanities* 82 (2004), pp. 11–35 (cit. on p. 88).

Arageorgis, A. "Fields, Particles, and Curvature: Foundations and Philosophical Aspects of Quantum Field Theory in Curved Space-Time". PhD thesis. University of Pittsburgh, 1995 (cit. on p. 196).

Arnold, V. I. *Mathematical Methods of Classical Mechanics*. Trans. by K. Vogtmann and A. Weinstein. 2nd ed. Vol. 60. New York: Springer-Verlag, 1989 (cit. on p. 147).

Ashtekar, A. and J. Lewandowski. "Background Independent Quantum Gravity: A Status Report". In: *Classical and Quantum Gravity* 21.15 (2004). URL: http://arxiv.org/abs/gr-qc/0404018 (cit. on p. 141).

Ashtekar, A. and T. A. Schilling. "Geometrical Formulation of Quantum Mechanics". In: *On Einstein's Path: Essays in Honor of Engelbert Schücking*. Ed. by A. Harvey. New York: Springer, 1997, pp. 23–65. URL: http://arxiv.org/abs/gr-qc/9706069 (cit. on pp. 165, 168, 172, 173, 175, 177, 179–182).

Awodey, S. "Structure in Mathematics and Logic: A Categorical Perspective". In: *Philosophia Mathematica* 4 (1996), pp. 209–237 (cit. on pp. 92, 93).

——— "An Answer to Hellman's Question: 'Does Category Theory Provide a Framework for Mathematical Structuralism?'" In: *Philosophia Mathematica* 12.1 (2004), pp. 54–64 (cit. on pp. 70, 91, 92, 97, 104).

——— "Structuralism, Invariance, and Univalence". In: *Philosophia Mathematica* 22.1 (2014), pp. 1–11 (cit. on pp. 89, 93).

Baez, J. C. "Quantum Quandaries: a Category-Theoretic Perspective". In: *The Structural Foundations of Quantum Gravity*. Ed. by S. French, D. Rickles, and J. Saatsi. New York: Oxford University Press, 2006. URL: http://arxiv.org/abs/quant-ph/0404040 (cit. on p. 262).

——— "Division Algebras and Quantum Theory". In: *Foundations of Physics* 42.7 (2012), pp. 819–855. URL: http://arxiv.org/abs/1101.5690 (cit. on p. 159).

Bargmann, V. "On Unitary Ray Representations of Continuous Groups". In: *Annals of Mathematics* 59.1 (1954), pp. 1–46 (cit. on pp. 264, 265).

Bell, J. S. "Against 'measurement'". In: *62 Years of Uncertainty: Erice, 5-14 August 1989*. Plenum Publishers, 1990. (Reprinted in: J. S. Bell. *Speakable and Unspeakable in Quantum*

*Mechanics.* 2nd ed. Cambridge: Cambridge University Press, 2004, pp. 213–231) (cit. on p. 6).

Bell, J. S. *Speakable and Unspeakable in Quantum Mechanics.* 2nd ed. Cambridge: Cambridge University Press, 2004 (cit. on pp. 6, 306).

Benacerraf, P. "What Numbers Could Not Be". In: *Philosophical Review* 74 (1965), pp. 47–73 (cit. on p. 123).

Bilu, Y., Y. Bugeaud, and M. Mignotte. *The Problem of Catalan.* Springer, 2014 (cit. on p. 96).

Bohr, N. "On the Constitution of Atoms and Molecules". In: *Philosophical Magazine* 26.151 (1913), pp. 1–25. (Reprinted in: N. Bohr. *Collected Works.* Ed. by U. Hoyer. Vol. 2. Amsterdam: Elsevier, 2008, pp. 161–185) (cit. on p. 11).

—— "On the Constitution of Atoms and Molecules (Part II)". In: *Philosophical Magazine* 26.153 (1913), pp. 476–502. (Reprinted in: N. Bohr. *Collected Works.* Ed. by U. Hoyer. Vol. 2. Amsterdam: Elsevier, 2008, pp. 188–214) (cit. on p. 11).

—— *Collected Works.* Ed. by U. Hoyer. Vol. 2. Amsterdam: Elsevier, 2008 (cit. on pp. 11, 307).

Boothby, W. M. "Transitivity of the Automorphisms of Certain Geometric Structures". In: *Transactions of the American Mathematical Society* 137 (1969), pp. 93–100 (cit. on p. 140).

Borges, J. L. "Funes the Memorious". In: *Ficciones.* Trans. by A. Kerrigan. Grove Press, 1962. (original Spanish title: *Funes el memorioso*, first published in the journal *La Nación* in June 1942) (cit. on p. 69).

Born, M., W. Heisenberg, and P. Jordan. "Zur Quantenmechanik II". In: *Zeitschrift für Physik* 35 (1926), pp. 557–615 (cit. on pp. 25, 185).

—— "On Quantum Mechanics II". In: *Sources of Quantum Mechanics.* Ed. by B. Van der Waerden. New York: Dover Publications, Inc., 1967, pp. 321–384 (cit. on pp. 21, 25, 33, 35, 163, 185).

Born, M. and P. Jordan. "Zur Quantenmechanik". In: *Zeitschrift für Physik* 34 (1925), pp. 858–888 (cit. on p. 22).

—— "On Quantum Mechanics". In: *Sources of Quantum Mechanics.* Ed. by B. Van der Waerden. New York: Dover Publications, Inc., 1967, pp. 277–306 (cit. on pp. 22, 23, 33, 34).

Born, M. and N. Wiener. "A New Formulation of The Laws of Quantization of Periodic and Aperiodic Phenomena". In: *Journal of Mathematics and Physics (MIT)* (1925–1926), pp. 84–98. (Reprinted in: N. Wiener. *Norbert Wiener: Collected Works. Volume III*. Ed. by P. Masani. Cambridge: The MIT Press, 1981, pp. 427–441) (cit. on p. 35).

Brading, K. and H. R. Brown. "Noether's Theorems and Gauge Symmetries". In: *arXiv preprint* (2000). URL: http://arxiv.org/abs/hep-th/0009058 (cit. on p. 145).

Bröcker, T. and T. tom Dieck. *Representations of Compact Lie Groups*. 1st ed. New York: Springer, 1985 (cit. on p. 288).

Brody, D. C. and L. P. Hughston. "Geometric Quantum Mechanics". In: *Journal of geometry and physics* 38.1 (2001), pp. 19–53. URL: http://arxiv.org/abs/quant-ph/9906086 (cit. on pp. 168, 188, 230, 238).

Burgess, J. P. "Review of *Philosophy of Mathematics: Structure and Ontology* by Stewart Shapiro". In: *Notre Dame Journal of Formal Logic* 40.2 (1999), pp. 283–291 (cit. on pp. 63, 105).

Button, T. "Realistic Structuralism's Identity Crisis: A Hybrid Solution". In: *Analysis* 66 (2006), pp. 216–222 (cit. on pp. 113, 114).

Cantoni, V. "Superposition of Physical States: a Metric Viewpoint". In: *Helvetica Physica Acta* 58 (1985), pp. 956–968 (cit. on p. 190).

Carnap, R. *Untersuchungen zur allgemeinen Axiomatik*. Darmstadt: Wissenschaftliche Buchgesellschaft, 2000 (cit. on p. 109).

Cartier, P. "Notion de spectre". In: *Première école d'été : Histoire conceptuelle des mathématiques - Dualité Algèbre-Géométrie*. Maison des Sciences de l'Homme. Universidade de Brasilia, 2008, pp. 232–242. URL: http://semioweb.msh-paris.fr/f2ds/docs/dualite_2008/dualite_doc_final_2008.pdf (cit. on p. 217).

Catren, G. "On Classical and Quantum Objectivity". In: *Foundations of Physics* 38.5 (2008), pp. 470–487. URL: http://philsci-archive.pitt.edu/4298/ (cit. on pp. 146, 150, 164).

——— "A Throw of the Quantum Dice Will Never Abolish the Copernican Revolution". In: *Collapse: Philosophical Research and Development* 5 (2009), pp. 453–500 (cit. on p. 136).

——— "On the Relation Between Gauge and Phase Symmetries". In: *Foundations of Physics* 44 (2014), pp. 1317–1335 (cit. on pp. 164, 250, 294).

Châtelet, G. *Figuring Space: Philosophy, Mathematics and Physics*. Trans. by R. Shore and M. Zagha. Dordrecht, The Netherlands: Springer Science & Business Media, 2000 (cit. on p. 230).

Chernoff, P. R. and J. E. Marsden. *Properties of Infinite Dimensional Hamiltonian Systems*. Lecture Notes in Mathematics. Heidelberg: Springer-Verlag, 1974 (cit. on p. 147).

Chihara, C. *A Structural Account of Mathematics*. Oxford: Oxford University Press, 2004 (cit. on p. 123).

Cirelli, R., M. Gatti, and A. Manià. "On the Nonlinear Extension of Quantum Superposition and Uncertainty Principles". In: *Journal of Geometry and Physics* 29.1 (1999), pp. 64–86 (cit. on p. 177).

———— "The Pure State Space of Quantum Mechanics as Hermitian Symmetric Space". In: *Journal of Geometry and Physics* 45.3 (2003), pp. 267–284. URL: http://arxiv.org/abs/quant-ph/0202076 (cit. on pp. 140, 168–170, 182, 185, 188, 190).

Cohen-Tannoudji, C., B. Diu, and F. Laloë. *Mécanique quantique*. Paris: Hermann, Collection Enseignement des Sciences, 1973 (cit. on p. 46).

Connes, A. *Noncommutative Geometry*. Trans. by S. Berberian. London: Academic Press, 1994 (cit. on p. 20).

Cook, R. T., ed. *The Arché Papers on the Mathematics of Abstraction*. Dordrecht: Springer, 2007 (cit. on p. 60).

Corfield, D. *Towards a Philosophy of Real Mathematics*. Cambridge: Cambridge University Press, 2003 (cit. on p. 250).

Corichi, A. "Quantum Superposition Principle and Geometry". In: *General Relativity and Gravitation* 38.4 (2006), pp. 677–687. URL: http://arxiv.org/abs/quant-ph/0407242 (cit. on pp. 189, 190).

Dalla Chiara, M. L. and G. Toraldo di Francia. "Individuals, Kinds and Names in Physics". In: *Bridging the Gap: Philosophy, Mathematics, Physics*. Dordrecht: Kluwer Academic Publishers, 1993, pp. 261–283 (cit. on p. 133).

Darrigol, O. *From c-Numbers to q-Numbers: The Classical Analogy in the History of Quantum Theory*. Berkeley: University of California Press, 1992 (cit. on pp. 19, 44, 48, 135).

Dell'Ambrosio, I. "Categories of $C^*$-algebras". Lecture Notes. URL: http://math.univ-lille1.fr/~dellambr/exercise_C_algebras.pdf (cit. on p. 216).

Dickson, M. "Non-relativistic Quantum Mechanics". In: *Philosophy of Physics (Handbook of the Philosophy of Science) 2 volume set*. Ed. by J. Butterfield and J. Earman. Vol. 1. Amsterdam: North-Holland Publishing Co., 2007, pp. 275–415. URL: http://philsci-archive.pitt.edu/3321/ (cit. on pp. 239, 244).

Dirac, P. A. M. "The Fundamental Equations of Quantum Mechanics". In: *Proceedings of the Royal Society of London* A109 (1925), pp. 642–653. (Reprinted in: P. A. M. Dirac. *The Collected Works of P.A.M. Dirac: 1924–1948*. Ed. by R. Dalitz. Cambridge: Cambridge University Press, 1995, pp. 65–78) (cit. on pp. 21, 24).

——— "Quantum Mechanics and a Preliminary Investigation of The Hydrogen Atom". In: *Proceedings of the Royal Society of London* A110 (1926), pp. 561–579. (Reprinted in: P. A. M. Dirac. *The Collected Works of P.A.M. Dirac: 1924–1948*. Ed. by R. Dalitz. Cambridge: Cambridge University Press, 1995, pp. 85–105) (cit. on pp. 25, 43).

——— "The Physical Interpretation of the Quantum Dynamics". In: *Proceedings of the Royal Society of London* 113 (1927), pp. 621–641. (Reprinted in: P. A. M. Dirac. *The Collected Works of P.A.M. Dirac: 1924–1948*. Ed. by R. Dalitz. Cambridge: Cambridge University Press, 1995, pp. 207–229) (cit. on pp. 40, 41, 43, 45, 47, 52, 54).

——— *The Principles of Quantum Mechanics*. 1st ed. Oxford: Clarendon Press, 1930 (cit. on pp. 42, 52, 54).

——— *The Principles of Quantum Mechanics*. 4th ed. Oxford: Oxford University Press, 1958 (cit. on pp. 166, 188).

——— *The Collected Works of P.A.M. Dirac: 1924–1948*. Ed. by R. Dalitz. Cambridge: Cambridge University Press, 1995 (cit. on pp. 21, 24, 25, 41, 310).

Dixmier, J. *Les C\*-algèbres et leurs représentations*. 2nd ed. Paris: Gauthiers-Villars, 1969 (cit. on pp. 198, 202).

Duke, G. *Dummett on Abstract Objects*. History of Analytical Philosophy. Hampshire: Palgrave MacMillan, 2012 (cit. on p. 59).

Dummett, M. *Frege: Philosophy of Mathematics*. London: Duckworth, 1991 (cit. on p. 59).

Duncan, A. and M. Janssen. "From Canonical Transformations to Transformation Theory, 1926–1927: The Road to Jordan's *Neue Bergündung*". In: *Studies In History and Philosophy of Science Part B: Studies In History and Philosophy of Modern Physics* 40.4 (2009), pp. 352–362 (cit. on p. 42).

Eckart, C. "The Solution of the Problem of the Single Oscillator by a Combination of Schrödinger's Wave Mechanics and Lanczos' Field Theory". In: *Proceedings of the National Academy of Science* 12 (1926), pp. 473–476 (cit. on p. 40).

Feferman, S. "Categorical Foundations and Foundations of Category Theory". In: *Logic, Foundations of Mathematics, and Computability Theory (Proc. Fifth Internat. Congr. Logic, Methodology and Philos. of Sci., Univ. Western Ontario).* Philos. Sci. Dordrecht, The Netherlands: University Western Ontario, 1977, pp. 149–169 (cit. on p. 97).

Fell, J. M. G. and R. S. Doran. *Representations of \*-Algebras, Locally Compact Groups, and Banach \*-Algebraic Bundles.* Vol. 1. San Diego: Academic press, 1988 (cit. on pp. 198, 218).

Fréchet, M. R. "Les ensembles abstraits et le calcul fonctionnel". In: *Rendiconti del Circolo Matematico di Palermo (1884–1940)* 30 (1910), pp. 1–26 (cit. on p. 124).

——— "Abstract Sets, Abstract Spaces and General Analysis". In: *Mathematics Magazine* 24.3 (1951), pp. 147–155 (cit. on p. 79).

French, S. and D. Krause. *Identity in Physics: A Historical, Philosophical, and Formal Analysis.* Oxford: Oxford University Press, 2006 (cit. on pp. 132–134).

Freyd, P. "Homotopy is Not Concrete". In: *The Steenrod Algebra and Its Applications: A Conference to Celebrate N.E. Steenrod's Sixtieth Birthday.* Ed. by F. P. Peterson. Springer, 1970, pp. 25–34. (Reprinted in: Reprints in Theory and Applications of Categories, 6 (2004) pp. 1-10.) URL: http://www.tac.mta.ca/tac/reprints/articles/6/tr6abs.html (cit. on p. 98).

Gelfand, I. "We Do Not Choose Mathematics as Our Profession, It Chooses Us: Interview with Yuri Manin". Trans. by M. Saul. In: *Notices of the AMS* 56.10 (2009), pp. 1268–1274 (cit. on p. 98).

Gelfand, I. and M. Naimark. "On the Imbedding of Normed Rings Into the Ring of Operators in Hilbert Space". In: *Matematicheskii Sbornik* 12 (1943), pp. 197–213 (cit. on pp. 210, 215).

Geroch, R. "A Method for Generating Solutions of Einstein's Equations". In: *Journal of Mathematical Physics* 12.6 (1971), pp. 918–924 (cit. on p. 180).

Ghirardi, G. C., A. Rimini, and T. Weber. "Unified Dynamics for Microscopic and Macroscopic Systems". In: *Physical Review D* 34.2 (1986), pp. 470–491 (cit. on p. 188).

Goodman, N. and W. V. O. Quine. "Steps Toward a Constructive Nominalism". In: *Journal of Symbolic Logic* 12 (1947), pp. 105–122 (cit. on p. 59).

Gracia-Bondía, J. M., J. C. Várilly, and H. Figueroa. *Elements of Noncommutative Geometry*. Boston: Birkhäuser, 2011 (cit. on p. 216).

Grothendieck, A. *Récoltes et semailles – Réflexions et témoignage sur un passé de mathématicien*. 1985 (cit. on p. 10).

Guillemin, V. and S. Sternberg. "Geometric Quantization and Multiplicities of Group Representations". In: *Inventiones Mathematicae* 67 (1982), pp. 515–538 (cit. on p. 294).

————— *Variations on a Theme by Kepler*. Vol. 42. American Mathematical Soc., 2006 (cit. on p. 145).

Haag, R. *Local Quantum Physics – Fields, Particles, Algebras*. 2nd ed. Heidelberg: Springer-Verlag, 1996 (cit. on pp. 194, 196).

Haag, R. and D. Kastler. "An Algebraic Approach to Quantum Field Theory". In: *Journal of Mathematical Physics* 5.7 (1964), pp. 848–861 (cit. on pp. 194, 195).

Hale, B. and C. Wright. "Logicism in The Twenty-First Century". In: *The Oxford Handbook of Philosophy of Mathematics and Logic*. Ed. by S. Shapiro. New York: Oxford University Press, 2005, pp. 166–202 (cit. on p. 60).

Heidegger, M. *What is a Thing?* Trans. by W. B. Barton and V. Deutsch. Indiana: Gateway Editions, Ltd., 1967 (cit. on p. 9).

Heisenberg, W. "Über quantentheoretische Umdeutung kinematischer und mechanischer Beziehungen". In: *Zeitschrift für Physik* 33 (1925), pp. 879–893 (cit. on pp. 6, 20).

————— "Quantum-theoretical Re-interpretation of Kinematic and Mechanical Relations". In: *Sources of Quantum Mechanics*. Ed. by B. Van der Waerden. New York: Dover Publications, Inc., 1967, pp. 261–276 (cit. on pp. 6, 20, 22, 163).

Hellman, G. "Three Varieties of Mathematical Structuralism". In: *Philosophia Mathematica* 9.3 (2001), pp. 184–211 (cit. on p. 90).

————— "Structuralism". In: *The Oxford Handbook of Philosophy of Mathematics and Logic*. Ed. by S. Shapiro. New York: Oxford University Press, 2005, pp. 536–562 (cit. on pp. 90, 94, 96, 101, 104).

Hepp, K. "Quantum Theory of Measurement and Macroscopic Observables". In: *Helvetica Physica Acta* 45 (1972), pp. 237–248 (cit. on p. 221).

Hermens, R. "Quantum Mechanics, From Realism to Intuitionism". MA thesis. Radboud University Nijmegen, 2010. URL: http://arxiv.org/abs/1002.1410 (cit. on p. 161).

Hooker, C. *The Logico-Algebraic Approach to Quantum Mechanics. Volume I: Historical Evolution*. Dordrecht, The Netherlands: Reidel Publishing Company, 1975 (cit. on p. 194).

Iglesias-Zemmour, P. *Symétries et moment*. Paris: Hermann, Éditeurs des Sciences et des Arts, 2000 (cit. on pp. 247, 262).

―――― *Aperçu des origines de la géométrie symplectique*. Actes du colloque "Histoire des géométries", vol. 1. 2004 (cit. on pp. 2, 3).

Isham, C. and J. Butterfield. "Topos Perspective on the Kochen-Specker Theorem: I. Quantum States as Generalized Valuations". In: *International Journal of Theoretical Physics* 37 (1998), pp. 2669–2733. URL: http://arxiv.org/abs/quant-ph/9803055 (cit. on p. 154).

Jammer, M. *The Conceptual Development of Quantum Mechanics*. 2nd ed. Los Angeles: Tomash Publishers, 1989 (cit. on pp. 18, 19, 27, 44).

Jauch, J.-M. "Systems of Observables in Quantum Mechanics". In: *Helvetica Physica Acta* 33 (1960), pp. 711–726 (cit. on p. 47).

―――― *Foundations of Quantum Mechanics*. Reading: Addison-Wesley, 1968 (cit. on p. 194).

Jauch, J.-M. and C. Piron. "What is Quantum Logic?" In: *Quanta, Essays in Theoretical Physics, dedicated to Gregor Wentzel*. Ed. by P. Freund, C. Goebel, and Y. Nambu. Chicago: University of Chicago Press, 1970, pp. 166–181 (cit. on p. 194).

Jordan, P. "Über eine neue Begründung der Quantenmechanik". In: *Zeitschrift für Physik* 40 (1927), pp. 809–838 (cit. on p. 42).

―――― "Über eine neue Begründung der Quantenmechanik II". In: *Zeitschrift für Physik* 44 (1927), pp. 1–25 (cit. on p. 42).

Jordan, P., J. von Neumann, and E. P. Wigner. "On an Algebraic Generalization of the Quantum Mechanical Formalism". In: *Annals of Mathematics* 35 (1934), pp. 29–64. (Reprinted in: J. von Neumann. *Collected Works*. Ed. by A. H. Taub. Oxford: Pergamon Press, 1961, Vol. II, pp. 409–444) (cit. on p. 193).

Kallmann, H. "Von der Anfängen der Quantentheorie—Eine persönliche Rückschau". In: *Physikalische Blätter* 22 (1966), pp. 489–500 (cit. on p. 31).

Kaplansky, I. "Algebras of Type I". In: *Annals of Mathematics* 56.3 (1952), pp. 460–472 (cit. on p. 237).

Keränen, J. "The Identity Problem for Realist Structuralism". In: *Philosophia Mathematica* 9.3 (2001), pp. 308–330 (cit. on pp. 105–108).

Kibble, T. "Geometrization of Quantum Mechanics". In: *Communications in Mathematical Physics* 65.2 (1979), pp. 189–201 (cit. on pp. 168, 176).

Knapp, A. W. *Lie Groups Beyond an Introduction.* 2nd ed. Boston: Birkhäuser, 2002 (cit. on pp. 277, 286).

Kobayashi, S. and K. Nomizu. *Foundations of Differential Geometry.* Vol. 1. New York: Wiley, 1963 (cit. on p. 170).

——— *Foundations of Differential Geometry.* Vol. 2. New York: Wiley, 1969 (cit. on p. 170).

Kosmann-Schwarzbach, Y. *Les Théorèmes de Noether. Invariance et lois de conservation au XXème siècle.* Palaiseau: Les éditions de l'école polytechnique, 2004 (cit. on p. 145).

Krömer, R. *Tool and Object: A History and Philosophy of Category Theory.* Vol. 32. Historical Studies. Berlin: Springer Science & Business Media, 2007 (cit. on p. 57).

Ladyman, J. "Mathematical Structuralism and The Identity of Indiscernibles". In: *Analysis* 65.3 (2005), pp. 218–221 (cit. on pp. 110, 111).

——— "Scientific Structuralism: On The Identity and Diversity of Objects in a Structure". In: *Aristotelian Society Supplementary Volume.* Vol. 81. 1. Wiley Online Library. 2007, pp. 23–43 (cit. on pp. 114, 119).

Lanczos, C. "Über eine feldmäßige Darstellung der neuen Quantenmechanik". In: *Zeitschrift für Physik* 35 (1926), pp. 812–830 (cit. on p. 185).

Landsman, N. P. "Quantization and Superselection Sectors I. Transformation Group $C^*$-algebras". In: *Rev. Math. Phys* 2 (1990), pp. 45–72 (cit. on p. 213).

——— "Classical and Quantum Representation Theory". In: *arXiv preprint* (1994). URL: http://arxiv.org/abs/hep-th/9411172 (cit. on pp. 209, 267).

——— "The Infinite Unitary Group, Howe Dual Pairs, and the Quantization of Constrained Systems". In: *arXiv preprint* (1994). URL: https://arxiv.org/abs/hep-th/9411171 (cit. on p. 177).

Landsman, N. P. "Poisson Spaces With a Transition Probability". In: *Review of Mathematical Physics* 9.1 (1997), pp. 29–57. URL: http://arxiv.org/abs/quant-ph/9603005 (cit. on pp. 218, 220, 221, 226, 227).

———— "Lecture Notes on $C^*$-algebras, Hilbert $C^*$-modules, and Quantum Mechanics". In: (1998). URL: http://arxiv.org/abs/math-ph/9807030 (cit. on p. 141).

———— *Mathematical Topics Between Classical and Quantum Mechanics*. New York: Springer, 1998 (cit. on pp. 5, 151, 157, 159, 173–175, 192, 197, 199–202, 207–213, 215, 217–220, 222–224, 226, 227, 229, 237, 247–250, 252, 253, 257–259, 267, 271–274, 291, 293, 294).

———— "Quantum Mechanics and Representation Theory: the New Synthesis". In: *Acta Applicandae Mathematica* 81.1 (2004), pp. 167–189 (cit. on p. 294).

———— "Lie Groupoids and Lie Algebroids in Physics and Noncommutative Geometry". In: *Journal of Geometry and Physics* 56.1 (2006), pp. 24–54. URL: http://arxiv.org/abs/math-ph/0506024 (cit. on pp. 252, 304).

———— "Between Classical and Quantum". In: *Philosophy of Physics (Handbook of the Philosophy of Science) 2 volume set*. Ed. by J. Butterfield and J. Earman. Vol. 1. Amsterdam: North-Holland Publishing Co., 2007, pp. 417–554. URL: http://arxiv.org/abs/quant-ph/0506082 (cit. on pp. 1, 55, 221).

Lang, S. *Algebra*. 3rd ed. New York: Springer GTM, 2002 (cit. on p. 245).

Lawvere, F. W. "The Category of Categories as a Foundation for Mathematics". In: *Proceedings of the Conference on Categorical Algebra*. Springer. Berlin, 1966, pp. 1–20 (cit. on p. 57).

———— "Variable Quantities and Variable Structures in Topoi". In: *Algebra, Topology and Category Theory - A collection of Papers in Honor of Samuel Eilenberg*. Ed. by A. Heller and M. Tierney. London: Academic Press, 1976, pp. 101–131 (cit. on pp. 124, 127, 131).

———— "Foundations and Applications: Axiomatization and Education". In: *The Bulletin of Symbolic Logic* 9.2 (2003), pp. 213–224 (cit. on p. 134).

Lawvere, F. W. and R. Rosebrugh. *Sets for Mathematics*. Cambridge: Cambridge University Press, 2003 (cit. on p. 2).

Leifer, M. S. "Is the Quantum State Real? An Extended Review of $\psi$-ontology Theorems". In: *Quanta* 3 (2014), pp. 67–155 (cit. on p. 37).

Leitgeb, H. and J. Ladyman. "Criteria of Identity and Structuralist Ontology". In: *Philosophia Mathematica* 16.3 (2008), pp. 388–396 (cit. on pp. 113, 116).

Lewis, D. *On the Plurality of Worlds*. New York: Basil Blackwell, 1986 (cit. on pp. 59, 143).

Linckelmann, M. "Alperin's weight conjecture in terms of equivariant Bredon cohomology". In: *Mathematische Zeitschrift* 250.3 (2005), pp. 495–513 (cit. on p. 242).

Linnebo, Ø. and R. Pettigrew. "Two Types of Abstraction for Structuralism". In: *The Philosophical Quaterly* 64.255 (2014), pp. 267–283 (cit. on pp. 65, 67).

Livine, E. "Covariant Loop Quantum Gravity?" In: *Approaches to Quantum Gravity*. Ed. by D. Oriti. Cambridge: Cambridge University Press, 2009, pp. 253–271 (cit. on p. 139).

London, F. "Über die Jacobischen Transformationen der Quantenmechanik". In: *Zeitschrift für Physik* 37 (1926), pp. 383–386 (cit. on p. 40).

Loux, M. J. *Metaphysics, a Contemporary Introduction*. 3rd. New York: Routledge, 2006 (cit. on p. 143).

Lowe, E. J. "Individuation". In: *The Oxford Handbook of Metaphysics*. Ed. by M. Loux and D. Zimmerman. Oxford: Oxford University Press, 2003, pp. 75–95 (cit. on p. 132).

Mac Lane, S. "Some Recent Advances in Algebra". In: *American Mathematical Monthly* 46 (1939), pp. 3–19 (cit. on p. 82).

—— *Mathematics, Form and Function*. New York: Springer, 1986 (cit. on pp. 62, 68, 69).

—— "Structure in Mathematics". In: *Philosophia Mathematica* 4.2 (1996), pp. 174–183 (cit. on p. 57).

Mac Lane, S. and I. Moerdijk. *Sheaves in Geometry and Logic: A First Introduction to Topos Theory*. New York: Springer-Verlag, 1992 (cit. on p. 281).

MacBride, F. "Speaking with Shadows: A Study of Neo-logicism". In: *British Journal for the Philosophy of Science* 54 (2003), pp. 103–163 (cit. on p. 60).

—— "Structuralism Reconsidered". In: *The Oxford Handbook of Philosophy of Mathematics and Logic*. Ed. by S. Shapiro. New York: Oxford University Press, 2005, pp. 563–589 (cit. on pp. 110, 118).

Majer, U. "The Axiomatic Method and the Foundations of Science: Historical Roots of Mathematical Physics in Göttingen (1900-1930)". In: *John von Neumann and the Foundations of Quantum Physics*. Ed. by M. Rédei and M. Stöltzner. Dordrecht: Kluwer Academic Publishers, 2001, pp. 11–31 (cit. on p. 48).

Makkai, M. "Towards a Categorical Foundation of Mathematics". In: *Logic Colloquium '95 (Haifa)*. Vol. 11. Lecture Notes Logic. Berlin: Springer, 1998, pp. 153–190 (cit. on pp. 57, 87, 88, 121, 122, 124, 125, 127, 132).

Manin, Y. I. "Georg Cantor and His Heritage". In: *Tr. Mat. Inst. Steklova* 246 (2004), pp. 208–216 (cit. on pp. 131, 133).

Marquis, J.-P. "Categories, Sets and the Nature of Mathematical Entities". In: *The Age of Alternative Logics: Assessing Philosophy of Logic and Mathematics Today*. Ed. by J. van Benthem et al. Dordrecht: Springer, 2006, pp. 181–192 (cit. on p. 80).

——— "Categorical Foundations of Mathematics – Or how to provide foundations for *abstract* mathematics". In: *The Review of Symbolic Logic* 6.1 (2013), pp. 51–75 (cit. on pp. 80, 81, 84, 87, 125).

——— "Mathematical Forms and Forms of Mathematics: Leaving the Shores of Extensional Mathematics". In: *Synthese* 190.12 (2013), pp. 2141–2164 (cit. on pp. 71, 78, 98).

——— "Mathematical Abstraction, Conceptual Variation and Identity". In: *Logic, Methodology and Philosophy of Science, Proceedings of the 14th International Congress (Nancy)*. Ed. by P. E. Bour et al. London: College Publications, 2014, pp. 299–322 (cit. on pp. 78, 80, 81, 86).

——— "Stairway to Heaven – The Abstract Method and Levels of Abstraction in Mathematics". In: (forthcoming) (cit. on pp. 80–84).

Marsden, J. E. and T. S. Ratiu. *Introduction to Mechanics and Symmetry. A Basic Exposition of Classical Mechanical Systems*. 2nd ed. New York: Springer, 1999 (cit. on pp. 147, 246–250, 252, 253, 262, 263, 269, 294).

Marsden, J. E. and A. Weinstein. "Reduction of Symplectic Manifolds With Symmetry". In: *Reports on Mathematical Physics* 5.1 (1974), pp. 121–130 (cit. on pp. 248, 292).

Mazur, B. "When is One Thing Equal to Some Other Thing?" In: *Proof and Other Dilemmas: Mathematics and Philosophy*. Ed. by B. Gold and R. A. Simons. Spectrum Series. Mathematical Association of America, 2008, pp. 221–242 (cit. on p. 74).

McCrimmon, K. *A Taste of Jordan Algebras*. New York: Springer-Verlag, 2004 (cit. on p. 157).

Mehra, J. and H. Rechenberg. *The Historical Development of Quantum Theory. Volume 3: The Formulation of Matrix Mechanics and Its Modifications*. New York: Springer-Verlag, 1982 (cit. on pp. 20–22, 185).

Mehra, J. and H. Rechenberg. *The Historical Development of Quantum Theory. Volumes 1 – 6*. New York: Springer-Verlag, 1982-2001 (cit. on p. 19).

———  *The Historical Development of Quantum Theory. Volume 5: Erwin Schrödinger and the Rise of Wave Mechanics*. New York: Springer-Verlag, 1987 (cit. on pp. 29, 31, 36).

———  *The Historical Development of Quantum Theory. Volume 6: The Completion of Quantum Mechanics. 1926–1941*. New York: Springer-Verlag, 2000 (cit. on p. 48).

Michor, P. W. and C. Vizman. "N–transitivity of Certain Diffeomorphism Groups". In: *Acta Math. Univ. Comenianae* 63.2 (1994), pp. 221–225. URL: http://arxiv.org/abs/dg-ga/9406005# (cit. on p. 140).

Mielnik, B. "Geometry of Quantum States". In: *Communications in Mathematical Physics* 9 (1968), pp. 55–80 (cit. on p. 168).

Mills, R. "Gauge fields". In: *100 Years of Gravity and Accelerated Frames: The Deepest Insights on Einstein and Yang-Mills*. Ed. by J.-P. Hsu and D. Fine. Vol. 9. Singapore: World Scientific, 2005, pp. 512–526 (cit. on p. 145).

Moore, G. H. "The Evolution of the Concept of Homeomorphism". In: *Historia Mathematica* (2007), pp. 333–343 (cit. on p. 76).

Muller, F. A. "The Equivalence Myth of Quantum Mechanics—Part I". In: *Studies in History and Philosophy of Science Part B: Studies in History and Philosophy of Modern Physics* 28 (1997), pp. 35–61 (cit. on p. 40).

———  "The Equivalence Myth of Quantum Mechanics—Part II". In: *Studies in History and Philosophy of Science Part B: Studies in History and Philosophy of Modern Physics* 28 (1997), pp. 219–247 (cit. on p. 40).

———  "How to Defeat Wüthrich's Abysmal Embarrassment Argument against Space-Time Structuralism". In: *Philosophy of Science* 78.5 (2011), pp. 1046–1057 (cit. on pp. 109, 112, 115, 133).

Ne'eman, Y. and S. Sternberg. "Internal Supersymmetry and Superconnections". In: *Symplectic Geometry and Mathematical Physics: Actes du colloque en l'honneur de Jean-Marie Souriau*. Ed. by P. Donato et al. Boston: Birkhäuser, 1991, pp. 326–354 (cit. on p. 277).

Norton, J. D. "General Covariance and the Foundations of General Relaitivity: Eight Decades of Dispute". In: *Reports on Progress in Physics* 56 (1993), pp. 791–858 (cit. on p. 300).

Parsons, C. "The Structuralist View of Mathematical Objects". In: *Synthese* (1990), pp. 303–346 (cit. on pp. 90, 95, 96, 103, 107).

Pauli, W. "Über das Wasserstoffspektrum vom Standpunkt der neuen Quantenmechanik". In: *Zeitschrift für Physik* 36 (1926), pp. 336–363 (cit. on p. 26).

——— "On The Hydrogen Spectrum From The Standpoint of The New Quantum Mechanics". In: *Sources of Quantum Mechanics.* Ed. by B. Van der Waerden. New York: Dover Publications, Inc., 1967, pp. 387–415 (cit. on p. 26).

Penrose, R. *The Road to Reality: A Complete Guide to the Laws of the Universe.* New York: Alfred A. Knopf, 2005 (cit. on p. 6).

Petitot, J. "Noncommutative Geometry and Transcendental Physics". In: *Constituting Objectivity. Trascendental Perspectives on Modern Physics.* Ed. by M. Bitbol, P. Kerszberg, and J. Petitot. Springer, 2009, pp. 415–455 (cit. on p. 20).

Piron, C. "Axiomatique quantique". In: *Helvetica Physica Acta* 37 (1964), pp. 439–468 (cit. on p. 194).

Procesi, C. *Lie Groups: An Approach through Invariants and Representations.* New York: Springer, 2007 (cit. on p. 276).

Prugovečki, E. *Quantum Mechanics in Hilbert Space.* 2nd ed. New York: Academic Press, 1981 (cit. on pp. 13, 14).

Puta, M. *Hamiltonian Mechanical Systems and Geometric Quantization.* Dordrecht, The Netherlands: Kluwer Academic Publishers, 1993 (cit. on pp. 139, 147).

Quine, W. V. O. *Word and Object.* Cambridge: Harvard University Press, 1960 (cit. on p. 110).

——— "Grades of Discriminability". In: *The Journal of Philosophy* 73 (1976), pp. 113–116 (cit. on p. 110).

Rédei, M. "Why John von Neumann Did Not Like The Hilbert Space Formalism of Quantum Mechanics (and What He Liked Instead)". In: *Studies In History and Philosophy of Science Part B: Studies In History and Philosophy of Modern Physics* 27.4 (1996), pp. 493–510 (cit. on p. 167).

Resnik, M. D. *Mathematics as a Science of Patterns.* New York: Oxford University Press, 1997 (cit. on pp. 95–97, 103, 123).

Rieffel, M. A. "Deformation Quantization and Operator Algebras". In: *Proceedings of Symposia in Pure Mathematics* 51 (1990), pp. 411–423 (cit. on p. 13).

Rodin, A. "Categories Without Structures". In: *Philosophia Mathematica* 19 (2011), pp. 20–46 (cit. on pp. 57, 61, 89, 94, 98).

Rovelli, C. *Quantum Gravity*. Cambridge: Cambridge University Press, 2004 (cit. on pp. 14, 139).

Ruetsche, L. *Interpreting Quantum Theories. The Art of the Possible*. Oxford: Oxford University Press, 2011 (cit. on p. 196).

Russell, B. "Logical Atomism". In: *Contemporary British Philosophers*. Ed. by J. Muirhead. London: Allen and Unwin, 1924, pp. 356–383. (Reprinted in: B. Russell. *Logic and Knowledge*. Ed. by R. Marsh. London: Allen and Unwin, 1956, pp. 323–343) (cit. on p. 62).

—— *Logic and Knowledge*. Ed. by R. Marsh. London: Allen and Unwin, 1956 (cit. on pp. 62, 320).

—— *The Principles of Mathematics*. London: Routledge, 1992. (First edition: 1903) (cit. on p. 88).

San Mauro, L. and G. Venturi. "Naturalness in Mathematics". In: *From Logic to Practice*. Ed. by G. Lolli, M. Panza, and G. Venturi. Springer, 2015, pp. 277–313 (cit. on p. 250).

Saunders, S. "Physics and Leibniz's Principles". In: *Symmetries in Physics: Philosophical Reflections*. Ed. by K. Brading and E. Castellani. Cambridge University Press, 2003, pp. 289–308 (cit. on p. 110).

Schiemer, G. and J. Korbmacher. "What Are Structural Properties?" Preprint available at http://www.jkorbmacher.com/ (cit. on p. 109).

Schilling, T. A. "Geometry of Quantum Mechanics". PhD thesis. The Pennsylvania State University, 1996 (cit. on pp. 168, 169, 174, 177, 179, 181, 185).

Schlimm, D. "Axioms in Mathematical Practice". In: *Philosophia Mathematica* 21 (2013), pp. 37–92 (cit. on p. 83).

Schrödinger, E. "An Undulatory Theory of the Mechanics of Atoms and Molecules". In: *The Physical Review* 28 (1926), pp. 1049–1070. (Reprinted in: E. Schrödinger. *Gesammelte Abhandlungen / Collected Papers. Volume 3*. Vienna: Austrian Academy of Science, 1984, pp. 280–301) (cit. on pp. 27, 28, 36).

—— "Quantisierung als Eigenwertproblem (I)". In: *Annalen der Physik* 79 (1926), pp. 361–376 (cit. on p. 26).

—— "Quantisierung als Eigenwertproblem (II)". In: *Annalen der Physik* 79 (1926), pp. 489–527 (cit. on pp. 26, 36).

Schrödinger, E. "Über das Verhältnis der Heisenberg-Born-Jordanschen Quantumemchanik zu der meinen". In: *Annalen der Physik* 79 (1926), pp. 734–756 (cit. on p. 32).

—— "Quantisierung als Eigenwertproblem (III)". In: *Annalen der Physik* 80 (1926), pp. 437–490 (cit. on p. 26).

—— "Quantisierung als Eigenwertproblem (IV)". In: *Annalen der Physik* 81 (1926), pp. 109–139 (cit. on p. 27).

—— *Collected Papers on Wave Mechanics*. Trans. by J. Shearer and W. Deans. 2nd ed. London and Glasgow: Blackie & Son, Ltd, 1928 (cit. on p. 27).

—— "La mécanique des ondes". In: *Electrons et Photons: Rapports et Discussions du Cinquième Conseil de Physique, tenu à Bruxelles du 24 au 29 Octobre 1927*. Paris: Gauthiers-Villars, 1928, pp. 185–213. (Reprinted in: E. Schrödinger. *Gesammelte Abhandlungen / Collected Papers. Volume 3*. Vienna: Austrian Academy of Science, 1984, pp. 302–323) (cit. on p. 27).

—— "On The Relation Between The Quantum Mechanics of Heisenberg, Born, and Jordan, and That of Schrödinger". In: *Collected Papers on Wave Mechanics*. Trans. by J. Shearer and W. Deans. London: Blackie & Son, 1928, pp. 45–61 (cit. on pp. 18, 31, 32, 37–39, 102, 137).

—— "Quantisation as a Problem of Proper Values. Part I". In: *Collected Papers on Wave Mechanics*. Trans. by J. Shearer and W. Deans. 2nd ed. London and Glasgow: Blackie & Son, Ltd, 1928, pp. 1–12 (cit. on pp. 26, 29).

—— "Quantisation as a Problem of Proper Values. Part II". In: *Collected Papers on Wave Mechanics*. Trans. by J. Shearer and W. Deans. 2nd ed. London and Glasgow: Blackie & Son, Ltd, 1928, pp. 13–40 (cit. on pp. 26, 28, 35, 36).

—— "Quantisation as a Problem of Proper Values. Part III". In: *Collected Papers on Wave Mechanics*. 2nd ed. London and Glasgow: Blackie & Son, Ltd, 1928, pp. 62–101 (cit. on p. 26).

—— "Quantisation as a Problem of Proper Values. Part IV". In: *Collected Papers on Wave Mechanics*. Trans. by J. Shearer and W. Deans. 2nd ed. London and Glasgow: Blackie & Son, Ltd, 1928, pp. 102–123 (cit. on pp. 27, 37).

—— *Mémoires sur la mécanique ondulatoire*. Trans. by A. Proca. Paris: Librairie Alcan, 1933 (cit. on p. 36).

Schrödinger, E. "Die gegenwärtige Situation in der Quantenmechanik". In: *Naturwissenschaften* 23.48 (1935), pp. 807–812 (cit. on p. 6).

—— "The Present Situation in Quantum Mechanics". Trans. by J. D. Trimmer. In: *Proceedings of the American Philosophical Society* 124.5 (1980), pp. 323–338 (cit. on p. 6).

—— *Gesammelte Abhandlungen / Collected Papers. Volume 3.* Vienna: Austrian Academy of Science, 1984 (cit. on pp. 27, 320, 321).

Segal, I. E. "Postulates for General Quantum Mechanics". In: *Annals of Mathematics* 48.4 (1947), pp. 930–948 (cit. on pp. 194, 195).

—— "Irreducible Representations of Operator Algebras". In: *Bulletin of the American Mathematical Society* 61 (1947), pp. 69–105 (cit. on p. 210).

—— "The Mathematical Meaning of Operationalism in Quantum Mechanics". In: *The Axiomatic Method. With Special Reference to Geometry and Physics.* Proceedings of an International Symposium held at UC Berkeley, Dec. 26 1957-Jan. 4, 1958. Ed. by L. Henkin, P. Suppes, and A. Tarski. Amsterdam: North-Holland Publishing Co., 1959, pp. 341–352 (cit. on p. 195).

—— "Mathematical Problems of Relativistic Physics". In: *Proceedings of the Summer Conference, Boulder, Colorado.* Ed. by M. Kac. American Mathematical Society, 1960 (cit. on pp. 63, 238).

Sepanski, M. R. *Compact Lie Groups.* New York: Springer, 2007 (cit. on pp. 287, 288).

Shapiro, S. "Mathematical Structuralism". In: *Internet Encyclopedia of Philosophy.* URL: http://www.iep.utm.edu/m-struct/ (cit. on pp. 90, 94, 100, 113).

—— *Philosophy of Mathematics: Structure and Ontology.* New York: Oxford University Press, 1997 (cit. on pp. 96, 99–101, 103, 106, 107, 113, 118, 122, 123, 126, 129).

—— "Identity, Indiscernibility, and *ante rem* Structuralism: The Tale of $i$ and $-i$". In: *Philosophia Mathematica* 16.3 (2008), pp. 285–309 (cit. on pp. 114, 116).

Shulman, M. "Homotopy Type Theory: A Synthetic Approach to Higher Equalities". In: *arXiv preprint* (2016). URL: http://arxiv.org/abs/1601.05035 (cit. on p. 300).

Shultz, F. W. "Pure States as Dual Objects for $C^*$-algebras". In: *Communications in Mathematical Physics* 82 (1982), pp. 497–509 (cit. on p. 218).

Souriau, J.-M. "Quantification géométrique. Applications". In: *Ann. Inst. Henri Poincaré* VI.4 (1967), pp. 311–341 (cit. on p. 248).

Souriau, J.-M. *Structure des systèmes dynamiques*. Paris: Dunod, 1970 (cit. on p. 147).

——— *Structure of Dynamical Systems. A Symplectic View of Physics*. Trans. by C. Cushman-de Vries. Boston: Birkhäuser, 1997 (cit. on p. 249).

Stachel, J. "Structural Realism and Contextual Individuality". In: *Hilary Putnam*. Ed. by Ben-Menahem. Cambridge: Cambridge University Press, 2005, pp. 203–219 (cit. on p. 119).

Steinitz, E. "Algebraische Theorie der Körper". In: *Journal für die reine und angewandte Mathematik* 137 (1910), pp. 167–309 (cit. on p. 117).

Strocchi, F. *An Introduction to the Mathematical Structure of Quantum Mechanics*. 2nd ed. Singapore: World Scientific, 2008 (cit. on pp. 16, 195, 197, 198, 200, 213, 216).

Summers, S. J. "On the Stone – von Neumann Uniqueness Theorem and Its Ramifications". In: *John von Neumann and the Foundations of Quantum Physics*. Ed. by M. Rédei and M. Stöltzner. Dordrecht: Kluwer Academic Publishers, 2001, pp. 135–152 (cit. on p. 196).

Takesaki, M. *Theory of Operator Algebras Vol. I*. New York: Springer, 2003 (cit. on pp. 198, 217).

The Univalent Foundations Program. *Homotopy Type Theory: Univalent Foundations of Mathematics*. Institute for Advanced Study: `http://homotopytypetheory.org/book`, 2013 (cit. on pp. 120, 121).

Timmermans, B. *Histoire philosophique de l'algèbre moderne – Les origines romantiques de la pensée abstraite*. Paris: Classiques Garnier, 2012 (cit. on p. 3).

Townsend, J. S. *A Modern Approach to Quantum Mechanics*. Sausalito: University Science Books, 2000 (cit. on p. 145).

Tuynman, G. M. and W. Wiegerinck. "Central Extensions and Physics". In: *Journal of Geometry and Physics* 4.2 (1987), pp. 207–258 (cit. on pp. 259, 273).

Van der Waerden, B. "From Matrix Mechanics and Wave Mechanics to Unified Quantum Mechanics". In: *The Physicist's Conception of Nature*. Ed. by J. Mehra. Dordrecht: D. Reidel Publishing Company, 1987, pp. 276–293 (cit. on pp. 30, 40).

Von Neumann, J. "Mathematische Begründung der Quantenmechanik". In: *Nachr. Ges. Wiss. Göttingen* (1927), pp. 1–57 (cit. on p. 49).

——— *Mathematische Grundlagen der Quantenmechanik*. Heidelberg: Springer-Verlag, 1932 (cit. on pp. 19, 49).

Von Neumann, J. "On an Algebraic Generalization of The Quantum Mechanical Formalism (Part I)". In: *Receuil Mathématique* 1.4 (1936), pp. 415–484. (Reprinted in: J. von Neumann. *Collected Works*. Ed. by A. H. Taub. Oxford: Pergamon Press, 1961, Vol. III, pp. 492–559) (cit. on p. 194).

—— *Mathematical Foundations of Quantum Mechanics*. Trans. by R. T. Beyer. Princeton: Princeton University Press, 1955 (cit. on pp. 19, 49–51, 53, 70, 82, 117, 140, 164).

—— *Collected Works*. Ed. by A. H. Taub. Oxford: Pergamon Press, 1961 (cit. on pp. 193, 194, 313, 324).

—— *Continuous Geometries with a Transition Probability*. Vol. 252. American Mathematical Society, 1981 (cit. on p. 220).

—— *John von Neumann: Selected Letters*. Ed. by M. Rédei. History of Mathematics. American Mathematical Society, 2005 (cit. on pp. 50, 167, 195).

Von Neumann, J. and G. Birkhoff. "The Logic of Quantum Mechanics". In: *Annals of Mathematics* 37.4 (1936), pp. 823–843. (Reprinted in: J. von Neumann. *Collected Works*. Ed. by A. H. Taub. Oxford: Pergamon Press, 1961, Vol. IV, pp. 105–125) (cit. on p. 193).

Weber, H. "Die allgemeinen Grundlagen der Galois'schen Gleichungstheorie". In: *Mathematische Annalen* 43 (1893), pp. 521–549 (cit. on p. 64).

Wegge-Olsen, N. E. *K-theory and C\*-algebras: a Friendly Approach*. New York: Oxford University Press, 1993 (cit. on p. 216).

Weinberg, S. *The Quantum Theory of Fields*. Vol. 1. New York: Cambridge University Press, 1996 (cit. on p. 264).

Weinstein, A. "Symplectic Categories". In: *Proceedings of Geometry Summer School, Lisbon*. 2009. URL: https://arxiv.org/pdf/0911.4133v1.pdf (cit. on p. 246).

Weyl, H. *Quantenmechanik und Gruppentheorie*. Leipzig: Teubner, 1928 (cit. on p. 243).

—— *The Theory of Groups & Quantum Mechanics*. Trans. by H. Robertson. New York: Dover Publications, Inc., 1931 (cit. on pp. 263, 286).

—— *The Classical Groups - Their Invariants and Representations*. 2nd ed. Princeton: Princeton University Press, 1946. (First edition: 1939) (cit. on p. 56).

—— *Philosophy of Mathematics and Natural Science*. Trans. by O. Helmer. Princeton: Princeton University Press, 1949 (cit. on pp. 61, 109, 120, 299).

Weyl, H. *Symmetry*. Princeton: Princeton University Press, 1952 (reprinted in 1989) (cit. on pp. 57, 131, 241).

Wiener, N. *Norbert Wiener: Collected Works. Volume III*. Ed. by P. Masani. Cambridge: The MIT Press, 1981 (cit. on pp. 35, 308).

Wigner, E. P. *Group Theory and Its Applications to the Quantum Mechanics of Atomic Spectra*. Trans. by J. J. Griffin. New York: Academic Press, 1959 (cit. on p. 264).

Woit, P. *Quantum Theory, Groups and Representations: An Introduction*. unpublished, 2015. URL: http://www.math.columbia.edu/~woit/QM/qmbook.pdf (cit. on p. 248).

Woodhouse, N. *Geometric Quantization*. 2nd. Oxford: Clarendon Press, 1991 (cit. on pp. 13, 139, 169).

Wright, C. *Frege's Conception of Numbers as Objects*. Aberdeen: Aberdeen University Press, 1983 (cit. on p. 60).

Wussing, H. *The Genesis of the Abstract Group Concept*. Trans. by A. Shenitze. Cambridge, MA: Dover Publications, Inc., 1984 (cit. on p. 64).

Wüthrich, C. "Challenging the Spacetime Structuralist". In: *Philosophy of Science* 76 (2010), pp. 1039–1051 (cit. on p. 105).

Zalamea, F. "The Mathematical Description of a Generic Physical System". In: *Topoi* 34.2 (2015), pp. 339–348. URL: http://dx.doi.org/10.1007/s11245-015-9322-7 (cit. on p. 275).

Zalamea, F. *Synthetic Philosophy of Contemporary Mathematics*. Trans. by Z. L. Fraser. Urbanomic/Sequence Press, 2012 (cit. on p. 260).

# Index

*The page in which the full definition of a concept is found appears in **bold typeface**.*

# Symbols

**Arrows**

| | |
|---|---|
| $E \longrightarrow F$ | Injection of $E$ in $F$ (monomorphism) |
| $E \longrightarrow F$ | Projection of $E$ onto $F$ (epimorphism) |
| $\mathrm{Hom}_{\mathcal{C}}(E, F)$ | Set of arrows from $E$ to $F$ in the category $\mathcal{C}$ |

**Groups**

| | |
|---|---|
| $G, H, \dots$ | Abstract (Lie) groups |
| $\mathfrak{g}, \mathfrak{h}, \dots$ | Abstract Lie algebras |
| $\mathfrak{g}^*, \mathfrak{h}^*, \dots$ | Dual space of $\mathfrak{g}, \mathfrak{h}$ (space of linear functionals) |
| $G \circlearrowright E$ (or $\mathfrak{g} \circlearrowright E$) | left action of $G$ (or $\mathfrak{g}$) on $E$ |
| $E \circlearrowleft G$ (or $E \circlearrowleft \mathfrak{g}$) | right action of $G$ (or $\mathfrak{g}$) on $E$ |
| $Ad$ | Adjoint action of $G$ on $\mathfrak{g}$ |
| $Co$ | Co-adjoint action of $G$ on $\mathfrak{g}^*$ |
| $\widehat{G}$ | Set of equivalence classes of irreducible unitary $G$-representations |

**Hilbert spaces**

| | |
|---|---|
| $\mathcal{H}$ | Abstract Hilbert space |
| $\mathcal{H}_G$ | Abstract Hilbert space equipped with a unitary $G$-representation |
| $\langle \cdot, \cdot \rangle$ | Hermitian product on $\mathcal{H}$ |
| $\mathbb{P}\mathcal{H}$ | Abstract projective Hilbert space |
| $\mathbb{S}\mathcal{H}$ | Unit sphere (vectors of norm 1 in $\mathcal{H}$) |
| $U(\mathcal{H})$ | Unitary operators on $\mathcal{H}$ |
| $\mathcal{B}(\mathcal{H})$ | $C^*$-algebra of bounded operators on $\mathcal{H}$ |
| $\mathcal{B}_{\mathbb{R}}(\mathcal{H})$ | Real Jordan-Lie algebra of bounded self-adjoint operators |
| $\mathcal{B}_{i\mathbb{R}}(\mathcal{H})$ | Real Jordan-Lie algebra of bounded anti-self-adjoint operators |
| $\circ$ | Composition of operators |
| $[\cdot, \cdot]_+$ (or $2\bullet$) | Anti-commutator (Jordan product) |
| $[\cdot, \cdot]$ (or $\frac{2}{i}\star$) | Commutator (Lie product) |

**$C^*$-algebras**

| | |
|---|---|
| $\mathcal{U}$ | Abstract $C^*$-algebra |
| $\mathcal{U}_{\mathbb{R}}$ | Abstract Jordan-Lie algebra (real part of a $\mathcal{U}$) |
| $\mathcal{S}(\mathcal{U})$ | Space of states |
| $\mathcal{P}(\mathcal{U})$ | Space of pure states |
| $\widehat{\mathcal{U}}$ | Set of equivalence classes of irreducible representations of $\mathcal{U}$ |
| $Cycl(\mathcal{U})$ | Set of equivalence classes of cyclic representations of $\mathcal{U}$ |

**Functions**

| | |
|---|---|
| $\mathcal{C}(M, \mathbb{C})$ | Continuous complex-valued functions on $M$ |
| $\mathcal{C}_0(M, \mathbb{C})$ | Continuous complex-valued functions vanishing at infinity |
| $\mathcal{C}^{\infty}(M, \mathbb{R})$ | Smooth real-valued functions on $M$ |

**Symplectic**

| | |
|---|---|
| $(S, \omega)$ | Abstract manifold $S$ with symplectic two-form $\omega$ |
| $S_{\mathfrak{g}}$ | Abstract symplectic manifold with a strongly Hamiltonian $\mathfrak{g}$-action |
| $f \cdot g$ | Pointwise multiplication of the functions $f$ and $g$ |
| $\{\cdot, \cdot\}_S$ or $\{\cdot, \cdot\}$ | Poisson bracket on $S$ |
| $\Gamma(TS)_\omega$ | Vector fields preserving the symplectic structure |
| $\Gamma(TS)_H$ | Hamiltonian vector fields |
| $\mathcal{C}^\infty(S, \mathbb{R}) \xrightarrow{v_-} \Gamma(TS)_H$ | Symplectic gradient |
| $\mathcal{C}^\infty(S, \mathbb{R})_{\mathcal{K}}$ | Smooth functions whose associated Hamiltonian vector field preserves the Kinematical structures |
| $\Omega^n(S)$ | Differential $n$-forms on $S$ |
| $J : S \to \mathfrak{g}^*$ | momentum map |
| $\widehat{J} : \mathfrak{g} \to \mathcal{C}^\infty(S, \mathbb{R})$ | co-momentum map |

**Varia**

| | |
|---|---|
| $S^n$ | Sphere in $n+1$-dimensions |
| $\partial S$ | boundary of the manifold $S$ |
| $V^*$ | Topological dual of the vector space $V$ |